\documentclass[11pt,a4paper]{article}

\usepackage[utf8]{inputenc}
\usepackage[T1]{fontenc}
\usepackage{amsmath,amssymb,amsfonts,mathtools}
\usepackage{bm}
\usepackage{physics}
\usepackage{geometry}
\usepackage{graphicx}
\usepackage{hyperref}
\usepackage{enumitem}
\usepackage{fancyhdr}
\usepackage{titlesec}
\usepackage{xcolor}
\usepackage{tcolorbox}
\usepackage{slashed}
\usepackage{cancel}
\usepackage{simplewick}
\usepackage{tikz}
\usetikzlibrary{decorations.markings,decorations.pathmorphing}

\newcommand{\be}{\begin{equation}}
\newcommand{\ee}{\end{equation}}

\hypersetup{colorlinks=true, linkcolor=blue!60!black, citecolor=green!50!black, urlcolor=blue!70!black}

\newcommand{\sv}{\vv{s}}

\newcommand{\vv}[1]{{\bm #1}}       % was \vec{#1}
\newcommand{\kv}{\vv{k}}
\renewcommand{\k}{\vv{k}}

\newcommand{\xv}{\vv{x}}
\newcommand{\x}{\vv{x}}

\newcommand{\rv}{\vv{r}}
\newcommand{\qv}{\vv{q}}
\newcommand{\Rv}{\vv{R}}
\newcommand{\Dv}{\vv{D}}
\newcommand{\Psiv}{\vv{\Psi}}
\newcommand{\vvv}{\vv{v}}
\renewcommand{\pv}{\vv{p}}          % momentum (used in Vlasov)

\renewcommand{\grad}{\boldsymbol{\nabla}}
\newcommand{\gradx}{\boldsymbol{\nabla}_{\!\xv}}
\newcommand{\gradr}{\boldsymbol{\nabla}_{\!\rv}}

\renewcommand{\dd}{\mathrm{d}}
\newcommand{\pder}[2]{\frac{\partial #1}{\partial #2}}

\newcommand{\ddq}{\frac{\dd^3 q}{(2\pi)^3}}

\newcommand{\avg}[1]{\langle #1 \rangle}

\newcommand{\Hc}{\mathcal{H}}        % conformal Hubble
\newcommand{\Hubble}{\Hc}            % alias (used throughout notes)
\newcommand{\Dp}{D_+}
\newcommand{\rhob}{\bar{\rho}}

\newcommand{\knl}{k_\mathrm{NL}}
\newcommand{\kosc}{k_\mathrm{osc}}
\newcommand{\kbao}{k_\mathrm{BAO}}
\newcommand{\keq}{k_\mathrm{eq}}
\newcommand{\kIR}{k_\mathrm{IR}}
\newcommand{\kUV}{k_\mathrm{UV}}
\newcommand{\Mpc}{\,h/\mathrm{Mpc}}
\newcommand{\hMpc}{\,h\mathrm{Mpc}^{-1}}

\newcommand{\Mpch}{\,\mathrm{Mpc}/h}

\newcommand{\lmfp}{\ell_\mathrm{mfp}}

\newcommand{\nbar}{\bar{n}}
\newcommand{\Phihat}{\hat{\Phi}}

\newcommand{\wickA}{% contraction (12)(34)
  \begin{tikzpicture}[baseline=(m.base)]
    \node[inner sep=0pt] (m) {$\delta_1(\kv_1)\;\delta_1(\kv_2)\;\delta_1(\kv_3)\;\delta_1(\kv_4)$};
    \node[inner sep=0pt] (d1) at (-1.62,0) {};
    \node[inner sep=0pt] (d2) at (-0.54,0) {};
    \node[inner sep=0pt] (d3) at ( 0.54,0) {};
    \node[inner sep=0pt] (d4) at ( 1.62,0) {};
    \draw[red,  thick] (d1.north) ++(0,0.15) to[bend left=60] ++(1.08,0);
    \draw[blue, thick] (d3.north) ++(0,0.15) to[bend left=60] ++(1.08,0);
  \end{tikzpicture}%
}
\newcommand{\wickB}{% contraction (13)(24)
  \begin{tikzpicture}[baseline=(m.base)]
    \node[inner sep=0pt] (m) {$\delta_1(\kv_1)\;\delta_1(\kv_2)\;\delta_1(\kv_3)\;\delta_1(\kv_4)$};
    \node[inner sep=0pt] (d1) at (-1.62,0) {};
    \node[inner sep=0pt] (d2) at (-0.54,0) {};
    \node[inner sep=0pt] (d3) at ( 0.54,0) {};
    \node[inner sep=0pt] (d4) at ( 1.62,0) {};
    \draw[red,  thick] (d1.north) ++(0,0.15) to[bend left=45] ++(2.16,0);
    \draw[blue, thick] (d2.north) ++(0,0.15) to[bend left=45] ++(2.16,0);
  \end{tikzpicture}%
}
\newcommand{\wickC}{% contraction (14)(23)
  \begin{tikzpicture}[baseline=(m.base)]
    \node[inner sep=0pt] (m) {$\delta_1(\kv_1)\;\delta_1(\kv_2)\;\delta_1(\kv_3)\;\delta_1(\kv_4)$};
    \node[inner sep=0pt] (d1) at (-1.62,0) {};
    \node[inner sep=0pt] (d2) at (-0.54,0) {};
    \node[inner sep=0pt] (d3) at ( 0.54,0) {};
    \node[inner sep=0pt] (d4) at ( 1.62,0) {};
    \draw[red,  thick] (d1.north) ++(0,0.15) to[bend left=40] ++(3.24,0);
    \draw[blue, thick] (d2.north) ++(0,0.15) to[bend left=60] ++(1.08,0);
  \end{tikzpicture}%
}

\title{ \vspace{-2.5cm}\begin{flushright}\normalfont\normalsize MIT-CTP/6084\end{flushright}
\vspace{1.2em}
\textbf{GGI Lectures on  
Large-Scale Structure Perturbation Theory (Effective Field Theory) }\\[10pt]
}
\vspace{-0.5em}
\author{Mikhail M. Ivanov\thanks{\texttt{ivanov99@mit.edu}}\\[6pt]
  \small\itshape Center for Theoretical Physics -- a Leinweber Institute, Massachusetts Institute of Technology,\\
  \small\itshape Cambridge, MA 02139, USA}
  \date{}

\begin{document}
\maketitle
\thispagestyle{empty}

\vspace{-0.9cm}
\begin{abstract}
\noindent
These notes are an introduction to non-linear perturbation 
theory for cosmological large-scale structure.
% an extended version of lectures 
% delivered at the 
% ``New Physics from Galaxy Clustering'' school at the Galileo Galilei Institute, Florence, in August 2025. 
They are aimed at undergraduate 
and beginning graduate students and do not require any cosmology or 
quantum field theory background. All necessary concepts are developed 
from scratch.
The lectures are intended to explain all key ingredients needed to 
model the observed clustering of galaxies in real and redshift spaces.
After a brief pedagogical introduction to the ideas of
effective field theory (EFT), we develop large-scale structure 
EFT in the context of Newtonian cosmology 
using symmetry principles. 
We discuss in detail 
the shortcomings of Standard Perturbation Theory, 
the non-linear evolution of baryon acoustic oscillations
and its relation to the equivalence principle,
counterterms and renormalization of the loop diagrams,
and stochastic effects. Then we develop EFT for galaxy bias 
and redshift space distortions. We also highlight 
some important facts about redshift-space stochasticity,
relevant for ongoing and future galaxy surveys. Finally, we 
introduce Lagrangian Perturbation Theory.
\end{abstract}

\tableofcontents
\newpage

%==========================================================================
\section{Chapter 1: Structure Formation and the EFT Ideology}
%==========================================================================

The basic empirical picture of structure formation in our Universe
goes as follows: small, linear dark-matter overdensities grow under
gravity, become non-linear at small scales, collapse to form halos,
and then complex baryonic physics inside halos produces the galaxies
we observe. Galaxies are extremely non-linear objects. There is
absolutely zero hope to understand this chain in full generality 
from first principles.
Nevertheless, on sufficiently large scales the \emph{statistical
distribution} of galaxies is predictable. This observation is the
central motivation for applying effective-field-theory (EFT) ideas
to large-scale structure: on large scales, perturbation theory and
the EFT go hand in hand, just as they do in fluid mechanics,
statistical physics, or particle physics.

What is an EFT? It is an approach in which one isolates the physics
on large scales from the small-scale complications. Specifically,
one builds a simplified description of the large-scale part of the
system. The EFT \emph{ideology} rests on the following assumptions:
\begin{itemize}
  \item There is a separation of scales: a long-wavelength scale
  $L_\mathrm{long}$, which we are interested in, and a short scale
  $L_\mathrm{short}$, where the complicated dynamics happens. They
  satisfy $L_\mathrm{long}\gg L_\mathrm{short}$.
  \item Some simplified dynamics emerges on scales
  $\sim L_\mathrm{long}$. It is captured by macroscopic,
  coarse-grained dynamical fields $\{F_i\}$, called ``degrees of
  freedom.''
  \item Observables on large scales, at a given finite accuracy,
  depend on the small-scale dynamics only through a \emph{finite}
  set of parameters and functions (``operators'') built from the
  macroscopic fields and their derivatives.
  \item This simplified description is organized as an analytic
  (derivative) expansion in the small parameters
  $L_\mathrm{short}/L_\mathrm{long}$ and $F_i$.
  \item The effective description is completely constrained by
  symmetries.
\end{itemize}
In short: the answer for any large-scale observable is an
\emph{analytic function} of $L_\mathrm{short}/L_\mathrm{long}$ in
the limit $L_\mathrm{short}/L_\mathrm{long}\to 0$, with a finite
number of coefficients encoding the short-scale physics.

%------------------------------------------------------------------
\subsection{A Simple Example: Multipole Expansion in Electrostatics}
%------------------------------------------------------------------

A textbook example is the multipole expansion in electrostatics.
For a localized (and arbitrarily complicated) charge distribution
of spatial size $a$, the potential at distance $R\gg a$ admits the
expansion
\be
  \Phi(\Rv) = \frac{Q}{R} + \frac{\Dv\cdot\hat{\vv{n}}}{R^2}
  + \frac{Q_{ij}\,\hat n^i \hat n^j}{R^3} + \cdots\,,\qquad
  \hat{\vv{n}} = \Rv/R\,,
\ee
with the monopole (total charge), dipole and quadrupole moments
\be
  Q = \int d^3x\;\rho_Q(\xv)\,,\qquad
  \Dv = \int d^3x\;\xv\,\rho_Q(\xv)\,,\qquad
  Q^{ij} = \int d^3x\left(x^i x^j - \frac{\delta^{ij}\xv^{\,2}}{3}\right)\rho_Q(\xv)\,,
\ee
built from the charge density $\rho_Q$. At a given order this
expansion involves only a few parameters. At very large distances
($R\to\infty$) it does not matter how complicated the distribution
of the charges is: the answer depends on a single parameter, the
system's total charge. At second order we need four parameters ---
the charge and the three components of the dipole moment $\Dv$;
the quadrupole adds five more (a symmetric $3\times 3$ tensor has
six components, minus one for tracelessness), and so on --- always
a \emph{finite} number. The power counting reads
\be
  \frac{\Dv\cdot\hat{\vv{n}}}{R^2}\sim \frac{Q\,a}{R^2}\ll \frac{Q}{R}\,,\qquad
  \frac{Q_{ij}\,\hat n^i\hat n^j}{R^3} \sim \frac{Q\,a^2}{R^3}\ll \frac{Q\,a}{R^2}\ll \frac{Q}{R}\,,
\ee
i.e.\ as long as $a/R\ll 1$ we expect the expansion to converge
rapidly. Note also the role of the symmetries: charge conservation
fixes the leading coefficient, and rotational invariance dictates
the tensor structures --- the $1/R^2$ term requires a vector, the
$1/R^3$ term a symmetric traceless tensor.

The key EFT lessons from the multipole expansion are:
\begin{enumerate}[nosep]
  \item \textbf{Analyticity:} the expansion is in powers of $a/R$.
  \item \textbf{Universality:} different microscopic realizations
  share the same long-distance description --- only the multipole
  moments matter, regardless of the small-scale details of the
  charge distribution.
  \item A \textbf{finite number of parameters} is sufficient at a
  given order.
\end{enumerate}
The multipole expansion is, however, too simple: the problem is
static, so we do not yet see the role of the dynamical degrees of
freedom. Let us consider a slightly more complicated EFT, which
will be directly relevant for structure formation.

%------------------------------------------------------------------
\subsection{Fluid Dynamics as an EFT: Navier--Stokes and Symmetry Constraints}
%------------------------------------------------------------------

Fluid dynamics is another canonical EFT. The short scale here is
the microscopic mean free path,
\be
  \lmfp \simeq \frac{1}{\bar{n}\,\avg{\sigma}}\,,
\ee
where $\bar n$ is the number density of particles and
$\avg{\sigma}$ the collision cross-section. On scales much larger
than $\lmfp$, a many-body system with arbitrarily complicated
interactions can be described by coarse-grained fields: the mass
density $\rho(\xv,t)$, the velocity $\vvv(\xv,t)$, and the pressure
$p(\xv,t)$ (for an adiabatic fluid, $p=c_s^2\rho$). The emergent
simplified dynamics is governed by the continuity and Navier--Stokes
equations,
\begin{align}
  &\frac{\partial \rho}{\partial t} + \nabla\cdot(\rho\,\vvv) = 0\,, \label{eq:continuity}\\
  &\rho\,\frac{D v^i}{D t} \equiv
  \rho\left(\frac{\partial v^i}{\partial t}+v^j\partial_j v^i\right)
  = -\partial^i p + \nu\rho\,\Delta v^i
  + \rho\left(\frac{\nu}{3}+\zeta\right)\partial^i \partial_k v^k\,, \label{eq:momentum}
\end{align}
where $\Delta=\partial^i\partial_i$, and $\nu$ and $\zeta$ are the
shear and bulk viscosity coefficients.

To decide which operators are important at a given scale, we
perform simple scaling estimates. Consider a state with constant
background density $\rho_0$ and velocity fluctuations characterized
by a wavenumber $k$, frequency $\omega$ and velocity amplitude
$v_k$, which we take to be of the order of the sound speed $c_s$.
These are just acoustic waves, with the dispersion relation
$\omega = c_s k$. Then:
\begin{itemize}[nosep]
  \item inertial (time-derivative) term:
  $\rho\,\partial_t v \sim \rho_0\,\omega\,v_k \sim \rho_0\,k\,c_s^2$\,,
  \item advection (non-linear) term:
  $\rho\,v\cdot\nabla v \sim \rho_0\,k\,c_s^2$\,,
  \item pressure gradient: $\nabla p \sim k\,c_s^2\,\rho_0$\,,
  \item viscous term: $\nu\rho\,\Delta v \sim \rho_0\,\nu\,k^2\,v_k$\,.
\end{itemize}
On dimensional grounds $\nu\sim c_s\,\lmfp$, hence the ratio of the
viscous to the perfect-fluid terms scales as
\be
  \frac{\nu k^2}{\omega} \sim k\,\lmfp \sim \frac{\lmfp}{\lambda} \ll 1\,.
\ee
This is the expansion parameter of the imperfect-fluid description:
the derivative expansion is controlled by $k\lmfp\ll 1$. In the
large-scale limit $k\to 0$ the viscous terms vanish and we are left
with the perfect-fluid equations. The viscous stress captures the
leading non-vanishing corrections to this description. The
structure is thus very similar to the multipole expansion: the
perfect-fluid equations are the analog of the monopole
approximation, while the viscosity terms in the Navier--Stokes
equation play the role of the dipole.

It is instructive to \emph{derive} the Navier--Stokes equation from
the EFT principles. This exercise shows that this equation is
inevitable if we stick to the EFT ideology. Once we have determined
that the density and the velocity are the relevant degrees of
freedom, mass conservation and momentum balance immediately give
\begin{align}
  &\frac{\partial \rho}{\partial t} + \nabla\cdot(\rho\,\vvv) = 0\,,\\
  &\rho\,\frac{D v^i}{D t} = \partial_j\tau^{ij}\,, \label{eq:momentum_EFT}
\end{align}
where $\tau^{ij}$ is the effective stress tensor that encodes the
forces generated by the complicated short-scale physics. On large
scales we can express it through $\rho$, $\vvv$ and their
derivatives. The symmetries then do all the work. $\tau^{ij}$ is a
rank-2 tensor with respect to 3D rotations, and its dependence on
the velocity must respect \textbf{Galilean symmetry}, i.e.\ it must
be invariant under constant velocity shifts
$\vvv\to\vvv+\vv{c}$. This implies that the velocity field can only
appear with a gradient acting on it --- terms like a bare $v^iv^j$
without derivatives are forbidden. The only options through order
$k\lmfp$ are
\begin{equation}
  \tau^{ij} = A\,\rho\,\delta^{ij}
  + B\,\lmfp\,\rho\,\partial^i v^j
  + C\,\lmfp\,\rho\,\delta^{ij}\,\partial_k v^k + \cdots\,,
\end{equation}
where $A\sim c_s^2$ and $B,C\sim c_s$ are constants fixed by the
microscopic dynamics, and the ellipsis stands for higher-order
terms. Symmetrizing this expression and renaming the constants, we
arrive at the familiar form
\begin{equation}\label{eq:stress_expansion}
  \tau^{ij} =
  \underbrace{-c_s^2\,\rho\,\delta^{ij}}_{\text{``monopole''}}
  \;+\;
  \underbrace{\rho\,\nu\left(\partial^i v^j+\partial^j v^i
  -\tfrac{2}{3}\,\delta^{ij}\,\partial_k v^k\right)
  + \rho\,\zeta\,\delta^{ij}\,\partial_k v^k}_{\text{``dipole''}}\,.
\end{equation}
The pressure is the only term allowed at zeroth order in gradients.
The shear and bulk viscous terms involve \emph{one} spatial
derivative acting on the velocity and describe momentum flux; they
are the leading derivative corrections and thus play the role of
the ``dipole'' in the organized expansion. If the precision of our
experiment is high enough, we may need to include higher-order
terms like $\partial_i v^j\,\partial_k v^k$, analogous to the
quadrupole.\footnote{Beyond the leading order two separate
expansions appear: one in the gradients, like
$\partial_i\partial_j\rho$ or $\partial_i\partial_j(\partial_k v^k)$,
and one in the number of fields, like the term
$\partial_i v^j\,\partial_k v^k$ above. New scales may also arise,
such as the relaxation time --- see e.g.\ the Burnett equations,
the M\"uller--Israel--Stewart theory, and the hydrodynamics of
heavy ions and holographic fluids.}

The upshot is that the symmetries (Galilean and rotational
invariance) together with the gradient-plus-field expansion
completely fix the leading form of $\tau^{ij}$: the Navier--Stokes
equation is determined by symmetries and dimensional analysis ---
it \emph{is} an EFT. Since the long-wavelength dynamics depends
only on a few coefficients ($c_s$, $\nu$, $\zeta$), many different
microscopic systems are described by the same macroscopic
equations: this is the statement of \emph{universality}. We have an
amazing equation that describes \emph{any} many-body interacting
dynamics with only three parameters. From the EFT point of view,
the only thing that distinguishes air from tar is the values of the
fluid coefficients; otherwise the dynamical equations of the two
systems are completely equivalent and fully determined by
symmetries.

\begin{tcolorbox}[colback=blue!5!white, colframe=blue!50!black, title=Key Concepts]
  \textbf{Simplicity:} 3 parameters characterise any fluid ($c_s$, $\nu$, $\zeta$).\\
  \textbf{Universality:} any fluid is described by 3 parameters, e.g.\ air \& tar.
\end{tcolorbox}

%==========================================================================
\section{Chapter 2: Recap of Linear Cosmology and Gaussian Random Fields}
%==========================================================================

\subsection{Linear Perturbations}

\noindent
We will review now some basic information about matter perturbations 
in linear cosmology. More details can be found in excellent textbooks such as
Dodelson, Schmidt, ``Modern Cosmology''~\cite{Dodelson:2020bqr}.
In what follows all numerical quantities and plots
are computed for the \textit{Planck2018} baseline $\Lambda$CDM cosmological 
model~\cite{Aghanim:2018eyx},
unless we fit data from a simulation, in which case 
we use the fiducial cosmology of that simulation.
The base $\Lambda$CDM parameters are: 
\begin{center}
\renewcommand{\arraystretch}{1.4}
\begin{tabular}{lll}
\hline\hline
Parameter & Symbol & Value \\
\hline
Baryon density & $\omega_b \equiv \Omega_b h^2$ & $0.02237$ \\
Cold dark matter density & $\omega_c \equiv \Omega_c h^2$ & $0.1200$ \\
Hubble constant & $h$ & $0.6736$ \\
Scalar spectral index & $n_s$ & $0.9649$ \\
Amplitude of scalar perturbations & $\ln(10^{10}A_s)$ & $3.044$ \\
Optical depth to reionisation & $\tau_{\rm reio}$ & $0.0544$ \\
\hline
\end{tabular}
\end{center}
The derived parameters relevant for large-scale structure are:
total matter fraction today $\Omega_m = 0.3153$,
the linear mass fluctuation amplitude 
$\sigma_8 = 0.8111$ (defined shortly).
The sum of neutrino masses is fixed to the minimal value
$\sum m_\nu = 0.06\;\mathrm{eV}$, with one massive and two
massless species.

The key object of interest is the overdensity contrast 
of matter defined as
\begin{equation}
  \delta = \frac{\rho(\xv,t) - \bar\rho(t)}{\bar\rho(t)}\,,
\end{equation}
where $\rho(t)$ is the background matter density that satisfies the Friedmann equations:
\be 
H^2\equiv \left(\frac{\dot{a}}{a}\right)^2=\frac{8\pi G}{3}(\rho(t)+\rho_\Lambda)\,,\quad \dot\rho+3H\rho = 0\,,
\ee 
where we have ignored the contributions of radiation and neutrinos.
Likewise, one can define the spatial fluctuations of galaxy number density $n_g$ through he 
via
\begin{equation}
  \delta_g = \frac{n_g(\xv,t) - \bar{n}_g(t)}{\bar{n}_g(t)}\,;\qquad
  \bar{n}_g = \frac{N}{V}\,,
\end{equation}
where $N$ is the total number of galaxies in an observed comoving volume $V$.
$\delta$, $\delta_g$ are \emph{random stochastic fields}. 
This means their observed phases do not carry any interesting information. 
To link these observables to fundamental physics, we need to study their \textbf{correlations}.
The simplest one is the 
two-point function of matter:
\begin{equation}
  \avg{\delta(\xv)\,\delta(\xv+\rv)} = \xi(|\rv|)\,,
\end{equation}
where the dependence only the module of the separation between the 
two points follows from the statistical homogeneity and isotropy of our Universe:
all directions look the same (hence no dependence on the vector $\rv$)
and different patches of the sky also look the same (hence no dependence on the exact position
of the galaxy $\xv$). Due to these arguments the one-point correlator can only be a constant which
we can absorb into the background density, which then ensures that
\be 
\langle \delta(\xv)\rangle = 0\,.
\ee

\begin{figure}
    \centering
    \includegraphics[width=0.49\linewidth]{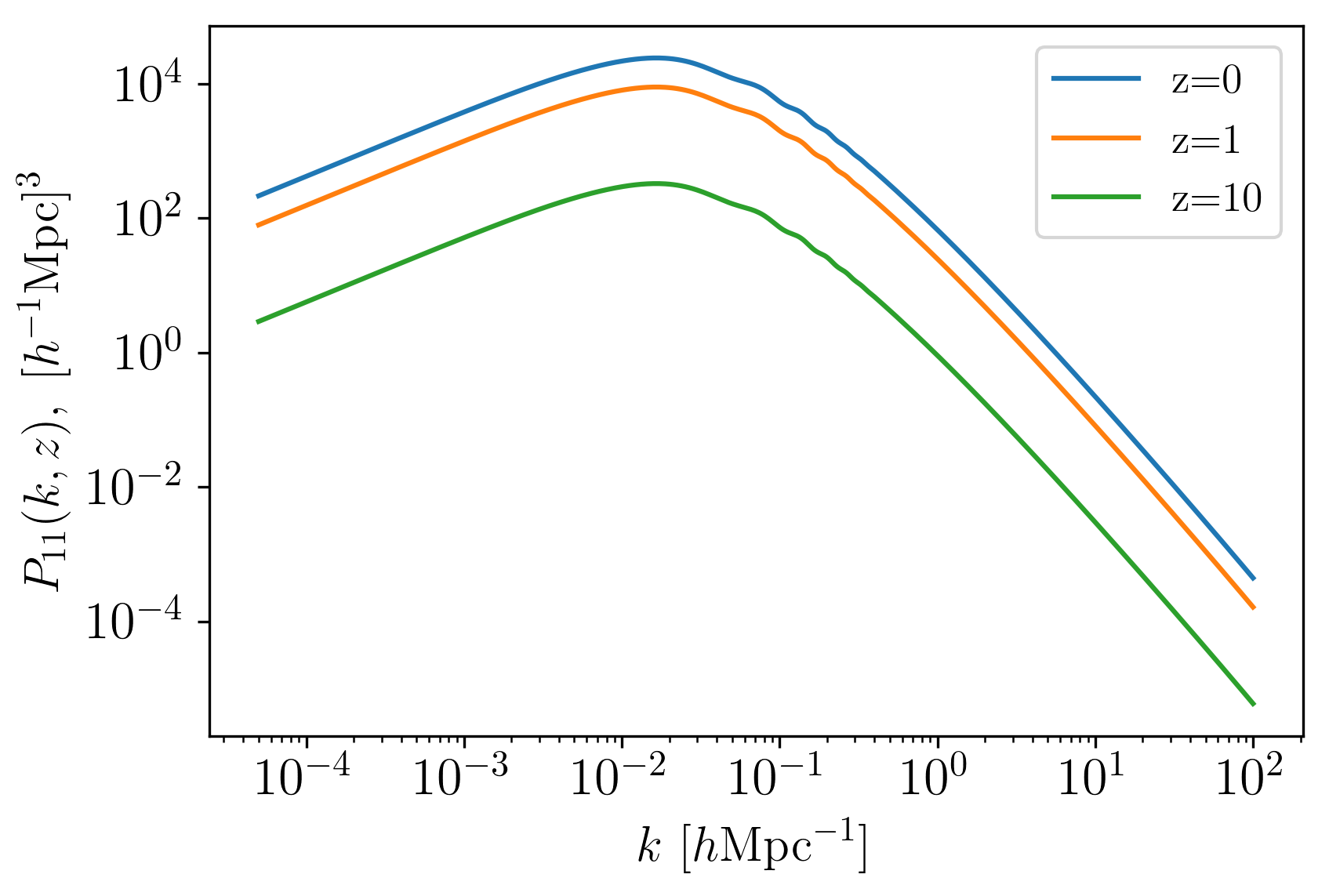}
        \includegraphics[width=0.49\linewidth]{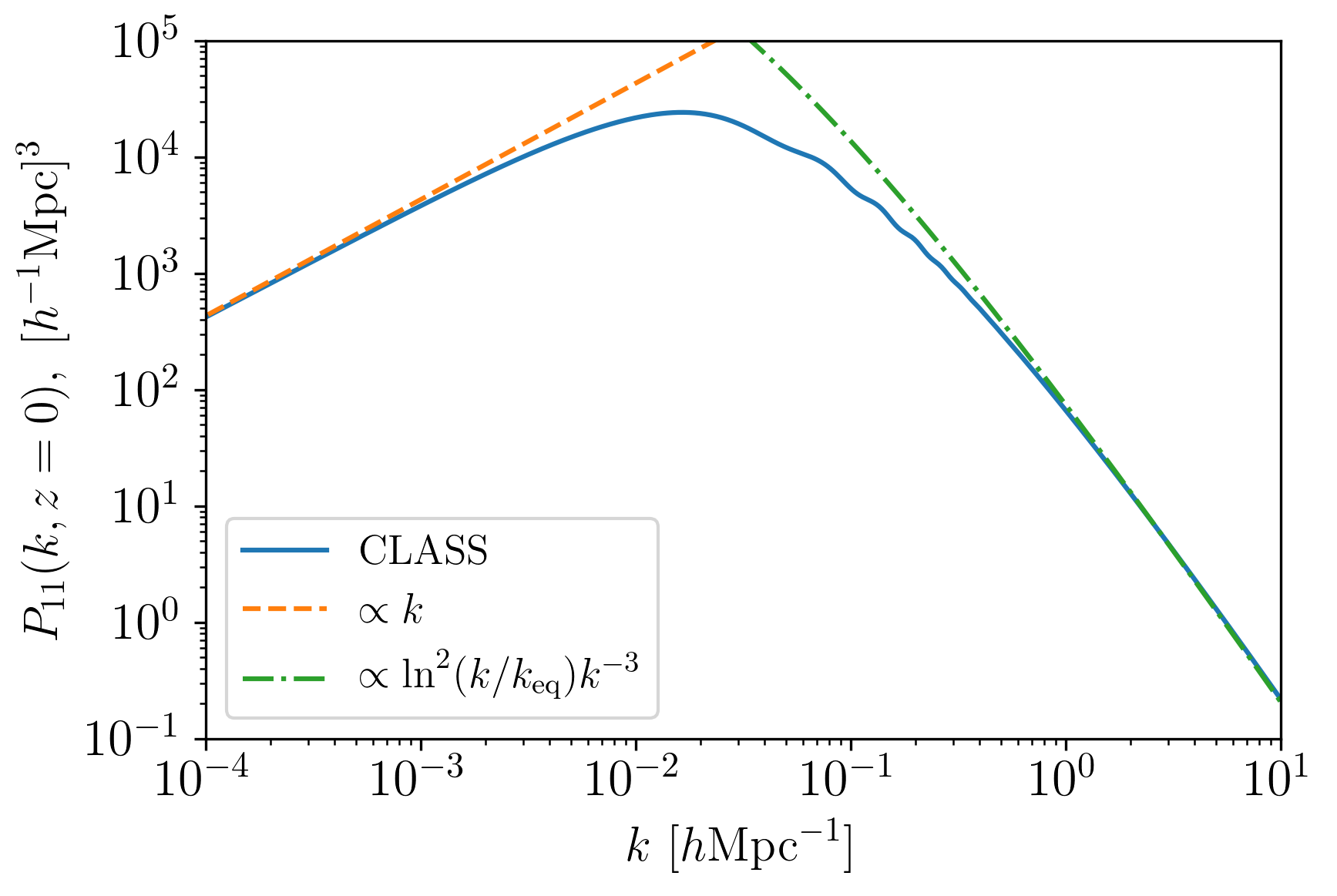}
    \caption{\textbf{Left:} the linear matter power spectrum as a function of wavenumber $k$ for several values of redshift, computed with the 
    \textsc{class} code~\cite{Lesgourgues:2011re}.  \textbf{Right:} the $z=0$ linear matter power spectrum and the asymptotic behaviours $\propto k$ as $k\to 0$ and $\propto k^{-3}\ln(k/k_{\rm eq})$,
    $k_{\rm eq}\approx 0.01~\Mpc$. We ignore the power spectrum tilt $n_s$. }
    \label{fig:P_of_k}
\end{figure}

It is convenient to work in Fourier space, where the Fourier image of the matter 
density contrast looks like this:
\begin{equation}\label{eq:deltak}
  \delta(\xv) = \int \frac{d^3 k}{(2\pi)^3}\;e^{i\kv\cdot\xv}\;\delta(\kv)\,,
  % \quad \delta(\kv)=\int \frac{d^3 k}{(2\pi)^3}\;e^{i\kv\cdot\xv} \delta(\xv)\;
  % \tag{1}
\end{equation}
The two correlation function of the Fourier-space density field reads
\begin{equation}\label{eq:pk}
  \avg{\delta(\kv)\,\delta(\kv')} = (2\pi)^3\,\delta_D^{(3)}(\kv+\kv')\,P(k)\,.
  % \tag{2}
\end{equation}
where $P(k)$ is called the
\textbf{power spectrum}. 
The power spectrum is a function of the wavenumber $k$ only. This is another consequence
of the statistical homogeneity and isotropy.
The Dirac delta function above reflects the momentum conservation
which is a consequence of the translational invariance suggested by the statistical homogeneity.
It is convenient to remove it (and $(2\pi)^3$) by introducing the 
primed correlator, which simply strips off the delta function:
\be 
\avg{\delta(\kv)\,\delta(\kv')}' \equiv P(k)\,.
\ee 
\medskip\noindent
From eq.~\eqref{eq:deltak} and eq.~\eqref{eq:pk} it is easy to show that the correlation function
is simply a Fourier transform of the power spectrum, which turns into a spherical Fourier transform 
thanks to the cosmological principle:
\begin{equation}
  \xi(r) = \int \frac{d^3 k}{(2\pi)^3}\;e^{i\kv\cdot\rv}\,P(k)
  = \int \frac{k^2\,dk}{2\pi^2}\;P(k)\;\frac{\sin(kr)}{kr}\,.
\end{equation}
The reverse us also true: the power spectrum is a spherical Fourier image of the two-point
correlation function:
\begin{equation}
  P(k) = 4\pi \int dr\;r^2\;\xi(r)\;\frac{\sin(kr)}{kr}\,.
\end{equation}
Let us discuss now the properties of the matter fluctuations 
in linear cosmological perturbation theory. For the usual baryons and cold dark matter 
the linear solution of the cosmological perturbation theory for the matter field after the recombination 
reads:
\begin{equation}
  \delta_1(\kv,z) = \Dp(z)\,\delta_0(\kv)\,,
\end{equation}
where $\delta_1(k,z)$ is the linear matter density, and $\delta_0$ is (up to a normalization) the initial matter density, 
which in the large-scale structure context means the matter density after recombination, when
baryons and dark matter form a single fluid to a very good approximation. A nice feature of linear matter 
perturbations is that their equation of motion does not depend on wavenumber, hence 
the time-evolution is simply captured by a function $\Dp(z)$, called the ``growth factor.''
In the standard cosmology matter grows uniformly on all scales.
%($\delta_1 \ll 1$ before nonlinear collapse).
Let us discuss now the \textit{linear matter power spectrum}:
\begin{equation}
  \avg{\delta_1(\kv,z)\;\delta_1(\kv',z)}' = P_{11}(k,z)\,.
\end{equation}
Cosmological evolution generates many interesting features in the $k$-dependence of $P_{11}$, see
the left panel of fig.~\ref{fig:P_of_k}. 
They can be summarized as follows. \textbf{Shape of $P_{11}(k,z)$:}
\noindent
\begin{itemize}[nosep]
   \item Peak at $k\sim \keq \approx 0.02\Mpc$.
  \item Low $k$ ($k\to 0$): $P_{11}(k) \propto k$ (goes to zero linearly).
  \item High $k$ ($k\to \infty$): $P(k) \sim k^{-3}\ln^2(k/\keq)$.
  \item For $k\sim 0.1~\Mpc$ there are wiggles produced by baryon acoustic oscillations (BAO). Their typical frequency is $\kbao = 1/(100\Mpch)\simeq 10^{-2}\Mpc$. The BAO is a relatively small feature $\sim 5\%$ of the total power, but it is incredibly important for cosmological constraints. 
\end{itemize}

The right panel of fig.~\ref{fig:P_of_k} demonstrates the above high-$k$ and low-$k$ asymptotics.
This funny shape is the consequence of different dynamics of the dark matter field 
during the radiation and matter domination. This is why it has a peak at the 
equality scale, which is imprinted during the transition between the two epochs.

\begin{figure}[htb!]
    \centering
    \includegraphics[width=0.49\linewidth]{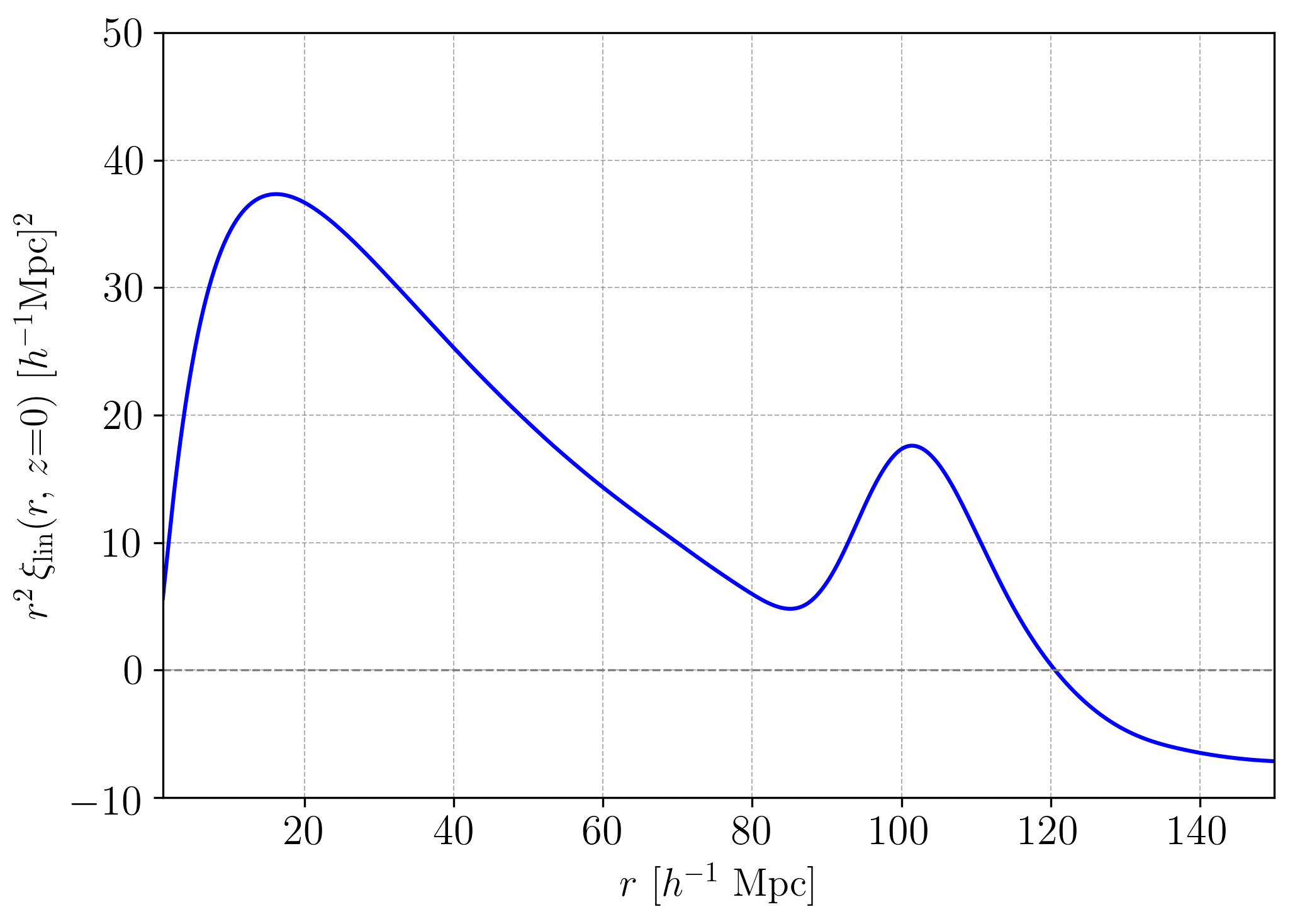}
    \caption{Linear correlation function $\xi_{\rm lin}(r)$ of the matter overdensity in position space. The bump at $r\simeq 100~\Mpch{}$ is the peak 
    of baryon acoustic oscillations (BAO).}
    \label{fig:xi_r}
\end{figure}
\begin{figure}
    \centering
    \includegraphics[width=0.49\linewidth]{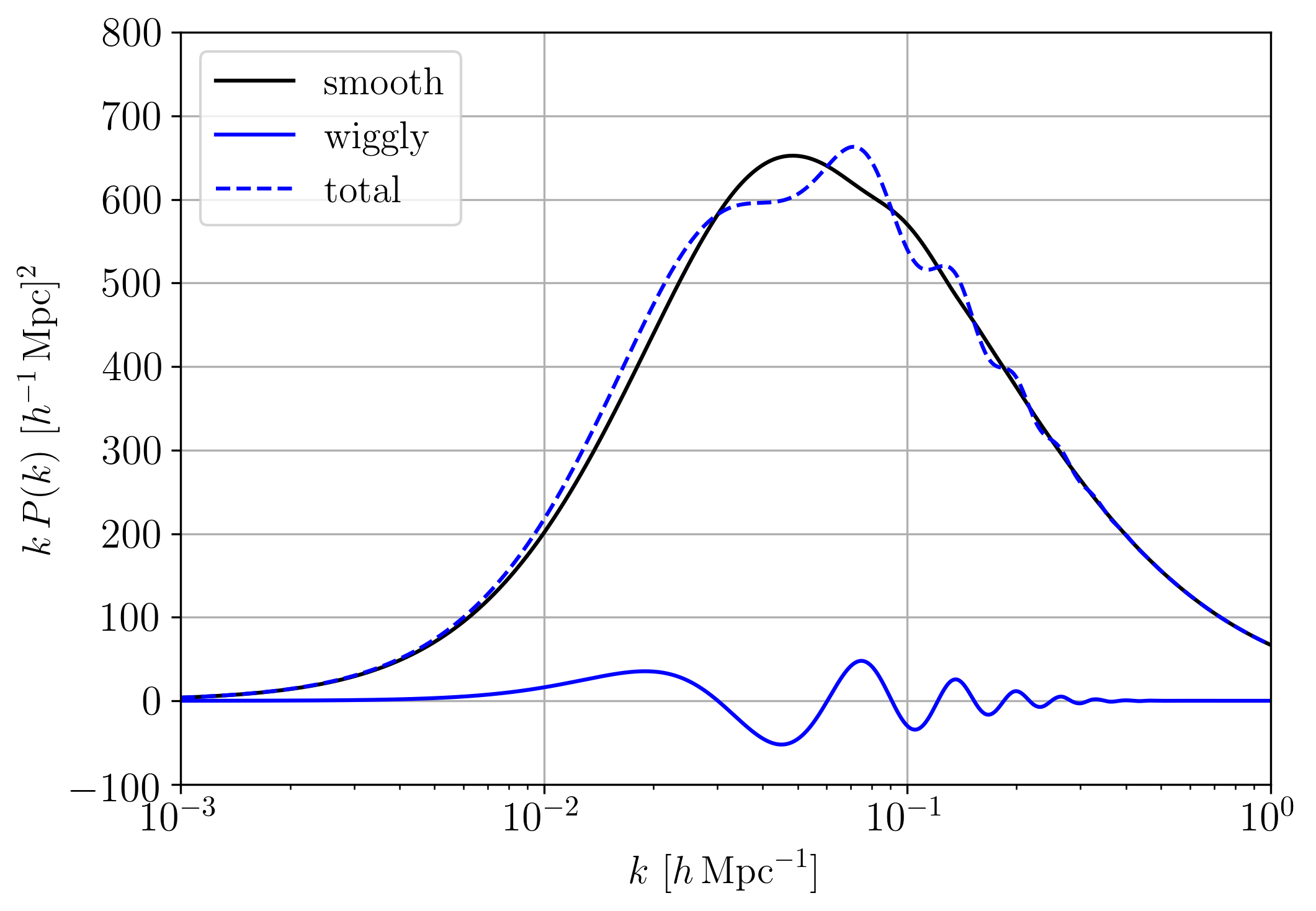}
        \includegraphics[width=0.49\linewidth]{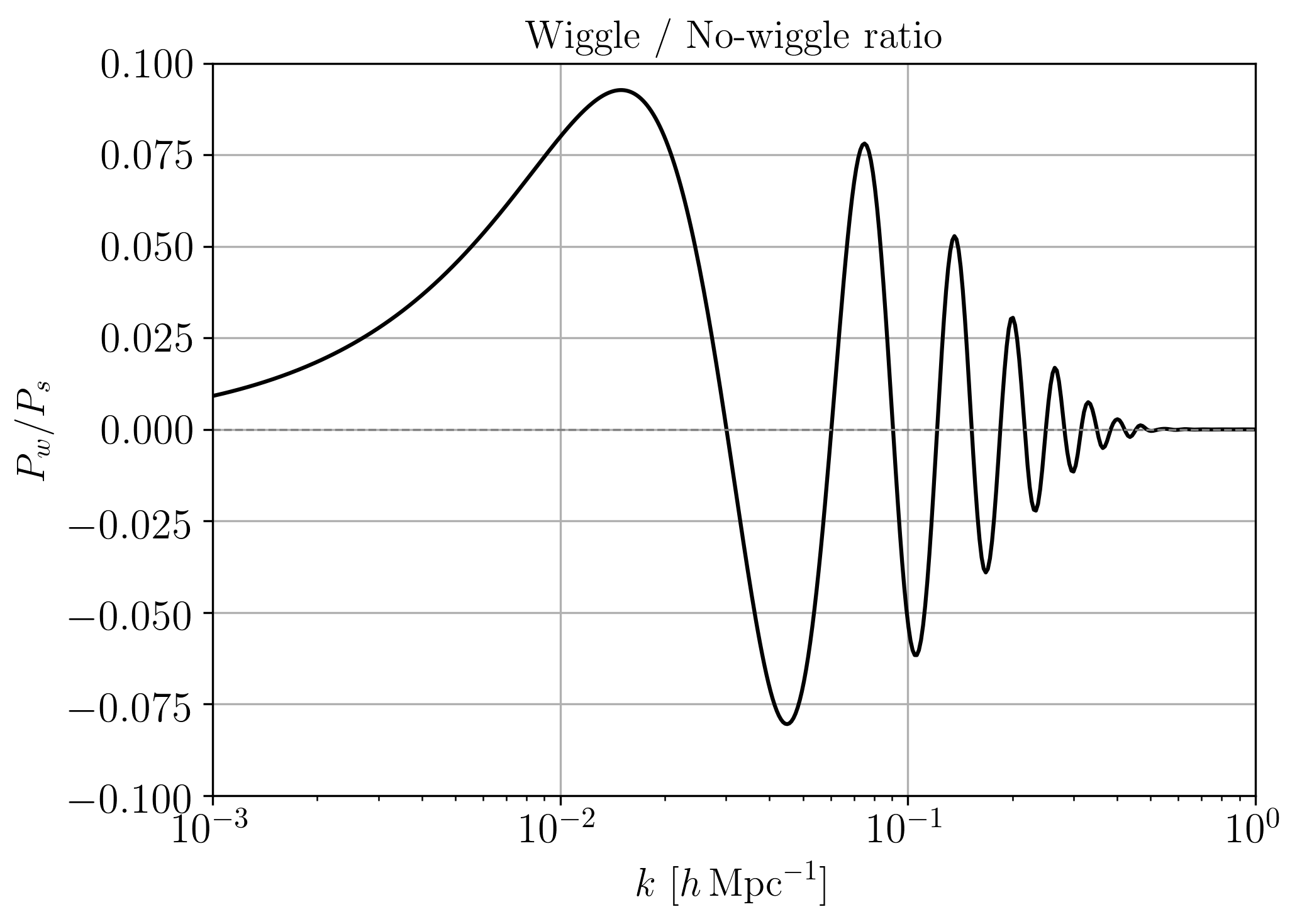}
    \caption{\textbf{Left:} the decomposition of the linear matter power spectrum into the ``wiggly'' and ``smooth'' components. The results are multiplied by $k$ to enhance the visibility of the wiggles. \textbf{Right:} the ratio of the wiggly to the 
    smooth power spectrum, which is about $\sim 5\%$.}
    \label{fig:Pw_nw}
\end{figure}

Fig. ~\ref{fig:xi_r} 
shows the linear correlation function (i.e. the Fourier transformed power spectrum). Crucially, one 
notices that all the BAO wiggles combine into a position-space
peak around  $\ell_{\rm BAO}=100\Mpch$. The BAO have a paramount 
significance in cosmology since they provide a standard ruler,
which can be observed at different redshifts and thus used 
to study the geometry of the Universe. The BAO also 
exhibit a peculiar evolution in non-linear perturbation theory,
which we will discuss shortly. To facilitate this discussion, 
it is useful to separate the total linear matter power spectrum
into the ``wiggly'' part $P_w$ that carries the wiggles and the
``smooth,'' i.e. ``non-wiggly'' part $P_{s}$, see Fig.~\ref{fig:Pw_nw}. 
Such separation 
is non-unique in practice since the slope of the smooth power 
spectrum varies considerably over the range of the wiggles.
Nevertheless, it is a nice tool that neatly highlights the important 
physics. Note that the ``wiggly'' power spectrum represents
only a small, $\sim 5\%$ fraction of the total power.

The linear matter power spectrum in $\Lambda$CDM has a very simple time-dependence: 
for a redshift $z$ it is simply the initial power spectrum rescaled by the growth factor squared
$D^2_+(z)$, as can be appreciated in the left panel of fig.~\ref{fig:P_of_k}. This simplicity allows one to choose $\delta_0(\kv)$  to be the linear matter 
density \textit{extrapolated} to redshift zero, in which case the growth factor now can be normalized
to unity, $D_+(z=0)=1$.

\begin{figure}
    \centering
    \includegraphics[width=0.49\linewidth]{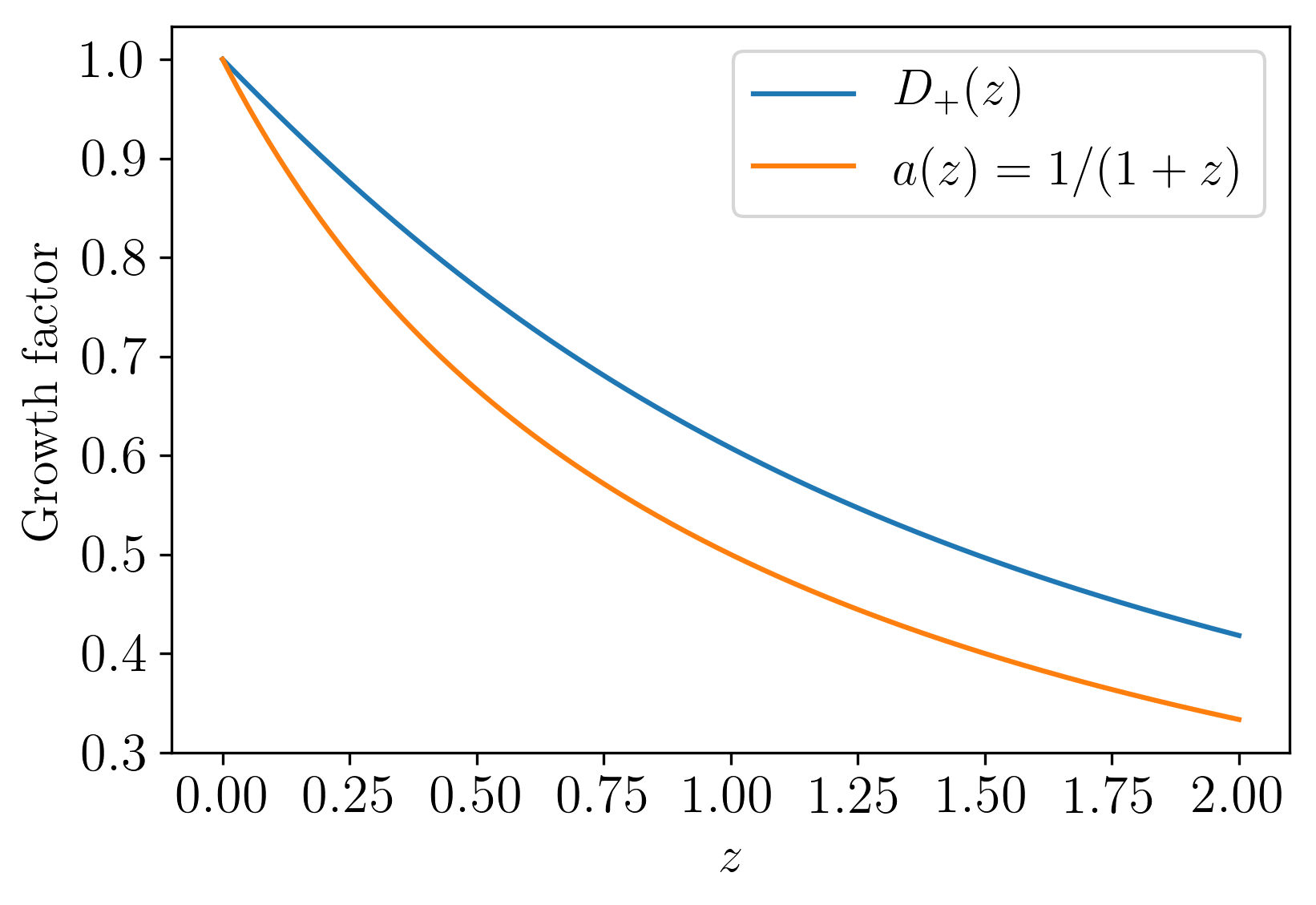}
        \includegraphics[width=0.49\linewidth]{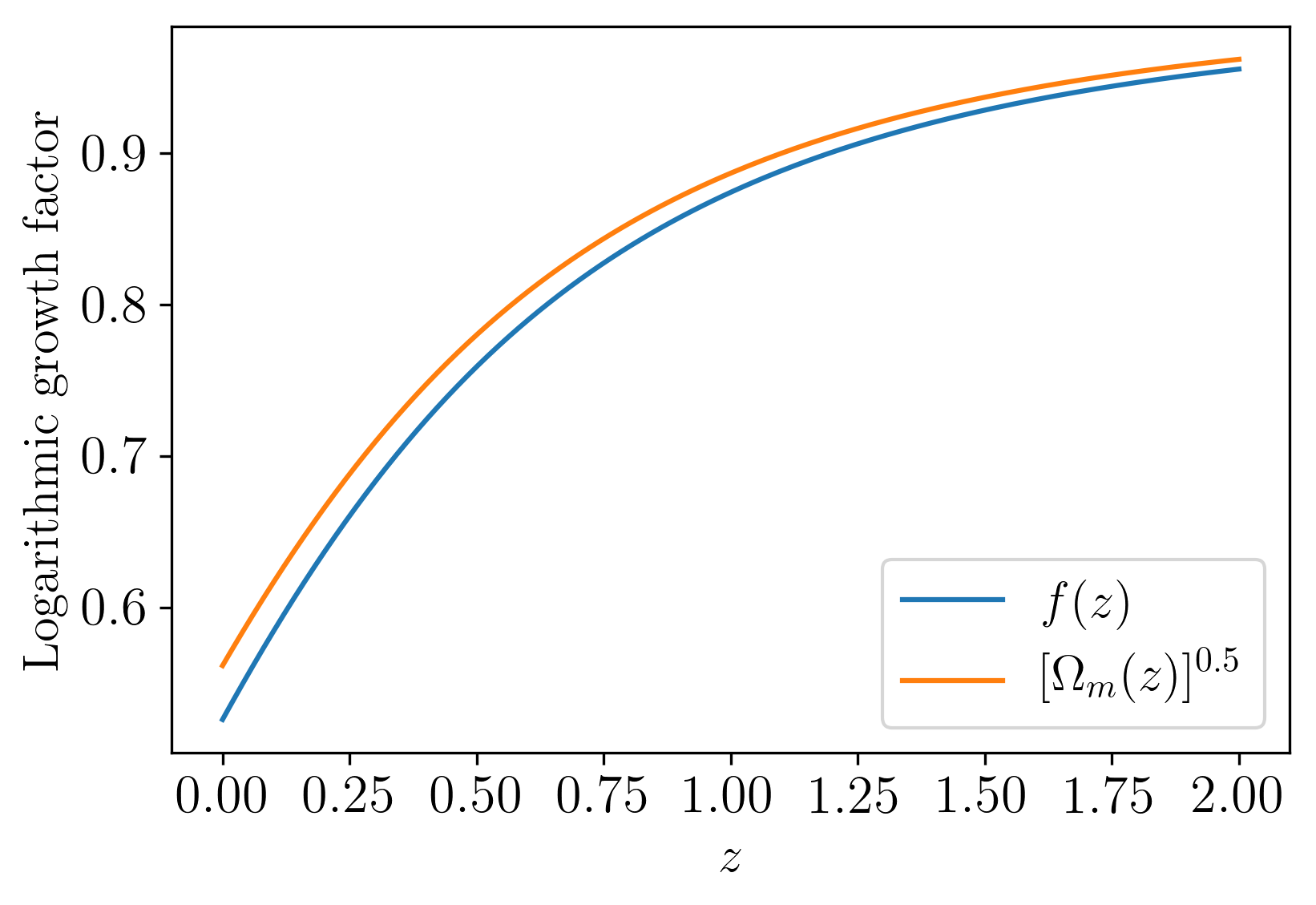}
    \caption{\textbf{Left:} the growth factor as a function of redshift $z$. We show the results for the fiducial $\Lambda$CDM cosmology and for a matter dominated Universe, where $D_+=1/(1+z)$. \textbf{Right:} the logarithmic growth factor $f(z)$ along with the approximate 
    function $[\Omega_m(z)]^{0.5}$. }
    \label{fig:D_of_z}
\end{figure}

In a matter-dominated Universe ($H^2 = \frac{8\pi G}{3}\,\rho_m(t)$) the growth factor is simply given by the scale factor $a(z)$ in the Friedman-Robertson-Walker metric, so we have
\begin{equation}
  P_{11}(k,z) = P_{11}(k,z\!=\!0)\;\Dp^2(z)\,,\qquad
  \Dp(z) = a(z)\,,\quad a = \frac{1}{1+z}\,.
\end{equation}
For $\Omega_m < 1$, i.e. in the presence of dark energy, $\Dp(z)$ deviates from $a(z)$.
This is shown in the left panel of figure~\ref{fig:D_of_z}. At  $z\approx 2$  the two functions are very 
different, which is a consequence of adopting the unit normalization at $z=0$. The scale factor 
grows with redshift much faster than $D_+(z)$, which reflects that dark energy slows down 
the growth of structure. 

Departures of the growth factor from the scale factor can be captured by the \textit{logarithmic}
\textit{growth factor} $f = d\ln\Dp/d\ln a$, see the right panel of figure~\ref{fig:D_of_z}. $f=1$ during matter domination, and when dark energy kicks in, it decreases to $f \approx 0.5$ at $z=0$. Approximately, the logarithmic growth factor 
behaves as $[\Omega_m(z)]^{0.5}$, where $\Omega_m(z)$ is the redshift-dependent abundance
of matter,
\be 
\Omega_m(z)\equiv \frac{\Omega_m(1+z)^3}{\Omega_m(1+z)^3+\Omega_\Lambda}\,.
\ee 
One can see that $\Omega_m(z)\to \Omega_m$ at $z=0$ and $\Omega_m(z)\to 1$ as $z\to \infty$.

Due to a simple time-dependence of linear theory, in what follows we will often suppress the explicit time 
dependence, assuming that unless stated otherwise, 
everything is computed at redshift zero, or at a specified redshift of 
interest.

%------------------------------------------------------------------
\subsection{Useful Quantities}
%------------------------------------------------------------------

\noindent\textbf{Useful quantities.}
Linear cosmological perturbation theory tells us a lot about the properties
of matter clustering in our Universe. Let's introduce some useful quantities 
that will prove instrumental in what follows.

\begin{figure}
    \centering
    \includegraphics[width=0.49\linewidth]{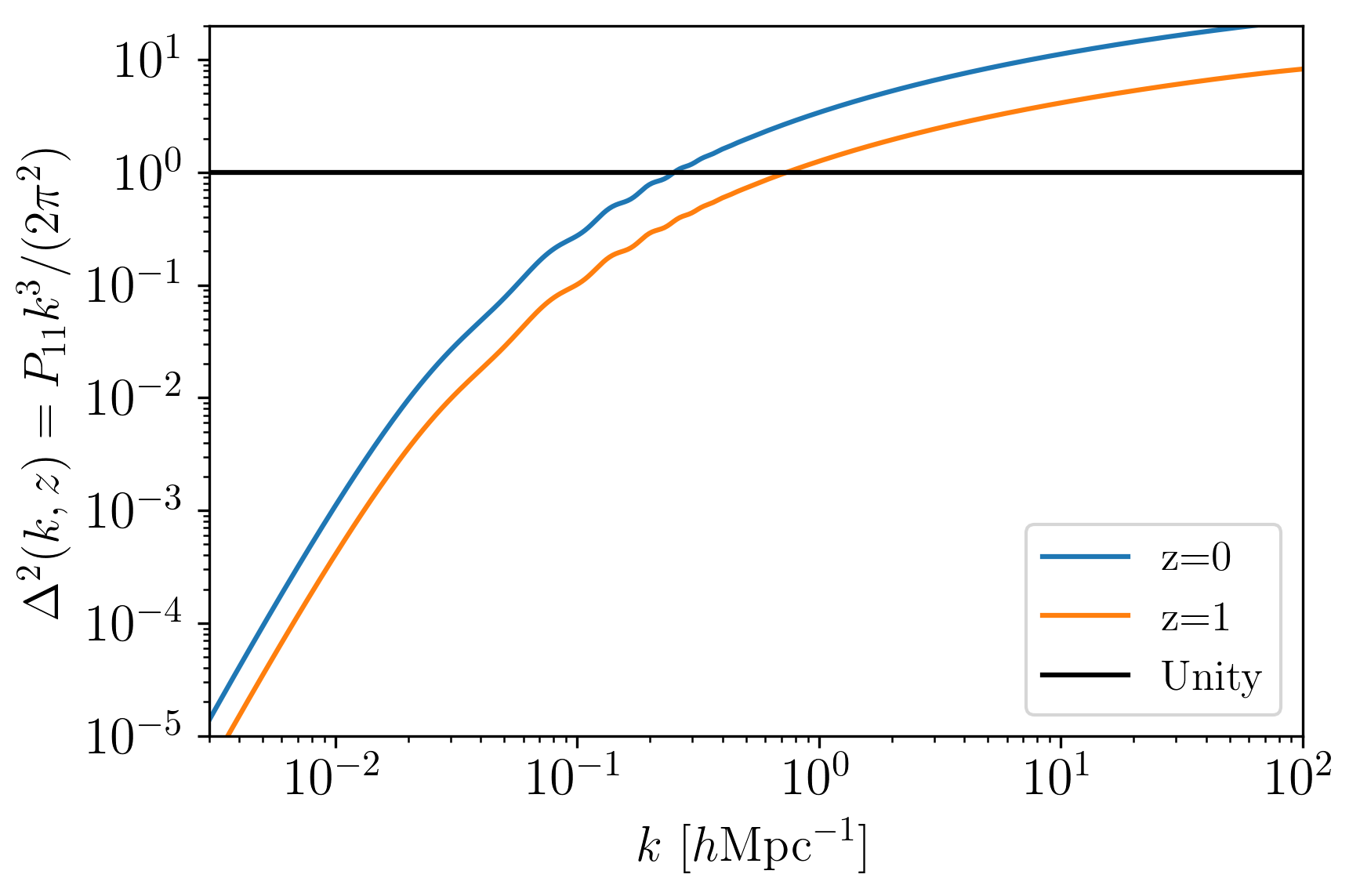}
        \includegraphics[width=0.49\linewidth]{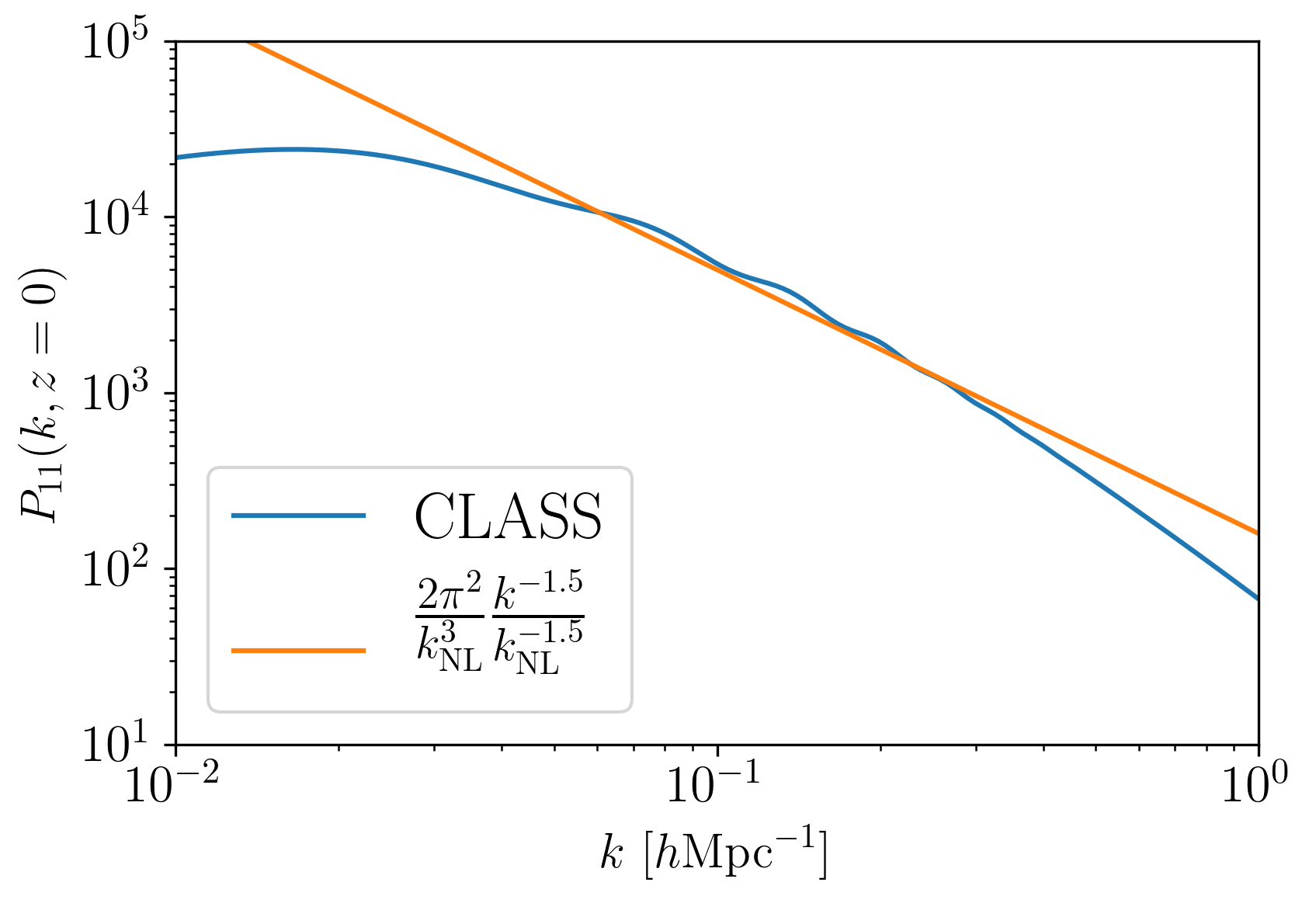}
    \caption{\textbf{Left:} The dimensionless power spectrum
  $\Delta^2(k,z) = P_{11}(k)\,k^3/(2\pi^2)$ at $z=0$ (blue) and $z=1$ (orange).
  The nonlinear scale $k_\mathrm{NL}(z)$ is defined by $\Delta^2(k_\mathrm{NL}) = 1$
  (black horizontal line);  one finds
  $k_\mathrm{NL} \approx 0.25\,h\,\mathrm{Mpc}^{-1}$ (at $z=0$) and  $k_\mathrm{NL} \approx 0.75\,h\,\mathrm{Mpc}^{-1}$ (at $z=0$).
  For $k \ll k_\mathrm{NL}$ the density field is perturbative
  ($\Delta^2 \ll 1$).
  \textbf{Right:} The linear power spectrum $P_{11}(k,z=0)$ computed
  with \textsc{class} (blue) compared to the power-law approximation
  $P_{11} \propto k^{-1.5}$ (orange), normalised to match at $k=\knl=0.25~\Mpc$.
  This scale-free approximation with effective slope
  $n_\mathrm{eff} \approx -1.5$ is used throughout these lectures for
  power-counting estimates (e.g.\ to relate the time dependence of
  counterterms to $c_s^2 \propto D^{4/(3+n)}$).
  The approximation is accurate to order one in the range
  $0.05 \lesssim k/(h\,\mathrm{Mpc}^{-1}) \lesssim 0.3$ relevant for
  one-loop calculations.}
    \label{fig:dimpk}
\end{figure}

\medskip\noindent
\textbf{a) Dimensionless power spectrum:}
\begin{equation}
  \Delta^2(k,z) = \frac{k^3}{2\pi^2}\,P(k,z)\,.
\end{equation}
This measures the contribution of a logarithmic interval of modes around $\sim k$ to the total 
variance of matter fluctuations:
\begin{equation}
  \avg{\delta^2(\xv)} = \xi(0) = \frac{1}{2\pi^2}\int d\ln k\;\Delta^2(k)\,.
\end{equation}
Roughly, the dimensional power spectrum tells how much a mode with wavenumber $k$ fluctuates
in position space. $ \Delta^2(k,z=0)$ in our $\Lambda$CDM model is shown in the left panel
of fig.~\ref{fig:dimpk}. It crosses unity on small scales, which means that the 
matter fluctuations there exceed unity, which invalidates perturbation theory.
To quantify this better we introduce 

\medskip\noindent
\textbf{b) Non-linear scale:}
$\knl$ is the momentum for which the fluctuations become $\mathcal{O}(1)$, i.e.\ $\delta_1 \sim 1$ (perturbation theory breakdown):
\begin{equation}
  \Delta^2(\knl) = 1\,,\qquad\text{i.e.}\quad
  \frac{\knl^3}{2\pi^2}\,P(\knl) = 1\,.
\end{equation}
For our fiducial Planck $\Lambda$CDM at $z=0$: $\knl \approx 0.25\Mpc$. Note that the non-linear scale
increases at high redshift, i.e. at $z=1$  one finds $\knl \approx 0.74\Mpc$.  This makes sense: 
matter fluctuations are smaller at high redshift and hence are more perturbative. 

\medskip\noindent
\textbf{c) Power-law universe (PLU) approximation.}  For a narrow range of scales the power spectrum 
can be well approximated as a power law,
\begin{equation}
  P_{11}^\mathrm{PLU}(k) \approx \frac{2\pi^2}{\knl^3}\left(\frac{k}{\knl}\right)^n\,.
\end{equation}
Note that this approximation automatically satisfies $\Delta_{11}^\mathrm{PLU}=1$ at $k=\knl$. If we want to match the actual power spectrum in the range $0.05 \lesssim k/(h\,\mathrm{Mpc}^{-1}) \lesssim 0.3$ we need to choose the following parameters:
\be 
\quad n \approx -1.5 \;\;\text{for }k\sim 0.1\Mpc\,,\quad \knl = 0.3\Mpc\,.
\ee 
Note that $\knl$ normalized to match a particular region
of $k$ might be somewhat different from the actual $\knl$ because the effective spectral index
$n$ reduces at larger $k$, producing a somewhat softer spectrum there. We will use the PLU spectrum 
for order-of-magnitude estimates where this small discrepancy would not matter.

\begin{figure}
    \centering
    \includegraphics[width=0.49\linewidth]{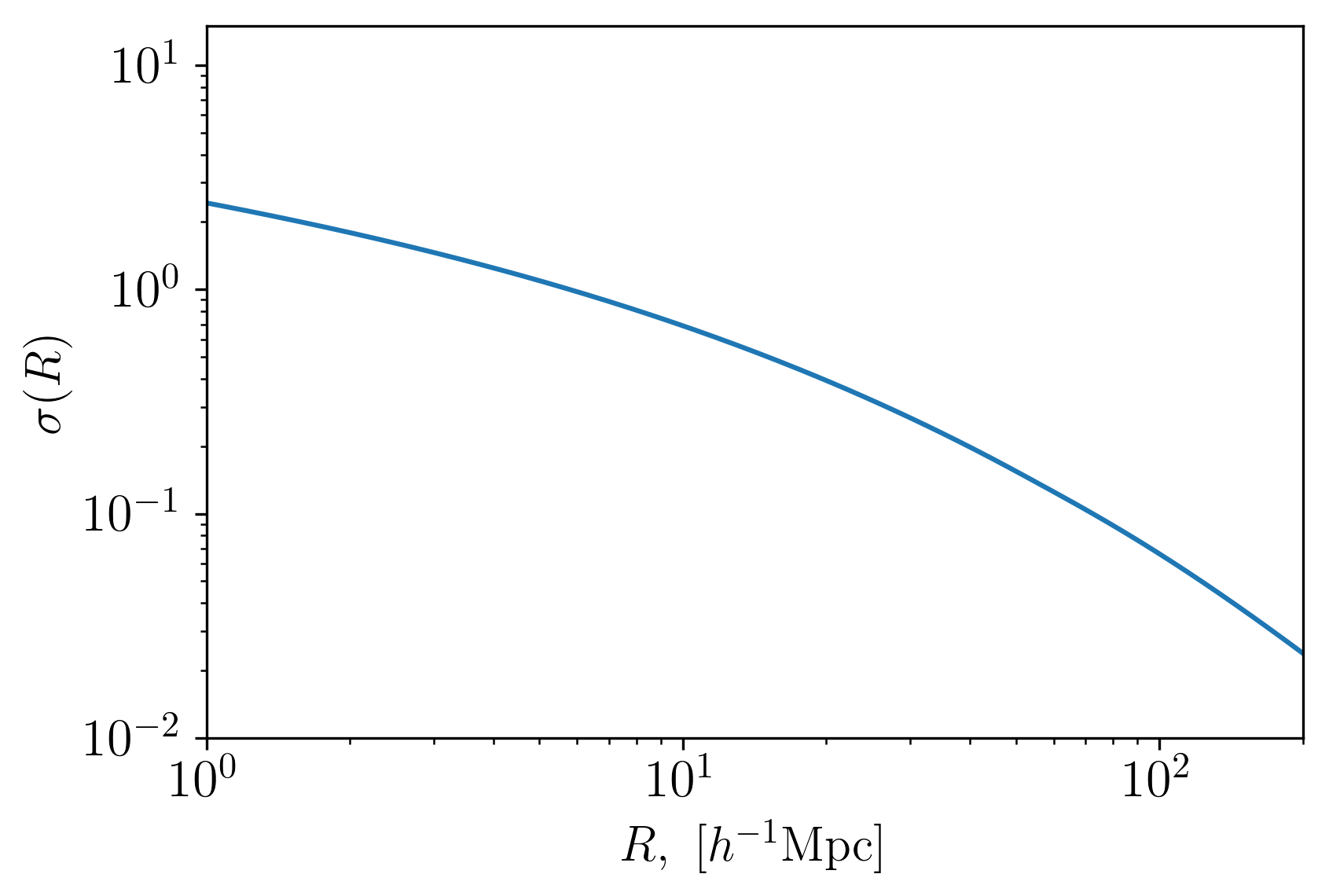}
        \includegraphics[width=0.49\linewidth]{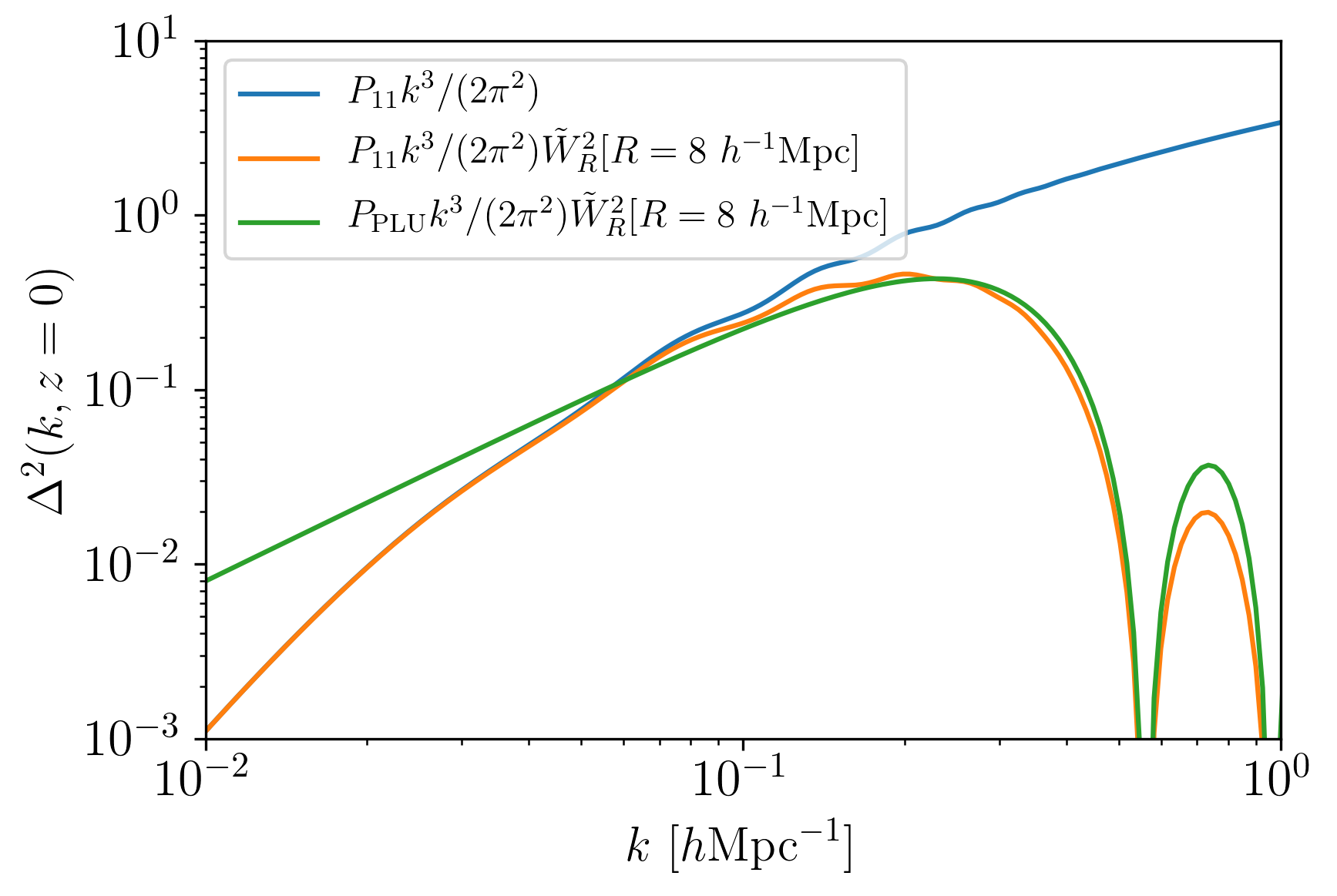}
    \caption{\textbf{Left:} The root-mean-square density fluctuation
  $\sigma(R) = \left[\int \frac{d^3q}{(2\pi)^3}\,P_{11}(q)\,
  \tilde{W}_R^2(q)\right]^{1/2}$ as a function of the smoothing scale $R$
  at $z=0$, using a top-hat window function. The familiar normalisation
  convention $\sigma_8 \equiv \sigma(R = 8\,h^{-1}\mathrm{Mpc}) \approx 0.8$
  is visible at $R = 8\,h^{-1}\mathrm{Mpc}$.
  As $R \to 0$, $\sigma(R)$ diverges, reflecting the UV sensitivity of
  the unsmoothed variance --- precisely the divergence that the EFT
  counterterms are designed to absorb.
  \textbf{Right:} The dimensionless power spectrum
  $\Delta^2(k) = P(k)\,k^3/(2\pi^2)$ at $z=0$ for three cases:
  the full linear spectrum $P_{11}$ from \textsc{class} (blue),
  and the smoothed spectrum
  $P_{11}(k)\,\tilde{W}_R^2(k)$ with a top-hat window at
  $R = 8\,h^{-1}\mathrm{Mpc}$ (green),
  and the smoothed
  power-law approximation $P_\mathrm{PLU} \propto k^{-1.5}$ (orange).
  The window function suppresses power on scales $k \gtrsim 1/R$,
  illustrating how spatial smoothing acts as a UV cutoff.
  The oscillatory features in the green curve are characteristic zeros
  of the Fourier-space top-hat window
  $\tilde{W}_R(k) = 3[\sin(kR) - kR\cos(kR)]/(kR)^3$.}
    \label{fig:sigma_of_R}
\end{figure}

\medskip\noindent
\textbf{d) Mass variance:}
\begin{equation}
  \avg{\delta^2(\xv)} = \xi(0) = \frac{1}{2\pi^2}\int d\ln q\;\Delta^2(q) \propto \lim_{q\to \infty}\ln^3(q) \to \infty\,,
\end{equation}
where in the last limit we took into account that the integrand $\Delta^2(q) \propto \ln^2(q)$
logarithmically
diverges in the UV.

We have already seen the above quantity. 
$\xi(0)$ measures a variance of density fluctuations 
at the same point, which can be used to estimate a typical amplitude of matter fluctuations. The problem
is  that this quantity diverges if we include contributions from arbitrarily 
small-scale (short wavelength) modes. In the particle physics speak
one says the mass variance diverges in the UV. A simple fix is to cut this off at 
$\sim \knl$, where linear theory breaks down. 
More systematically, one can \textbf{smooth the density field}, say using a spherical top-hat 
filter,
\begin{equation}
  W_R(\xv) = \frac{3}{4\pi R^3}\;\Theta_H(R - |\xv|)\,,
\end{equation}
where $\Theta_H$ is the Heaviside step function. The smoothed field $\delta_R$ 
\begin{equation}
  \delta_R(\xv) = \int d^3 y\;\delta(\xv+\vv{y})\,W_R(\vv{y})\,
\end{equation}
does not have fluctuations on scales smaller
than $R$.
In Fourier space we have $\delta_R(\kv) = \delta(\kv)\,\widetilde{W}_R(kR)$ with 
\begin{equation}
  \widetilde{W}_R = \frac{3\,j_1(kR)}{kR}\equiv  3\frac{\sin(kR) - kR\cos(kR)}{(kR)^3}\,.
\end{equation}
The smoothed variance reads:
\begin{equation}\label{eq:sig2_def}
  \sigma^2(R) \equiv \avg{\delta_R^2(\xv)}
  = \frac{1}{2\pi^2}\int d\ln k\;k^3\,P_{11}\,|\widetilde{W}_R|^2\,.
\end{equation}
The r.m.s. $\sigma(R)$  is shown in the left panel of fig.~\ref{fig:sigma_of_R}. We see that 
$\sigma(R)<1$ for $R\lesssim 5~\Mpch$. As long as we are doing the calculations 
for distances larger than this, perturbation theory should work. 
Finally, a familiar choice from textbook cosmology is to pick $R=8~\Mpc$, 

\medskip\noindent
\textbf{Example:} $\sigma(R\!=\!8) \approx 0.8$, \; $\sigma^2 \approx 0.64$ $\to$ small-ish. Next-to-leading order: $\sigma^4 \approx 0.41$.

It is also instructive to take a look 
at the integrand $k^3\,P_{11}\,|\widetilde{W}_R|^2$, shown in the right panel of fig.~\ref{fig:sigma_of_R} for $R=8~\Mpc$. 
First, we see
that the low-pass filter $W_R$  indeed suppresses the small scale power so that we get a finite 
answer for  the mass variance. 
Second, numerically, the integrand peaks at $k \approx 1.22/R = \pi/(2R)$, which tells us how to 
convert position space distances to the Fourier space wavenumbers.
In addition, and more importantly, because of this sharp peak around $1.22/R$, 
we see that even the very crude PLU actually provides a very good  approximation to the smoothed density variance integrand, which can be contrasted with the broadband power in fig.~\ref{fig:dimpk}.

\medskip\noindent
\textbf{e) Displacement field.}
It is useful to know how much a typical matter particle moves during the age of the Universe. 
Since we have a huge number of particles, they form a fluid with a continuous density, 
so it makes sense to talk about displacement field as opposed to displacements
of individual particles.
In conformal time $\tau$, the comoving displacement of a fluid element with the initial position 
$\xv_\mathrm{ini}$  is defined as
\begin{equation}
  \Psiv = \xv - \xv_\mathrm{ini} = \int_{\tau_\mathrm{ini}}^{\tau} d\tau_1\;\vvv(\xv_\mathrm{ini},\tau_1)\,,
\end{equation}
where $\vvv(\xv_\mathrm{ini},\tau_1)$ is the velocity of the fluid element 
with the initial position $\xv_\mathrm{ini}$ along its trajectory parameterized by
conformal 
time $\tau_1$.

In linear theory for the adiabatic mode initial conditions 
the velocity is curl-free, and its divergence is proportional to the density field by the continuity equation, 
i.e. we have 
\begin{equation}
  v^i = -a\,H\,f\;\frac{\partial^i \delta}{\Delta}\,,\qquad
  \vvv(\kv) = a H f\,\Dp \frac{i\kv}{k^2}\;\delta_0(\kv)=a H f\,\frac{i\kv}{k^2}\;\delta_1(\tau,\kv)\,,
\end{equation}
(note the appearance of the \textit{conformal} Hubble parameter $\mathcal{H} =\partial_\tau a/a\equiv aH$),
which implies
\begin{equation}\label{eq:displacement}
  \boxed{\Psiv(\kv) = \frac{i\kv}{k^2}\,\delta_1(\tau,\kv)\,.}
\end{equation}
From this we compute the displacement variance:
\begin{equation}
  \avg{\Psi_i(\xv)\,\Psi_j(\xv)}
  = \frac{\delta_{ij}}{3}\int \frac{P_{11}(k)}{k^2}\;\frac{d^3k}{(2\pi)^3}
  = \frac{\delta_{ij}}{6\pi^2}\int d\ln k\;k\,P_{11}(k)
  \equiv \delta_{ij}\,\sigma_d^2\,,
\end{equation}
where we introduced the 1D displacement variance
\be 
\sigma_d^2 \equiv \frac{1}{6\pi^2}\int d\ln k\;k\,P_{11}(k)~\,.
\ee 
Since the integrand $k\,P_{11}(k)\propto k^{-2}\ln^2 k$ as $k\to  \infty$,
$\sigma_d^2$  converges in the UV.
At $z=0$ one finds $\sigma_d^2 \approx 34\;(\Mpch)^2$.
Can we trust this result? No, because 
the integration above $\knl$ is not legitimate. 
Instead, we should stop around $\knl$, which gives us an estimate
\be
\frac{1}{6\pi^2} \int_0^{\knl} d\ln k\;k\,P_{11}(k) = 29\;(\Mpch)^2\,.
\ee 
From the 1D variance we can estimate a typical 3D r.m.s. displacement as
$\avg{|\Delta x|^2} = \avg{\Psi^i\Psi_i} = 3\,\sigma_d^2$, implying 
$\Delta x \sim \sqrt{3}\,\sigma_d \sim 9.3~\Mpch$. Thus
a typical matter particle travels by about $10~\Mpch$ during the age of the Universe.

\medskip\noindent
Note that the displacement variance is closely related to the 
velocity dispersion: 
\be 
\avg{v_i v_j} = \mathcal{H}^2 f^2\,\delta_{ij}\,\sigma_d^2\,.
\ee
 % hence $\avg{|\Delta x|^2} = \avg{\Psi^i\Psi^j} = 3\,\sigma_d^2$, so 

% \medskip\noindent
\subsection{Gaussian Random Fields and the Wick theorem}
The final important ingredient of our lectures is the Wick theorem. 
Observational data suggests that the initial conditions in our 
Universe are Gaussian, i.e. 
$\delta_1$ is a \textbf{Gaussian random field}. This implies that all odd-point  
correlation functions of this field vanish, while the even-point correlation 
functions can be expressed through the two point functions. At the lowest orders we have:
\begin{align}
  \avg{\delta_1(\kv_1)\delta_1(\kv_2)} &= (2\pi)^3\delta_D^{(3)}(\kv_1+\kv_2)\;P_{11}(k_1)\,,\\
  \avg{\delta_1\delta_1\delta_1} &= 0\,.
  % \avg{\delta_1\delta_1\delta_1\delta_1} &\sim P_{11}\,P_{11} \,.
\end{align}
\textbf{Wick's theorem}  precisely tells us how to compute higher order correlators. 
In general, an even $2N$-point correlator of a Gaussian random field $\phi$ is given by 
\begin{equation}
  \avg{\phi_{i_1}\cdots\phi_{i_{2N}}}
  = \sum_{\text{ordered pairings }P_i}
  \prod_{\substack{\text{pairs }ij\\ \text{in }P_i}} C_{ij}\,,\qquad
  C_{ij} = \avg{\phi_i\,\phi_j}\,.
\end{equation}
A derivation of this theorem from the Gaussian generating functional
is given in Appendix~\ref{app:GRF}.
In the cosmological context we have:
\be 
\begin{split}
  &\avg{\delta_1(\kv_1)\,\delta_1(\kv_2)\,\delta_1(\kv_3)\,\delta_1(\kv_4)}
  \;=\;\big\langle\wickA\big\rangle
  \;\\
  &+\;\big\langle\wickB\big\rangle
  \;+\;\big\langle\wickC\big\rangle\,,
  \end{split}
\ee
where we explicitly show different pairings that contract two fields
into a power spectrum, yielding
\begin{align}
  &= (2\pi)^3\delta_D^{(3)}(\kv_1+\kv_2)\;P_{11}(k_1)
     \times (2\pi)^3\delta_D^{(3)}(\kv_3+\kv_4)\;P_{11}(k_3)
     \nonumber\\[4pt]
  &\quad+ (2\pi)^3\delta_D^{(3)}(\kv_1+\kv_3)\;P_{11}(k_1)
     \times (2\pi)^3\delta_D^{(3)}(\kv_2+\kv_4)\;P_{11}(k_2)
     \nonumber\\[4pt]
  &\quad+ (2\pi)^3\delta_D^{(3)}(\kv_1+\kv_4)\;P_{11}(k_1)
     \times (2\pi)^3\delta_D^{(3)}(\kv_2+\kv_3)\;P_{11}(k_2)\,.
\end{align}

%==========================================================================
\section{Chapter 3: EFT of Dark Matter Clustering}
%==========================================================================

After this solid build-up, we are ready to develop non-linear perturbation theory
for matter clustering. As before, we are working in the approximation 
where dark matter and baryons form a single fluid (total matter).
In this approximation there is no distinction between total matter and dark matter, 
so we will use these terms interchangeably.

\subsection{Equations of Motion}

Let's derive the equation of motion for matter fluctuations. 
We will follow the symmetry-based derivation of fluid equations from Chapter 1. 
On scales much inside the horizon ($k/a \gg H$) we can work in the regime of 
\textbf{Newtonian cosmology}.
In physical coordinates and time, matter denvity and velocity fields satisfy the 
\textbf{continuity equation:}
\begin{equation}
  \frac{\partial\rho}{\partial t} + \nabla_{\bm r}\cdot(\rho\,{\bm V}) = 0~\,,
\end{equation}
\medskip\noindent
\textbf{the Euler equation:}
\begin{equation}
  \rho\;\frac{DV^i}{Dt} = -\rho\, \frac{\partial}{\partial r^i}\Phi + \frac{\partial}{\partial r^j}\tau^i_{~j}\,,
\end{equation}
where 
\begin{equation}
  \frac{D}{Dt} = \frac{\partial}{\partial t}\bigg|_{\rv}
    + V^j\frac{\partial}{\partial r^j}
\end{equation}
is the material (convective) derivative and $\tau^i_{~j}$ is a general stress-energy tensor. Gravity is captured by the
\textbf{Poisson equation:}
\begin{equation}\label{eq:poisson_phys}
\nabla_{\bm r}^2 \Phi  = 4\pi G \sum_a\left(\rho_a + 3p_a\right)
= 4\pi G\rho - \Lambda\,,
\end{equation}
where the sum runs over all components of the Universe, and
$\rho_a+3p_a$ is the  the
correct source of the Newtonian potential in the 
appropriate (sub-horizon) limit
of general relativity. For non-relativistic matter $p=0$ and one recovers the
usual source $4\pi G\rho$; a cosmological constant,
$p_\Lambda=-\rho_\Lambda=-\Lambda/8\pi G$, contributes the constant
$-\Lambda$ displayed in the second equality. The crucial point for
what follows is that at late times all components other than matter
are \emph{smooth}: they affect the background expansion, but carry
no fluctuations.
In what follows we will use 
conformal time $\tau$: $a\,d\tau = dt$, and comoving coordinates $x^i$ defined by the Friedmann metric
\begin{equation}
  ds^2 = -dt^2 + a^2(t)\,d\xv^{\,2} = a^2\!\left(-d\tau^2 + d\xv^{\,2}\right)\,,
\end{equation}
i.e. $ \rv = a(\tau)\,\xv$. We will keep using the conformal Hubble parameter
$\Hubble \equiv a'/a$, with primes denoting $\dd/\dd\tau$. 
The usual dots will denote the derivatives w.r.t.  cosmic time $t$. 
 It is
convenient to express the peculiar velocity in terms of conformal-time
derivatives.  
Since $\dd t = a\,\dd\tau$, one has
$\dot\xv = \xv'/a$, so that ${\bm v} = \xv'$.  The peculiar velocity in
our convention therefore has dimensions of velocity (comoving
displacement per conformal time).
The gradient operators are related by
$\gradr = a^{-1}\gradx$, and a time derivative at fixed $\rv$
transforms as
\begin{equation}\label{eq:dt-chain}
  \pder{}{t}\bigg|_{\rv}
    = \frac{1}{a}\pder{}{\tau}\bigg|_{\xv}
      - H \,\xv\cdot\gradx\,,
\end{equation}
where the second term accounts for the fact that a point of fixed~$\xv$
corresponds to a time-dependent $\rv$.

We decompose now the velocity, density, and gravitational potentials into the background
and fluctuations as 
\be
\label{eq:bg_split}
\begin{split}
& \vv{V} = H\rv + \vvv(\xv,\tau)\,,\quad \rho = \bar\rho\,(1+\delta(\xv,\tau))\,,\quad \Phi=-\frac{1}{2}\ddot{a}a|\xv|^2+\phi(\xv,\tau)\,.
\end{split}
\ee
The background density satisfies
$\dot{\bar{\rho}} +3H  \bar\rho=0$.
% \be 
% \begin{split}
% & \,,\\
% &
% \end{split}
% \ee 
The full velocity can be decomposed into the Hubble flow $+$ peculiar velocity via $\vv{V} = H\rv + \vvv$. Note that this is always possible to do 
by means of simple ensemble averaging.

It is straighforward to show that the perturbations satisfy the following 
equations (comoving coordinates, conformal
  time): 
\begin{tcolorbox}[colback=green!5!white, colframe=green!50!black]
\textbf{Perturbed continuity equation:}
\begin{equation}\label{eq:cont}
  \boxed{\dfrac{\partial\delta}{\partial\tau} + \partial_i\!\left[(1+\delta)\,v^i\right] = 0\,,}
\end{equation}
\medskip\noindent
\textbf{Perturbed Euler equation:}
\begin{equation}\label{eq:E-euler-conf}
  \boxed{ \dfrac{\partial v^{i}}{\partial\tau} 
  % \;v^{i\prime} 
  + \Hubble\,v^i + v^j\partial_jv^i
  = -\partial_i\phi
    + \frac{1}{\rho}\,\partial_j\tau^i{}_j\,,}
\end{equation}
\medskip\noindent
\textbf{Perturbed Poisson equation:}
\begin{equation}\label{eq:poisson}
  \boxed{ \nabla^2_x \phi = \frac{3}{2}\Hubble^2 \Omega_m(\tau)\delta\,.}
\end{equation}
\end{tcolorbox}
When the physical stress-tensor is set to zero, $\tau_{ij}=0$, the above equations
are simply equations for pressureless perfect fluid 
in cosmological perturbation theory. 
These equations are appropriate for a fluid of non-relativistic 
non-interacting or weakly interacting particles, like dark matter.
That's why histrionically, people assumed that this was the equation that describes
structure formation in our Universe. 

Our Universe, however, looks very different from a fluid of free
dark matter particles. We know that most of dark matter actually resides in bound structures
such as dark matter halos, so the actual small-scale behavior of the matter fluid
is strongly influenced by these non-linear bound objects. It is not obvious then
that $\tau_{ij}=0$ for a gas of dark matter halos. In fact it's not. 
In addition, dark matter in our Universe can be described by scenarios
different from that of non-relativistic collisionless particles. For instance, 
dark matter can be an axion-like field with a quantum pressure,
which then will have a non-zero $\tau_{ij}$ even if we ignore
the halos. But even in the absence of these physical scenarios, 
simply from symmetries and dimensional analysis (EFT--!) we must conclude 
that the most general description
must involve a non-zero $\tau_{ij}$. 

One can also obtain justification in favor of non-zero $\tau_{ij}$
from the UV point of view. This is the approach adopted in the original EFT papers~\cite{Baumann:2010tm,Carrasco:2012cv}. If we assume that 
dark matter is just a collection of non-relativistic particles, 
we'd get a velocity dispersion $\langle v_i v_j \rangle-\langle v_i \rangle\langle v_j \rangle$ 
in the right hand side of
the Euler equation following from
the taking the moments of the Boltzmann equation for the DM particles. Even if we start from 
collisionless CDM particles with well defined velocities, a dispersion
will be generated dynamically through orbit crossing (also known as shell-crossing).
Alternatively, one could start from the collisionless Boltzmann equation, coarse-grain
it over non-linear small scale modes, take moments, and end up with a nonzero $\tau_{ij}$
in the r.h.s. of the Euler equation. In either case $\tau_{ij}$ is interpreted
as an effective stress-tensor generated by non-linear dynamics.

Historically, people started building the non-linear theory for matter clustering assuming $\tau_{ij}=0$.
This happened to be a good starting point because the contribution of $\tau_{ij}$
is suppressed by gradients, just like in the case of usual fluids. 
This approximation went down in the literature as ``standard perturbation theory''  (SPT)~\cite{Bernardeau:2001qr}.
SPT represents the leading order approximation, while the terms coming from $\tau_{ij}$
capture a non-trivial gradient expansion. 
These terms are often called ``EFT corrections,''
because they appear when you start taking dimensional analysis and 
symmetries seriously. 
Thus, SPT can be thought of as an analog of the \textbf{``monopole''} approximation, while the gradient contributions can be thought of as higher order moments, i.e. \textbf{dipole, etc.}  Schematically we have:
\begin{equation}
 \underbrace{ \frac{\partial v^i}{\partial\tau} + \mathcal{H}v^i + v^j\partial_j v^i+\partial^i\Phi}_{\text{``monopole,'' SPT}}
  = 
  \quad\underbrace{\partial^j \tau^i_{~j}}_{\text{``dipole+higher orders,'' EFT}}\,.
\end{equation}
Let us start with the ``monopole''  approximation. Unlike the monopole of
electrostatic, it captures a highly non-trivial and rich dynamics. Importantly, 
unlike the case of the perfect fluid, we will see that this dynamics will \textit{require}
including the higher order EFT corrections. So let's start with SPT.

%------------------------------------------------------------------
\subsection{Standard Perturbation Theory (SPT)}
%------------------------------------------------------------------

To simplify our life going forward, let's introduce a new time variable:
\begin{equation}
  \eta = \ln\Dp\,,\quad(\Dp = e^\eta)\,.
\end{equation}
In addition, let's use a new variable for the velocity divergence:
\be 
\Theta = -\frac{\partial_i v^i}{f\Hubble}\,.
\ee 
Transforming eqs.~(\ref{eq:cont},\ref{eq:E-euler-conf},\ref{eq:poisson})
into 
Fourier space and using the Poisson equation to solve for $\phi$  (and dropping $\tau_{ij}$)
we get a closed system of equations for $\delta$ and $\Theta $ only:
\begin{align}\label{eq:SPT_cont}
  \partial_\eta \delta_{\kv} - \Theta_{\kv}
  &= \int_{\qv_1}\int_{\qv_2}(2\pi)^3 \delta_D^{(3)}(\kv-\qv_{12})\;\alpha(\qv_1,\qv_2)\;\Theta_{\qv_1}\,\delta_{\qv_2}\,,\\[4pt]
\label{eq:SPT_euler}
  \partial_\eta\Theta_{\kv} - \frac{3\Omega_m}{2f^2}\,\delta_{\kv}
  + \left(\frac{3\Omega_m}{2f^2} - 1\right)\Theta_{\kv}
  &=  \int_{\qv_1}\int_{\qv_2}(2\pi)^3 \delta_D^{(3)}(\kv-\qv_{12})\;\beta(\qv_1,\qv_2)\;\Theta_{\qv_1}\,\Theta_{\qv_2}\,,
\end{align}
with the coupling kernels:
\begin{equation}\label{eq:alpha_beta}
  \alpha(\qv_1,\qv_2) = \frac{\qv_{12}\cdot\qv_1}{q_1^2}\,,\qquad
  \beta(\qv_1,\qv_2) = \frac{q_{12}^2\;(\qv_1\cdot\qv_2)}{2\,q_1^2\,q_2^2}\,.
\end{equation}
The presence of time-dependent coefficients $\propto \frac{\Omega_m}{f^2}$ makes life
somewhat hard for us. The equation  simplifies in the two 
% \medskip\noindent
\textbf{cases:}
\begin{enumerate}[nosep]
  \item $\Omega_m = 1$, $f=1$: matter-dominated universe (MDU), also known in the general cosmology literature as an Einstein--de Sitter (EdS) Universe. Since we know we have a significant fraction
  of dark energy in our Universe, it would be a terrible approximation if used naively.
  \item $\Omega_m = f^2 \neq 1$: EdS in LSS. We've seen in the previous section
  that this is a relatively good approximation even in the presence of dark energy. This is what in the LSS literature called the EdS approximaiton because 
  effectively it produces the same answer as $\Omega_m = 1$, $f=1$. In this case time \& space dependence factorize.
\end{enumerate}
We will use the latter approximation in what follows.
It is easy to see now that the linearized EdS equations
\begin{align} 
  \partial_\eta \delta_{\kv} - \Theta_{\kv}
  &= 0\,,\\[4pt]
  \partial_\eta\Theta_{\kv} - \frac{3}{2}\,\delta_{\kv}
  + \frac{1}{2}\Theta_{\kv}
  &=  0\,,
\end{align}
supplemented with the initial condition 
\be 
\Theta = \delta = e^{\eta}\delta_0\,, \quad as \quad \eta \to -\infty\,,
\ee  
easily recovers the linear theory adiabatic mode solution: 
\be 
\delta_1 = \Dp\,\delta_0\,; \; \Theta_1 = \delta_1\,; \; v^i = -f\Hubble\;\dfrac{\partial^i\delta}{\Delta}\,.
\ee 
Note that one can also easily derive the equation for the curl part the velocity field ${\bm \omega}=\nabla\times {\bm v}$:
\be 
\partial_\eta {\bm \omega}+ \frac{1}{f}{\bm \omega}= 0\,.
\ee  
In SPT the curl decays as $\propto 1/a$ since it does not have a source. The gravitational potential is scalar so it supports
only gradient modes. 
The situation, however, is different if we include $\tau_{ij}$, which 
generates voracity at the non-linear level~\cite{Pueblas:2008uv,Mercolli:2013bsa}. This is, however, a higher order effect irrelevant for 
our lectures.
We can focus on the scalar component $\Theta$
and develop a general perturbative solution to the non-linear equations.

\bigskip\noindent
\textbf{Perturbative expansion.}  It turns out that the following expansion solves the SPT equations 
perturbatively: 
\be 
\begin{split}
  &\delta = \Dp\,\delta_0 + \Dp^2(\tau)\int d^3q\;F_2(\kv-\qv,\qv)\;(2\pi)^3\delta_0{}({\kv-\qv})\,\delta_0({\qv})
  + \;\\
  &+
 \Dp^3(\tau)\int d^3q_1\,d^3q_2\,d^3q_3\;\delta_D^{(3)}(\kv-\qv_{123}) F_3(\qv_1,\qv_2,\qv_3)\;(2\pi)^3\delta_0(\qv_1)\delta_0(\qv_2)\delta_0(\qv_3)
  + \cdots \\
    &\Theta = \Dp\,\delta_0 + \Dp^2(\tau)\int d^3q\;G_2(\kv-\qv,\qv)\;(2\pi)^3\delta_0({\kv-\qv})\,
    \delta_0({\qv})
  + \;\\
  &+
 \Dp^3(\tau)\int d^3q_1\,d^3q_2\,d^3q_3\;\delta_D^{(3)}(\kv-\qv_{123}) G_3(\qv_1,\qv_2,\qv_3)\;(2\pi)^3\delta_0(\qv_1)\delta_0(\qv_2)\delta_0(\qv_3)
  + \cdots
  \end{split}
\ee
the the kernels $F_n$, $G_n$ can be recursively
defined by plugging the above ansatz into eqs.~\eqref{eq:SPT_cont}--\eqref{eq:SPT_euler}, with the result: 
\be 
\begin{split}
F_n(\qv_1,...\qv_n) = &  
\sum_{m-1}^{n-1}\frac{G_m(\qv_1,...,\qv_m)}{(2n+3)(n-1)}
[(2n+1)\alpha(\qv_{1m},\qv_{(m+1)...n})F_{n-m}(\qv_{m+1},...\qv_n)\\
&+2\beta(\qv_{1m},\qv_{(m+1)n})G_{n-m}(\qv_{m+1},...\qv_n)]\,,\\
G_n(\qv_1,...\qv_n) = &  \sum_{m-1}^{n-1}\frac{G_m(\qv_1,...,\qv_m)}{(2n+3)(n-1)}
[3\alpha(\qv_{1m},\qv_{(m+1)...n})F_{n-m}(\qv_{m+1},...,\qv_n)\\
&+2n\beta(\qv_{1m},\qv_{(m+1)...n})G_{n-m}(\qv_{m+1},...,\qv_n)]\,,
\end{split}
\ee
and $F_1=G_1=1$ (recovering the linear growing mode). 
In particular, at first non-trivial order we find
\be\label{eq:F2}
\begin{split}
F_2(\qv_1,\qv_2) = \frac{5}{7}+\frac{(\qv_1\cdot \qv_2)}{q_1q_2}\left(
\frac{q_1}{q_2}+\frac{q_2}{q_1}
\right) +\frac{2}{7}\frac{(\qv_1\cdot\qv_2)^2}{q_1^2 q_2^2}\,,\\
G_2(\qv_1,\qv_2)  = \frac{3}{7}+\frac{(\qv_1\cdot \qv_2)}{q_1q_2}\left(
\frac{q_1}{q_2}+\frac{q_2}{q_1}
\right) +\frac{4}{7}\frac{(\qv_1\cdot\qv_2)^2}{q_1^2 q_2^2}\,.
\end{split}
\ee 
Note that all kernels by default are symmetrized w.r.t. their arguments.
An instructive alternative derivation of $F_2$ and $G_2$, based on the
retarded Green's functions of the fluid equations in the variable
$\eta$, is presented in Appendix~\ref{app:green-eta}.
Another useful way to write down the above expansion is
\be 
\delta=\sum_{n=1}\delta^{(n)}\,,\quad \Theta=\sum_{n=1}\Theta^{(n)}\,,
\ee 
where $\delta^{(n)}$ terms explicitly absorb the time dependence:
\begin{align}
  \delta^{(n)}(\kv,a) &= \int\!\frac{\dd^3 q_1}{(2\pi)^3}
    \cdots\frac{\dd^3 q_n}{(2\pi)^3}\;
    (2\pi)^3\delta_D^{(3)}(\kv-\kv_{1\cdots n})\;
    F_n(\kv_1,\ldots,\kv_n)\;\delta_1(\kv_1)\cdots\delta_1(\kv_n)\,,
    \label{eq:deltan-kernel}\\[4pt]
  \Theta^{(n)}(\kv,a) &= \int\!\frac{\dd^3 q_1}{(2\pi)^3}
    \cdots\frac{\dd^3 q_n}{(2\pi)^3}\;
    (2\pi)^3\delta_D^{(3)}(\kv-\kv_{1\cdots n})\;
    G_n(\kv_1,\ldots,\kv_n)\;\delta_1(\kv_1)\cdots\delta_1(\kv_n)\,,
    \label{eq:thetan-kernel}
\end{align}
with the identification $\delta^{(1)}=\delta_1$.

We can see that the time dependence of  the solution 
factorizes in front of each term in the series 
thanks to the EdS approximation.
% \bigskip\noindent
% \textbf{$F_2$ kernel:}
% \begin{equation}\label{eq:F2}
%   \boxed{F_2(\kv_1,\kv_2) = \frac{5}{7}
%   + \frac{\kv_1\cdot\kv_2}{2k_1^2}
%   + \frac{\kv_1\cdot\kv_2}{2k_2^2}
%   + \frac{2}{7}\,\frac{(\kv_1\cdot\kv_2)^2}{k_1^2\,k_2^2}\,.}
% \end{equation}
Using $\qv_1\cdot\qv_2 = q_1 q_2\cos\theta_{q_1 q_2}$ we observe a re-appearance of the 
gravitational 
multipole expansion in the $F_2$ kernel. In particular we see:
\begin{itemize}[nosep]
  \item $5/7$: monopole,
  \item terms $\propto \cos\theta$: dipole,
  \item term $\propto \cos^2\theta$: quadrupole.
\end{itemize}
\textbf{Einstein--de Sitter approximation in LSS:}
The coefficients $5/7$, $1/2$, and $2/7$ etc. are specific to the EdS approximation. 
In general they will be time-dependent functions. 
In the LSS context the term ``EdS'' approximation refers to the procedure when one solves
the equations of motion perturbatively, as above, and then replaces $\Dp$ in the above answer by the actual 
linear growth factor
$\Dp$ computed by Boltzmann codes. In other words, one 
uses the \emph{exact} $\Dp$, $f$ from the linear theory equations, but the $F_n$, $G_n$ kernels obtained in the limit $\Omega_m = f^2 = 1$.
This version of the EdS approximation actually gives an 
exact answer in linear theory (i.e. at the leading approximation), and significantly suppresses the error at higher orders. 
That's why it has become standard in the field.  

Before we proceed, let us mention that it is possible to solve the pressure-less perfect fluid 
equations even without the EdS approximation~\cite{Takahashi:2008yk,
Carrasco:2012cv,
Fasiello:2016qpn,Garny:2020ilv,Steele:2020tak,Fasiello:2022lff}, 
but the exact result matches the approximate one to
better than 1\%. Given that this $\lesssim 1\%$ mismatch appears only 
starting at the one-loop order,
they are likely to be 
totally negligible in all 
practical applications both for current and future galaxy surveys, 
though there are 
efficient ways to compute 
these effects if needed~\cite{Fidler:2026wqg}.

\bigskip\noindent
\textbf{One-loop power spectrum.} We are in the position to compute the \textit{non-linear}
matter power spectrum in SPT. The fluid equations have the following perturbative solution for the density field:
\begin{equation}
  \delta_\mathrm{NL}=\delta = \delta^{(1)} + \delta^{(2)} + \delta^{(3)} +\cdots\,, %\delta_1 + \int F_2\,\delta_1^2 + \int F_3\,\delta_1^3 + \cdots\,,
\end{equation}
which we can square now to get the matter power spectrum,
$\avg{\delta_{\rm NL}(\kv)\,\delta_{\rm NL}(\kv')}
= (2\pi)^3\delta_D(\kv+\kv')\,P(k)$.
Keeping terms through order $O(\delta_1^3)$
we get, 
\begin{align}\label{eq:Pexpand}
  \avg{\delta_{\rm NL}(\kv)\,\delta_{\rm NL}(\kv')}'
  &= \underbrace{\avg{\delta^{(1)}\,\delta^{(1)}}'}_{P_{11}}
    + \underbrace{2\avg{\delta^{(1)}\,\delta^{(2)}}'}_{\text{= 0 (odd)}}
    + \underbrace{\avg{\delta^{(2)}\delta^{(2)}}'}_{P_{22}}
    + \underbrace{2\avg{\delta^{(3)}\delta_1}'}_{2P_{13}}
    + \ldots
\end{align}
% \begin{align}
%   \avg{\delta_\mathrm{NL}(\kv)\,\delta_\mathrm{NL}(\kv)}'
%   &= \avg{\delta_1\,\delta_1}'
%   + 2\avg{\delta_1\;\int F_2\,\delta_1^2}' \quad(\text{$=0$ for Gaussian ICs})\nonumber\\
%   &\quad + \avg{\int F_2\,\delta_1^2\;\int F_2\,\delta_1^2}'
%   + 2\avg{\int F_3\,\delta_1^3\;\delta_1}'\,,
% \end{align}
where the factor of 2 above comes from the fact that there are two symmetric terms 
from each $\delta_\mathrm{NL}$ above.  All expectation values are
computed via \textbf{Wick's theorem}: for Gaussian initial conditions,
every $n$-point function of $\delta_1$ factorizes into a sum over all
distinct pairings (contractions), with each pair giving a linear
power spectrum:
\begin{equation}\label{eq:wick-rule}
  \contraction{}{\delta_1}{(\kv_a)\;}{\delta_1}
  \delta_1(\kv_a)\;\delta_1(\kv_b)
  \;\equiv\;
  \avg{\delta_1(\kv_a)\,\delta_1(\kv_b)}
  = (2\pi)^3\delta_D(\kv_a+\kv_b)\,P_{11}(k_a)\,.
\end{equation}
$\avg{\delta_1\,\delta^{(2)}}$ is the vanishing term.
It involves three $\delta_1$ fields (one external, two from
$F_2$).  An odd number of Gaussian fields cannot be fully paired, so we have
\begin{equation}\label{eq:odd-vanish}
  \avg{\delta^{(1)}(\kv)\;\delta^{(2)}(\kv')}'
  = \int\!\frac{\dd^3 q}{(2\pi)^3}\;F_2({\bm q},\kv'{-}{\bm q})\;
  \avg{\delta^{(1)}(\kv)\;\delta^{(1)}({\bm q})\;\delta^{(1)}(\kv'{-}{\bm q})}
  = 0\,.
\end{equation}
Obviously the above term is non-zero for \textit{non-Gaussian} initial conditions.
This case will be discussed elsewhere.

The next term is called the ``22'' term or $P_{22}$. 
This correlator involves two copies of $\delta^{(2)}$:
\begin{align}
  &\avg{\delta^{(2)}(\kv)\;\delta^{(2)}(\kv')}'= \int\!\frac{\dd^3 q_1}{(2\pi)^3}\!\int\!\frac{\dd^3 q_2}{(2\pi)^3}\;
    F_2({\bm q}_1,\kv{-}{\bm q}_1)\;
    F_2({\bm q}_2,\kv'{-}{\bm q}_2)
  \nonumber\\[2pt]
  &\hspace{4em}\times\;
    \avg{\delta_1({\bm q}_1)\;\delta_1(\kv{-}{\bm q}_1)\;
         \delta_1({\bm q}_2)\;\delta_1(\kv'{-}{\bm q}_2)}'\,.
  \label{eq:P22-setup}
\end{align}
Wick's theorem produces three pieces: 
Contraction $(12)(34)$:
\begin{equation}
  \contraction{}{\delta_1}{({\bm q}_1)\;}{\delta_1}
  \contraction[2ex]{\delta_1({\bm q}_1)\;\delta_1(\kv{-}{\bm q}_1)\;}
    {\delta_1}{({\bm q}_2)\;}{\delta_1}
  \delta_1({\bm q}_1)\;\delta_1(\kv{-}{\bm q}_1)\;
  \delta_1({\bm q}_2)\;\delta_1(\kv'{-}{\bm q}_2)
\end{equation}
generates terms contributing only at $\kv=0$, $\kv'=0$. This is the \textbf{disconnected} piece
that vanishes once we take into account that $F_2(\qv,-\qv)=0$. Then there are two identical 
connected pieces from  $(13)(24)$ and $(14)(23)$ contractions, giving 
\begin{equation}\label{eq:P22}
  \boxed{\;P_{22}(k)
  = 2\int\!\frac{\dd^3 q}{(2\pi)^3}\;
    \bigl[F_2({\bm q},\kv{-}{\bm q})\bigr]^2\;
    P_{11}(q)\;P_{11}(|\kv{-}{\bm q}|)\;}\,.
\end{equation}
\bigskip\noindent
The $P_{13}$ term is computed in the same fashion. Starting from
\begin{align}
  &\avg{\delta^{(3)}(\kv)\;\delta_1(\kv')}'= \int\!\frac{\dd^3 q_1}{(2\pi)^3}\!\int\!\frac{\dd^3 q_2}{(2\pi)^3}\;
    F_3({\bm q}_1,{\bm q}_2,\kv{-}{\bm q}_1{-}{\bm q}_2)
  \nonumber\\[2pt]
  &\hspace{2em}\times\;
    \avg{\delta_1({\bm q}_1)\;\delta_1({\bm q}_2)\;
         \delta_1(\kv{-}{\bm q}_1{-}{\bm q}_2)\;
         \delta_1(\kv')}'\,,
  \label{eq:P13-setup}
\end{align}
and collecting three identical pieces connected by the Wick pairing we arrive at 
\begin{equation}\label{eq:P13}
  \boxed{\;P_{13}(k)
  = 3\,P_{11}(k)\int\!\frac{\dd^3 q}{(2\pi)^3}\;
    F_3(\kv,{\bm q},-{\bm q})\;P_{11}(q)\;}
\end{equation}
Summing the two we end up with  
\begin{equation}\label{eq:P1loop}
\begin{split}
& P_{\rm NL}(k) = P_{\rm lin}(k) +  \Delta P_{1\text{-loop}}(k)  = P_{11}(k) + P_{22}(k) + 2\,P_{13}(k)\,.\\
\end{split}
\end{equation}
The sum of the two terms on the right are called the ``one-loop matter power spectrum correction:'' 
\be 
\boxed{\Delta P_{\rm 1-loop}(k) =  P_{22}(k) + 2\,P_{13}(k)}\,.
\ee 
The Wick contractions computed above can be organised systematically
using a diagrammatic technique analogous to Feynman diagrams in
quantum field theory.

\subsubsection*{Feynman rules}

\tikzset{
  Fvertex/.style={draw,fill=white,minimum size=7pt,
    inner sep=0pt,rectangle},
  dot/.style={circle,fill=black,inner sep=1.5pt},
  pdot/.style={circle,fill=black,inner sep=2.2pt},
  arr/.style={thick,decoration={markings,
    mark=at position #1 with {\arrow{>}}},postaction={decorate}},
  arr/.default=0.55,
  arrL/.style={thick,decoration={markings,
    mark=at position #1 with {\arrow{<}}},postaction={decorate}},
  arrL/.default=0.55,
  plain/.style={thick},
}

\begin{figure}[h]
\centering
\begin{tikzpicture}[scale=1.0]
  \node[Fvertex,minimum size=9pt] (V) at (0,0) {};
  \node[below=4pt] at (V) {$F_n$};
  \draw[arrL] (-2.5,0) -- (V);
  \node[above=2pt] at (-1.6,0) {$\kv$};
  \draw[arr=0.65,thick] (V.east) -- ++(35:2.2)
    node[right]{${\bm q}_n$};
  \draw[arr=0.65,thick] (V.east) -- ++(-12:2.2)
    node[right]{${\bm q}_2$};
  \draw[arr=0.65,thick] (V.east) -- ++(-35:2.2)
    node[right]{${\bm q}_1$};
  \draw[dashed,thick,arr=0.65] (V.east) -- ++(20:2.0);
  \draw[dashed,thick,arr=0.65] (V.east) -- ++(8:2.0);
\end{tikzpicture}

\vspace{10pt}

\begin{tikzpicture}[scale=0.85]
  % === P11 ===
  \begin{scope}[xshift=0cm]
    \node at (-1.8,1.8) {$P_{11}$};
    \node[Fvertex] (A) at (-0.9,0) {};
    \node[Fvertex] (B) at (0.9,0) {};
    \node[below=4pt] at (A) {\footnotesize$F_1$};
    \node[below=4pt] at (B) {\footnotesize$F_1$};
    \node[pdot] (D) at (0,0) {};
    \node[below=5pt] at (D) {\footnotesize$P_{\rm lin}$};
    \draw[arr] (A.west) -- (-2.3,0);
    \draw[arr] (B.east) -- (2.3,0);
    \draw[arr=0.6] (A.east) -- (D);
    \draw[arrL=0.4] (D) -- (B.west);
  \end{scope}

  % === P22 ===
  \begin{scope}[xshift=6cm]
    \node at (-1.6,1.8) {$P_{22}$};
    \node[Fvertex] (L) at (-1.0,0) {};
    \node[Fvertex] (R) at (1.0,0) {};
    \node[below=5pt] at ([xshift=-2pt]L.south) {\footnotesize$F_2$};
    \node[below=5pt] at ([xshift=2pt]R.south) {\footnotesize$F_2$};
    \draw[arr] (L.west) -- (-2.4,0);
    \draw[arr] (R.east) -- (2.4,0);
    \node[pdot] (TU) at (0,1.05) {};
    \draw[arr=0.55] (L.north) to[out=75,in=180] (TU);
    \draw[arrL=0.45] (TU) to[out=0,in=105] (R.north);
    \node[below=4pt] at (TU) {\footnotesize$P_{\rm lin}$};
    \node[pdot] (TD) at (0,-1.05) {};
    \draw[arr=0.55] (L.south) to[out=-75,in=180] (TD);
    \draw[arrL=0.45] (TD) to[out=0,in=-105] (R.south);
    \node[above=4pt] at (TD) {\footnotesize$P_{\rm lin}$};
  \end{scope}

  % === P13 ===
  \begin{scope}[xshift=12cm]
    \node at (-1.9,1.8) {$P_{13}$};
    \node[Fvertex] (F3) at (-0.8,0) {};
    \node[below=4pt] at (F3) {\footnotesize$F_3$};
    \node[Fvertex] (Fo) at (1.4,0) {};
    \node[below=4pt] at (Fo) {\footnotesize$F_1$};
    \draw[arr] (F3.west) -- (-2.3,0);
    \draw[arr] (Fo.east) -- (2.9,0);
    \node[pdot] (BD) at (0.3,0) {};
    \draw[arr=0.6] (F3.east) -- (BD);
    \draw[arrL=0.4] (BD) -- (Fo.west);
    \node[below=5pt] at (BD) {\footnotesize$P_{\rm lin}$};
    \node[pdot] (TT) at (-0.8,1.22) {};
    \draw[arr=0.55] (-0.8,0.12) arc(270:90:0.55);
    \draw[arrL=0.45] (-0.8,1.22) arc(90:-90:0.55);
    \node[below=4pt] at (TT) {\footnotesize$P_{\rm lin}$};
  \end{scope}
\end{tikzpicture}
\caption{\textbf{Top:} the SPT vertex $F_n(\qv_1,\ldots,\qv_n)$ with
  outgoing momentum $\kv=\qv_1+\cdots+\qv_n$ (dashed lines stand for
  the remaining legs). \textbf{Bottom:} the tree-level and one-loop power
  spectrum diagrams. Each fat dot is a Wick contraction of two
  linear fields, i.e.\ an insertion of $P_{\rm lin}$, with the
  momentum of both lines flowing into it; the open squares are the
  kernels $F_n$, with $F_1=1$ the trivial vertex.}
\label{fig:powerspec}
\end{figure}
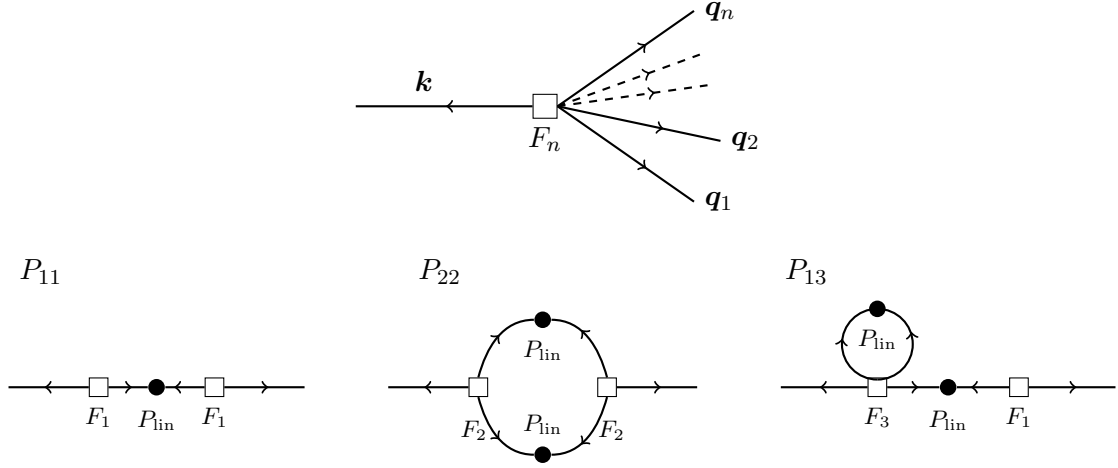

\begin{enumerate}[nosep]
  \item \textbf{Vertex:} an open square (Fig.~\ref{fig:powerspec},
    top) denotes the symmetrized kernel
    $F_n(\qv_1,\ldots,\qv_n)$, with one outgoing line of momentum
    $\kv=\qv_1+\cdots+\qv_n$ and $n$ incoming lines, each carrying
    one factor of the linear field $\delta_1(\qv_i)$.
  \item \textbf{Line ($P_{11}$ insertion):} a filled circle
    (Fig.~\ref{fig:powerspec}, bottom) joining two legs is a Wick
    contraction of two linear fields,
    $\avg{\delta_1(\qv)\,\delta_1(-\qv)}' = P_{11}(q)$. The arrows
    on the two joined legs both point \emph{into} the circle: the
    contracted fields carry momenta $\qv$ and $-\qv$, so momentum
    flows into the insertion from both sides. Each internal momentum
    left free by momentum conservation is integrated with
    $\int \dd^3q/(2\pi)^3$; the number of such integrals is the
    number of loops $L$, and an $L$-loop diagram scales as
    $P_{11}^{L+1}$.
  \item \textbf{Multiplicity:} multiply each diagram by the number
    of distinct Wick pairings that produce the same topology.
\end{enumerate}

% \begin{enumerate}[nosep]
%   \item \textbf{Vertex:} an open square (Fig.~\ref{fig:powerspec},
%     top) denotes the symmetrized kernel
%     $F_n(\qv_1,\ldots,\qv_n)$, with one outgoing line of momentum
%     $\kv=\qv_1+\cdots+\qv_n$ and $n$ incoming lines, each carrying
%     one factor of the linear field $\delta_1(\qv_i)$.
%   \item \textbf{Line:} every line joining two legs is a Wick
%     contraction, $\avg{\delta_1(\qv)\,\delta_1(-\qv)}' = P_{11}(q)$.
%     Each internal momentum left free by momentum conservation is
%     integrated with $\int \dd^3q/(2\pi)^3$; the number of such
%     integrals is the number of loops $L$, and an $L$-loop diagram
%     scales as $P_{11}^{L+1}$.
%   \item \textbf{Multiplicity:} multiply each diagram by the number
%     of distinct Wick pairings that produce the same topology.
% \end{enumerate}

\noindent
Reading the two one-loop diagrams is a matter of routing the momenta.
In $P_{22}$ (the ``sunset''), momentum conservation at the two $F_2$
vertices fixes the arcs to carry $\qv$ and $\kv-\qv$, leaving one
free integral; in $P_{13}$ (the ``tadpole''), two legs of $F_3$ close
into a loop carrying $\qv$ and $-\qv$, while the third leg connects
to the external point through the bridge $P_{11}(k)$.

\textbf{Homework:} use the rules to recover the boxed results
\eqref{eq:P22} and \eqref{eq:P13}, including the multiplicities
($2$ for $P_{22}$; $3$ per ordering for $P_{13}$, hence $6$ in
$2P_{13}$).

The two topologies play very different roles, and this distinction
will organize much of what follows. In $P_{13}$ one factor of
$P_{11}(k)$ sits \emph{outside} the loop: $P_{13}$-type diagrams are
corrections to the \emph{propagation} of the mode $k$, multiplicative
in the linear spectrum. In $P_{22}$ both $P_{11}$'s are evaluated at
loop momenta: $P_{22}$-type diagrams describe genuine
\emph{mode coupling} --- power generated at $\kv$ by pairs of modes
$\qv$, $\kv-\qv$. We will meet this dichotomy again in the IR limits
below, and in the EFT renormalization: the UV sensitivity of the
$13$-type diagrams is absorbed by the $k^2 P_{11}(k)$ counterterm
(Chapter 3), while that of the $22$-type diagrams requires the
stochastic $k^4$ term (Chapter 4).

Finally, let us compute the one-loop integral.
One may notice that 
our propagators $P_{11}$ are complicated numerical functions
of momenta, in contrast to the usual QFT diagrams (e.g. the propagators 
typically scale as $k^{-2}$ in the massless case). Because of this we need to do the loop
integrals numerically  for a given set of $\Lambda$CDM model parameters. 

\begin{figure}
    \centering
    \includegraphics[width=0.79\linewidth]{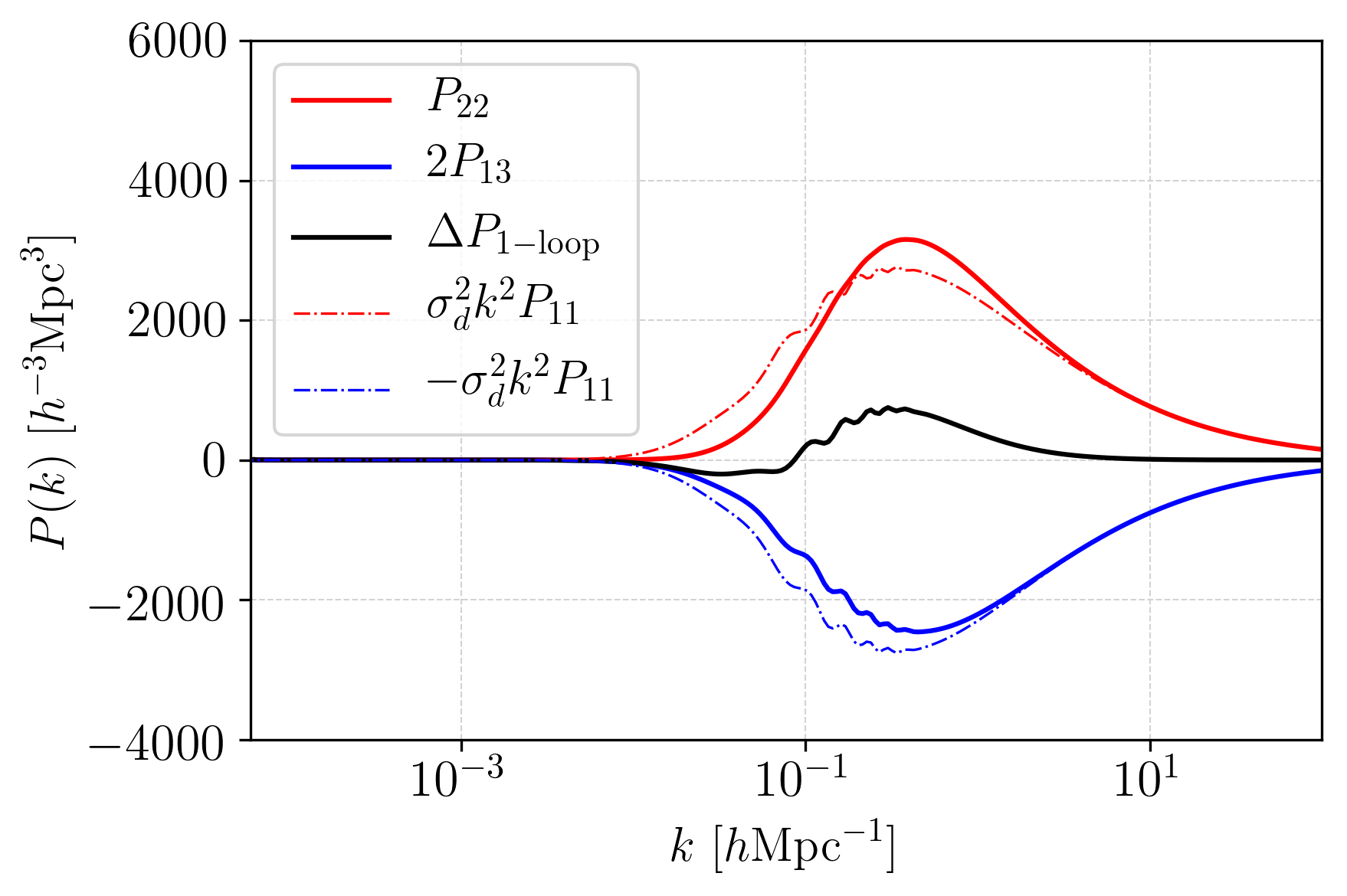}
    \caption{The one-loop matter power spectrum $\Delta P_{\rm 1-loop}$, which is a sum of the $2P_{13}$ and $P_{22}$ diagrams. 
    Dot-dashed lines show the leading infrared (IR) asymptotics of the loop integrals proportional to the displacement variance $\sigma_d^2$.
    The computation is carried out numerically with \textsc{class-pt} code~\cite{Chudaykin:2020aoj}. }
    \label{fig:p22-p13}
\end{figure}

The numerical evaluation for our fiducial cosmology yields the result
displayed in Fig.~\ref{fig:p22-p13}.
One important observation is that the total power spectrum
is a result of a large cancellation between $P_{13}$ and $P_{22}$
contributions. 
Each of this diagrams individually is a factor of ten greater than
the residual.
This, in fact, is symptomatic of the problems with the 
convergence in the
infrared limit, which we will discuss in detail shortly.

\begin{figure}[htb!]
    \centering
    \includegraphics[width=0.79\linewidth]{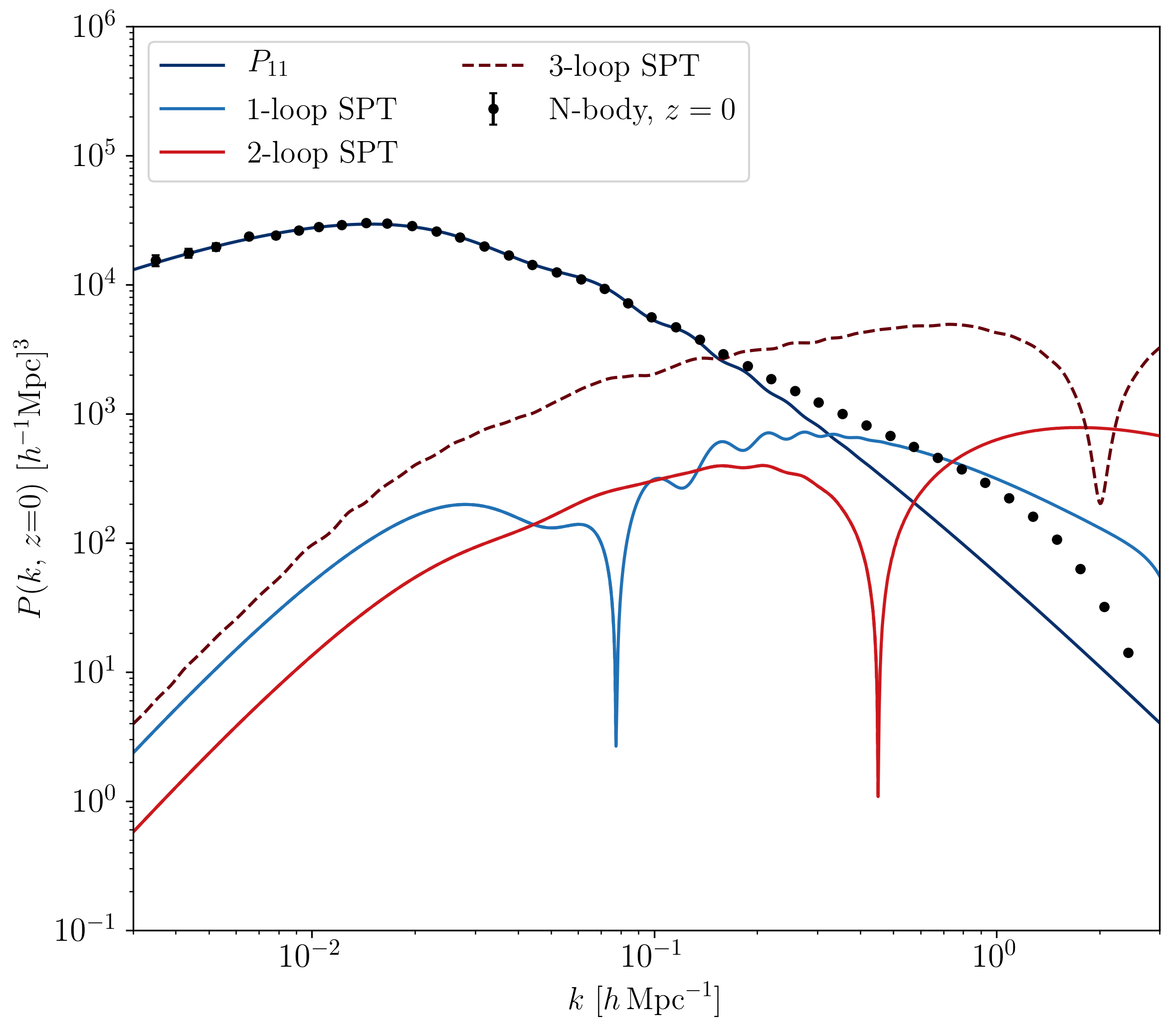}
    \caption{Matter power spectrum at $z=0$: SPT loop corrections versus Horizon Run N-body simulation data~\cite{Kim:2011ab}. The three-loop SPT power spectrum is adapted from Ref.~\cite{Blas:2013aba}.}
    \label{fig:spt_loops}
\end{figure}

In principle, nothing stops us from computing more SPT loops: at $L$
loops the diagrams involve the correlators
$\avg{\delta^{(m)}\delta^{(n)}}$ with $m+n=2L+2$ and kernels up to
$F_{2L+1}$; at two loops these are $P_{33-I}$, $P_{33-II}$ (there are two distinct topologies in the 3-3 sector), $P_{24}$ and $P_{15}$,
with two free loop momenta. 
Such calculations have been done through the three loop  
order even before the full development of EFT~\cite{Scoccimarro:1996jy,Scoccimarro:1999kp,Blas:2013bpa,Blas:2013aba} (see Figure~\ref{fig:spt_loops}).
Back then the pioneers used to think that SPT is a full solution
to non-linear perturbation theory. In this logic, computing higher
loops will lead to a better match to data.
To test this,
Figure~\ref{fig:spt_loops} shows the SPT performance 
against the N-body data. There we show the linear theory prediction
and SPT corrections through the three loop order.
We see that computing more loops is simply not helpful 
-- the two-loop corrections 
happened to be of the same order of magnitude as the one-loop,
and the three-loop SPT power spectrum is greater than the first two
corrections combined, on all scales. 
In addition, these corrections don't look like the data, e.g. 
the tree-loop power spectrum grows at $k\sim 0.1~\hMpc$, while the 
N-body data actually shows the decrease of $P(k)$.
Clearly, the naive SPT
expansion is in trouble. 

Historically, the pioneers thought that SPT diverges like an
asymptotic series, and that some kind of resummation might improve
its convergence properties. This logic often works for asymptotic
series, as e.g.\ in the case of the anharmonic oscillator in
ordinary quantum mechanics. This line of thinking produced a vast
literature on resummations of SPT. These resummations, however,
were a red herring --- the real issue was the fundamental
inconsistency of the perfect fluid equations. This is nicely
demonstrated, e.g., in Ref.~\cite{McQuinn:2015tva}, which focused
on matter clustering in 1+1 dimensions. The perfect pressureless
fluid equations can be solved exactly in this case, but the exact
solution still fails to describe the data --- a clear demonstration
that no resummation of diagrams within the wrong theory can
compensate for the absence of the correct physical ingredients.

Now we know that the breakdown of the SPT loop expansion
happens due to the absence of a proper renormalization
of the corresponding loop diagrams. 
This renormalization is provided by the effective stress tensor,
which we have argued must be present in the equations of motion
purely on symmetry grounds. 
It's instructive nevertheless
to study the IR and UV issues of the SPT expansion more 
carefully in order to understand how the full EFT treatment fixes them.

%------------------------------------------------------------------
\subsection{IR Limit of Loop Corrections and the Equivalence Principle}
%------------------------------------------------------------------
Consider the limits of the loop integrals when the loop momenta
are parametrically smaller than the external momentum $k$. Formally this can be done by introducing an IR scale $\kIR$ such that
$q \leq \kIR$ and $k\gg \kIR$.
These are called
\textbf{IR limits:}
\begin{itemize}[nosep]
  \item $P_{13}$: $k \gg q$, %\; $q \leq \kIR$.
  \item $P_{22}$: $k \gg q$ \emph{and} $k \gg |\k-\qv|$ 
\end{itemize}
In the $P_{22}$ case the two IR domains are related by symmetry $\qv\leftrightarrow \k-\qv$, so it sufficient to just consider one and multiply the answer by 2.
The
\textbf{leading-order IR limits} read:
\begin{align}
  P_{22}^{\mathrm{IR,LO}} &\approx \frac{k^2}{3}\,P_{11}(k)\int_0^{\kIR}
  d^3q\;\frac{P_{11}(q)}{q^2}
  = k^2\,\sigma_{d,\mathrm{IR}}^2\;P_{11}(k)\,,\\[4pt]
 2 P_{13}^{\mathrm{IR,LO}} &= -\frac{k^2}{3}\,P_{11}(k)\int_0^{\kIR}
  d^3q\;\frac{P_{11}(q)}{q^2}
  = -k^2\,\sigma_{d,\mathrm{IR}}^2\;P_{11}(k)\,.
\end{align}
When summed together, the two leading IR corrections cancel:
\begin{equation}\label{eq:IR_cancel}
  \boxed{2P_{13}^{\mathrm{IR,LO}} + P_{22}^{\mathrm{IR,LO}} = 0\,.}
  \qquad\text{(Leading-order IR cancellation)}
\end{equation}
This is illustrated in Fig.~\ref{fig:p22-p13}, where one can see 
that the IR corrections actually
dominate the individual $P_{13}$ and $P_{22}$ diagrams. To produce figure Fig.~\ref{fig:p22-p13} we had to take the formal limit $\kIR \to \infty$ yielding $\sigma_{d,\mathrm{IR}}^2\to \sigma_{d}^2$. This might seem to be at odds with the IR limit $q/k\to 0$.
However, the displacement variance integrals actually saturate at $q\sim k_{\rm eq}\simeq 0.02~\hMpc$, so extrapolation 
of $\kIR$ to small scales produces only a tiny error.

The first important consequence of the above cancellation 
result is that
the physical non-linear coupling of 
a mode $k$ to large scales \textbf{is not}
controlled by the large-scale displacement field, as might be naively
inferred from the SPT expansion. 
We will see momentarily that such coupling, in fact, is forbidden by the equivalence principle.
The second important consequence of this cancellation is 
that in practice we need to keep high precision 
of our calculation of individual SPT integrals 
if we want to compute the total $\Delta P_{\rm 1-loop}$ accurately.
Indeed, if the target 
precision on $\Delta P_{\rm 1-loop}$ is $O(1\%)$, the
error on our individual $P_{22},P_{13}$ computations 
must be $O(0.1\%)$. While this is not a big problem at the one-loop
order, the situation becomes more difficult at higher orders where
the infrared domains become more complicated and the computations
become entangled by sub-divergences. In this case the development
of infrared-safe integrands (i.e. integrands with offending divergences subtracted) becomes vital~\cite{Blas:2013bpa,Blas:2013aba,Carrasco:2013sva,Carrasco:2013mua}.

\textbf{The origin of IR cancellation: a toy model.}
To dive deeper into the nature of this cancellation, let us consider a toy model of a shift 
of a density field $\delta_1(\kv)$ from its initial 
position $\x$ by a homogeneous displacement
$\Delta x^i_L$ 
% ($\x\equiv \x_0+\Delta \x_L$), 
sourced by the large-scale (long-wavelength) modes. In Fourier space the shifted field can be written as
\be
\tilde\delta_1(\kv)=\int d^3x~ e^{-i\kv (\xv+\Delta \xv_L)} \delta_1(\xv)=e^{-i\kv \Delta \xv_L} \delta_1(\kv)~\,.
\ee
Let us now Taylor expand this in $\Delta \xv_L$, as one does in
SPT:
\be
\begin{split}
& \tilde\delta_1(\kv)=
\delta_1(\kv)\left(1-i\kv\Delta \xv_L-\frac{1}{2}k_ik_j
\Delta x^i_L\Delta x^j_L+...\right)\equiv \delta_1(\kv)+\delta_2(\kv)+\delta_3(\kv)+...\,,\\
& \delta_2(\kv)\equiv -i
\delta_1(\kv) k_i \Delta x^i_L\,,\quad
\delta_3(\kv)\equiv -\frac{1}{2}
\delta_1(\kv)k_ik_j
\Delta x^i_L\Delta x^j_L~\,.
\end{split}
\ee
Let's compute now the power spectrum of the shifted field 
$\tilde \delta_1$ to $O(\Delta x_L^2k^2)$, equivalent to 
a one-loop order. We have:
\be 
\langle \tilde \delta_1(\k)\tilde \delta_1(-\k)\rangle'=
\langle \delta_1(\k)  \delta_1(-\k)\rangle'+
\langle \delta_2(\k)  \delta_2(-\k)\rangle'
+2\langle \delta_1(\k)  \delta_3(-\k)\rangle'\,,
\ee 
where one can see the emergence of the baby versions of the $P_{22}$
and $P_{13}$ integrals. Specifically, we have:
\be 
\begin{split}
& \tilde{P}_{22}(k)\equiv \langle \delta_2(\k)  \delta_2(-\k)\rangle'=
\Delta x^i_L \Delta x^j_L k^ik^jP_{11}(k)\,, \\
& 
\tilde{P}_{13}(k)\equiv \langle \delta_1(\k)  \delta_3(-\k)\rangle'=-\frac{1}{2}
\Delta x^i_L \Delta x^j_L k^ik^jP_{11}(k)\,.
\end{split}
\ee 
$\tilde{P}_{22}(k)$ is a positive contribution
that describes the increase of power due to 
a coherent shift of both $\delta_1$'s. 
The displacement of each $\delta_1$ at this order is reminiscent 
of a gravitational pull, which increases clustering.
In contrast, 
$\tilde{P}_{13}(k)$ describes the apparent reduction of power
when one shifts one $\delta_1$ at a time. At this order in
perturbation it looks as if the matter 
particles were moved away from
each other, hence the suppression of clustering.
The two effects cancel each other exactly, $\tilde{P}_{22}+2\tilde{P}_{13}=0$. This happens for a very good reason: the position
space correlation function depends only on a relative distance between the two particles:
\begin{equation}
  \avg{\delta(\xv+\Delta\xv_L)\,\delta(\xv+\Delta\xv_L+\rv)} = \xi(|\rv|)\,,
\end{equation}
hence a constant displacement can't change the power spectrum.
In our toy model the cancellation of the displacement contributions
happened, in effect, thanks to the homogeneity of the Universe,
related to momentum conservation. In the actual problem
$\Delta x^j$ itself is sourced by the underlying density field
$\delta_1$, and the cancellation happens due to the
equivalence principle. While our toy model can't show this,
it exhibits the key mechanism for this cancellation: the
immunity of the power spectrum to homogeneous shifts of matter by virtue of underlying symmetries.
We now sharpen this statement and show that the cancellation
follows from an exact symmetry of the equations of motion.

\begin{figure}
    \centering
    \includegraphics[width=0.99\linewidth]{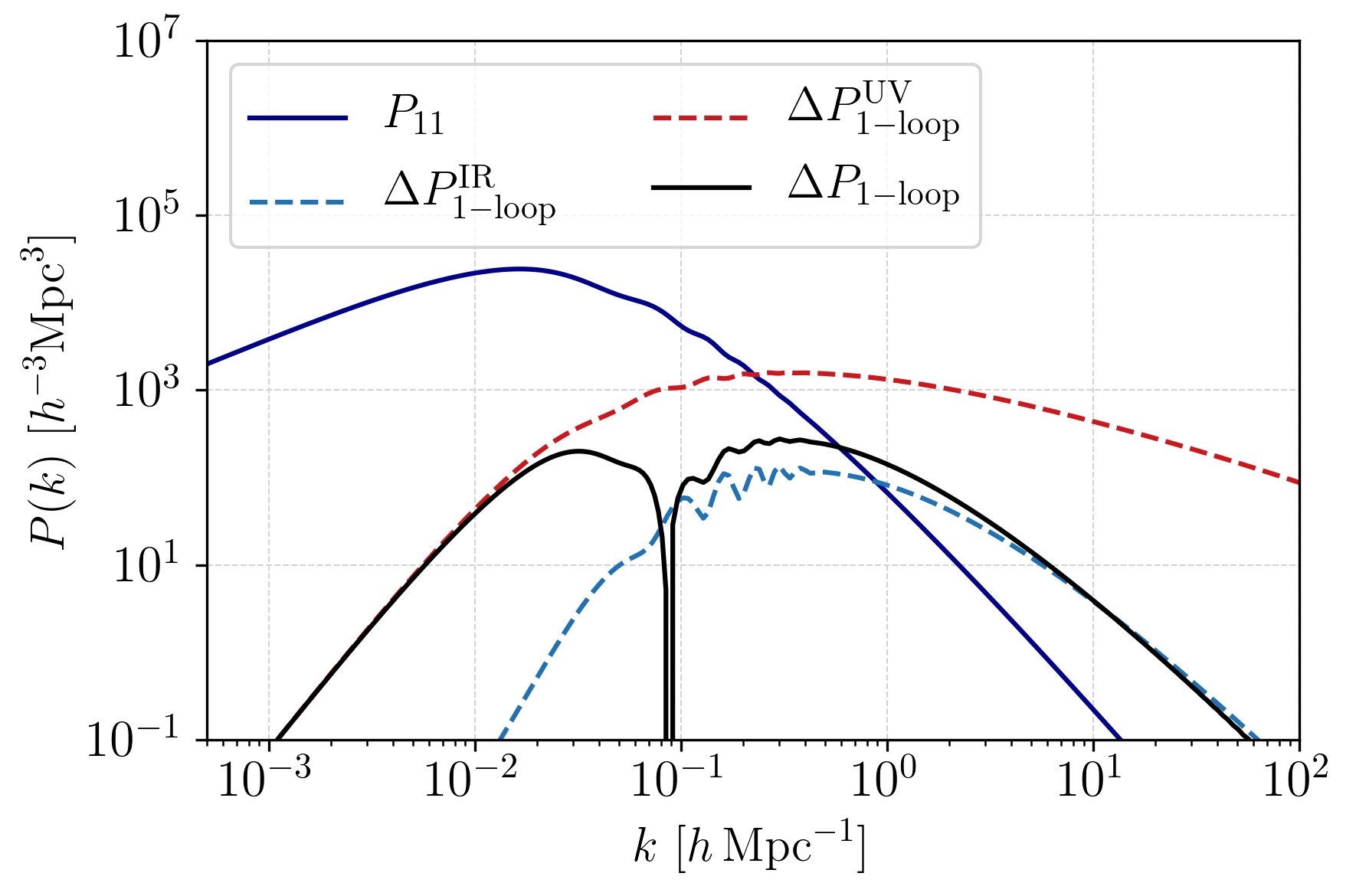}
    \caption{One-loop power spectrum and its asymptotic IR and UV limits at $z=0$.}
    \label{fig:1-loop_asymp}
\end{figure}

\bigskip\noindent
\textbf{The boost symmetry and the equivalence principle.}
The toy model relied on a \emph{constant} shift, which is removed by
ordinary translation invariance. 
The displacement generated by a
physical long mode grows in time, so we need to use a 
more physical expression
for $\Delta \x_L$. 
In addition, we need to prove that the displaced
density is actually a solution of the fluid equations.
The crucial
observation is that the full non-linear system
(eqs.~\ref{eq:cont},~\ref{eq:E-euler-conf},~\ref{eq:poisson}) is
\emph{exactly} invariant under a time-dependent uniform boost~\cite{Scoccimarro:1996jy,Kehagias:2013yd,Creminelli:2013mca,Creminelli:2013poa,Creminelli:2013nua,Baldauf:2015xfa},
\be\label{eq:boost}
  \tilde\xv = \xv + \vv{n}(\tau)\,,\qquad
  \tilde\vvv(\tilde\xv,\tau) = \vvv(\xv,\tau) + \vv{n}'(\tau)\,,\qquad
  \tilde\Phi(\tilde\xv,\tau) = \Phi(\xv,\tau)
  - \left(\vv{n}'' + \Hubble\,\vv{n}'\right)\cdot\tilde\xv\,,
\ee
with $\delta$ and $\tau_{ij}$ transforming without extra terms, and
$\vv{n}(\tau)$ an \emph{arbitrary} function of time. The check takes
two lines: in the Euler equation the frame acceleration
$\vv{n}''+\Hubble\vv{n}'$ generated on the left-hand side is exactly
compensated by the uniform force from the shifted potential on the
right; the continuity equation involves only quantities that
transform as scalars; and the Poisson equation is untouched because
the extra piece of $\tilde\Phi$ is linear in
$\tilde\xv$.\footnote{Strictly speaking, a potential growing linearly
with $\xv$ is not a standard perturbation. The transformation should
be understood as the $q\to 0$ limit of a physical long-wavelength
mode --- an ``adiabatic mode'' in the terminology of Weinberg~\cite{Weinberg:2003sw}.}

Let us now use this symmetry to construct the solution in the
presence of a long mode. Take a solution $\delta_S$, $\vvv_S$ of the
non-linear equations without the long mode, and apply the
boost~\eqref{eq:boost} with $\vv{n}(\tau) = \Delta\xv_L(\tau)$, the
displacement of a long-wavelength linear growing mode
$\delta_L(\qv)$,
\be
  \Delta\xv_L(\qv,\tau) = \frac{i\qv}{q^2}\,\delta_L(\qv,\tau)\,,
\ee
cf.~Eq.~\eqref{eq:displacement}. Since this mode is long-wavelength, it is appropriate to use a linear solution for it. The transformed configuration then
carries the extra uniform velocity $\vvv_L = \Delta\xv_L'$.
Crucially, since the long growing mode obeys its own linearized
Euler equation,
\be
  \Delta\xv_L'' + \Hubble\,\Delta\xv_L' = -\grad\Phi_L\,,
\ee
the shift of the potential in Eq.~\eqref{eq:boost} is precisely the
long-mode potential in the $q\to 0$ limit: the transformation has
generated \emph{both} the velocity and the gravitational force of
the physical long mode. This is where the equivalence principle does
real work --- a single change of frame compensates the
long-wavelength force for every fluid element at once only because
they all fall identically. We conclude that, to leading order in
$q$, the solution in the presence of the long mode is the translated
short-scale solution:
\be\label{eq:soft_translation}
  \delta_S(\kv,\tau)\Big|_{L}
  = e^{-i\kv\cdot\Delta\xv_L(\tau)}\;\delta_S(\kv,\tau)
  + O\!\left(\partial_i\partial_j\Phi_L\right)\,,
\ee
where the corrections --- the density and tidal couplings
$\partial_i\partial_j\Phi_L\sim\delta_L$ --- are \emph{not} enhanced
by $1/q$. Note that Eq.~\eqref{eq:soft_translation} is
non-perturbative in $\Delta\xv_L$; the
only expansion we've been using is $q\to 0$.
The cancellation is now manifest. At equal times, and for
\emph{each realization} of the long mode,
\be
  P(k,\tau)\Big|_L
  = \left|e^{-i\kv\cdot\Delta\xv_L(\tau)}\right|^2 P_S(k,\tau)
  = P_S(k,\tau)\,:
\ee
the phases cancel between $\delta$ and $\delta^*$ before any
averaging over $\delta_L$ is performed, so no IR-enhanced
contribution can survive --- at any order in perturbation theory. To
make contact with the diagrams, expand
Eq.~\eqref{eq:soft_translation} in $\Delta\xv_L$, treat the latter
as a Gaussian random field with variance
$\avg{\Delta x_L^i\,\Delta x_L^j} = \delta^{ij}\sigma^2_{d,\mathrm{IR}}$
(built from the modes $q\leq\kIR$), and correlate. The term with one
power of $\Delta\xv_L$ on each side is the IR limit of $P_{22}$,
while the term with $(\Delta\xv_L)^2$ acting on one side is the IR
limit of $2P_{13}$:
\be
\begin{split}
  P_{22}^{\mathrm{IR,LO}}
  &= \avg{(\kv\cdot\Delta\xv_L)^2}\,P_{11}(k)
  = +\,k^2\,\sigma^2_{d,\mathrm{IR}}\,P_{11}(k)\,,\\
  2P_{13}^{\mathrm{IR,LO}}
  &= 2\cdot\left(-\frac{1}{2}\right)\avg{(\kv\cdot\Delta\xv_L)^2}\,P_{11}(k)
  = -\,k^2\,\sigma^2_{d,\mathrm{IR}}\,P_{11}(k)\,,
\end{split}
\ee
reproducing the direct diagrammatic computation above and summing to
zero, Eq.~\eqref{eq:IR_cancel}.
Likewise, matching
the same expansion stemming from Eq.~\eqref{eq:soft_translation} to the
SPT solution fixes the soft limits of the kernels --- a non-trivial
prediction of the symmetry:
\be\label{eq:soft_theorems}
  \lim_{q\to 0}\; 2\,F_2(\qv,\kv-\qv) = \frac{\kv\cdot\qv}{q^2}
  + O(q^0)\,,\qquad
  \lim_{q\to 0}\; 3\,F_3(\kv,\qv,-\qv) = -\frac{(\kv\cdot\qv)^2}{2\,q^4}
  + \text{less singular}\,,
\ee
where the factor of $2$ counts the two orderings of the arguments in
$\delta^{(2)}$, and the factor of $3$ the Wick contractions in
$\avg{\delta_1\,\delta^{(3)}}$. 

\textbf{Homework:} verify from the
explicit kernels of \S3.2 that
$F_2(\qv,\kv-\qv)\to (\kv\cdot\qv)/(2q^2)$ and
$F_3(\kv,\qv,-\qv)\to -(\kv\cdot\qv)^2/(6q^4)$, in agreement with
Eq.~\eqref{eq:soft_theorems}.

The argument comes with a set of corollaries:
\begin{enumerate}
  \item \textbf{Equal times only.} At unequal times the phases
  combine into
  \mbox{$e^{-i\kv\cdot[\Delta\xv_L(\tau_1)-\Delta\xv_L(\tau_2)]}\neq 1$}:
  the \emph{relative} displacement of the two time slices is
  frame-independent and physical. Accordingly, unequal-time
  correlators (propagators, response functions) are not protected by
  the symmetry --- they are exponentially damped by the large-scale
  displacements.
  \item \textbf{Adiabaticity and the equivalence principle.} The
  construction requires the long mode to be the adiabatic growing
  mode (velocity locked to the gravitational force) and all species
  to fall identically. A long-range fifth force with
  species-dependent couplings, isocurvature velocity perturbations,
  or the relative baryon--dark-matter velocity cannot be removed by
  a single change of frame and leave residual IR effects in the
  (cross-)correlations. Searching for the corresponding violations
  of the IR structure (``consistency relations'') is a test of the
  equivalence principle on cosmological scales.
  \item \textbf{Only the displacement-enhanced terms are
  protected.} The symmetry removes the terms enhanced by
  $\sigma^2_{d,\mathrm{IR}}$ and nothing else: the $O(q^0)$
  couplings --- the separate-universe response to the long-mode
  density and tides --- are physical. These are precisely the
  subleading $\sigma_\ell^2$ effects that we compute next.
  \item \textbf{Coherence.} The symmetry removes only displacements
  that are coherent across the scales probed by the observable. For
  an observable that compares points separated by
  $\ell_\mathrm{BAO}$ --- the BAO wiggles --- modes with
  $q\gtrsim\kosc$ produce \emph{relative} displacements of the two
  points, which are physical and, as we will see, non-perturbative.
  This is the origin of the IR resummation discussed below; the
  equivalence principle still leaves its imprint there, forcing the
  contribution of modes $q\ll\kosc$ to the BAO damping to be
  suppressed as $(q/\kosc)^2$.
\end{enumerate}

%------------------------------------------------------------------
\subsection{IR resummation of the BAO}
%------------------------------------------------------------------

\textbf{Physical IR effects.}
Let us now compute the sub-leading IR limit that one gets after the 
leading order cancellations. To that end one needs to Taylor expand
$P(|\k-\qv|)$ in $P_{22}$, so we assume at this point that this expansion
is convergent.
We find the following
\textbf{subleading contributions:}
\begin{equation}\label{eq:IR_sub}
 \Delta P^{\rm IR}_{1\text{-loop}}(k)
= \left(\frac{569}{735}\,P_{11}(k) - \frac{47}{105}\,k\,P'_{11}(k) + \frac{1}{10}\,k^2\,P''_{11}(k)\right)\cdot\sigma_\ell^2\,,
\end{equation}
where we observe the appearance of the large-scale mass variance:
\be \label{eq:sig2ell_def}
\sigma_\ell^2 = \int_0^{\kIR} d^3q\;P_{11}(q)~\,,
\ee 
which is essentially the same object as
a smoothed mass variance introduced in Eq.~\eqref{eq:sig2_def}.
The only formal difference is that the smoothing in~\eqref{eq:sig2_def}
was implemented by a position-space top hat filter, while 
Eq.~\eqref{eq:sig2ell_def} uses a sharp cutoff in momentum space. 
We see now that a mode $k$ couples to IR modes $q\ll k$
via the mass variance. Physically, a large-scale over-density
mode $q$
behaves like a separate Universe which boosts the growth of 
smaller overdensities of wavenumber $k$ in its background. 
Likewise, a large-scale under-density will locally slow 
down the growth of structure. 
The large-scale coupling should then be proportional 
to the average mass variance, i.e. $\sigma_\ell^2$.
For reasonable values $\kIR\sim 0.02~\hMpc$, the mass fluctuation
is very small, $\sigma_\ell^2\sim 10^{-2}$. However, 
as $k$ increases, more and more modes fall into the 
IR domain, so that $\kIR$ should increase as well.
In this case
$\sigma_\ell^2$ grows rapidly and the IR limit~\eqref{eq:IR_sub} starts to dominate the entire
one-loop integral. For instance, using $\kIR=k/10$ we
see that the IR modes account to
the bulk of $\Delta P_{\rm 1-loop}$ for $k\gtrsim 0.1~\hMpc$,
see Fig.~\ref{fig:1-loop_asymp}.

\bigskip\noindent
\textbf{BAO and IR resummation.}
The subleading IR result~\eqref{eq:IR_sub} contains derivatives
of the linear power spectrum, so the actual size of the IR
corrections depends on how featureful $P_{11}$ is.
To quantify this, let us decompose the linear power spectrum
into a smooth (broadband) part and a wiggly part that contains
the baryon acoustic oscillations:
\be\label{eq:Pw_decomp}
P_{11}(k)= P_s(k)+P_w(k)\,,\qquad
P_w(k)= f_s(k)\,\sin\left(\frac{k}{\kosc}\right)\,,\qquad
\kosc=(100\Mpch)^{-1}\simeq 10^{-2}\hMpc\,,
\ee
where $f_s(k)$ is a smooth envelope function.
An important remark is in order. The derivation of
Eq.~\eqref{eq:IR_sub} relied on the Taylor expansion of
$P_{11}(|\k-\qv|)$ in $q/k$. This expansion
has a different convergence rate for the smooth and wiggly parts.

For the smooth part the logarithmic derivatives are of the order
of the spectral tilt,
\be
  \frac{d\ln P_s}{d\ln k} \sim n = O(1)\,,
\ee
so all three terms in Eq.~\eqref{eq:IR_sub} are comparable and
the IR correction is safely small,
\be
\Delta P^{s}_{1\text{-loop}}(k)\Big|_{\rm IR}\sim P_s(k)\,\sigma_\ell^2\ll P_s(k)\,.
\ee
Hence, for the smooth part of the power spectrum the IR
corrections are perturbative.

The situation is very different for the wiggly part: every
derivative acting on the rapidly oscillating factor brings
down an enhancement $k/\kosc\gg 1$,
\be
  k\,P_w' \approx f_s\,\frac{k}{\kosc}\,\cos\left(\frac{k}{\kosc}\right)\,,\qquad
  k^2 P_w'' \approx -f_s\,\frac{k^2}{\kosc^2}\,\sin\left(\frac{k}{\kosc}\right)\,,
\ee
where we differentiate only the sine, as the envelope $f_s$ is smooth.
Plugging these expressions into Eq.~\eqref{eq:IR_sub}, we obtain
\be\label{eq:Pw_1loop}
\Delta P^{w}_{1\text{-loop}}(k)\Big|_{\rm IR} = \sigma_\ell^2\,f_s\left[\frac{569}{735}\,\sin\!\left(\frac{k}{\kosc}\right)
  - \frac{47}{105}\,\frac{k}{\kosc}\,\cos\!\left(\frac{k}{\kosc}\right)
  - \frac{1}{10}\,\frac{k^2}{\kosc^2}\,\sin\!\left(\frac{k}{\kosc}\right)\right]\,.
\ee
To interpret this result it is convenient to combine the tree-level
and one-loop wiggly contributions using
$\sin X-\epsilon \cos X=\sin\big(X(1-\epsilon)\big)+O(\epsilon^2)$,
valid for $\epsilon\ll 1$. This gives
\be\label{eq:Pw_combined}
P_w(k) + \Delta P^w_{1\text{-loop}}(k)\Big|_{\rm IR} =
f_s\,\sin\!\left(\frac{k}{\kosc}\Big(1-\frac{47}{105}\,\sigma_\ell^2\Big)\right)
  \left(1 + \frac{569}{735}\,\sigma_\ell^2 - \frac{1}{10}\,\frac{k^2}{\kosc^2}\,\sigma_\ell^2\right)\,.
\ee
We observe two distinct physical effects. First, the long-wavelength
modes shift the BAO phase by $\Delta=\frac{47}{105}\sigma_\ell^2\sim 10^{-2}$.
This effect is small and fully perturbative.\footnote{The smallness of this effect is foundation of the BAO method of distance measurements  
in cosmology. $\sigma_\ell^2\ll 1$ is, however, somewhat of a coincidence
of our $\Lambda$CDM-like universe. If the matter power spectrum had been enhanced on large scales, $\sigma_\ell^2$ could have been much larger, 
and the BAO scale would have been strongly shifted by the non-linear evolution.}
Second, the oscillations get suppressed
by the last term in the brackets, and this term is dangerous.
Indeed, already at $k=0.1~\hMpc$ we have
$k/\kosc \simeq 0.1\times 100=10$, so that
\be
  \frac{k^2}{\kosc^2}\,\sigma_\ell^2 \sim 100\times 10^{-2} \sim 1\,,
\ee
i.e. the ``correction'' to the wiggly part is of order one.
Perturbation theory breaks down for the BAO part of the power
spectrum even at wavenumbers where the broadband part is
perfectly under control. 

Two comments are in order here.
The result for the phase shift above is exact.
The result for the suppression, however, is not: in deriving
Eq.~\eqref{eq:Pw_1loop} we have Taylor-expanded
$P_{11}(|\k-\qv|)$ and hence
retained only the modes with $q\leq \kosc$, while
the modes $\kosc\lesssim q\ll k$ also contribute at $O(1)$.
To capture them one has to replace the Taylor expansion
of the wiggly component with a finite difference, which yields the full IR limit
% NB: check the overall normalization/factor of 2 against your conventions
\be\label{eq:Pw_full}
\Delta P^{w}_{1\text{-loop}}(k)\Big|_{\rm IR,~full}=2\int_{q\leq \kIR} d^3q\;
\frac{(\k\cdot\qv)^2}{q^4}\,P_{11}(q)
\left[P_w(|\k-\qv|)-P_w(k)\right]\,,
\ee
whose structure makes two points manifest: (a) the non-linear 
coupling of an oscillating mode $\k$ to mode $q\gtrsim \kosc$ is actually 
$k^2\sigma_d^2$ and (b) a strictly homogeneous
displacement ($P_w(|\k-\qv|)\to P_w(k)$ as $q\to 0$) drops out,
in agreement with the equivalence principle. 

Since the offending coupling is $O(1)$, the IR-enhanced
contributions must be resummed to all orders in perturbation
theory. This is called \textbf{IR resummation}.
There are two ways to carry it out:
\begin{enumerate}[nosep]
  \item Resum the IR-enhanced parts of loop diagrams
  diagram by diagram directly in Eulerian PT
  (systematically implemented in time-sliced perturbation theory, TSPT~\cite{Blas:2015qsi,Blas:2016sfa,Ivanov:2018gjr,Vasudevan:2019ewf});
  \item Use Lagrangian PT~\cite{Matsubara:2007wj,Vlah:2015sea,Vlah:2016bcl,Senatore:2014via,Baldauf:2015xfa,Chen:2020zjt}, based on
  $\xv = \xv_\mathrm{ini} + \Psiv(\tau,\xv_\mathrm{ini})$,
  where $\xv_\mathrm{ini}$ is an initial position and $\Psiv$ is the dynamical
  displacement field. There the displacement is kept non-perturbatively from the start and never Taylor-expanded.\footnote{A nice feature of LPT is that it does not require a ``wiggly-smooth'' split of the matter power spectrum, thereby removing any potential uncertainty associated with it.}
\end{enumerate}
Either way, the net result is the exponentiation of the
damping factor: at leading order the IR-resummed wiggly
spectrum reads
\be\label{eq:Pw_IRres_LO}
  P^w_{\mathrm{IR\text{-}res,\,LO}}(k) = e^{-\Sigma^2 k^2}\;P_w(k)\,,
\ee
with the BAO damping factor
\be\label{eq:Sigma2_def}
  \Sigma^2 = \frac{1}{6\pi^2}\int_0^{\kIR} dq\,P_{11}(q)\,
  \left[1 - j_0\!\left(\frac{q}{\kosc}\right) + 2\,j_2\!\left(\frac{q}{\kosc}\right)\right]\,,
\ee
where $j_\ell$ are spherical Bessel functions.
For $q\gg \kosc$ the expression in the square brackets
approaches unity and we recognize in $\Sigma^2$ the dispersion
of the relative displacements: only the motions that do not
reduce to a homogeneous shift across one BAO wavelength
damp the wiggles. In the opposite limit
$q \ll \kosc$ the brackets reduce to $\frac{3}{10}(q/\kosc)^2$,
so that this part of the integral gives
\be
\Sigma^2 \approx \frac{1}{10}\,\frac{\sigma_\ell^2}{\kosc^2}\,,
\ee
reproducing the suppression term in Eq.~\eqref{eq:Pw_combined}:
the very long modes couple to the BAO through the mass
variance $\sigma_\ell^2$, and not through their displacements ---
the equivalence-principle protection (corollary~4 of the boost-symmetry
argument above) at work inside the damping integral.

Finally, the one-loop power spectrum with IR resummation is
assembled as follows: the tree-level wiggly part is damped
(and corrected for the damping already generated at one loop),
while the loop integrals are evaluated on the leading-order
resummed input spectrum:
\begin{equation}\label{eq:IR_resum}
  \boxed{\;P_{\rm 1-loop}^{\rm IR-res}(k)
  = P_s + P_w\,e^{-\Sigma^2 k^2}\!\left(1 + \Sigma^2 k^2\right)
  + P_{1\text{-loop}}\!\left[P_s + P_w\,e^{-\Sigma^2 k^2}\right]\,,\;}
\end{equation}
where $P_{1\text{-loop}}[P]$ denotes the one-loop integrals
computed with the input spectrum $P$. The factor
$(1+\Sigma^2k^2)$ removes the double counting of the damping:
expanding Eq.~\eqref{eq:IR_resum} to first order in $\Sigma^2$
reproduces the fixed-order one-loop result, so the difference
is a two-loop effect.

\begin{figure}
    \centering
        \includegraphics[width=0.49\linewidth]{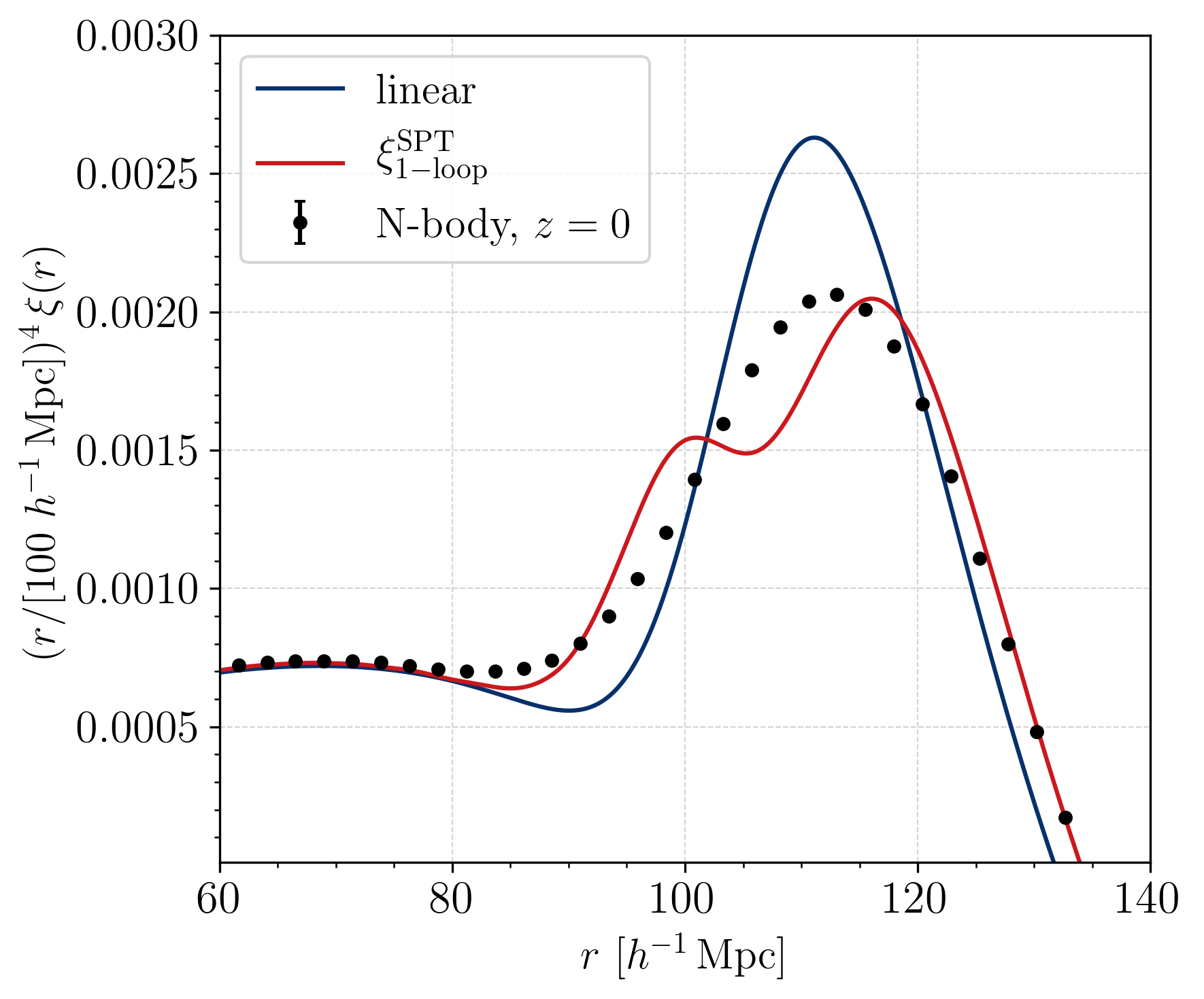}
    \includegraphics[width=0.49\linewidth]{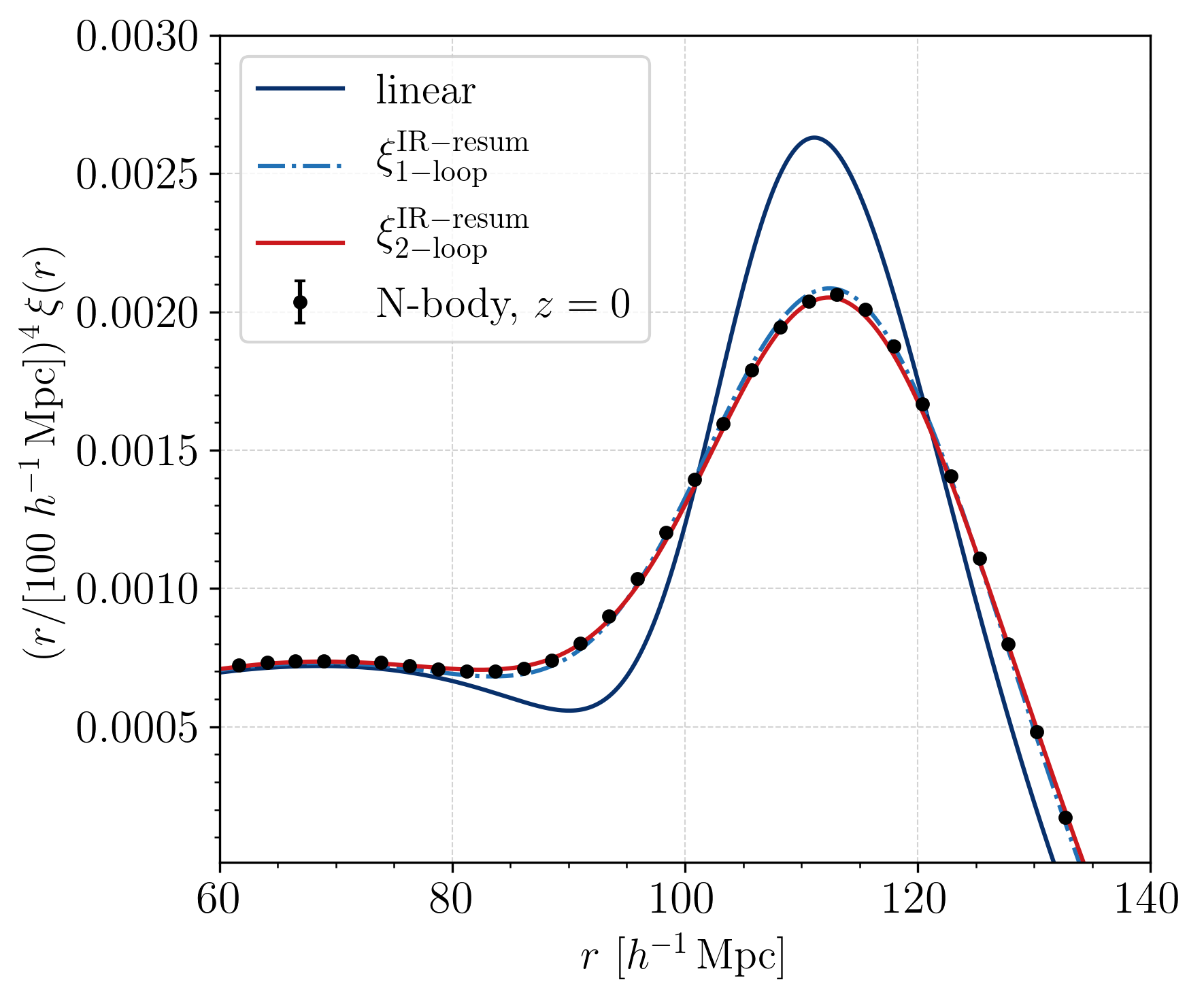}
    \caption{BAO in the position space correlation function at $z=0$ multiplied by $r^4$ to enhance the visibility of the BAO peak.
    The data is from the Horizon Run N-body simulation~\cite{Kim:2011ab,Blas:2016sfa}.
    \textbf{Left:} failure of 1-loop SPT to capture the shape of the BAO peak. \textbf{Right:} IR-resummed EFT captures the non-linear distortion of the peak. Note that the EFT predictions also include the UV counterterms (discussed below). }
    \label{fig:fit_xi_nbody}
\end{figure}

Figure~\ref{fig:fit_xi_nbody} illustrates the results
of this section at the level of the position-space two-point 
correlation function $\xi(r)$. The left panel displays the Horizon Run 
N-body data
at $z=0$~\cite{Kim:2011ab}. We see that the linear theory significantly over-predicts 
the amplitude of the BAO peak in the correlation function. 
The SPT one-loop correction, however, predicts a totally 
incorrect shape and spurious oscillations. 
These are produced by $O(1)$ SPT 1-loop corrections 
to $P_w$, proportional to its logarithmic derivatives. 
The IR-resummation formula~\eqref{eq:IR_resum} kills the spurious distortions, and provides 
an accurate prediction for the shape of the BAO peak
at one-loop order seen in the right panel of Figure~\ref{fig:fit_xi_nbody}. This prediction
slightly improves upon including the two-loop corrections.

%------------------------------------------------------------------
\subsection{UV Limit of Loop Corrections}
%------------------------------------------------------------------

Let us now consider the opposite regime, in which the loop momentum
is parametrically larger than the external one, $q\gg k$. In analogy
with the IR case, we introduce a scale $\kUV$ such that the loop
momenta satisfy $q\geq \kUV\gg k$, and expand the loop integrands
accordingly. These are the \textbf{UV limits} of the loop
corrections.
For the $P_{22}$ diagram the expansion of the $F_2$ kernels yields
\begin{equation}\label{eq:P22_UV}
  P_{22}(k)\Big|_{q \gg k}
  = \frac{9}{196\pi^2}\;k^4 \int_{\kUV}^\infty \frac{dq}{q^2}\;
  [P_{11}(q)]^2\,.
\end{equation}
This result has two remarkable features. First, it scales as $k^4$
and carries no dependence on $P_{11}(k)$ whatsoever: the
\emph{small-scale modes decouple} from the long-wavelength mode.
Whatever the short modes do, their imprint on the large-scale power
spectrum is a smooth, noise-like contribution $\propto k^4$. We will
see later (via Peebles' momentum-conservation argument in the chapter on
stochastic effects) that this scaling is not an accident of SPT, but
is dictated by mass and momentum conservation. Second, we can ask
when this integral actually converges. For a power-law universe,
$P_{11}(q)\propto q^\nu$, we get
$P_{22}^{\mathrm{UV}} \propto k^4\int dq\; q^{2\nu-2}$, which
diverges for $\nu \geq 1/2$. This, however, does not happen 
in our Universe, where the above integral converges quite quickly.

For the $P_{13}$ diagram one finds in the same limit
\begin{equation}\label{eq:P13_UV}
  2P_{13}(k)\Big|_{q\gg k}
  = -\frac{61}{630\pi^2}\;k^2\,P_{11}(k)\int_{\kUV}^\infty dq\;P_{11}(q)\,.
\end{equation}
In contrast to $P_{22}$, this contribution is proportional to
$P_{11}(k)$, i.e.\ it modulates the linear power spectrum. Recalling
the displacement dispersion $\sigma_d^2$ introduced in Chapter~2, we
can rewrite it as
\begin{equation}
  2P_{13}(k)\Big|_{q\gg k}
  = -\frac{61}{105}\,k^2\,\sigma_{d,\mathrm{UV}}^2\,P_{11}(k)\,,\qquad
  \sigma_{d,\mathrm{UV}}^2 \equiv \frac{1}{6\pi^2}\int_{\kUV}^\infty dq\;P_{11}(q)\,,
\end{equation}
i.e.\ $2P_{13}^{\rm UV}\sim k^2\,P_{11}(k)\,\avg{|\Psiv|^2}_\mathrm{UV}$:
the linear spectrum is modulated by the displacements generated by
the small-scale modes. This is the same physics we encountered in the
IR discussion --- random displacements smear the correlations of 
mode $k$ --- except that now
the displacements are sourced by the short modes, so there is no
symmetry that forces this effect to cancel. 
The upshot is that the leading coupling to small scale modes  
is via the displacement variance. This leading asymptotic
displayed in Fig.~\ref{fig:1-loop_asymp}, 
in which we set $\kUV=0$ appropriate in the $k\to 0$ limit. 
We see
that this UV asymptotic dominates $\Delta P_{\rm 1-loop}$
for $k\lesssim 10^{-2}~\hMpc$. 

As far as convergence is concerned, 
for a power-law universe
the above integral diverges for $\nu \geq -1$. 
In $\Lambda$CDM, where $P_{11}(q)$ decays as $q^{-3}\ln^2 q$, 
this UV
integral happened to converge, just like $P_{22}$. 
Historically, the cosmological community at this point 
made the same mistake as particle physicists some fifty years
earlier.
Whether the loop integrals converge or diverge 
does not matter -- their small scale part is not reliably 
described by the underlying theory anyway. Therefore, one needs to
renormalize these integrals in any case. 
% Even though the l these integrals converge to a wrong answer.
Indeed, 
the region $q\gtrsim \knl$ contains $O(1)$ fluctuations, for which
neither the perturbative expansion nor the underlying pressureless
perfect fluid equations are valid.
Whatever SPT returns for this part of the integral is simply wrong.\footnote{There were signs though: it was known that
the SPT predictions blow up for a power-law power spectrum with $\nu \geq -1$,
while the N-body simulations produce a perfectly normal structure for these spectra, see e.g. Fig.~3 of~\cite{Peacock:2003hh} for $\nu=0$.} 

The crucial observation is that we do not need to know the details 
of the small-scale dynamics to understand its effects: as
Eqs.~\eqref{eq:P22_UV}--\eqref{eq:P13_UV} show, the small-scale
region contributes to the large-scale power spectrum only through
terms with an analytic, universal $k$-dependence, $k^2 P_{11}(k)$ and
$k^4$, multiplied by some unknown numbers that depend on the small scale power spectrum. 
We can therefore cut off the loop
integrals at some scale $\Lambda$ and replace the untrusted
contribution by a term of exactly this form with a free coefficient,
e.g.\ for $P_{13}$:
\begin{equation}\label{eq:UV_split}
  k^2 P_{11}(k)\!\int_{\kUV}^\infty dq\;P_{11}(q)
  \;\longrightarrow\;
  k^2 P_{11}(k)\!\int_{\kUV}^{\Lambda} dq\;P_{11}(q)
  \;+\; C_\Lambda\,k^2\,P_{11}(k)\,.
\end{equation}
The coefficient $C_\Lambda$ encodes the true (non-perturbative)
small-scale physics; it cannot be computed within the theory and
must depend on $\Lambda$ in such a way that the total result is
cutoff-independent. In other words:
\begin{center}
  \textbf{We need counterterms!}
\end{center}
Where do such terms come from in the EFT? From the effective stress
tensor $\tau_{ij}$ --- the subject of the next chapter.

%==========================================================================
\section{Chapter 4: Effective Stress Tensor for Dark Matter }
%==========================================================================

We concluded the previous chapter with the observation that the SPT
loop integrals receive contributions from UV modes $q\gtrsim\knl$,
for which perturbation theory is meaningless, and that this UV
sensitivity shows up as cutoff-dependent terms of the form
$C_\Lambda\,k^2\,P_{11}(k)$. In this chapter we show that the
effective stress tensor $\tau_{ij}$, which we introduced on symmetry
grounds in the equations of motion but have ignored so far, supplies
precisely the counterterms needed to cure this problem. Along the way
we will encounter an important structural novelty of the EFT of LSS:
the stress tensor is non-local in time.

%------------------------------------------------------------------
\subsection{The Effective Sound Speed}
%------------------------------------------------------------------

Our starting point is the Euler (Navier--Stokes) equation with the
effective stress tensor,
\begin{equation}
  \rho\frac{Dv^i}{D\tau} = -\rho\,\partial^i\Phi + \partial^j\tau_{ij}\,.
\end{equation}
What is $\tau_{ij}$? To build up some intuition, let us start
(naively) with the expression suggested by ordinary fluid dynamics,
\begin{equation}\label{eq:tau_fluid}
  \tau_{ij} = -p\,\delta_{ij}
  + \rho\,\zeta\,\delta_{ij}\,(\partial_k v^k)
  + \rho\,\mu\!\left(\partial_i v_j + \partial_j v_i - \tfrac{2}{3}\,\delta_{ij}\,\partial_k v^k\right)\,,
\end{equation}
i.e.\ pressure plus bulk ($\zeta$) and shear ($\mu$) viscosity terms.
For an adiabatic fluid $p=c_s^2\rho$. Let us now evaluate this
expression on the linear-theory growing mode, for which
\begin{equation}
  v^{(1)}_i = -f\Hubble\,\frac{\partial_i\delta_1}{\Delta}
  = -f\Hubble\,\partial_i\Phihat^{(1)}\,,\qquad
  \Delta\Phihat \equiv \delta\,,
\end{equation}
where we used the Poisson equation
$\Delta\Phi = \frac{3}{2}\Omega_m\Hubble^2\delta$ and introduced the
rescaled potential $\Phihat = 2\Phi/(3\Omega_m\Hubble^2)$. The
velocity gradients then reduce to second derivatives of the
potential, and Eq.~\eqref{eq:tau_fluid} takes the form
\begin{equation}\label{eq:tau_lin}
  \frac{\tau_{ij}}{\rhob}
  \simeq -c_s^2\,\delta_{ij}\,(1+\delta)
  + \tilde\zeta\,\delta_{ij}\,\delta
  + \tilde\mu\;\partial_i\partial_j\Phihat\,,
\end{equation}
where we absorbed the background factors ($f$, $\Hubble$, etc.) into
the redefined constants $\tilde\zeta$, $\tilde\mu$. This simple
exercise teaches us an important structural lesson: at leading order
in perturbations and gradients, the symmetries allow exactly two
operators, $\delta_{ij}\,\delta$ and the \textbf{tidal tensor}
$\partial_i\partial_j\Phihat$. 
At the same time $\delta\equiv \Delta \hat\Phi$,
so the right building block, i.e. the degree of freedom, 
is only $\partial_i\partial_j\Phihat$ (at the lowest order).
Taking the divergence and using the
Poisson equation once again, both collapse onto a single structure,
\begin{equation}\label{eq:cseff_def}
  \frac{\partial^j \tau_{ij}}{\rhob}
  = \left(-c_s^2 + \tilde\zeta + \tilde\mu\right)\partial_i\delta
  \equiv -c_{s,\mathrm{eff}}^2(\tau)\;\partial_i\delta\,.
\end{equation}
Only one combination of the fluid parameters is observable at this
order: the \textbf{effective sound speed} $c_{s,\mathrm{eff}}^2$. This
is the universality of the EFT at work --- whatever the microscopic
physics generating $\tau_{ij}$, its leading-order effect is captured
by a single number.

Inserting Eq.~\eqref{eq:cseff_def} into the Euler equation and
repeating the steps that led to the SPT system
\eqref{eq:SPT_cont}--\eqref{eq:SPT_euler} (Fourier transform, time
variable $\eta=\ln\Dp$, EdS approximation), we obtain
\begin{equation}\label{eq:euler_cs}
  \partial_\eta\Theta_{\kv} - \frac{3}{2}\,\delta_{\kv} + \frac{1}{2}\,\Theta_{\kv}
  = \int\beta\,\Theta_{\qv_1}\Theta_{\qv_2}
  - \frac{k^2\,c_{s,\mathrm{eff}}^2}{\Hubble^2}\,\delta_{\kv}\,,
\end{equation}
where the factors of $f$, $\Omega_m$ have been absorbed into
$c_{s,\mathrm{eff}}^2$; they will play no role in what follows.

The new term is a linear source, so its effect can be computed
exactly with the retarded Green's function of the linearized
fluid equations. Writing $\delta = \delta_\mathrm{SPT} + \Delta\delta^{(c_s^2)}$
and using the linear continuity equation
$\partial_\eta\delta_{\kv}=\Theta_{\kv}$ to eliminate $\Theta$,
Eq.~\eqref{eq:euler_cs} reduces, at leading order in perturbations,
to a second-order equation with \emph{constant} coefficients and a
source evaluated on the linear solution:
\begin{equation}\label{eq:growth_source}
  \left(\partial_\eta^2 + \frac{1}{2}\,\partial_\eta - \frac{3}{2}\right)
  \Delta\delta^{(c_s^2)}_{\kv}(\eta)
  = -\,k^2\,\frac{c_{s,\mathrm{eff}}^2(\eta)}{\Hubble^2(\eta)}\;
  \delta_1(\kv,\eta)\,.
\end{equation}
The retarded Green's function of this operator is derived in
Appendix~\ref{app:green-eta}: thanks to the constant coefficients in the EdS regime, it
depends only on $s=\eta-\eta'$,
\begin{equation}
  g_\eta(s) = \frac{2}{5}\,\Theta_H(s)\left[e^{s} - e^{-3s/2}\right]\,,
\end{equation}
where $\Theta_H$ is the Heaviside function. Hence, using
$\delta_1(\kv,\eta')=e^{\eta'}\delta_0(\kv)$,
\begin{equation}\label{eq:ddelta_cs_int}
  \Delta\delta^{(c_s^2)}_{\kv}(\eta)
  = -\,k^2\int_{-\infty}^{\eta} d\eta'\;g_\eta(\eta-\eta')\;
  \frac{c_{s,\mathrm{eff}}^2(\eta')}{\Hubble^2(\eta')}\;e^{\eta'}\,\delta_0(\kv)
  \equiv -k^2\gamma(\eta)\delta_1(\kv)\,,
\end{equation}
where we made the factorization of time and scale dependence manifest.
In principle, we do not need to compute the time-dependence 
of $\gamma(\eta)$, which is \textit{a priori} not calculable in EFT. 
We can simply treat it as a non-perturbative
input to be extracted from data. However, it is 
instructive to compute it explicitly in some simple cases.
% To carry out the time integral $\gamma(\eta)$ 
% we need the time dependence of the
% sound speed. 
Since $c_{s,\mathrm{eff}}^2$ is generated by the
small-scale fluctuations, whose amplitude grows with time, it is
natural to parameterize it as a power of the growth factor,
\begin{equation}\label{eq:cs_ansatz}
  \frac{c_{s,\mathrm{eff}}^2(\eta)}{\Hubble^2(\eta)}
  = \ell_0^2\,e^{m\eta} = \ell_0^2\,\Dp^m\,,\qquad m>0\,,
\end{equation}
where positivity of $m$ guarantees that the source switches off at
early times and the integral converges. The integral is then
elementary:
\begin{equation}
  \int_{-\infty}^{\eta} d\eta'\;\frac{2}{5}
  \left[e^{\eta-\eta'} - e^{-\frac{3}{2}(\eta-\eta')}\right]
  e^{(m+1)\eta'}
  = \frac{2}{5}\left[\frac{1}{m}-\frac{1}{m+5/2}\right]e^{(m+1)\eta}
  = \frac{e^{(m+1)\eta}}{m\left(m+\tfrac{5}{2}\right)}\,,
\end{equation}
so that the correction is proportional to the linear field at the
\emph{same} time,
\begin{equation}\label{eq:ddelta_cs}
  \Delta\delta^{(c_s^2)}_{\kv}(\eta) = -\,k^2\,\gamma(\eta)\;\delta_1(\kv,\eta)\,,
  \qquad
  \boxed{\;\gamma(\eta) = \frac{2}{m\,(2m+5)}\;
  \frac{c_{s,\mathrm{eff}}^2(\eta)}{\Hubble^2(\eta)}\;}
\end{equation}
Thus $\gamma$ is the instantaneous value of
$c_{s,\mathrm{eff}}^2/\Hubble^2$ (which has dimensions of length
squared), suppressed by an $O(1)$ number that encodes the memory of
the growth history.

Which value of $m$ should we use? Renormalization provides an
answer. As we will see below, $\gamma$ must cancel a
cutoff-dependent term proportional to
$\int_{\knl}^{\Lambda} dq\,P_{11}(q;\eta)\propto \Dp^2(\eta)$, so
at least part of the counterterm must carry $m=2$, for which
\begin{equation}
  \gamma(\eta) = \frac{1}{9}\,\frac{c_{s,\mathrm{eff}}^2(\eta)}{\Hubble^2(\eta)}\,.
\end{equation}
Other values of $m$ are relevant too, e.g. the finite part of the 
counterterm $\gamma_{\rm fin}$ (discussed shortly) capturing the physical 
small scale response has a different scaling.
As a byproduct, the matrix propagator of
Appendix~\ref{app:green-eta} also delivers the \emph{velocity}
counterterm generated by the same source, with
$\Delta\Theta^{(c_s^2)} = (m+1)\,\Delta\delta^{(c_s^2)}$: the ratio
$(m+1)\neq 1$ signals the decaying-mode admixture induced by the
stress tensor (a pure growing mode has $\Theta/\delta = 1$).

The correction to the power
spectrum is
\begin{equation}\label{eq:P_ctr}
  \avg{\left(\delta_\mathrm{SPT}+\Delta\delta^{(c_s^2)}\right)
  \left(\delta_\mathrm{SPT}+\Delta\delta^{(c_s^2)}\right)}'
  = P_{11} + P_{22} + 2P_{13}
  - 2k^2\gamma\,P_{11}(k) + k^4\gamma^2\,P_{11}(k)\,.
\end{equation}
The last term is of the size of a two-loop contribution and can be
dropped at the one-loop order we work. The crucial observation is that the
\textbf{counterterm} $-2k^2\gamma\,P_{11}(k)$ has exactly the same
$k$-dependence as the UV limit~\eqref{eq:P13_UV} of $2P_{13}$ --- precisely
what is needed to absorb the UV sensitivity of the loop. We will
complete this renormalization program below, but first we have to
deal with an important subtlety.

\begin{figure}
    \centering
        \includegraphics[width=0.49\linewidth]{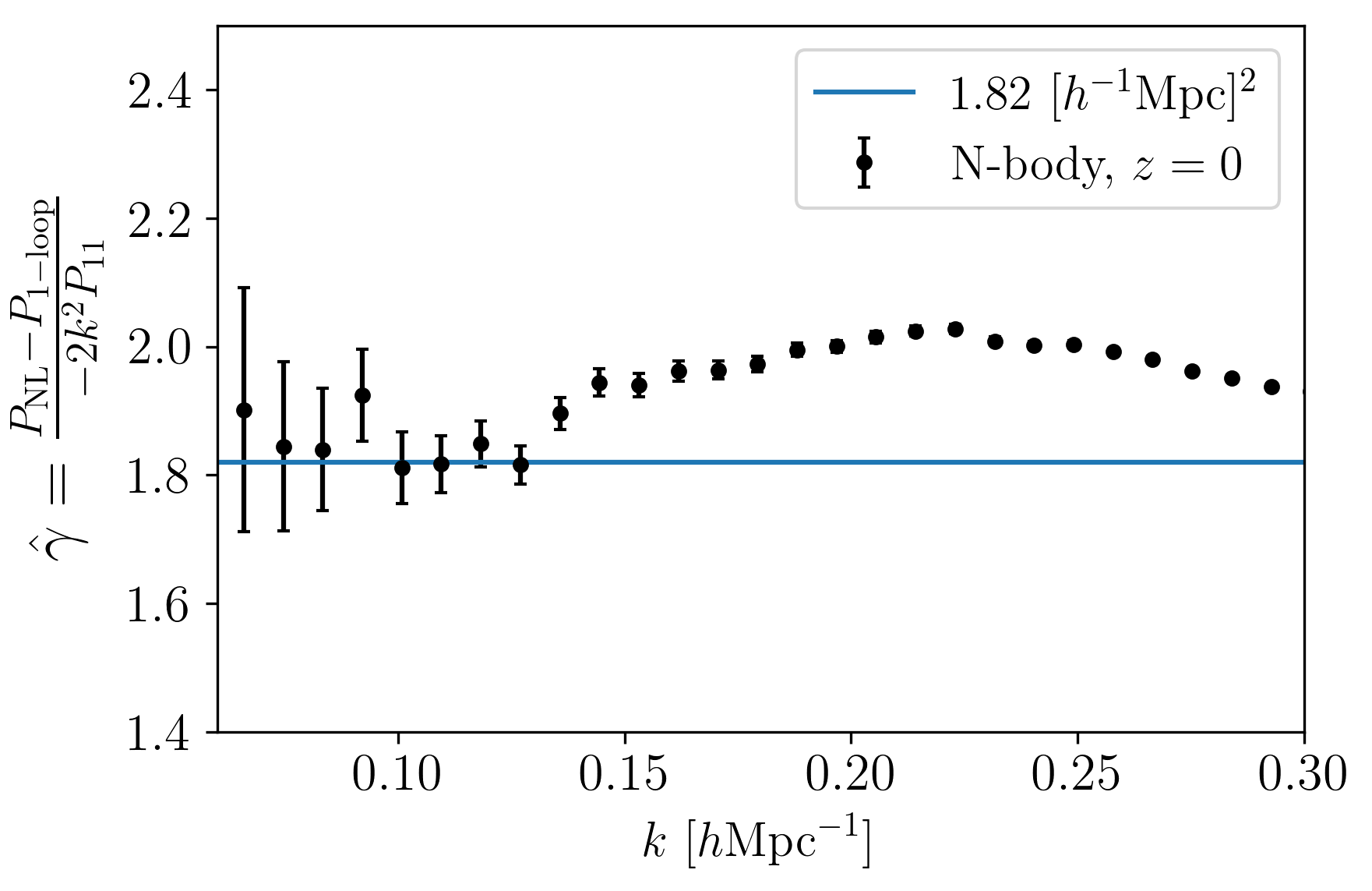}
    \includegraphics[width=0.49\linewidth]{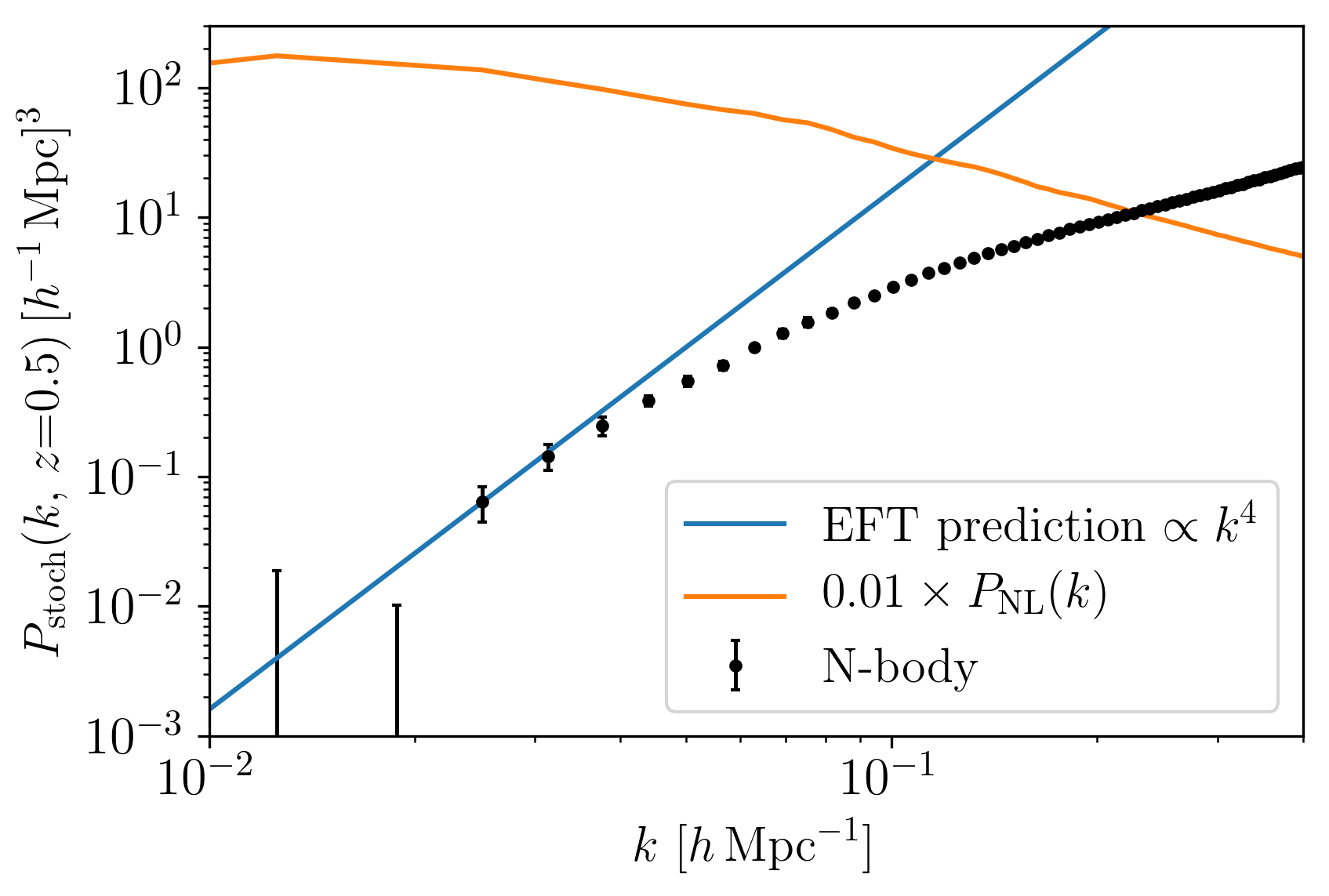}
    \caption{\textbf{Left:} effective sound speed \textit{estimator} $\hat \gamma$, showing the prediction for this parameter from individual (independent) $k$ bins from the N-body data of~\cite{Kim:2011ab}. The sound speed is constant $\gamma\sim 1.8~[\hMpc]^2$ on large scales. The apparent scale-dependence
    at $k\gtrsim 0.13~\hMpc$ 
    is the effect of two-loop corrections not included in the estimator.
    \textbf{Right:} the stochastic power spectrum from the
    \texttt{Quijote} HR simulation suite~\cite{Villaescusa-Navarro:2019bje}. On large scales $P_{\rm stoch}$ follows the EFT $k^4$
    asymptotic behavior. Note that $P_{\rm stoch}$ gets shallower around $k\sim 0.1~\hMpc$ relevant for observations, 
    which implies a breakdown of the power-law expansion. }
    \label{fig:cs_and_k4}
\end{figure}

%------------------------------------------------------------------
\subsection{Time Non-Locality}
%------------------------------------------------------------------

Hold on! In writing Eq.~\eqref{eq:tau_fluid} we tacitly assumed that
$\tau_{ij}$ at time $\tau$ depends on the fields at the same time
$\tau$ only. For ordinary fluids this is justified: the derivative
expansion is controlled by
\begin{equation}
  \lmfp\,\frac{\partial}{\partial x} \sim k\,\lmfp \ll 1\,,\qquad
  \tau_\mathrm{coll}\,\frac{\partial}{\partial t} \sim \omega\,\tau_\mathrm{coll} \ll 1\,,
\end{equation}
and since $\omega = c_s k$ while $c_s\tau_\mathrm{coll} = \lmfp$, the
two expansion parameters coincide: time derivatives and gradients are
equivalent, and both expansions converge. Physically, the fluid
relaxes to local equilibrium on the collision time
$\tau_\mathrm{coll}$, which is much shorter than the period of the
sound wave --- the fluid response is effectively instantaneous.
That's why we can use a local in time expansion for the stress tensor
from the get-go.

In LSS the analogue of the relaxation time is the dynamical time of
the small-scale structure formation, which is of the order of the Hubble time,
i.e.\ precisely the timescale on which the long-wavelength
perturbations evolve:
\begin{equation}
  \Delta\tau\;\partial_\tau\,\delta_1 \sim \frac{1}{\Hubble}\cdot\Hubble\,\delta_1
  \sim O(1)\cdot\delta_1\,.
\end{equation}
A Taylor expansion in time derivatives therefore \emph{does not
converge}: the response of the small-scale structure to a
long-wavelength perturbation is spread over the entire history of the
Universe. The stress tensor must be written as a functional of the
fields on the past trajectory~\cite{Carrasco:2013mua,Baldauf:2014qfa},\footnote{A remark on notation: the spatial derivatives inside the
memory integral are taken with respect to the spatial argument at
time $\tau'$ and only then evaluated on the trajectory, i.e.\
$\Delta\delta(\xv_\mathrm{fl}[\xv,\tau;\tau'],\tau') \equiv
\left(\Delta_{\vv{y}}\,\delta\right)(\vv{y},\tau')\big|_{\vv{y}=\xv_\mathrm{fl}(\tau')}$,
as dictated by the equivalence principle: the response is built from
quantities measurable in the local frame of the fluid element at the
past time. In contrast, the outer derivatives in
$\partial^i\partial^j\tau_{ij}(\xv,\tau)$ act at the field point
$\xv$ at time $\tau$. Differentiating the composite function with
respect to $\xv$ instead would generate Jacobian factors
$\partial x^k_\mathrm{fl}/\partial x^i = \delta^k_i + O(\partial\Psi)$;
since $\partial_i\Psi^j$-type terms are themselves allowed operators,
the two conventions differ only by a reshuffling of higher-order
Wilson coefficients, and the distinction drops out entirely at the
linear order used below. The same convention is understood in the
bias functional of \S6.2.}
\begin{equation}\label{eq:tau_nonlocal}
  \frac{\partial^i\partial^j\tau_{ij}}{\rhob}(\xv,\tau)
  = \int_{\tau_i}^{\tau} d\tau'\;K(\tau,\tau')\;
  \Delta\delta\!\left(\xv_\mathrm{fl}[\xv,\tau;\tau'],\tau'\right)\,,
\end{equation}
where $K(\tau,\tau')$ is a memory kernel with support over a few
Hubble times. Crucially, by the equivalence principle the response is
carried along by the fluid element, so the density must be evaluated
on the fluid trajectory $\xv_\mathrm{fl}$ defined by
\begin{equation}
  \frac{d\xv_\mathrm{fl}(\tau')}{d\tau'} = \vvv[\xv_\mathrm{fl}(\tau'),\tau']\,,\qquad
  \xv_\mathrm{fl}(\tau) = \xv
  \quad\Longrightarrow\quad
  \xv - \xv_\mathrm{fl}(\tau') = \int_{\tau'}^{\tau} d\tau_1\;\vvv[\xv_\mathrm{fl}(\tau_1),\tau_1]\,.
\end{equation}
Since the displacements are perturbatively small, we can expand:
\begin{align}\label{eq:tau_nonlocal_exp}
  \frac{\partial^i\partial^j\tau_{ij}}{\rhob}
  &= \int_{\tau_i}^{\tau} d\tau'\;K(\tau,\tau')
  \left[\Delta\delta(\xv,\tau')
  - \partial_i\Delta\delta(\xv,\tau')\int_{\tau'}^\tau d\tau_1\;v^i(\xv,\tau_1) + \cdots\right]\,,
\end{align}
where the second term is $O(\delta_1^2)$ and contributes only at
higher orders in perturbation theory. At linear order we can use
$\delta_1(\xv,\tau') = \Dp(\tau')\,\delta_0(\xv)$, and the time
integral factorizes:
\begin{equation}\label{eq:tau_nonlocal_LO}
  \int_{\tau_i}^{\tau} d\tau'\;K(\tau,\tau')\;\Delta\delta_1(\xv,\tau')
  = \left[\frac{1}{\Dp(\tau)}\int_{\tau_i}^{\tau} d\tau'\;K(\tau,\tau')\,\Dp(\tau')\right]
  \Dp(\tau)\,\Delta\delta_0(\xv)
  \equiv c_{s,\mathrm{eff}}^2(\tau)\;\Delta\delta_1(\xv,\tau)\,.
\end{equation}
The whole memory kernel has collapsed into a single time-dependent
coefficient: at this order the non-local-in-time stress tensor is
completely indistinguishable from the local sound speed ansatz of the
previous section. \textbf{Locality in time emerges order by order in
perturbation theory}, with the non-locality absorbed into the (time
dependence of the) EFT parameters. 
Can we keep doing this simplifications forever?
\textbf{Yes,} as long as perturbation theory is applicable.  
Is it messy at higher order? \textbf{Yes}, but there exist efficient 
algorithms to deal with it systematically~\cite{Abolhasani:2015mra},
which do not distinguish between the local and non-local in time
operators. 
In addition, 
there is a Lagrangian formulation of perturbation
theory and EFT, in which time non-locality 
is automatically implemented. More on this in Chapter 7. 

Notice that the term
$\sim \partial_i\delta\cdot v^i$ in the
expansion~\eqref{eq:tau_nonlocal_exp} naively ``breaks'' Galilean
symmetry (correlation functions are of course fine); it can be
eliminated by a coordinate redefinition, i.e.\ by passing to
Lagrangian coordinates. 

%------------------------------------------------------------------
\subsection{UV Matching 
}
%------------------------------------------------------------------

\noindent
\textbf{One-loop renormalization.}
Let us see how the sound speed fixes the UV problem
of SPT.
Consider the one-loop integral $I_{13} \equiv \int_\qv F_3(\kv,\qv,-\qv)\,P_{11}(q)$.
It receives contributions from all scales.
Let's split the integration domain:
\begin{equation}
  I_{13} = \underbrace{\int_0^{\knl} \ddq\; F_3\,P_{11}(q)}_{\text{calculable (perturbative)}}
  + \underbrace{\int_{\knl}^{\Lambda} \ddq\;F_3\,P_{11}(q)}_{\text{not trustworthy}}\,.
\end{equation}
The first part of the integral is called ``calculable'' since
it depends on the perturbative modes for which our 
theory is correct. 
Hence, this part of the integral we can trust. 
Note, however, that it's not entirely clear if the right 
scale there should be $\knl$, or perhaps $\knl/5$. Any
uncertainty in this scale will be compensated by the
counterterm. At this point it is important to stress that
even divergent integrals have trustworthy pieces, which can be 
roughly estimated by setting $\kUV\sim \knl$. These kinds of estimates
are known as ``naturalness arguments'' in particle physics, and 
they can be used to estimate the sizes of various effects, e.g.
the masses of the charm quark and the charged pions.

The UV part ($q > \knl$) involves non-perturbative physics: the
modes there do not obey the perturbative equations of motion, so
this part of the integral is wrong and, on top of that, cutoff-dependent.  
The cutoff is an artifact of our computation. 
The physical prediction should not depend on it.
Using the UV
limit~\eqref{eq:P13_UV}, the untrusted part of $2P_{13}$ is
\begin{equation}\label{eq:P13_UV_part}
  2P_{13}(k)\Big|_{\knl}^{\Lambda}
  \approx -\frac{61}{630\pi^2}\;k^2\,P_{11}(k)\int_{\knl}^{\Lambda} dq\;P_{11}(q)\,,
\end{equation}
which has exactly the $k$-dependence of the counterterm in
Eq.~\eqref{eq:P_ctr}. (The UV part of $P_{22}$ scales as $k^4$ and
will be dealt with by the stochastic counterterm in the next
chapter.) We therefore split the EFT parameter as
\begin{equation}
  \gamma = \gamma_\mathrm{inf}(\Lambda) + \gamma_\mathrm{fin}\,,\qquad
  \gamma_\mathrm{inf}(\Lambda) = -\frac{61}{1260\pi^2}\int_{\knl}^{\Lambda} dq\;P_{11}(q)\,,
\end{equation}
so that the $\Lambda$-dependent parts cancel, leaving
\begin{equation}\label{eq:PEFT_1loop}
  P_\mathrm{EFT}^{1\text{-loop}}(k)
  = P_{11}(k) + P_{22}(k) + 2P_{13}(k)\Big|_0^{\knl}
  - 2\gamma_\mathrm{fin}(\knl)\,k^2\,P_{11}(k)\,.
\end{equation}
The finite part $\gamma_\mathrm{fin}$ is a genuinely new input: it
encodes the true contribution of the non-perturbative small-scale
physics, which the EFT cannot predict. It has to be fitted to data or
measured in $N$-body simulations. 
This procedure is called \textit{UV matching}.
Note also that
$\gamma_\mathrm{fin}$ depends on the arbitrary split scale $\knl$
(such dependence is called the scheme-dependence); only the total, renormalized result is physical.

In practice one often does not bother introducing the intermediate scale at
all: the loop integrals are evaluated with $\Lambda\to\infty$, and
the fitted counterterm automatically absorbs both the UV garbage and
the true finite contribution. This is nothing but a particular
(and convenient) choice of scheme.

\medskip\noindent
\textbf{Higher-loop scaling problem.}
At each loop order, the SPT UV integrals introduce factors of the mass variance:
\begin{equation}
  \sigma^2(Q) = \int_0^{Q} \ddq\;P_{11}(q)\,,
\end{equation}
which is $\mathcal{O}(1)$ for $Q \sim \knl$. Specifically we have the leading scaling:
\be 
\Delta P^{\rm SPT}_{\rm L-loop}\sim k^2 P_{11}(k) \int_{\knl}^{\Lambda} dq~P_{11}(q)\sigma^{2L-2}(q)~\,.
\ee 
While $\sigma^{2}(q)$ diverges $\propto \ln^3(q)$, the above
integrals converge, but to progressively larger numbers,
which  grow as $(3L-1)!/2^{3L}$~\cite{Blas:2013aba}.
So the leading UV contributions at each loop order are \emph{all the same size}, which explains the numerical result of Fig.~\ref{fig:spt_loops}:
\begin{equation}
  P_\mathrm{2\text{-loop}}^\mathrm{UV} \sim P_\mathrm{1\text{-loop}}^\mathrm{UV}\,,\qquad
  P_\mathrm{3\text{-loop}}^\mathrm{UV} > P_\mathrm{1\text{-loop}}^\mathrm{UV} + P_\mathrm{2\text{-loop}}^\mathrm{UV}\,.
\end{equation}
The SPT perturbative series \textbf{does not converge} in the UV.
The EFT counterterms must absorb this UV sensitivity order by order.
Note that every new SPT loop simply re-defines the counterterm $\gamma$.
Indeed, the $k$-dependence of all of them is exactly $k^2P_{11}$.
Since we already have this operator included at the one-loop order,
the new UV corrections are totally redundant -- they do not require new 
counterterms. In other words, these re-definitions are completely unobservable,
and hence we can simply choose a scheme in which these parts of higher loop 
corrections are identically set to zero. One says that the 
redundant parts are subtracted from the loop integrals.
What remains calculable after this subtractions 
are the IR-dominated parts of the loops,
which, as we will see in the next chapter, are suppressed by powers
of $k/\knl \ll 1$. Thus, the strong UV sensitivity of SPT
loops is an artifact of the perfect fluid description. 
After renormalization the loops organize a well behaved 
parametric series. This agrees well with 
simulation-based results that found that the actual 
power spectrum
is very weakly sensitive to small scale dynamics~\cite{Nishimichi:2014rra,Garny:2015oya}, in full agreement with the EFT-based expectations.

\medskip\noindent
\textbf{EFT fitting in practice.}
The one-loop EFT power spectrum in the $\Lambda\to\infty$ scheme is:
\begin{equation}
  P_\mathrm{EFT}^{1\text{-loop}}(k) = P_{11}(k) + P_{22}^\mathrm{SPT}(k) + 2P_{13}^\mathrm{SPT}(k)
  - 2\gamma\,k^2\,P_{11}(k)\,,
\end{equation}
where $\gamma$ is a free parameter (the time-averaged effective sound
speed squared, in units of $(\mathrm{Mpc}/h)^2$).
The counterterm $-2\gamma\,k^2\,P_{11}$ cancels the $\Lambda$-dependent part of $P_{13}$.
Fitting to Horizon Run $N$-body simulation~\cite{Kim:2011ab,Blas:2013bpa,Blas:2016sfa} at $z=0$ gives $\gamma \approx 1.8~\,(\mathrm{Mpc}/h)^2$, see the right panel of~\ref{fig:cs_and_k4}, where we show
the \textit{estimator} for $\gamma$ for individual $k$ bins. We see
that the estimator is scale-independent up to $k \sim 0.15\Mpc$,
after which it picks up a scale dependence due to the two-loop
corrections omitted in the theory prediction~\cite{Baldauf:2015aha}. Nevertheless, the 
consistency of  $\gamma$ across independent bins is an important
test of EFT: once the scheme is fixed, this parameter
should take a well determined constant value.

\begin{figure}
    \centering
    \includegraphics[width=0.49\linewidth]{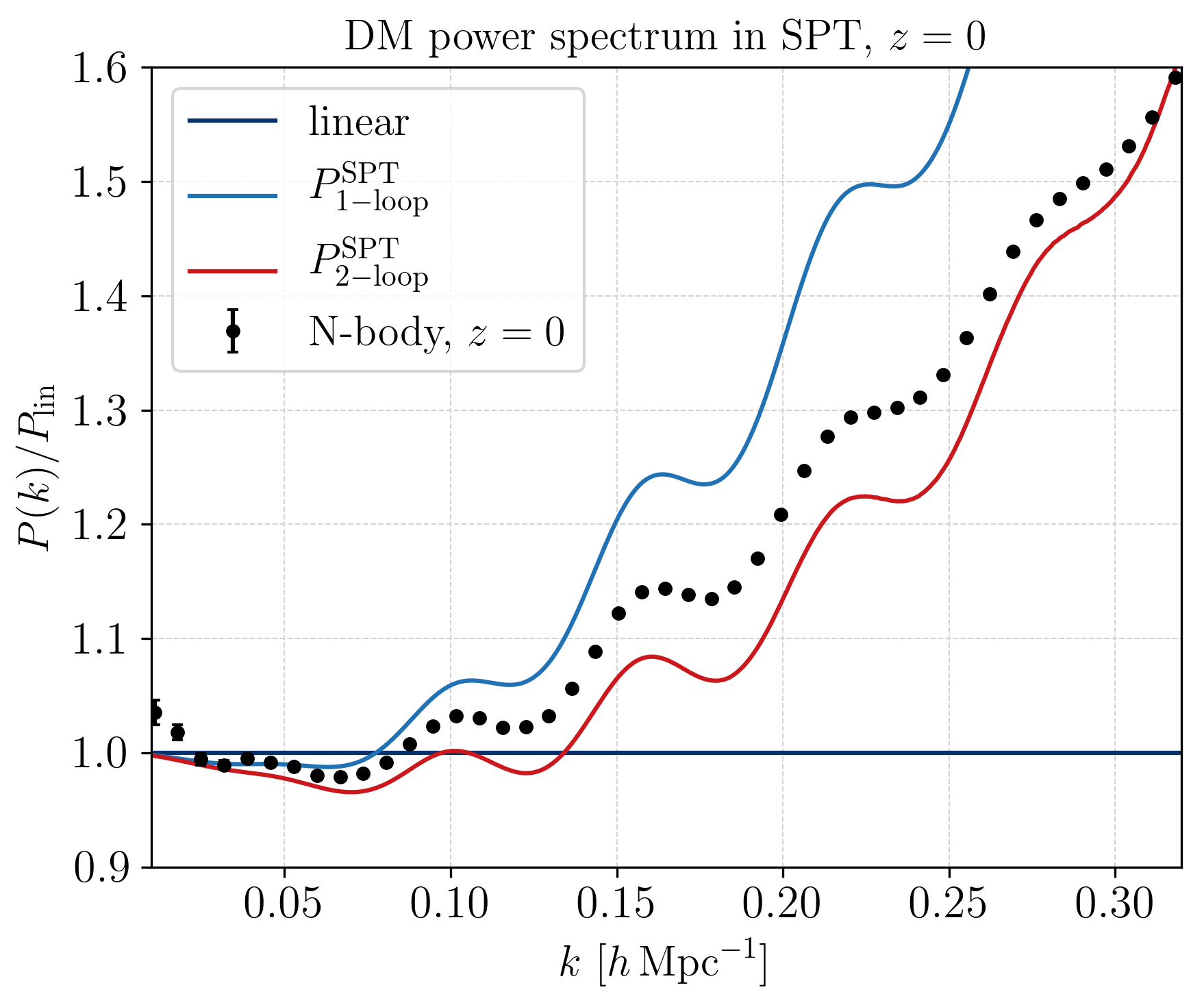}
    \includegraphics[width=0.49\linewidth]{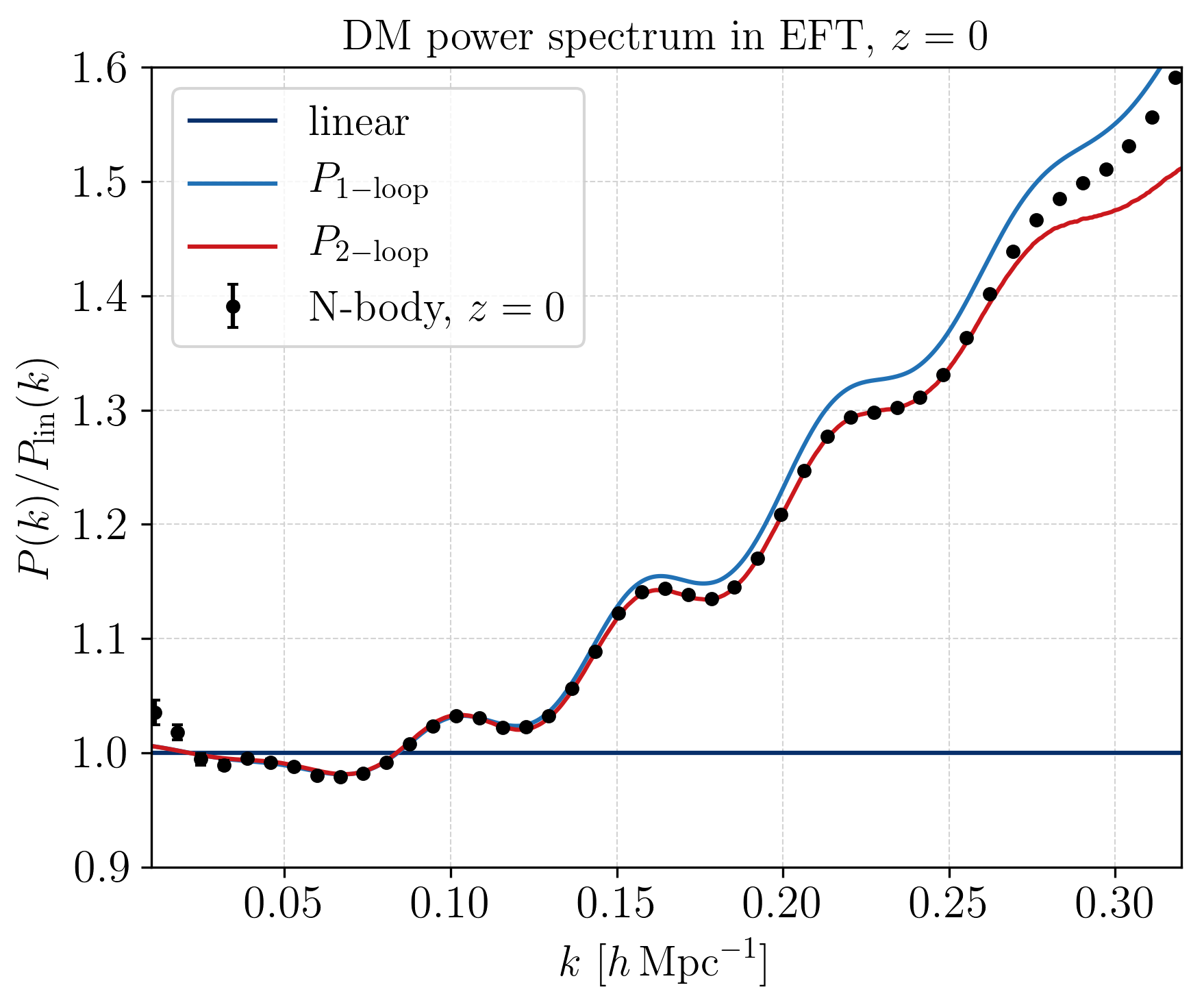}
    \caption{Dark-matter power spectrum at $z=0$, divided by the linear
  power spectrum $P_{\rm lin}(k)$.
  \textbf{Left:} Standard Perturbation Theory (SPT).  The one-loop
  correction (light blue) overshoots the N-body data (black points)
  already at $k \gtrsim 0.05\,h\,\mathrm{Mpc}^{-1}$, and the
  two-loop result (red) develops unphysical oscillations without
  converging toward the data.
  \textbf{Right:}  The one-loop EFT result, which includes the
  counterterm $-2\gamma\,k^2 P_{\rm lin}(k)$ with $\gamma$
  fit to simulations, agrees with the N-body data to sub-percent
  accuracy up to $k \simeq 0.15\,h\,\mathrm{Mpc}^{-1}$.  The
  two-loop EFT result extends the agreement to
  $k \simeq 0.25\,h\,\mathrm{Mpc}^{-1}$.  The failure of SPT at
  high~$k$ is a direct consequence of the missing UV counterterms:
  the loop integrals receive uncontrolled contributions from
  short-scale modes that are absorbed into the EFT parameters.}
    \label{fig:fit_nbody}
\end{figure}

Figure~\ref{fig:fit_nbody} summarizes this section. 
The left panel shows the failure of SPT to fit the N-body data
on all scales: the one-loop SPT prediction overshoots the data, 
while the two-loop one undershoots.  The right panel displays 
the EFT results: the power spectrum prediction is $\Lambda$-independent 
and agrees with simulations up to $k \simeq 0.15\Mpc$. 
Once upgraded to two-loops~\cite{Carrasco:2013mua,Baldauf:2015aha,Foreman:2015lca,Bakx:2025jwa,Saraivanov:2026sxc,Chen:2026usz}, EFT fits the data up to $k_{\rm max}\simeq 0.25~\hMpc$.

\medskip\noindent
\textbf{Baryonic effects.} It is worth mentioning that 
baryonic feedback on dark matter (i.e. backreaction due 
to star formation etc.) can also be included 
in EFT by considering baryons and dark matter as two separate fluids
with different stress tensors~\cite{Lewandowski:2014rca}.
The net effect is simply a shift of the dark matter counterterm
$\gamma\to \gamma+\gamma_b$, where $\gamma_b$ is the baryonic 
contribution. Intuitively, this can
be understood as coming from a simple Taylor expansion of the 
Jeans suppression in linear theory
\[
  \delta_1 e^{-\frac{k^2}{k_J^2}}\approx 
\delta_1\left(1-\frac{k^2}{k_J^2}\right)\,.
\]

\paragraph{A historic remark.} At the beginning, the presence of fitting 
parameters for EFT dark matter predictions was considered a tragedy
by many non-EFT researches. How come a theory can't predict everything? 
In the EFT framework it is quite OK: we acknowledge 
that small-scales are difficult, but in the large-scale limit 
all of their complexity can be 
captured by a few parameters only. 
For pure N-body dynamics one can develop
more sophisticated UV models that allow one to estimate
the sound speed fairly accurately~\cite{Nascimento:2024hle,Nascimento:2024rbv,Garny:2025zlq}.
The point, however, is that for more complicated situations, 
i.e. dynamics beyond N-body (e.g. ultralight axion dark matter),
or in the presence of baryonic feedback, 
such estimates are much harder 
to make, so one has to resort to fitting EFT parameters from the data.
In addition, once we start looking at the galaxies, there will be even 
more EFT parameters whose accurate 
semi-analytic predictions are very challenging. Thus, fitting 
them from the data is universally acknowledged as a normal practice.

%==========================================================================
\subsection{Stochastic Effects}
%==========================================================================

In the previous chapter we renormalized the $k^2 P_{11}(k)$ UV
sensitivity of the one-loop power spectrum using the response of the
effective stress tensor to the long-wavelength fields. This, however,
cannot be the whole story: the UV limit of $P_{22}$,
Eq.~\eqref{eq:P22_UV}, produced a $k^4$ contribution completely
independent of $P_{11}(k)$, for which no counterterm has appeared so
far. The missing ingredient is that the symmetries allow $\tau_{ij}$
to contain pieces that are not determined by the long-wavelength
cosmological fields at all:
\be\label{eq:tau_stoch}
  \frac{\tau_{ij}}{\rhob} \supset \varepsilon\,\delta_{ij} + \varepsilon_{ij}\,,\qquad
  \avg{\varepsilon\,\delta_1} = 0\,,\quad
  \avg{\varepsilon_{ij}\,\delta_1} = 0\,.
\ee
These \textbf{stochastic} fields represent the contributions of the
fully non-linear, virialized modes, which have decoupled from the
cosmological fields: their evolution has erased the memory of the
initial conditions, so from the point of view of the long modes they
behave as pure noise, uncorrelated with $\delta_1$. For the one-loop
power spectrum the scalar and tensor components in
Eq.~\eqref{eq:tau_stoch} are degenerate, so for convenience we keep
only the scalar piece, $\tau_{ij}=\rhob\,\varepsilon\,\delta_{ij}$,
for which
\be
  \frac{1}{\rhob}\,\partial^i\partial^j\tau_{ij} = \Delta\varepsilon\,.
\ee
The equation of motion then reads
\be
  \partial_\eta\Theta_{\kv} - \frac{3}{2}\,\delta_{\kv} + \frac{1}{2}\,\Theta_{\kv}
  = \int\beta\,\Theta_{\qv_1}\Theta_{\qv_2}
  - \frac{k^2 c_{s,\mathrm{eff}}^2}{\Hubble^2}\,\delta_{\kv}
  - \frac{k^2}{\Hubble^2}\,\varepsilon_{\kv}\,,
\ee
and solving it perturbatively with the same Green's function as in
the previous chapter, we obtain
\be\label{eq:ddelta_stoch}
  \Delta\delta^{(\tau)}_{\kv} = -k^2\,\gamma\,\delta_1(\kv) - k^2\,\varepsilon(\kv)\,,
\ee
where, with a slight abuse of notation, $\varepsilon(\kv)$ in the
last expression denotes the time integral of the stochastic source
against the Green's function. Since $\avg{\varepsilon\,\delta_1}=0$,
all cross terms vanish, and the power spectrum receives a new,
purely additive contribution:
\be
  \avg{|\delta_\mathrm{SPT} + \Delta\delta^{(\tau)}|^2}'
  = P_{11} + P_{22} + 2P_{13}
  - 2k^2\gamma\,P_{11}(k) + k^4\,\avg{|\varepsilon(\kv)|^2}'\,.
\ee
$k^4\avg{|\varepsilon(\kv)|^2}'$ above is the 
stochastic power spectrum of the matter
field 
$P_{\rm stoch}$.
What can we say about the underlying power spectrum
$\avg{|\varepsilon(\kv)|^2}'$? Symmetries and
dimensional analysis are enough. The field $\varepsilon$ is generated
by the small-scale dynamics, with support at $q\gtrsim\knl$, so its
correlations extend only over distances $\sim 1/\knl$. For
$k\ll\knl$ its power spectrum must therefore be analytic in $k$, and
rotational invariance forbids odd powers,  yielding:
\be\label{eq:eps_power}
  \avg{|\varepsilon(\kv)|^2}' = A_0 + A_1\,k^2 + \cdots
\ee
Combining everything, we arrive at the
\textbf{final one-loop EFT power spectrum:}
\begin{equation}\label{eq:P_EFT}
  \boxed{P_{1\text{-loop}}^\mathrm{EFT}
  = P_{11} + P_{22} + 2P_{13}
  - 2k^2\gamma\,P_{11}(k) + A_0\,k^4 + \cdots}
\end{equation}

\bigskip\noindent
\textbf{Renormalization.}
Let us verify that the new parameter completes the renormalization
program of the previous chapter. Isolating the $\Lambda$-dependent
UV parts of the loops with the help of the asymptotics of \S3.4,
\be
  i_{22} \equiv \frac{9}{196\pi^2}\int_{\knl}^{\Lambda}\frac{dq}{q^2}\;P_{11}^2(q)\,,\qquad
  i_{13} \equiv \frac{61}{1260\pi^2}\int_{\knl}^{\Lambda}dq\;P_{11}(q)\,,
\ee
we can write
\be
  P_{1\text{-loop}}^\mathrm{EFT}
  = \Big[P_{11} + P_{22} + 2P_{13}\Big]_0^{\knl}
  + k^4\left(A_0 + i_{22}\right)
  - 2k^2\left(\gamma + i_{13}\right)P_{11}(k)\,.
\ee
The UV parts of \emph{both} loop integrals now sit next to an EFT
parameter with exactly the same $k$-dependence. Splitting
$A_0 = A_0^\mathrm{inf}(\Lambda) + A_0^\mathrm{fin}$ and
$\gamma = \gamma^\mathrm{inf}(\Lambda) + \gamma^\mathrm{fin}$ as
before, the cutoff dependence is completely removed --- this is
renormalization at work. The finite part of the stochastic
contribution scales as
\be
  P_\mathrm{stoch}(k) = A_0^\mathrm{fin}\,k^4\,.
\ee
It is instructive to compare this with the popular halo model, whose
one-halo term (the analog of $P_\mathrm{stoch}$) approaches a \emph{constant} at small $k$~\cite{Seljak:2000gq,Cooray:2002dia}. 
If you are reading this notes, you probably already know 
which prediction is the correct one. 
Precision measurements of $P_\mathrm{stoch}$ from N-body 
simulations
support the $k^4$ scaling in the $k\to 0$ limit 
predicted by the EFT,
see the 
right panel of
Fig.~\ref{fig:cs_and_k4} (more detail in~\cite{Baldauf:2015zga,Ivanov:2024hgq,Ivanov:2024xgb}). This is not a coincidence. 
As we now
show, the constant $P_\mathrm{stoch}$
is in conflict with basic conservation laws.

\bigskip\noindent
\textbf{Peebles' argument.}
The $k^4$ scaling of the stochastic power is not an accident of
perturbation theory --- it is fixed by mass and momentum
conservation. Following Peebles~\cite{1980lssu.book.....P} (see also~\cite{Abolhasani:2015mra}), let's model the small-scale structure as
a collection of particles of masses $m_n$ at positions
$\xv_n$, confined to a region of size $R\ll 1/k$. 
These particles can be thought of as dark 
matter particles inside a halo.
The small-scale non-perturbative
dynamics rearranges the particles, $\xv_n\to\xv_n+\Delta\xv_n$,
generating a density perturbation
\be
  \delta\rho(\kv)\Big|_\mathrm{stoch}
  = \sum_n m_n\,e^{i\kv\cdot(\xv_n+\Delta\xv_n)} - \sum_n m_n\,e^{i\kv\cdot\xv_n}\,.
\ee
Expanding in $\kv\cdot\Delta\xv_n\ll 1$,
\be
  \delta\rho(\kv)\Big|_\mathrm{stoch}
  = \sum_n i\,\kv\cdot\Delta\xv_n\;m_n\,e^{i\kv\cdot\xv_n}
  - \frac{1}{2}\sum_n \left(\kv\cdot\Delta\xv_n\right)^2 m_n\,e^{i\kv\cdot\xv_n}
  + \cdots
\ee
In the long-wavelength limit we can set $e^{i\kv\cdot\xv_n}\to 1$
inside the region. There is no $O(k^0)$ term because the total mass
of the region is conserved by the internal dynamics. The $O(k)$
(dipole) term is proportional to $\sum_n m_n\,\Delta\xv_n$, i.e.\ to
the displacement of the center of mass --- and this vanishes as
well, because internal gravitational forces cannot generate a net
momentum. The leading surviving contribution is the quadrupole:
\be
  \frac{\delta\rho}{\rho}\Big|_\mathrm{stoch} \propto k^2
  \qquad\Longrightarrow\qquad
  \avg{\Big|\frac{\delta\rho}{\rho}\Big|^2}_\mathrm{stoch} \propto k^4\,.
\ee
The EFT recovers this symmetry-based result automatically: the
stochastic source enters the equations of motion only through
$\partial^i\partial^j\tau_{ij}$, which supplies the two powers of $k$
at the level of the field.

Before closing this Section let us mention
that if a small-scale reshuffling is correlated with a 
long-wavelength density field $\delta_1(\k)$, it will
produce an effective sound speed contribution $\sim k^2P_{11}(k)$.
In fact, this correlation must be present since the reshuffling 
happens inside the halos, whose distribution
is modulated by the linear density fluctuations $\delta_1$
via the peak-background split arguments which we will use in the next Chapter to motivate galaxy bias.
Therefore, the effective sound speed could have been discovered
by Peebles already in the 1980's! 

%------------------------------------------------------------------
\subsection{Power Counting in a Scaling Universe}
%------------------------------------------------------------------

Having collected all the ingredients of the EFT at one loop, let us
estimate their relative sizes~\cite{Pajer:2013jj}. For this purpose it is convenient to
use the power-law universe of Chapter 2,
\be
  P_{11}^\mathrm{PLU}(k) = \frac{2\pi^2}{\knl^3}\left(\frac{k}{\knl}\right)^{n}\,,
\ee
with $n=-1.5$, which mimics the slope of the $\Lambda$CDM spectrum
around the non-linear scale. In this universe the linear field
scales as
$\delta_1\sim (k/\knl)^{(3+n)/2} = (k/\knl)^{0.75}$,
and each loop brings an extra factor of the dimensionless power:
\begin{align}
  P_{22},\;P_{13} &\sim \int d^3q\,P_{11}(q)\,P_{11}(k)
  \sim \left(\frac{k}{\knl}\right)^{3+n} P_{11}(k)\,,\\
  P_{2\text{-loop}} &\sim \left(\frac{k}{\knl}\right)^{2(3+n)} P_{11}(k)\,.
\end{align}
Collecting all contributions to the dimensionless power spectrum
$\Delta^2(k)\sim k^3 P(k)$:
\begin{equation}
  \Delta^2(k) \sim \underbrace{\left(\frac{k}{\knl}\right)^{n+3}}_\text{tree}
  + \underbrace{\left(\frac{k}{\knl}\right)^{2(3+n)}}_{1\text{-loop}}
  + \underbrace{\left(\frac{k}{\knl}\right)^{3(3+n)}}_{2\text{-loop}}
  + \underbrace{\left(\frac{k}{\knl}\right)^{n+5}}_{k^2 P_{11}\text{ ctr}}
  + \underbrace{\left(\frac{k}{\knl}\right)^{7}}_{k^4\text{ stoch}}
  + \cdots
\end{equation}
For $n=-1.5$ the exponents evaluate to
\begin{equation}
  \Delta^2(k) \sim \left(\frac{k}{\knl}\right)^{1.5}
  + \left(\frac{k}{\knl}\right)^{3}
  + \left(\frac{k}{\knl}\right)^{4.5}
  + \left(\frac{k}{\knl}\right)^{3.5}
  + \left(\frac{k}{\knl}\right)^{7}
  + \cdots
\end{equation}
The hierarchy is now manifest. The counterterm is only slightly
smaller than the one-loop correction (the difference of the
exponents, $0.5$, reflects the mild $k^2/\knl^{n+2}$ suppression),
so the two must always be included together. The stochastic term,
by contrast, is strongly suppressed by its $k^4$ scaling:
\begin{equation}
  \boxed{P_\mathrm{tree} \gg P_{1\text{-loop}} \simeq P_\mathrm{ctr}
  \gg P_{2\text{-loop}} \gtrsim P_\mathrm{stoch}\,.}
\end{equation}
The scaling Universe estimates apply to the 
physical, \textit{renormalized} quantities.
Note that the scaling arguments can be used to estimate the time-dependence
of the finite parts of the counterterms. Indeed, from the linear scaling 
$\Delta^2(k)\propto D_+^2(z)$ we extract 
\[
  \knl^{-(3+n)}\propto D^2_+(z)\quad \quad \Rightarrow \quad \quad 
  \frac{\knl(z)}{\knl(z=0)}=D^{-\frac{2}{n+3}}_+(z)\sim D^{-\frac{4}{3}}_+(z)~\,.
\]
This implies $\gamma_{\rm fin}\sim 1/(\knl)^2\propto D^{\frac{8}{3}}_+$,
which matches simulations quite well~\cite{Lazeyras:2019dcx,Chudaykin:2022sdl,Chen:2026usz}.

%==========================================================================
\section{Chapter 5: Galaxy Bias}
%==========================================================================

So far we have been describing the clustering of dark matter. What
we actually observe, however, are galaxies: highly non-linear
objects that form at special locations of the matter field --- in
the peaks of the density, inside virialized halos. 
Thanks to this formation mechanism,
on large scales galaxies trace the distribution of matter
in the Universe. This effect is known as \textbf{galaxy bias}.
Remarkably, even
though galaxy formation itself is hopelessly complicated, the
\emph{large-scale distribution} of galaxies can be described
analytically by the same EFT logic: all the complications are
absorbed into a finite set of bias coefficients.

Before going for a full-blown EFT treatment of the problem, let
us consider phenomenological models of halo formation
and their correlations with the cosmological initial 
conditions. These models are not as accurate as the full EFT, 
but they provide some nice intuition behind \textbf{halo bias}, 
a relationship between 
the matter halos and the underlying matter field. 

\subsection{Phenomeological Models for Halo Bias}

The first relevant ingredient is \textbf{spherical collapse}.
Consider, in an EdS background, a spherical top-hat region of
initial radius $R_i$ and overdensity $\bar\delta_i>0$. By the
Newtonian shell theorem the sphere evolves as a closed Friedmann
universe, independently of its environment: its radius obeys
\begin{equation}
  \ddot{R} = -\frac{GM}{R^2}\,,\qquad
  M = \frac{4\pi}{3}\,\bar\rho_{m,i}\left(1+\bar\delta_i\right)R_i^3
  = \mathrm{const}\,.
\end{equation}
For a bound (overdense) region the solution is the famous cycloid,
\begin{equation}
  R = A\,(1-\cos\theta)\,,\qquad
  t = B\,(\theta-\sin\theta)\,,\qquad
  A^3 = GM B^2\,,
\end{equation}
while the background density in EdS is
$\bar\rho_m = 1/(6\pi G t^2)$. The overdensity of the sphere is
therefore known exactly at all times:
\begin{equation}\label{eq:sph_collapse_exact}
  1+\delta = \frac{3M}{4\pi R^3}\cdot\frac{1}{\bar\rho_m}
  = \frac{9}{2}\,\frac{(\theta-\sin\theta)^2}{(1-\cos\theta)^3}\,.
\end{equation}
The milestones of the evolution: at \emph{turnaround},
$\theta=\pi$, the sphere decouples from the expansion and reaches
its maximal radius $R_\mathrm{ta}=2A$, with
$1+\delta_\mathrm{ta} = 9\pi^2/16 \simeq 5.55$; at $\theta=2\pi$
the sphere formally \emph{collapses} to a point, at the time
$t_\mathrm{col}=2\pi B$.

Now compare with linear theory. Expanding
Eq.~\eqref{eq:sph_collapse_exact} at small $\theta$,
\begin{equation}
  \delta \simeq \frac{3}{20}\,\theta^2
  = \frac{3}{20}\left(\frac{6t}{B}\right)^{2/3}
  \propto t^{2/3} \propto a\,,
\end{equation}
i.e.\ we recover the linear growing mode, as we must. The key trick
is to ask: what value does the \emph{formal linear extrapolation} of
the initial overdensity reach at the moment when the true non-linear
solution collapses? Setting $t=t_\mathrm{col}=2\pi B$,
\begin{equation}\label{eq:deltac}
  \boxed{\;\delta_c = \frac{3}{20}\,(12\pi)^{2/3} \simeq 1.686\,.\;}
\end{equation}
This number is the bridge between the (Gaussian, calculable) initial
conditions and the (non-perturbative) formation of halos: a region
collapses by time $\tau_0$ if its \emph{linearly
evolved} smoothed density exceeds $\delta_c$. In EdS the result is
exact and independent of the halo mass; in $\Lambda$CDM $\delta_c$
acquires only a weak cosmology dependence. In reality the collapse
to a point is halted by shell crossing and violent relaxation: the
object settles into virial equilibrium at roughly half the
turnaround radius, with a characteristic density contrast
$1+\delta_\mathrm{vir}\simeq 18\pi^2 \simeq 178$ --- the origin of
the familiar $\Delta\simeq 200$ conventions used to define the halo
mass and radius.

\medskip\noindent
\textbf{Homework:} using energy conservation and the virial theorem,
show that virialization at half the turnaround radius implies
$1+\delta_\mathrm{vir} = 18\pi^2$ if the collapse time is taken as
$2\pi B$.

%------------------------------------------------------------------
\paragraph{The Press--Schechter Halo Mass Function.}
%------------------------------------------------------------------

The end result of the non-linear collapse of matter is the
formation of halos. Let us study how the halos are distributed with
respect to their masses. This distribution is called the
\textbf{halo mass function} (HMF), $d\bar n/d\ln M$, where $\bar n$
is the halo number density in units of $(\Mpch)^{-3}$. The
Press--Schechter approach~\cite{Press:1973iz} 
provides a simple analytic model for it.
We have just seen that in the spherical collapse model a region of
initial comoving radius $R$ collapses when its linearly extrapolated
overdensity reaches $\delta_c=1.686$. Following Press and Schechter,
let us assume that the same criterion applies to different patches of our Universe:
a spherical region collapses into a halo of mass
$M(R)=\frac{4\pi}{3}\bar\rho_m R^3$ within the lifetime of the
Universe if its linear overdensity, extrapolated to today, satisfies
$\bar\delta_1(R)\geq\delta_c$; regions with
$\bar\delta_1(R)<\delta_c$ do not make it. Whether a region
collapses or not is thus determined entirely by the initial
conditions, i.e.\ by the linear power spectrum $P_{11}(k)$.

To compute the probability of collapse we first need the one-point
PDF of the smoothed linear density $\delta_R$ (as in Chapter 2, we
work with linear fields extrapolated to redshift zero). 
This is best to do using the functional integral techniques, introduced
in Appendix~\ref{app:GRF}.
Since the
initial field is Gaussian, the PDF is given by the (normalized)
Gaussian functional integral
\begin{equation}
  P(\delta_R)= \mathcal{N}\int \mathcal{D}\delta_1\;
  \exp\Bigg\{-\int_\kv \frac{\delta_{1}(\kv)\delta_{1}(-\kv)}{2P_{11}(k)}
  \Bigg\}\;
  \delta_D^{(1)}\!\left(\delta_R-\int d^3x\;W_R(\xv)\,\delta_{1}(\xv)\right)\,,
\end{equation}
where the delta function selects the field configurations with the
prescribed smoothed density, and in Fourier space the constraint
reads $\int_\kv \delta_1(\kv)\tilde{W}_R(kR)$ with the top-hat
window of Chapter 2. Using the integral representation of the delta
function,
\begin{equation}
  P(\delta_R)=\mathcal{N}\int_{-i\infty}^{i\infty}\frac{d\lambda}{2\pi}\,
  e^{-\lambda\delta_R}
  \int\mathcal{D}\delta_1\,
  \exp\Bigg\{-\int_\kv \frac{\delta_{1}(\kv)\delta_{1}(-\kv)}{2P_{11}(k)}
  +\lambda\int_\kv \delta_1(\kv)\tilde{W}_R(kR)\Bigg\}\,,
\end{equation}
the functional integral is Gaussian and is computed by evaluating
the exponent on the saddle point,
$\delta_1(-\kv) = \lambda\,P_{11}(k)\tilde{W}_R(kR)$, which leaves
an ordinary Gaussian integral over $\lambda$:
\begin{equation}\label{eq:PS_lambda}
  P(\delta_R)=\mathcal{N}'\int_{-i\infty}^{i\infty}\frac{d\lambda}{2\pi}\,
  \exp\Bigg\{-\lambda\delta_R
  +\frac{\lambda^2}{2}\int_\kv P_{11}(k)\,|\tilde{W}_R(kR)|^2\Bigg\}\,.
\end{equation}
The coefficient of $\lambda^2$ is nothing but the smoothed mass
variance $\sigma^2(R)$ of Eq.~\eqref{eq:sig2_def}. Performing the
last integral, we get
\begin{equation}\label{eq:PS_pdf}
  P(\delta_R)=\frac{1}{\sqrt{2\pi\sigma^2(R)}}\,
  \exp\Bigg\{-\frac{\delta_R^2}{2\sigma^2(R)}\Bigg\}\,.
\end{equation}
The probability for a region of radius $R$ to collapse is then
\begin{equation}\label{eq:PsP}
  \mathcal{P}(\delta_R\geq\delta_c)
  =\int_{\delta_c}^\infty d\delta_R\;P(\delta_R)
  =\frac{1}{2}\,\mathrm{erfc}\!\left(\frac{\nu}{\sqrt{2}}\right)\,,\qquad
  \nu\equiv\frac{\delta_c}{\sigma(R)}\,.
\end{equation}
In the limit $R\to 0$ the variance diverges and we expect \emph{all}
matter to end up in halos of some (small) mass; yet
Eq.~\eqref{eq:PsP} saturates at $1/2$. The mismatch is corrected by
an ad hoc factor of two, $\mathcal{P}\to 2\mathcal{P}$ ---
famously introduced by hand by Press and Schechter, and explained
rigorously in the excursion-set formulation, where it accounts for
the underdense regions embedded in larger collapsing ones (the
cloud-in-cloud problem). By definition, this probability equals the
fraction of the total mass contained in halos with masses above
$M(R)$,
\begin{equation}
  2\mathcal{P}(\delta_R\geq\delta_c)
  =\frac{1}{\bar\rho_m}\int_{M(R)}^{\infty} M'\,\frac{d\bar n}{dM'}\,dM'\,.
\end{equation}
Differentiating both sides with respect to $M$ we obtain the
\textbf{Press--Schechter mass function}:
\begin{equation}\label{eq:PSHMF}
  \boxed{\;M\frac{d\bar n}{dM}
  =\frac{\bar\rho_m}{M}\,
  f_\mathrm{PS}\!\left(\nu\right)
  \left|\frac{d\ln\sigma(R[M])}{d\ln M}\right|\,,\qquad
  f_\mathrm{PS}(\nu)=\sqrt{\frac{2}{\pi}}\;\nu\,e^{-\nu^2/2}\,.\;}
\end{equation}
Note the structure of this result: the mean halo abundance is
\emph{non-perturbative} in $P_{11}$ --- the variance appears in an
exponent. This is an example of a quantity that is beyond the reach
of perturbation theory, just as the mean free path or the size of
molecules can be estimated in quantum mechanics but cannot be
computed within fluid dynamics. In the EFT context, the HMF is the
kind of input we can use for order-of-magnitude estimates of the
non-perturbative EFT parameters, but never for controlled
predictions.

%------------------------------------------------------------------
\paragraph{Thresholding: Peak-Background Split and Halo Bias.}
%------------------------------------------------------------------

\begin{figure}[h!]
\begin{center}
\includegraphics[width=0.7\textwidth]{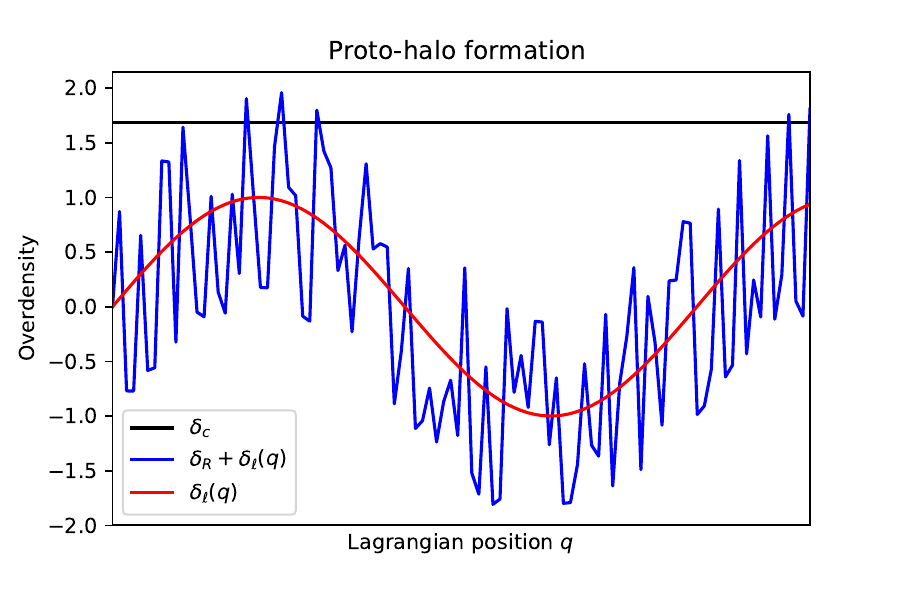}
\end{center}
\caption{\label{fig:thresh}
1D cartoon explaining the mechanism of correlating
proto-halo positions with long-wavelength fluctuations in the
thresholding model. Long-wavelength modes locally act as separate
universes that assist halo formation by pushing small-scale
fluctuations closer to (or farther from) the critical threshold
$\delta_c=1.686$.}
\end{figure}

So far we have considered only the mean abundance of halos. In the
spherical collapse model every region evolves independently; in the
actual Universe the modes interact, and these interactions modulate
halo formation and induce inhomogeneities in the halo distribution.
Imagine a small region sitting in the background of a
long-wavelength mode $\delta_\ell$. By the equivalence principle
(cf.\ Chapter 3), the large-scale fluctuation acts as a tiny
separate universe with slightly more matter in it than average:
locally, the small-scale fluctuations grow a bit faster, and faster
growth means earlier collapse. The long mode therefore \emph{assists}
halo formation, see the cartoon in Fig.~\ref{fig:thresh}. The most
naive way to account for this is to weaken the collapse criterion in
such a region:
\begin{equation}\label{eq:thresh_crit}
  \delta_R(\qv) \geq \delta_c - \delta_\ell(\qv)\,,
\end{equation}
where $\qv$ is the initial (Langrangian) position.
Plugging this into the HMF~\eqref{eq:PSHMF} and expanding in
$\delta_\ell$,
\begin{equation}
\begin{split}
  M\frac{d\bar n}{dM}\bigg|_{\delta_\ell}
  &=\frac{\bar\rho_m}{M}\,
  f_\mathrm{PS}\!\left(\frac{\delta_c-\delta_\ell}{\sigma(R)}\right)
  \left|\frac{d\ln\sigma}{d\ln M}\right|
  = M\frac{d\bar n}{dM}\bigg|_{0}
  \times\left(1-\frac{d\ln f_\mathrm{PS}}{d\nu}\,
  \frac{\delta_\ell(\qv)}{\sigma(R)}+O(\delta_\ell^2)\right)\,.
\end{split}
\end{equation}
The coefficient of $\delta_\ell$ is the (Lagrangian) linear bias:
\begin{equation}\label{eq:b1L_thresh}
  b_1^L
  =-\frac{1}{\sigma(R)}\,\frac{d\ln f_\mathrm{PS}}{d\nu}
  \bigg|_{\nu=\delta_c/\sigma(R)}
  =\frac{\delta_c^2-\sigma^2(R)}{\delta_c\,\sigma^2(R)}\,,
\end{equation}
so that the Lagrangian-space overdensity of would-be halos
(proto-halos) is
\begin{equation}\label{eq:deltah_L}
  \delta_h(\qv)\equiv\frac{\bar n(\qv)-\bar n}{\bar n}
  = b_1^L\,\delta_1(\qv)\,.
\end{equation}
This is the first appearance of the tracer--matter relation. Note
that this is precisely the Lagrangian bias coefficient of
Chapter~7: the observed (Eulerian) tracers are obtained by advecting
the proto-halos with the flow, whence $b_1 = 1+b_1^L$,
Eq.~\eqref{eq:b1_lag}. Thus the thresholding model gives us an estimate
of $b_1^L$. High-mass ($\nu\gg1$) objects have
$b_1^L\simeq\delta_c/\sigma^2\gg 1$: rare peaks are strongly
biased.

This result can be derived more rigorously from the same
thresholding picture used for the mean HMF, by studying the
clustering of the regions above threshold. The two-point function of
these regions --- our proxy for the halo two-point function ---
measures the excess probability of finding \emph{two} spheres, at
Lagrangian separation $\rv$, both above the threshold:
\begin{equation}
  p_2\big(\delta_R(\qv+\rv)>\delta_c,\;\delta_R(\qv)>\delta_c\big)
  = p_1^2\left(1+\xi_h^L(r)\right)\,,
\end{equation}
where the superscript $L$ reminds us that we work in Lagrangian
coordinates. We need the joint PDF of the smoothed field at two
points. Repeating the steps that led to Eq.~\eqref{eq:PS_pdf} with
two Lagrange multipliers, and noting that the second constraint,
$\delta_{R_2}=\int d^3q\,\delta_1(\qv)W_{R_2}(\qv+\rv)$, brings a
factor $e^{-i\kv\cdot\rv}$ in Fourier space, the Gaussian integrals
give
\begin{equation}\label{eq:pdf2joint}
  P(\delta_{R_1},\delta_{R_2})=
  \frac{1}{2\pi}\,
  \frac{1}{\sqrt{\sigma_{R_1}^2\sigma_{R_2}^2-\sigma^4_{R_1R_2}(\rv)}}\,
  \exp\Bigg\{-\frac{\sigma^2_{R_2}\delta^2_{R_1}
  +\sigma^2_{R_1}\delta^2_{R_2}
  -2\,\sigma^2_{R_1R_2}(\rv)\,\delta_{R_1}\delta_{R_2}}
  {2\left(\sigma_{R_1}^2\sigma^2_{R_2}-\sigma^4_{R_1R_2}(\rv)\right)}
  \Bigg\}\,,
\end{equation}
with the smoothed cross-correlation
\begin{equation}\label{eq:sigma12}
  \sigma^2_{R_1R_2}(\rv)\equiv
  \int_\kv P_{11}(k)\,e^{-i\kv\cdot\rv}\,
  \tilde{W}_{R_1}(kR_1)\,\tilde{W}_{R_2}(kR_2)\,.
\end{equation}
For equal radii, $\sigma^2_{RR}(\rv)=\xi_R(r)$ is simply the
filtered linear correlation function of matter. Integrating the
joint PDF over both densities above threshold and dividing by
$p_1^2$, with $p_1$ from Eq.~\eqref{eq:PsP}, we arrive at
\begin{equation}\label{eq:xihL}
  1+\xi_h^L(r)=\sqrt{\frac{2}{\pi}}\,
  \frac{1}{\left[\mathrm{erfc}(\nu/\sqrt{2})\right]^2}
  \int_{\nu}^\infty d\nu'\;e^{-\nu'^2/2}\,
  \mathrm{erfc}\!\left[\frac{\nu-\nu'\hat\xi_R(r)}
  {\sqrt{2\left(1-\hat\xi^2_R(r)\right)}}\right]\,,\qquad
  \hat\xi_R(r)\equiv\frac{\xi_R(r)}{\sigma^2(R)}\,.
\end{equation}
A formal Taylor expansion in $\xi_R(r)$ organizes this result as a
series of tracer--matter relations,
\begin{equation}\label{eq:bNL}
  \xi_h^L(r)=\sum_{N=1}^\infty \frac{1}{N!}\,(b_N^L)^2\,
  \left[\xi_R(r)\right]^N\,,\qquad
  b_N^L=\sqrt{\frac{2}{\pi}}\,
  \frac{1}{\mathrm{erfc}(\nu/\sqrt{2})}\,
  \frac{e^{-\nu^2/2}}{\sigma^N(R)}\,
  \mathrm{He}_{N-1}(\nu)\,,
\end{equation}
where $\mathrm{He}_N(x)$ are the Hermite
polynomials, $\mathrm{He}_0=1$, $\mathrm{He}_1(x)=x$,
$\mathrm{He}_2(x)=x^2-1$, etc. In the high-mass limit
$\sigma(R)\to 0$ the bias parameters simplify,
\begin{equation}
  b_N^L\Big|_{\nu\gg 1}=\frac{\delta_c^N}{\sigma^{2N}(R)}\,,
\end{equation}
which correctly reproduces the naive
estimate~\eqref{eq:b1L_thresh} in the same limit. Importantly, the above 
result, while being statistical in nature, can be reproduced
by promoting the Lagrangian proto-halo overdensity field $\delta_g$
to be a power-law series in the local dark matter overdensity field $\delta_1$~\cite{Ferraro:2012bd}:
\begin{equation}
  \delta_h^L = f(\delta_m=\delta_1) = b_1\,\delta_1 + \frac{b_2}{2}\,\delta_1^2 + \cdots\,.
\end{equation}
This forms a basis for the deterministic galaxy bias. 

Let us finish with a historic remark. Simple as it is, the
thresholding model provided the first mechanism correlating
small-scale structure --- halos and galaxies --- with the
long-wavelength cosmological fluctuations. This was a breakthrough:
it explained for the first time why galaxy
clustering should depend on cosmology~\cite{Kaiser:1984sw,Bardeen:1985tr}.
At the same time the model is very naive. Tidal fields
should participate as well --- they act as an \emph{anisotropic}
separate universe modulating proto-halo formation --- so the halo
density must respond to the tidal tensor too. A fixed threshold and
a specific window function are further simplifications that are hard
to justify. On large scales, the lesson we retain is the structure
of the answer: the tracer density is a function of the long-wavelength
environment,
\begin{equation}
  \delta_g = f(\delta_m) = b_1\,\delta_m + \frac{b_2}{2}\,\delta_m^2 + \cdots\,,
  \qquad
  P_{gg}^\mathrm{tree}(k) = b_1^2\,P_{11}(k)\,,
\end{equation}
with a single parameter capturing \emph{any} galaxy selection at
linear order. This is the simplification principle 
of EFT at play again: just like the potential of any 
charge distribution on large scales is characterized only by a single parameter, 
its charge,
the galaxy overdensity is charachterized by \textbf{linear bias} $b_1$.
In what follows we promote this observation to 
systematic, symmetry-based EFT expansion~\cite{McDonald:2009dh,Assassi:2014fva,Mirbabayi:2014zca,Senatore:2014eva} (see~\cite{Desjacques:2016bnm} for an excellent review).
The galaxy overdensity
is a function of the underlying degrees of freedom sourced by 
the matter field. 
EFT writes down a symmetry-based Taylor approximation 
for this function.

%------------------------------------------------------------------
\subsection{The General EFT Bias Expansion}
%------------------------------------------------------------------

The local PBS ansatz $\delta_g=f(\delta_m)$ cannot be the whole
story. Guided by the symmetries, let us ask what the galaxy density
is actually allowed to depend on. By the equivalence principle, a
uniform potential or a uniform gravitational acceleration is
locally unobservable, so $\Phi$ and $\partial_i\Phi$
cannot appear: the allowed building blocks are the second
derivatives, i.e. $\partial_i\partial_j\Phi$ (the tidal tensor), the
velocity gradients $\partial_i v^j$, and the stochastic fields.
Moreover, galaxy formation takes a Hubble time, so, exactly as for
the effective stress tensor of Chapter 4, the dependence is
non-local in time and must be evaluated along the past fluid
trajectory. The full EFT expression is a functional
\begin{equation}\label{eq:bias_functional}
  \delta_g(\xv,\tau) = \int^{\tau} d\tau'\;
  F\!\left[\partial_i\partial_j\Phi(\xv_\mathrm{fl},\tau'),\;
  \partial_i v^j(\xv_\mathrm{fl},\tau'),\;
  \varepsilon(\xv_\mathrm{fl},\tau');\,\tau,\tau'\right]\,.
\end{equation}
$\varepsilon$ now is the stochastic field of galaxies which
is different from that of matter.
In perturbation theory this expression collapses, order by order,
to a \emph{local-in-time} expansion --- the same mechanism we saw in
\S4: each $\delta^{(n)}$ carries a fixed time dependence
$\propto\Dp^n$, so all the memory integrals reduce to constant
coefficients. At the one-loop order, we can write
$\delta_g = f\big(\partial_i\partial_j\Phihat,\,
\partial_i\partial_j\Phihat_v,\varepsilon\big)(\xv,\tau)$,
where $\Phihat$ and $\Phihat_v$ are the rescaled gravitational and
velocity potentials ($\Delta\Phihat = \delta$,
$\Delta\Phihat_v = \Theta$) --- the bias
expansion is \emph{effectively local in time}.

Organizing the expansion by the symmetries, up to third order (plus
the leading higher-derivative term) we get
\begin{equation}\label{eq:bias_expansion}
  \boxed{\delta_g = b_1\,\delta + \frac{b_2}{2}\,\left(\delta^2-\langle \delta^2\rangle \right) + b_{ \mathcal{G}_2}\, \mathcal{G}_2
  + \frac{b_3}{3!}\,\delta^3 + b_{ \mathcal{G}_2\delta}\, \mathcal{G}_2\,\delta + b_{ \mathcal{G}_3}\, \mathcal{G}_3 + b_{\Gamma_3}\,\Gamma_3
  + b_{\nabla^2\delta}\,\nabla^2\delta + \varepsilon + \cdots}
\end{equation}
with the tidal operators defined as
\begin{equation}
  \mathcal{G}_2(\Phihat) \equiv \left(\partial_i\partial_j\Phihat\right)^2 - \left(\Delta\Phihat\right)^2\,,\qquad
  \Gamma_3 \equiv  \mathcal{G}_2(\Phihat) -  \mathcal{G}_2(\Phihat_v)\,,
\end{equation}
and the cubic Galileon operator, built from the Hessian matrix
$(\mathcal{H}_{\Phihat})_{ij}\equiv\partial_i\partial_j\Phihat$:
\begin{equation}\label{eq:G3_def}
   \mathcal{G}_3(\Phihat) \equiv
  \left(\Delta\Phihat\right)^3
  - 3\,\Delta\Phihat\,\left(\partial_i\partial_j\Phihat\right)^2
  + 2\,\partial_i\partial_j\Phihat\;\partial^j\partial^k\Phihat\;\partial_k\partial^i\Phihat
  = 6\,\det\left(\partial_i\partial_j\Phihat\right)\,.
\end{equation}
The last equality follows from Newton's identity
$\det M = \frac{1}{6}\big[(\tr M)^3 - 3\tr M\,\tr M^2 + 2\tr M^3\big]$
for a $3\times 3$ matrix. Together with $\delta^3=(\Delta\Phihat)^3$
and $ \mathcal{G}_2\,\delta$, the operator $\mathcal{G}_3$ completes the basis of cubic
rotational invariants of the tidal tensor: at this order one can
build only $\tr\mathcal{H}^3_{\Phihat}$, $\tr\mathcal{H}^2_{\Phihat}\tr\mathcal{H}_{\Phihat}$
and $(\tr\mathcal{H}_{\Phihat})^3$, and $\{\delta^3, \mathcal{G}_2\delta, \mathcal{G}_3\}$
is a convenient repackaging of these three structures.
Several comments are in order:
\begin{itemize}[nosep]
  \item The \textbf{tidal bias} $b_{\mathcal{G}_2}$ is the first qualitatively
  new ingredient beyond PBS: the equivalence principle allows the
  galaxy field to respond to the full tidal tensor
  $\partial_i\partial_j\Phihat$, not just to its trace $\delta$.
  $\mathcal{G}_2$ starts at second order in perturbations.
  \item $\Gamma_3$ starts at \emph{third} order, since
  $\Phihat=\Phihat_v$ at linear order; it is the operator through
  which the velocity potential first enters.
  \item $b_{\nabla^2\delta}\,\nabla^2\delta$ is the leading
  \textbf{higher-derivative} bias: the coefficient scales as the
  square of the spatial scale of galaxy formation (e.g.\ the halo radius), $b_{\nabla^2\delta}\sim R_*^2$.
  \item $\varepsilon$ is the \textbf{stochastic} field --- the part of
  the galaxy field uncorrelated with the long-wavelength
  cosmological fields (see \S5.4).
  \item  The above expression is a nested expansion
  in $\delta_1$. The bias operators are functions of the non-linear density
  field $\delta$ and its tidal potential, which themselves need to be expanded 
  in $\delta_1$ using the matter EFT. This is a Eulerian bias expansion
  formulated in terms of the finite-time matter field.
\end{itemize}

%------------------------------------------------------------------
\subsection{Galaxy Kernels, One-Loop Power Spectrum, and Redundant Operators}
%------------------------------------------------------------------

Plugging the perturbative solutions for $\delta$ and $\Theta$ into
the bias expansion defines the \textbf{galaxy kernels} $K_n$,
analogous to the SPT kernels $F_n$:
\begin{align}
  \delta_g^\mathrm{det}(\kv) &= b_1\,\delta_1(\kv) + \int_\qv K_2(\kv-\qv,\qv)\;\delta_1(\kv-\qv)\,\delta_1(\qv)\nonumber\\
  &\quad + \int_{\qv_1,\qv_2,\qv_3}(2\pi)^3\delta_D^{(3)}(\kv-\qv_{123})\;K_3(\qv_1,\qv_2,\qv_3)\;\delta_1(\qv_1)\delta_1(\qv_2)\delta_1(\qv_3) + \cdots
\end{align}
At second order,
\begin{equation}\label{eq:K2}
  K_2(\kv_1,\kv_2) = b_1\,F_2(\kv_1,\kv_2) + \frac{b_2}{2}
  + b_{\mathcal{G}_2}\left[\frac{(\kv_1\cdot\kv_2)^2}{k_1^2\,k_2^2} - 1\right]\,,
\end{equation}
while $K_3$ combines $b_1 F_3$ with the SPT-evolved quadratic and cubic bias operators.
Repeating the steps of \S3.3, the deterministic part of the
\textbf{one-loop galaxy power spectrum} is
\begin{equation}\label{eq:Pgg_1loop}
  \boxed{P_{gg}^{1\text{-loop}}(k)\Big|_{\rm det} = b_1^2\,P_{11}(k)
  + 2\int_\qv \left[K_2(\kv-\qv,\qv)\right]^2 P_{11}(|\kv-\qv|)\,P_{11}(q)
  + 6\int_\qv K_3(\kv,\qv,-\qv)\;P_{11}(k)\,P_{11}(q)\,.}
\end{equation}

\medskip\noindent
\textbf{Redundant operators.}
An important simplification takes place in
Eq.~\eqref{eq:Pgg_1loop}: the cubic operators $\delta^3$, $\mathcal{G}_3$ and
$G_2\delta$ \emph{do not contribute} to the one-loop power spectrum.
To see why, consider the $P_{13}$-type contribution of the
$\delta^3$ term:
\begin{equation}
  2\,\avg{b_1\delta_1\,,\;\tfrac{b_3}{3!}\,\delta_1^3}'
  = b_1 b_3\;\sigma_\Lambda^2\;P_{11}(k)\,,\qquad
  \sigma_\Lambda^2 \equiv \int^{\Lambda} d^3q\;P_{11}(q)\,,
\end{equation}
which diverges as $\Lambda\to\infty$. Crucially, its $k$-dependence
is \emph{exactly} that of the linear bias term, so it is completely
absorbed by the renormalization. Indeed, combining this $P_{13}$
contribution with the linear one we get
\be
b_1^2P_{11} + b_1 b_3\;\sigma_\Lambda^2\;P_{11}(k)=\left(b_1+
\frac{b_3}{2}\sigma^2_\Lambda
\right)^2P_{11}(k) + O(\sigma^4_\Lambda P_{11}(k))\,,
\ee 
where $O(\sigma^4_\Lambda)$  are two-loop corrections which we 
can ignore at the  one-loop order. 
We see that the physical predictions depend on a combination
of $b_1$ and $b_3\sigma^2_\Lambda$. Hence $b_1$ should be interpreted
as a ``bare'' parameter, while the relevant combination
is the physical renormalized linear bias:
\begin{equation}
  b_1^\mathrm{ren} = b_1 + \frac{b_3}{2}\,\sigma_\Lambda^2+\cdots \,.
\end{equation}
Similarly including $\mathcal{G}_2\delta$ and the \textit{SPT-evolved}
$b_2\delta^2$ (i.e. plugging the SPT formula 
$\delta=\delta_1+\int F_2\delta_1^2$ into
$\delta^2$),
we end up with the following net expression:
\be \label{eq:b1_ren_full}
  b_1^\mathrm{ren} = b_1+\left(\frac{b_3}{2}+\frac{34b_2}{21}-\frac{4b_{\mathcal{G}_2\delta}}{3}\right)\sigma^2_\Lambda\,.
\ee  
% \medskip\noindent
% \textbf{Homework.} Derive the above expression. 

\medskip\noindent
As far as the cubic Galileon  $\mathcal{G}_3$  is concerned, 
its one-loop power spectrum contribution vanishes identically.
The parameter measured from the data is always the renormalized one;
$\delta^3$, $\mathcal{G}_3$, $\mathcal{G}_2\delta$ add nothing new and are
\textbf{redundant} at this order. 
In practice, it is convenient to use \textit{dimensional regularization}
scheme, in which all convergent integrals are evaluated at $\Lambda\to \infty$,
but all divergent integrals are 
set to zero~\cite{Pajer:2013jj,Simonovic:2017mhp}.\footnote{See~\cite{Rubira:2023vzw,Bakx:2026rmd} for an explicit cutoff scheme.} 
In this scheme $\sigma^2_\Lambda=0$,
so $b_1$ measured from a linear bias fit is the same as $b_1$ measured 
at the one-loop order.

The only cubic operator that
survives in the one-loop power spectrum is $\Gamma_3$. 
The
independent deterministic parameters at one loop are therefore
\begin{equation}
  b_1\,,\quad b_2\,,\quad b_{\mathcal{G}_2}\,,\quad b_{\Gamma_3}\,,\quad b_{\nabla^2\delta}\,,
\end{equation}
supplemented by the stochastic contributions to which we now turn.

%------------------------------------------------------------------
\subsection{Galaxy Stochasticity and Shot Noise}
%------------------------------------------------------------------

Just like matter, the galaxy field contains a component that is not determined by the
long-wavelength fields at all:
\begin{equation}
  \delta_g = \delta_g^\mathrm{det} + \varepsilon\,,\qquad
  \avg{\delta_g^\mathrm{det}\,\varepsilon} = 0\,.
\end{equation}
This stochastic field is required both physically --- galaxy
formation depends on the small-scale environment, which is
decoupled from the long modes --- and by consistency of the EFT.
Indeed, consider the $P_{22}$-type contribution of the quadratic
bias:
\begin{equation}
  P_{gg}(k) \supset \frac{b_2^2}{4} P_{\delta^2\delta^2}(k)
  = \frac{b_2^2}{2}\int d^3q\;P_{11}(q)\,P_{11}(|\kv-\qv|)\,.
\end{equation}
In the UV limit $k\to 0$:
\begin{equation}\label{eq:k0Pd2d2}
  P_{\delta^2\delta^2}\Big|_{k\to 0}
  \approx \frac{b_2^2}{2}\left[\int_0^{\knl} d^3q\;P_{11}^2(q)
  + \int_{\knl}^{\Lambda} d^3q\;P_{11}^2(q)\right]
  = \text{calculable constant} + C_0^\Lambda\,.
\end{equation}
The small scales decouple and produce \emph{constant},
$k$-independent power --- including a cutoff-dependent piece
$C_0^\Lambda$ that must be cancelled by a constant counterterm.
This is supplied by the stochastic field: since $\varepsilon$ is
generated at $q\gtrsim\knl$, its power spectrum is analytic for
$k\ll\knl$,
\begin{equation}
  \avg{\varepsilon\,\varepsilon}'_{\kv} = \alpha_0 + \alpha_1\,k^2 + \cdots\,,
\end{equation}
and the total power
$\avg{\delta_g\delta_g}' = \avg{\delta_g^\mathrm{det}\delta_g^\mathrm{det}}' + \avg{\varepsilon\varepsilon}'$
is rendered cutoff-independent by the split
$\alpha_0 = \alpha_0^\Lambda + \alpha_0^\mathrm{fin}$, with
$C_0^\Lambda + \alpha_0^\Lambda$ finite. Note the contrast with the
matter field: there, mass and momentum conservation (Peebles'
argument of Chapter 5) forced the stochastic power to scale as
$k^4$. The galaxy \emph{number} density obeys no such conservation
laws: the galaxies keep forming by collapsing matter 
and being destroyed via merging throughout the cosmic history.
Thus a constant $\alpha_0$ is allowed and is indeed present.

\medskip\noindent
\textbf{Shot noise.}
How large is $\alpha_0$? A useful estimate comes from a set of $N$
randomly placed points (Poisson sampling) with mean density
$\nbar=N/V$. The density fluctuations then 
\begin{equation}
  \delta_g(\xv) = \frac{1}{\nbar}\sum_{i=1}^{N}\delta_D^{(3)}(\xv-\xv_i) - 1
  \qquad\Longrightarrow\qquad
  \delta_g(\kv) = \frac{1}{\nbar}\sum_i e^{i\kv\cdot\xv_i}\,,\quad \kv\neq 0\,.
\end{equation}
Noting that $\delta^{(3)}_D(\kv=0)=V$, the power spectrum is then
\begin{equation}
  P(k) = \frac{\avg{\delta_g(\kv)\,\delta_g(-\kv)}}{V}
  = \frac{1}{V\nbar^2}\sum_{i,j}\avg{e^{i\kv\cdot(\xv_i-\xv_j)}}
  = \frac{1}{V\nbar^2}\Big(\underbrace{\sum_{i=j}1}_{=\,N}
  + \underbrace{\sum_{i\neq j}\avg{e^{i\kv\cdot(\xv_i-\xv_j)}}}_{=\,0\;\text{(no correlations)}}\Big)
  = \frac{1}{\nbar}\,.
\end{equation}
This is the \textbf{Poisson shot noise}. It sets the natural
normalization of the stochastic power,
\begin{equation}\label{eq:shot_noise}
  \avg{\varepsilon\,\varepsilon}'_{\kv}
  = \frac{1}{\nbar}\left(\tilde\alpha_0
  + \tilde\alpha_1\,\frac{k^2}{k_\mathrm{stoch}^2} + \cdots\right)\,,
\end{equation}
with $\tilde\alpha_0 = 1$ for a pure Poisson process. In practice
$O(1)$ deviations are expected, e.g.\ from \textbf{halo exclusion}~\cite{Baldauf:2013hka,Desjacques:2016bnm}
(halos have finite size and cannot overlap) and from the non-linear
environment dependence of galaxy formation, so $\tilde\alpha_0$ must
be treated as a free parameter.

Finally, let us make a comment 
on the calculable part of the $P_{\delta^2\delta^2}$
integral in Eq.~\eqref{eq:k0Pd2d2},
\be \label{eq:constPd2d2}
C_{\delta^2\delta^2}=\frac{b_2^2}{2} \int_0^{\knl} d^3q\;P_{11}^2(q)~\,.
\ee 
Based on naturalness, this part is physical, and hence should be
present in theory predictions. However, since it is fully degenerate with the shot noise, in practice it is often
subtracted from the deterministic power spectrum. 
This does not change 
the estimate~\eqref{eq:shot_noise} because 
$C_{\delta^2\delta^2}$ in~\eqref{eq:constPd2d2} is typically 
smaller than the true shot noise of the realistic galaxy samples.\footnote{As a point of comparison, we can consider DESI DR1 LRG2 sample at $z=0.7$,  
which has $b_2\approx 0.3$, $C_{\delta^2\delta^2}\approx 42~[\Mpch]^3$ and $\bar n^{-1}\simeq 4.7\cdot 10^3~[\Mpch]^3$~\cite{Chudaykin:2025aux}.}
However, there are important exceptions: the neutral hydrogen (HI) field~\cite{Obuljen:2022cjo}
and the Lyman-$\alpha$ forest~\cite{Ivanov:2023yla,deBelsunce:2025bqc}. 
These tracers have tiny 
physical stochastic power spectrum, so $C_{\delta^2\delta^2}\gg \avg{\varepsilon\,\varepsilon}'_{\kv}$. Absorbing $C_{\delta^2\delta^2}$ into 
$\avg{\varepsilon\,\varepsilon}'_{\kv}$ is a bad idea in this case because 
this inflates the amplitude of $\tilde{\alpha}_0$
that one has to allow in the fit. In such cases it is more practical
to keep $C_{\delta^2\delta^2}$  as part of the deterministic power
spectrum.

\begin{figure}
    \centering
    \includegraphics[width=0.89\linewidth]{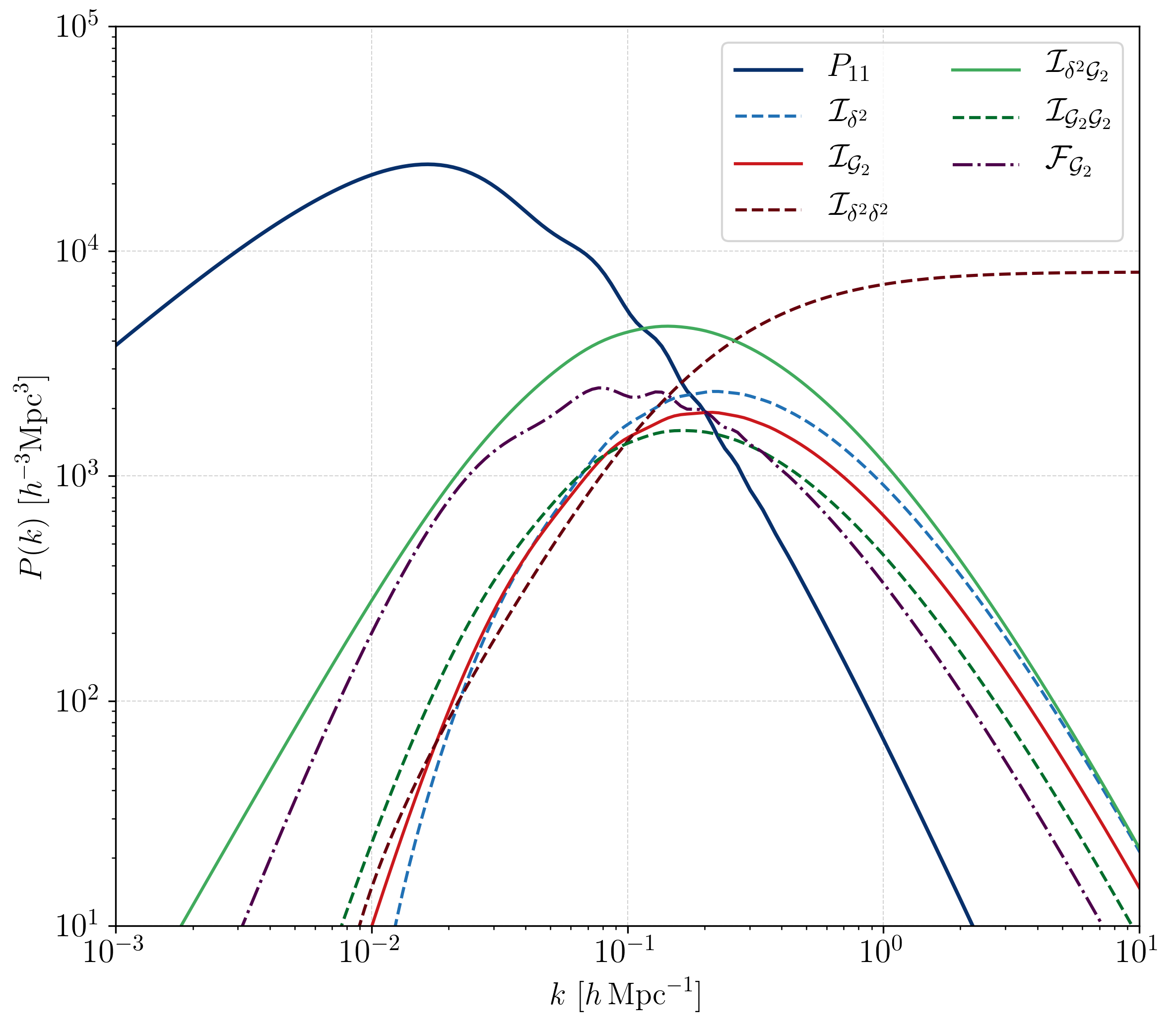}
    \caption{One-loop galaxy power spectrum basis shapes for the fiducial cosmology at $z=0$.}
    \label{fig:gal-shapes}
\end{figure}

\begin{figure}
    \centering
    \includegraphics[width=0.49\linewidth]{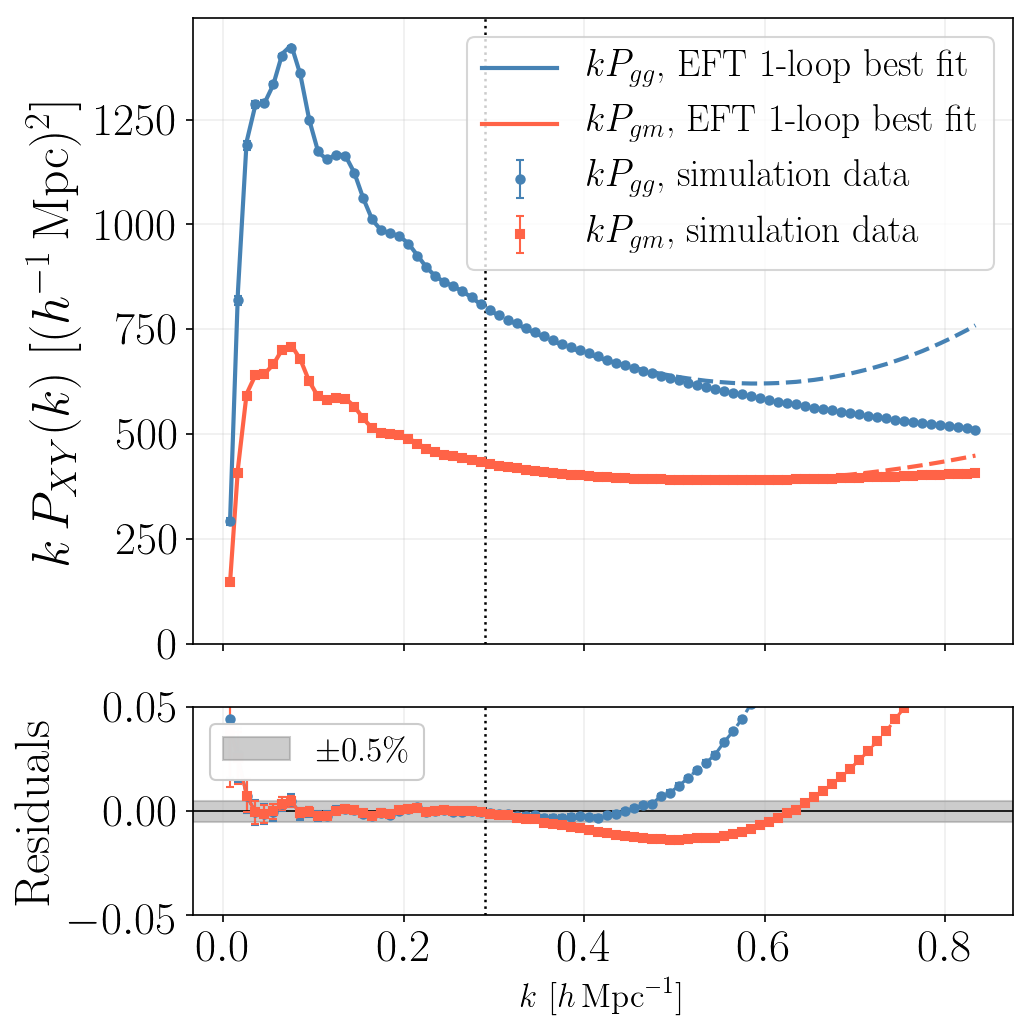}
    \includegraphics[width=0.49\linewidth]{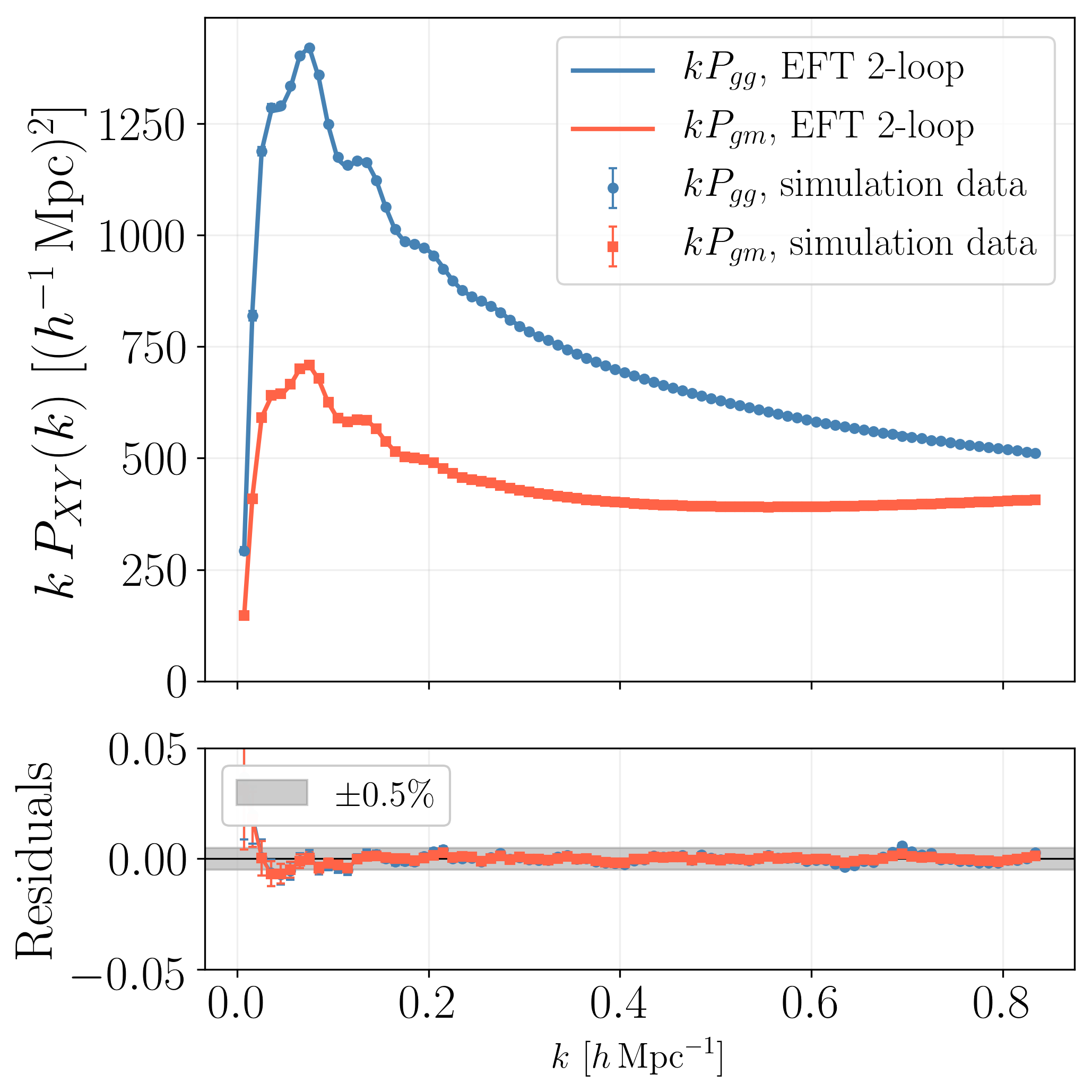}
    \caption{EFT fits to galaxy power spectrum and galaxy-matter 
    cross-spectrum of the mock luminous red galaxy sample from PT Challenge N-body simulation at $z=0.61$.
  \textbf{Left:} One-loop predictions discussed in the main text. 
  The EFT fit is accurate up to $k_{\rm max}=0.29~\hMpc$ (depicted by the vertical dashed line).
  \textbf{Right:}  Two-loop EFT predictions from Ref.~\cite{Ivanov:2026zos}, accurate up to $k_{\rm max}=0.85~\hMpc$.}
    \label{fig:fit_gal}
\end{figure}

\subsection{Comparison with Simulation Data}

\medskip\noindent
\textbf{The Final Theory Prediction.}
Collecting all the terms, the one-loop galaxy auto power spectrum reads
\be
\begin{split}
\label{eq:Pgg_explicit}
P_{gg}(k)=&~b^2_1\left(P_{11}(k)+\Delta P_{\rm 1\text{-}loop}(k)\right)
+b_1 b_2\,\mathcal{I}_{\delta^2}(k)
+2\,b_1 b_{\mathcal{G}_2}\,\mathcal{I}_{\mathcal{G}_2}(k)\\
&+b_1\left(2\,b_{\mathcal{G}_2} + \frac{4}{5}\,b_{\Gamma_3}\right)\mathcal{F}_{\mathcal{G}_2}(k)
+\frac{1}{4}\,b^2_2\,\mathcal{I}_{\delta^2\delta^2}(k)
+b^2_{G_2}\,\mathcal{I}_{\mathcal{G}_2 \mathcal{G}_2}(k)\\
&+\frac{1}{2}b_2\, b_{\mathcal{G}_2}\,\mathcal{I}_{\delta^2 \mathcal{G}_2}(k)
+P_{\nabla^2\delta}(k)+P_{\varepsilon\varepsilon}(k)\,,
\end{split}
\ee
where all spectra are evaluated at the same redshift (the time
dependence is left implicit), $P_{\varepsilon\varepsilon}$ is the
stochastic power spectrum of \S5.4, and we abbreviated the Fourier-space
kernel of the tidal operator $\mathcal{G}_2$ (the same combination that appears
in the quadratic galaxy kernel~\eqref{eq:K2}):
\be
\mathcal{S}(\kv_1,\kv_2) \equiv \frac{(\kv_1\cdot\kv_2)^2}{k_1^2\,k_2^2}-1\,.
\ee
The basis loop integrals are defined as\footnote{Note we have an extra factor of 2 in $\mathcal{I}_{\delta^2 \mathcal{G}_2}$ compared to~\cite{Simonovic:2017mhp,Chudaykin:2020aoj}. }
\begin{subequations}\label{eq:loop_integrals}
\begin{align}
&\mathcal{I}_{\delta^2}(k) \equiv 2\int_\qv F_2(\qv,\kv-\qv)\,
  P_{11}(|\kv-\qv|)\,P_{11}(q)\,,\\
&\mathcal{I}_{\mathcal{G}_2}(k) \equiv 2\int_\qv \mathcal{S}(\qv,\kv-\qv)\,
  F_2(\qv,\kv-\qv)\,P_{11}(|\kv-\qv|)\,P_{11}(q)\,,\\
&\mathcal{F}_{\mathcal{G}_2}(k) \equiv 4\,P_{11}(k)\int_\qv
  \mathcal{S}(\qv,\kv-\qv)\,F_2(\kv,-\qv)\,P_{11}(q)\,,\\
&\mathcal{I}_{\delta^2\delta^2}(k) \equiv 2\int_\qv
  P_{11}(|\kv-\qv|)\,P_{11}(q) - 2\int_\qv P^2_{11}(q)\,,\\
&\mathcal{I}_{\mathcal{G}_2 \mathcal{G}_2}(k) \equiv 2\int_\qv \mathcal{S}^2(\qv,\kv-\qv)\,
  P_{11}(|\kv-\qv|)\,P_{11}(q)\,,\\
&\mathcal{I}_{\delta^2 \mathcal{G}_2}(k) \equiv 4\int_\qv \mathcal{S}(\qv,\kv-\qv)\,
  P_{11}(|\kv-\qv|)\,P_{11}(q)\,,\\
&P_{\nabla^2\delta}(k) \equiv
  -2\,b_1\left(b_{\nabla^2\delta} + \gamma\,b_1\right) k^2\,P_{11}(k)\,,
\end{align}
\end{subequations}
where $\gamma$ is the matter counterterm of Chapter~4. Note that the
second term in $\mathcal{I}_{\delta^2\delta^2}$ subtracts the constant
$k\to 0$ limit of the first: this is precisely the UV constant of the
shot-noise renormalization discussed in \S5.4, which is degenerate with
(and absorbed into) $P_{\varepsilon\varepsilon}$. 
The one-loop bias shapes for our fiducial cosmology
at $z=0$ are displayed in Fig.~\ref{fig:gal-shapes}.

The galaxy--matter
cross spectrum relevant for lensing surveys 
is obtained in the same way:
\be
\begin{split}
\label{eq:Pgm_explicit}
P_{gm}(k)=&~b_1\left(P_{11}(k)+\Delta P_{\rm 1\text{-}loop}(k)\right)
+\frac{1}{2}\,b_2\,\mathcal{I}_{\delta^2}(k)
+b_{\mathcal{G}_2}\,\mathcal{I}_{\mathcal{G}_2}(k)\\
&+\left(b_{\mathcal{G}_2} + \frac{2}{5}\,b_{\Gamma_3}\right)\mathcal{F}_{\mathcal{G}_2}(k)
-\left(b_{\nabla^2\delta} + 2\,\gamma\,b_1\right) k^2\,P_{11}(k)+P_{\varepsilon\varepsilon_m}(k)\,.
\end{split}
\ee
An important point is that the stochastic cross-spectrum $P_{\varepsilon\varepsilon_m}(k)$ involves a matter field now, which scales as $k^2$ at $k\to 0$. Therefore,
its EFT expansion starts at $O(k^2)$:
\begin{equation}\label{eq:shot_noise_gm}
  P_{\varepsilon\varepsilon_m}
  = \frac{1}{\nbar}\tilde\alpha'_1\,\frac{k^2}{k_\mathrm{stoch}^2} + \cdots\,.
\end{equation}

\medskip\noindent
\textbf{IR resummation.} The bias expansion \eqref{eq:bias_expansion}
is built from operators that satisfy the equivalence principle, and thus
they do not involve locally unobservable displacements, which 
violate the perturbative expansion for the BAO. All the displacements
in the Eulerian bias expansion,
which do not cancel for modes $q\gtrsim \kosc$ and require resummation,
are produced by the SPT evolution. These can be resummed in much the same way
as the offending contributions in the matter case. The result is the same expression
\eqref{eq:IR_resum} but applied now to the galaxy power spectrum.

\medskip\noindent
\textbf{Comparison with simulations.} 
The EFT fits to the galaxy power spectrum 
and galaxy-matter cross-spectra 
from the PT Challenge simulation at $z=0.61$~\cite{Nishimichi:2020tvu}
are shown in Figure~\ref{fig:fit_gal}.
The left panel shows the one-loop fit, while the right panel 
displays the 
two-loop EFT model~\cite{Bakx:2025cvu,Ivanov:2026zos,Bakx:2026rmd}, which is currently the state-of-the-art. 
Note that at the two-loop order the one-loop operators $\delta^3,\delta\mathcal{G}_2$
and $\mathcal{G}_3$ are not redundant anymore.

\medskip\noindent
\textbf{A comment on scales.}
The EFT of galaxies contains several \textit{a priori} different expansion
scales: the matter non-linear scale $\knl$, the higher-derivative
bias scale ($b_{\nabla^2\delta}^{-1/2}\sim R_*^{-1}$), and the
stochastic scale $k_\mathrm{stoch}$ controlling the
scale-dependence of the stochasticity. For a generic galaxy sample
these may be \emph{parametrically different}, and which one fails
first depends on the sample.
For instance, massive halos and galaxies residing in them at $z\sim 0.5$
have large $R_*\simeq 5~[\Mpch]^2$, so the gradient expansion 
breaks down at $k \simeq 0.4~\hMpc$ for them~\cite{Ivanov:2024xgb,Ivanov:2025qie}. 
At the same time the stochasticity
scale 
is typically quite low even for massive halos, $k_\mathrm{stoch}\sim 1~\hMpc$.

\medskip\noindent
\textbf{Power counting.}
As a result of having different scales in the expansion, the power counting 
rules are somewhat different for galaxies. Whilst the galaxy bias coefficients 
do not change the power counting parametrically, e.g. $b_1\sim O(1)$
for realistic samples, the counting of the stochasticity is quite different. 
Galaxy samples used in real spectroscopic surveys satisfy\footnote{See e.g. DESI DR1 samples~\cite{Chudaykin:2025aux,DESI:2024aax}.}
$P_{gg}(k=0.2~\hMpc) \bar n\simeq 1$, so that at the BAO scales
$P_{\epsilon\epsilon}\sim \bar n^{-1}$ has the same counting as the linear power spectrum.
Then the $k^2$ stochasticity scales as one-loop deterministic power
spectrum. We stress, however, that the counting is different for other 
galaxy samples. For instance, photometric galaxies used in 
imaging surveys have  $P_{gg}(k=0.2~\hMpc) \bar n\gg 1$~\cite{Chen:2024vvk,Ivanov:2026dvl}. On the opposite part of the spectrum  we have
quasars, which live in the regime 
$P_{gg}(k=0.2~\hMpc) \bar n\ll 1$~\cite{Chudaykin:2022nru}.
The great advantage of EFT is that it provides the power counting rules,
which we can use to estimate what terms are important for any 
galaxy sample in question.

%==========================================================================
\section{Chapter 6: Redshift-Space Distortions}
%==========================================================================

There is one more distortion separating us from the observations.
A galaxy survey does not measure positions: it measures two angles
and a \emph{redshift}, $(\alpha,\delta,z)$, which are converted into
Cartesian coordinates assuming an FRW background. The observed
redshift, however, contains a Doppler contribution from the peculiar
velocity of the galaxy on top of the Hubble flow,
\begin{equation}
  z_\mathrm{obs} = H\,r_\mathrm{phys} + v_\mathrm{pec,\parallel}
  \equiv H\,s_\mathrm{RSD}\,,
\end{equation}
so the inferred (``redshift-space'') distance $s_\mathrm{RSD}$
differs from the true one. These are the
\textbf{redshift-space distortions} (RSD).

\subsection{The Mapping from Real to Redshift Space}

In comoving units and in the plane-parallel (distant-observer)
approximation, with the line of sight along $\hat{z}=(0,0,1)$, the
mapping reads
\begin{equation}\label{eq:RSD_mapping}
  \boxed{\sv = \xv + \frac{v_z}{\Hubble}\;\hat{z}\,,}
\end{equation}
where $v_z = \vvv\cdot\hat{z}$ is the line-of-sight peculiar
velocity and $\Hubble$ is the conformal Hubble rate. The mapping
merely \emph{reshuffles} the galaxies along the line of sight, so
their number is conserved:
\begin{equation}
  \left(1+\delta_g^{(s)}(\sv)\right) d^3s = \left(1+\delta_g(\xv)\right) d^3x\,.
\end{equation}
Multiplying by $e^{-i\kv\cdot\sv}$ and integrating, we obtain an
\emph{exact} expression for the redshift-space density field in
Fourier space:
\begin{equation}\label{eq:delta_s_exact}
  \delta_g^{(s)}(\kv) = \delta_g(\kv)
  + \int d^3x\;e^{-i\kv\cdot\xv}
  \left(e^{-ik_z v_z(\xv)/\Hubble} - 1\right)\left(1+\delta_g(\xv)\right)\,.
\end{equation}
Note that this expression is fully non-perturbative in the velocity
field --- all the approximations will enter only when we start
expanding the exponential.

%------------------------------------------------------------------
\subsection{Kaiser Formula (Linear Theory)}
%------------------------------------------------------------------

Let us first linearize Eq.~\eqref{eq:delta_s_exact}. It is
convenient to introduce the rescaled velocity
$u_z \equiv v_z/(f\Hubble)$, in terms of which the expansion
parameter is $f k_z u_z$. To linear order,
\begin{equation}
  \delta_g^{(s)}(\kv) = b_1\,\delta_{\kv}
  + \int d^3x\;e^{-i\kv\cdot\xv}\left(-if k_z\,u_z(\xv)\right)\,.
\end{equation}
For the linear growing mode $u^i = -\partial^i\delta_1/\Delta$
(cf.\ \S3.2), so in Fourier space
$u_z(\kv) = i\,(k_z/k^2)\,\delta_1(\kv)$ and
\begin{equation}
  \delta_g^{(s)}(\kv) = b_1\,\delta_1(\kv) + f\,\frac{k_z^2}{k^2}\,\delta_1(\kv)\,.
\end{equation}
We arrive at the \textbf{Kaiser formula}~\cite{Kaiser:1987qv}:
\begin{equation}\label{eq:Kaiser}
  \boxed{\delta_g^{(s)}(\kv) = \left(b_1 + f\mu^2\right)\delta_1(\kv)\,,\qquad
  \mu \equiv \frac{k_z}{k}\,,}
\end{equation}
and the linear (tree-level) redshift-space power spectrum
\begin{equation}\label{eq:tree_kaiser}
  \avg{\delta_g^{(s)}\,\delta_g^{(s)}}'_\mathrm{lin}
  = \left(b_1 + f\mu^2\right)^2 P_{11}(k)\,.
\end{equation}

\begin{tcolorbox}[colback=green!5!white, colframe=green!50!black, title=Physical Interpretation]
  \textbf{Squashing effect:} Coherent infall velocities towards overdensities enhance the apparent clustering along the line of sight. This is captured by the by $f\mu^2$ term in the Kaiser formula.
  This term also breaks isotropy: the power spectrum depends on $\mu = \cos\theta_{k,\hat{z}}$.
\end{tcolorbox}

\medskip\noindent
\textbf{Fingers of God.}
Virialized motions within clusters create elongation of structures along the line of sight in redshift space~\cite{Jackson:2008yv}.
This is a \emph{non-perturbative} effect: the velocities inside virialized halos are large and random.
In the EFT framework, Fingers of God are absorbed into the counterterms and stochastic contributions, as we will see below.

%------------------------------------------------------------------
\subsection{Multipole Expansion}
%------------------------------------------------------------------

The angular dependence is conveniently expanded in Legendre
polynomials,
\begin{equation}
  P_g^{(s)}(k,\mu) = \sum_\ell P_\ell(k)\,\mathcal{L}_\ell(\mu)\,,\qquad
  P_\ell(k) = \frac{2\ell+1}{2}\int_{-1}^{1} d\mu\;P_g^{(s)}(k,\mu)\;\mathcal{L}_\ell(\mu)\,.
\end{equation}
Expanding the Kaiser result
$(b_1+f\mu^2)^2 = b_1^2 + 2b_1 f\mu^2 + f^2\mu^4$
with the help of
$\mathcal{L}_2 = (3\mu^2-1)/2$ and
$\mathcal{L}_4 = (35\mu^4-30\mu^2+3)/8$,
we find that only $\ell = 0,2,4$ are non-zero at tree level:
\begin{align}
  P_0(k) &= \left(b_1^2 + \frac{2}{3}\,b_1 f + \frac{1}{5}\,f^2\right)P_{11}(k)\,,\\
  P_2(k) &= \left(\frac{4}{3}\,b_1 f + \frac{4}{7}\,f^2\right)P_{11}(k)\,,\\
  P_4(k) &= \frac{8}{35}\,f^2\,P_{11}(k)\,.
\end{align}
Non-linear evolution generates higher multipoles at loop level, but
the bulk of the information resides in $\ell=0,2,4$.
The reason this basis is convenient to use 
is that the multipoles can be
easily estimated from the data~\cite{Scoccimarro:2015bla}.

%------------------------------------------------------------------
\subsection{Beyond Linear Order --- RSD Kernels}
%------------------------------------------------------------------

To go beyond linear theory, we expand the exponential in
Eq.~\eqref{eq:delta_s_exact} in powers of $fk_z u_z$. For the
one-loop power spectrum we need terms up to cubic order:
\begin{equation}\label{eq:delta_s_expanded}
  \delta_g^{(s)}(\kv) = \delta_g(\kv)
  - if k_z\left[(1+\delta_g)\,u_z\right]_{\kv}
  + \frac{(if k_z)^2}{2}\left[(1+\delta_g)\,u_z^2\right]_{\kv}
  + \frac{(if k_z)^3}{6}\left[(1+\delta_g)\,u_z^3\right]_{\kv}\,,
\end{equation}
where $[f]_{\kv} \equiv \int d^3x\;e^{-i\kv\cdot\xv} f(\xv)$.
Substituting the perturbative expansions of $\delta_g$ (bias
kernels $K_n$) and of the velocity (SPT kernels $G_n$) defines the
\textbf{redshift-space kernels} $Z_n$:
\begin{equation}
  \delta_g^{(s)}(\kv) = \sum_{n=1}^\infty (2\pi)^3\int_{\qv_1\cdots\qv_n}
  \delta_D^{(3)}(\kv-\qv_{1\cdots n})\;
  Z_n(\qv_1,\ldots,\qv_n)\;\delta_1(\qv_1)\cdots\delta_1(\qv_n)\,.
\end{equation}
The first two kernels are
\begin{align}
  Z_1(\kv) &= b_1 + f\mu^2\,,\\
  Z_2(\qv_1,\qv_2) &= K_2(\qv_1,\qv_2)
  + f\mu_{12}^2\,G_2(\qv_1,\qv_2)
  + \frac{f\mu_{12}k_{12}}{2}\left[\frac{\mu_1}{q_1}\left(b_1+f\mu_2^2\right)
  + \frac{\mu_2}{q_2}\left(b_1+f\mu_1^2\right)\right]\,,
\end{align}
where $\mu_i = q_{i,z}/q_i$ and $\mu_{12} = (q_{1,z}+q_{2,z})/|\qv_1+\qv_2|$;
the expression for $Z_3$ is analogous but lengthy (see e.g.~\cite{Ivanov:2019pdj}).
The deterministic one-loop redshift-space power spectrum is then built exactly as
in real space,
\begin{equation}
\begin{split}
  P_{gg,\,1\text{-loop}}^{(s)}(k,\mu) = & Z_1^2\,P_{11}(k)
  + 2\int_\qv Z_2^2(\kv-\qv,\qv)\,P_{11}(|\kv-\qv|)P_{11}(q) \\
  & + 6\,Z_1(\kv)\int_\qv Z_3(\kv,\qv,-\qv)\,P_{11}(k)P_{11}(q)\,.
  \end{split}
\end{equation}

%------------------------------------------------------------------
\subsection{Renormalization of Contact Operators}
%------------------------------------------------------------------

The building blocks of the RSD expansion~\eqref{eq:delta_s_expanded}
are \textbf{contact operators}: products of fields evaluated at the
same point, such as $\left[\delta_g(\xv)\,u^i(\xv)\right]_{\kv}$.
In Fourier space such a product is a convolution over \emph{all}
momenta, so it is UV-sensitive: short modes of both fields
contribute to it even when the external momentum $k$ is small.
These operators must be renormalized~\cite{Senatore:2014vja,Perko:2016puo,Bakx:2025pop}.

The procedure follows the same logic as everywhere in this course.
Split every field into long- and short-wavelength parts,
\begin{equation}
  \delta_g = \delta_g^{\ell} + \delta_g^{s}\,,\qquad
  u_i = u_i^{\ell} + u_i^{s}\,,
\end{equation}
and average over the short modes with the long modes held fixed;
we denote this average by $\avg{\ldots}_\mathrm{s.s.}$. This is
exactly the \textbf{Born--Oppenheimer} logic of molecular physics:
the fast (short-scale) degrees of freedom are integrated out in the
background of the slow (long-wavelength) ones, and the result is an
effective operator built from the slow fields.

As a warm-up, consider the galaxy density itself:
\begin{equation}
  \avg{\delta_g}_\mathrm{s.s.} = \delta_g^{\ell} + \avg{\delta_g^{s}}_\mathrm{s.s.}\,.
\end{equation}
The short-mode average in the presence of the long fields is itself
a function of the long fields,
$\avg{\delta_g^{s}}_\mathrm{s.s.} = \sum_{\tilde O} z_{\tilde O}\,\tilde O
= b_1'\,\delta_\ell + b_2'\,\delta_\ell^2 + \cdots$,
so it merely reshuffles the bias coefficients:
$\avg{\delta_g}_\mathrm{s.s.} \equiv \delta_g^\mathrm{R}$. Nothing
new here. The novelty appears for the velocity-weighted operators:
\begin{equation}\label{eq:BO_u}
  \avg{(1+\delta_g)\,u_i}_\mathrm{s.s.}
  = (1+\delta_g^{\ell})\,u_i^{\ell}
  + \avg{\delta_g^{s}}_\mathrm{s.s.}\,u_i^{\ell}
  + \avg{(1+\delta_g^{\ell}+\delta_g^{s})\,u_i^{s}}_\mathrm{s.s.}
  \equiv \left(1+\delta_g^\mathrm{R}\right)u_i^{\ell} + O_u^{i}\,,
\end{equation}
where the last term defines a genuinely new operator $O_u^i$: the
response of the short-scale ``atmosphere'' to the long-wavelength
environment. Note that $O_u^i$ is a \emph{scalar under Galilean
transformations}: a boost is carried entirely by $u_i^\ell$, while
the short modes only know about locally observable quantities.
Analogously,
\begin{align}
  \avg{(1+\delta_g)\,u_i u_j}_\mathrm{s.s.}
  &= \left(1+\delta_g^\mathrm{R}\right)u_i^{\ell}u_j^{\ell}
  + u_i^{\ell}\,O_u^{j} + u_j^{\ell}\,O_u^{i} + O_{u^2}^{ij}\,,\\
  \avg{(1+\delta_g)\,u_i u_j u_k}_\mathrm{s.s.}
  &
  % = [u^i_{\ell} u^j_{\ell} u^k_{\ell} (1+\delta_g) ]_R 
  = 
(1+\delta^R_{g})
u_i^{\ell} 
u_j^{\ell} 
u_k^{\ell}   
+ (u^{\ell}_i u^{\ell}_j O_{u}^{k} +\text{2 perm.})
+ (u^{\ell}_i O_{u^2}^{jk} +\text{2 perm.})
+ O_{u^3}^{ijk} \,.
  % \cdots + O_{u^3}^{ijk}\,.
\end{align}
The EFT now tells us what these operators can be. By symmetries
(Galilean invariance, the equivalence principle, rotational
invariance) and dimensional analysis, at leading order in
perturbations and derivatives:
\begin{equation}\label{eq:O_operators}
  O_u^{i} = c_1\,\partial^i\delta + \cdots\,,\qquad
  O_{u^2}^{ij} = c_2\,\delta^{ij}\,\delta + c_3\,\partial^i\partial^j\Phihat + \cdots
\end{equation}
A Galilean-invariant vector must be a gradient of a scalar, and a
symmetric tensor admits exactly the two structures shown.

Let us now insert the renormalized operators into the
mapping~\eqref{eq:delta_s_expanded} and project onto the line of
sight. Using $\partial^i\partial^j\Phihat \to (k^ik^j/k^2)\,\delta$
in Fourier space, we find
\begin{align}
  -if k_z\,\hat{z}_i\left[\avg{(1+\delta_g)u^i}_\mathrm{s.s.}\right]_{\kv}
  &= -if k_z\,\hat{z}_i\left[(1+\delta_g^\mathrm{R})\,u^i_{\ell}\right]_{\kv}
  + c_1 f\,k_z^2\,\delta_1(\kv)\,,\\
  \frac{(if k_z)^2}{2}\,\hat{z}_i\hat{z}_j\left[\avg{(1+\delta_g)u^iu^j}_\mathrm{s.s.}\right]_{\kv}
  &= -\frac{f^2 k_z^2}{2}\left[(1+\delta_g^\mathrm{R})\,u_\ell^iu_\ell^j\,\hat{z}_i\hat{z}_j\right]_{\kv}
  -\frac{f^2 k_z^2}{2}\left(c_2 + c_3\,\mu^2\right)\delta_1(\kv)\,,
\end{align}
where all higher-order terms in $\delta_1$ have been dropped (they
contribute beyond one loop). Collecting the new pieces together
with the real-space higher-derivative bias, the renormalized
redshift-space field takes the form
\begin{equation}
  \delta_g^{(s)} = \delta^{(s)}_\mathrm{SPT}
  - b_{\nabla^2\delta}\,k^2\,\delta_1
  + k_z^2\left(c_1 f - \frac{1}{2}\,f^2 c_2 + \frac{1}{2}\,f^2 c_3\,\mu^2\right)\delta_1\,,
\end{equation}
(the signs of the arbitrary constants are conventional). Using
$k_z^2 = k^2\mu^2$, all counterterms combine into
\begin{equation}\label{eq:RSD_ctr}
  \boxed{\Delta\delta^{(s)}_\mathrm{ctr}(\kv)
  = k^2\,\delta_1(\kv)\left[C_0 + \tilde{C}_1\,f\mu^2 + \tilde{C}_2\,f^2\mu^4\right]\,,}
\end{equation}
where $C_0 = -b_{\nabla^2\delta}$ is inherited from real space, while
$\tilde{C}_1$ and $\tilde{C}_2$ are the two genuinely new RSD
counterterms. Correlating with the tree-level field, the
counterterm contribution to the power spectrum is
$P^{(s)}_\mathrm{ctr} = 2\left(b_1+f\mu^2\right)
\left[C_0 + \tilde{C}_1 f\mu^2 + \tilde{C}_2 f^2\mu^4\right] k^2 P_{11}(k)$.

\medskip\noindent
\textbf{Renormalization.} This structure is enough to renormalize the
one-loop power spectrum in redshift space. Note that there is
\emph{no} $\mu^6 k^2$ term at the level of the field: the symmetries
allow only the two tensor structures in
Eq.~\eqref{eq:O_operators}, so the angular dependence of the
counterterms truncates at $\mu^4$. This is a non-trivial structural
prediction of the EFT --- a purely phenomenological fit would have
no reason to stop at $\mu^4$.

\medskip\noindent
\textbf{Non-renormalization of the growth factor.}
Importantly, the condition $\langle \delta_g^{(s)}\rangle=0$
and the linear bias renormalization Eq.~\eqref{eq:b1_ren_full}
dictate that RSD loops \textbf{do not} generate power spectrum
corrections $\propto \mu^2 P_{11}\sigma^2_\Lambda, \mu^4 P_{11}\sigma^2_\Lambda$. Therefore, the cosmology-dependent growth factor 
in the Kaiser formula~\eqref{eq:tree_kaiser} is protected
from complications of the galaxy formation physics.~\footnote{A contamination,
however, may happen if the galaxy sample is selected in an orientation-dependent way~\cite{2009MNRAS.399.1074H}. However, a 
chain of unlucky events need to happen for such ``selection effects''
to get generated. First, galaxies must have a strong enough alignment with the ambient tidal field. And second, 
they should be preferentially selected based on their orientation. 
Selection effects are believed to be small for ongoing surveys 
like DESI, though see~\cite{Obuljen:2020ypy,Singh:2020cvu} for BOSS.
Also see~\cite{Desjacques:2018pfv} for galaxy EFT in the presence
of selection effects.
}

%------------------------------------------------------------------
\subsection{Stochasticity in Redshift Space}
%------------------------------------------------------------------

Exactly as in real space, the operators generated by the short-mode
averaging contain stochastic components, uncorrelated with the
long-wavelength fields:
\begin{equation}
  O_u^{i} \supset \varepsilon^{i}\,,\qquad
  O_{u^2}^{ij} \supset \varepsilon^{ij}\,.
\end{equation}
Retracing the steps of the previous section, the stochastic part of
the redshift-space galaxy field is
\begin{equation}
  \delta^{(s)}_{g,\mathrm{stoch}} = \varepsilon
  - if k_z\,\hat{z}_i\,\varepsilon^{i}
  - \frac{f^2 k_z^2}{2}\,\hat{z}_i\hat{z}_j\,\varepsilon^{ij}\,.
\end{equation}
The correlators of the stochastic fields are fixed by rotational
invariance of the underlying dynamics (the line of sight is not a
preferred direction of the \emph{dynamics}) and analyticity:
\begin{equation}
  \avg{\varepsilon^{i}\,\varepsilon^{j}}' = \delta^{ij}\,B_1 + \cdots\,,\qquad
  \avg{\varepsilon^{i}\,\varepsilon}' = i k^{i}\,B_2 + \cdots\,,\qquad
  \avg{\varepsilon^{ij}\,\varepsilon}' = \delta^{ij}\,B_3 + \cdots
\end{equation}
Assembling the stochastic power spectrum, every new term comes with
two powers of $k_z$ (one from each factor of the mapping or one
from the gradient in the cross-correlator), so
\begin{equation}\label{eq:Pstoch_RSD}
  \avg{\delta^{(s)}_{g,\mathrm{stoch}}\,\delta^{(s)}_{g,\mathrm{stoch}}}'
  = \avg{\varepsilon\varepsilon}' + A_2\,k_z^2
  = A_0 + A_1\,k^2 + A_2\,k^2\mu^2\,,
\end{equation}
where $A_2$ is a combination of $B_1$, $B_2$, $B_3$ (with the
appropriate powers of $f$). Normalizing to the Poisson shot noise
as in \S5.4,
\begin{equation}\label{eq:Pstoch_rsd}
  P^{(s)}_\mathrm{stoch}(k,\mu)
  = \frac{1}{\nbar}\left(\tilde\alpha_0
  + \tilde\alpha_1\,\frac{k^2}{\knl^2}
  + \tilde\alpha_2\,\frac{k^2\mu^2}{k_{\mathrm{RSD,stoch}}^2}\right)\,.
\end{equation}

%====================================================================
\subsection{The Alcock--Paczy\'nski effect}\label{sec:AP}
%====================================================================

Another important effect arises when we convert observed angles and 
redshifts to distances, which is called the \textit{The Alcock--Paczy\'nski 
(AP) distortions}. The spatial part of the flat FRW
metric in comoving spherical coordinates $(\chi,\theta,\phi)$ is
\begin{equation}
  \dd\ell^2 = \dd\chi^2 + \chi^2\,\dd\Omega^2\,,
\end{equation}
where $\chi$ is comoving distance.
A small radial separation in redshift maps to a comoving distance via
$\dd\chi = c\,\dd z/H(z)$, while a small angular separation
$\dd\theta_\perp$ maps to a transverse distance
$\dd\ell_\perp = D_M(z)\,\dd\theta_\perp$, where
$D_M(z)=(1+z)\,D_A(z)$ is the comoving angular diameter distance.
Defining Cartesian-like comoving coordinates aligned with the line of
sight,
\begin{equation}\label{eq:AP-dxobs}
  \dd x_\parallel = \frac{c}{H(z)}\,\dd z\,,\qquad
  \dd x_\perp = D_M(z)\,\dd\theta_\perp\,,
\end{equation}
the line element becomes $\dd\ell^2 = \dd x_\parallel^2 +
\dd x_\perp^2$, as it should in comoving space.  Both conversion
factors depend on the cosmology.

\paragraph{The AP distortion.}
In practice one must adopt a \emph{fiducial} cosmology
$(H^{\rm fid},\,D_M^{\rm fid})$ to turn $(z,\theta_\perp)$ into
distances.  If the true cosmology differs, the fiducial and true
coordinates are related by
\begin{equation}\label{eq:AP-rescaling}
  \dd x_\parallel^{\rm fid}
    = \tilde{q}_\parallel\,\dd x_\parallel^{\rm true}\,,\qquad
  \dd x_\perp^{\rm fid}
    = \tilde{q}_\perp\,\dd x_\perp^{\rm true}\,,
\end{equation}
with the dilation parameters
\begin{equation}\label{eq:AP-qs}
  {\;\tilde{q}_\parallel
    = \frac{H^{\rm true}(z)}{H^{\rm fid}(z)}\,,\qquad
  \tilde{q}_\perp
    = \frac{D_M^{\rm fid}(z)}{D_M^{\rm true}(z)}\;}\,.
\end{equation}
A sphere of radius $r$ in true comoving space maps to an ellipsoid
with semi-axes $\tilde{q}_\parallel\,r$ (radial) and $\tilde{q}_\perp\,r$
(transverse).  The anisotropy is governed by the AP ratio
\begin{equation}\label{eq:qAP}
  \tilde{q}_{\rm AP} \equiv \frac{\tilde{q}_\parallel}{\tilde{q}_\perp}
  = \frac{D_M^{\rm fid}(z)\,H^{\rm true}(z)}
         {D_M^{\rm true}(z)\,H^{\rm fid}(z)}\,,
\end{equation}
while the isotropic dilation is
$\tilde{q}_{\rm iso} = (\tilde{q}_\perp^2\,\tilde{q}_\parallel)^{1/3}$.

\paragraph{Units and the $h$-rescaling.}
In practice, Boltzmann codes such as \textsc{class} output
distances in Mpc and $H(z)$ in km\,s$^{-1}$Mpc$^{-1}$,
while the data analysis pipeline quotes wavenumbers in
$h_{\rm fid}/\text{Mpc}$ and power spectra in
$[\mathrm{Mpc}/h_{\rm fid}]^3$, having converted angles and
redshifts using the fiducial Hubble constant $h_{\rm fid}$.
The natural distance unit for the AP parameters is therefore
$h^{-1}\mathrm{Mpc}$, in which the $h$-dependence drops out
entirely.

The line-of-sight distance element in $h^{-1}\mathrm{Mpc}$ is
$\dd x_\parallel = c\,\dd z/(H/h) = c\,\dd z/(100\,E(z))$,
where $E(z)\equiv H(z)/H_0$ is the dimensionless Hubble
parameter.  Since $H/h = 100\,E(z)\;\mathrm{km\,s^{-1}Mpc^{-1}}$
is independent of $h$, the line-of-sight dilation depends only on
$E(z)$.  Similarly, the comoving angular diameter distance in
$h^{-1}\mathrm{Mpc}$ is
$D_M^{[h]} \equiv h\,D_M = (c/100)\int_0^z \dd z'/E(z')$,
which is also $h$-independent.  The AP dilation parameters are
therefore defined as
\begin{equation}\label{eq:AP-qs}
  \boxed{\;q_\parallel
    = \frac{E^{\rm true}(z)}{E^{\rm fid}(z)}\,,\qquad
  q_\perp
    = \frac{D_M^{{\rm fid},\,[h]}(z)}
           {D_M^{{\rm true},\,[h]}(z)}\;}\,,
\end{equation}
with the AP ratio
\begin{equation}\label{eq:qAP}
  q_{\rm AP} \equiv \frac{q_\parallel}{q_\perp}
  = \frac{D_M^{{\rm fid},\,[h]}(z)\;E^{\rm true}(z)}
         {D_M^{{\rm true},\,[h]}(z)\;E^{\rm fid}(z)}\,,
\end{equation}
and the isotropic dilation
$q_{\rm iso} = (q_\perp^2\,q_\parallel)^{1/3}$.
Since both $E(z)$ and $D_M^{[h]}$ are $h$-independent, the
parameters $q_\parallel$, $q_\perp$, and $q_{\rm AP}$ depend only on
background quantities such as $\Omega_m$, $\Omega_\Lambda$, the dark energy equation of state.

\paragraph{Remapping of wavevectors.}
Since Fourier modes are conjugate to position, wavevectors
(in $h\Mpc$) transform inversely to coordinates
(in $h^{-1}\mathrm{Mpc}$):
\begin{equation}\label{eq:AP-kmap}
  k_\parallel^{\rm true} = q_\parallel\,k_\parallel^{\rm fid}\,,\qquad
  k_\perp^{\rm true} = q_\perp\,k_\perp^{\rm fid}\,,
\end{equation}
where $k^{\rm fid}$ is in $h_{\rm fid}/\text{Mpc}$ and $k^{\rm true}$ is
in $h_{\rm true}/\text{Mpc}$.
Writing $k_\parallel = k\mu$ and $k_\perp = k\sqrt{1-\mu^2}$
(with $\mu\equiv\hat\k\cdot\hat{\vv{n}}$ the cosine to the line of
sight), the true wavenumber and angle expressed in terms of fiducial
quantities are
\be
  k^{\rm true}
  = k^{\rm fid}\,q_\perp\,F(\mu^{\rm fid})\,,
 \quad \quad \quad
  \mu^{\rm true}
  = \frac{q_{\rm AP}\,\mu^{\rm fid}}{F(\mu^{\rm fid})}\,,
  \label{eq:AP-mutrue}
\ee
where
\begin{equation}\label{eq:AP-F}
  F(\mu) \equiv \sqrt{1 + (q_{\rm AP}^2-1)\,\mu^2}\,.
\end{equation}
For $q_{\rm AP}=1$, $F=1$ and
$\mu^{\rm true}=\mu^{\rm fid}$; the remapping reduces to
$k^{\rm true}=q_\perp\,k^{\rm fid}$.

\paragraph{Effect on the power spectrum.}
The overdensity $\delta(\x)$ is a scalar: it does not change under
the coordinate remapping, only the volume element does.  The
Fourier-space volume transforms as
$\dd^3 k^{\rm fid} = (q_\parallel\,q_\perp^2)^{-1}\,
\dd^3 k^{\rm true}$,
and the Dirac delta function picks up the inverse Jacobian.
The operational formula used in data analyses is thus
\begin{equation}\label{eq:AP-Pk-practical}
  \boxed{\;P^{\rm obs}(k,\mu)
  = \frac{1}{q_\perp^2\,q_\parallel}\;
    P^{\rm true}\!\bigl(k\,q_\perp\,F(\mu),\;
      q_{\rm AP}\,\mu/F(\mu)\bigr)\;}\,,
\end{equation}
Note that the power spectra on the two sides carry different
$h$-units: $P^{\rm obs}$ is in $[{\rm Mpc}/h_{\rm fid}]^3$
(matching $k$ in $h_{\rm fid}/{\rm Mpc}$), while $P^{\rm true}$ from
\textsc{class} is in $[{\rm Mpc}/h_{\rm true}]^3$.
This unit difference is accounted for by the Jacobian. 
The three physical effects encoded in this formula are:
remapping of $k$ ($k^{\rm true}=q_\perp F\,k$),
remapping of $\mu$ ($\mu^{\rm true}=q_{\rm AP}\mu/F$), and the
volume prefactor ($1/q_\perp^2 q_\parallel$).

\paragraph{Combining with redshift-space distortions.}
In redshift space the true power spectrum is already anisotropic.
In the Kaiser approximation,
\begin{equation}\label{eq:Kaiser}
  P^{\rm true}(k,\mu) = b_1^2\bigl(1+\beta\,\mu^2\bigr)^2\,
    P_{\rm lin}(k)\,,\qquad \beta\equiv f/b_1\,,
\end{equation}
where $f=\dd\ln\Dp/\dd\ln a$ is the logarithmic growth rate and
$b_1$ the linear bias (with the $h$-rescaling absorbed).
Substituting Eq.~\eqref{eq:AP-mutrue} into Eq.~\eqref{eq:Kaiser}
and applying Eq.~\eqref{eq:AP-Pk-practical}:
\begin{equation}\label{eq:Pobs-full}
  P^{\rm obs}(k,\mu) = \frac{b_1^2}{q_\perp^2\,q_\parallel}\;
    \biggl(1 + \beta\,
      \frac{q_{\rm AP}^2\,\mu^2}{F^2(\mu)}\biggr)^{\!2}\;
    P_{\rm lin}\!\bigl(k\,q_\perp\,F(\mu)\bigr)\,.
\end{equation}
The observed multipoles are obtained by projecting onto Legendre
polynomials:
\begin{equation}
  P_\ell^{\rm obs}(k) = \frac{2\ell+1}{2}\int_{-1}^{1}\!\dd\mu\;
    P^{\rm obs}(k,\mu)\;\mathcal{L}_\ell(\mu)\,.
\end{equation}

\medskip\noindent
\textbf{Are we making a mistake?}
Finally, let is comment on a confusing point about the AP distortions. 
Often in the literature it is presented as if we are intentionally making
a mistake by assuming some wrong fiducial cosmology, which then need to be iterated. The correct interpretation
in the EFT-based full-shape analyses 
is that the AP distortions is just a convenient way to 
measure distances in galaxy clustering by normalizing them to some 
fiducial baseline. Using a fiducial cosmology we do not make a mistake 
\textit{per se}. The distance dependence is always fully forward-modeled,
and even if the fiducial cosmology happened to be different from the one
inferred from the data, this is totally fine as long as the 
AP distortions are modeled consistently (e.g. without additional approximations like a Taylor expansion of $H_{\rm fid}/H_{\rm true}$ around unity).

\subsection{Summary and Discussion}

\medskip\noindent
\textbf{IR resummation.} The last effect we need to include is a generalization of IR resummation in the case of RSD. 
Unlike the bias case, the RSD mapping explicitly depends on the 
velocity of the tracer w.r.t the observer, which is not a
frame-invariant quantity. As such, it is not protected by the equivalence 
principle. However, the correlation functions are, so one can still
derive the generalizations of the IR cancellation arguments 
in the RSD case. Just like in the real space case, the cancellation
does not happen for
modes $q\sim \kosc$, which require IR resummation. 
This is done 
by promoting the damping factor $\Sigma^2$ in Eq.~\eqref{eq:Sigma2_def}
to be direction-dependent: 
\be
\label{eq:sigmatot}
\begin{split}
\Sigma^2_{\rm tot}(z,\mu)& =(1+f(z)\mu^2(2+f(z)))D^2_+(z)\Sigma^2
+f^2(z)\mu^2(\mu^2-1)D^2_+(z)\delta\Sigma^2 \\
\delta\Sigma^2& \equiv \frac{1}{2\pi^2}\int_0^{\kIR}dq\,P_{11}(q)j_2\left(\frac{q}{\kosc}\right) \,.
\end{split}
\ee
The direction dependence of the damping factor is a serious
practical issue, which can be resolved either by using further
approximations~\cite{Ivanov:2018gjr} or going into Lagrangian space~\cite{Chen:2020zjt}. 
The latter also provides some insight into the physical 
origin of the damping factor, which will be discussed in the 
Chapter 7.

\medskip\noindent
\textbf{The scales of RSD.}
The scales $k_\mathrm{RSD,stoch}$ (and the analogous
$k_\mathrm{RSD,det}$ controlling the size of $\tilde{C}_1$,
$\tilde{C}_2$) may be \emph{parametrically suppressed} with respect
to $\knl$ for some galaxy samples:
\begin{equation}
  k_\mathrm{RSD} \ll \knl\,,
\end{equation}
because of the small-scale velocity dispersion of virialized
objects, and the fact that the velocity field is, in general,
more non-linear than the density field. This is the EFT
manifestation of the \textbf{Fingers of God}: their leading 
deterministic effect is captured by the $k^2\mu^2 P_{11}$
and $k^2 \mu^4 P_{11}$ counterterms, whilst the stochastic 
effect
is captured by the $\mu^2 k^2$ stochastic term.
The corresponding coefficients may be 
enhanced by $(\knl/k_\mathrm{RSD})^2$, and
the derivative expansion in redshift space typically breaks down at
$k\sim k_\mathrm{RSD}$ rather than $\knl$~\cite{Ivanov:2021fbu,Ivanov:2021zmi}.

If fingers-of-God are present, 
the breakdown of EFT is manifest in large values
of the inferred counterterms $\tilde C_1$ and
$\tilde C_2$ if one is using the normalization 
$k_\mathrm{RSD} = \knl$. 
For instance, luminous red galaxies have $k_\mathrm{RSD}\simeq 0.25~\hMpc$,
implying that
at $k_{\rm max}=0.2~\hMpc$ the next-to-leading order counterterms $\sim k^4 P_{11}$
are not negligible. A good policy is to include some of these counterterms 
in the theory model even though formally they are higher order just because
their contribution is not negligible in power counting.
This practice is common in particle physics, see e.g.~\cite{Tackmann:2024kci,Chang:2025ohh}.
In this vein, Ref.~\cite{Ivanov:2019pdj} suggested to include the following higher-derivative contribution
\be 
P_{\rm RSD,~\nabla^4_{\hat z}\delta}=\tilde{c}k^4\mu^4 f^4 P_{11}(k)\,.
\ee 
Once the cosmological constraints are marginalized over such extra counterterms,
the are more robust and conservative w.r.t. small-scale non-linearity. For the $\Lambda$CDM power
spectrum it happened that $\sim k^4P_{11}\sim k^2$ at $k_{\rm max}\simeq 0.2~\hMpc$,
so that the $k^2\mu^2$ stochastic contributions are degenerate
with the higher derivative corrections and can
effectively absorb them in the fit~\cite{Chudaykin:2024wlw,Ivanov:2024xgb,Chudaykin:2025aux}. However,
the fit then returns incorrect values of $\tilde{\alpha}_2$, which are 
inconsistent with precision measurements of these parameters from simulations 
(discussed shortly). 
In addition, one can propose a data-driven argument: as long 
as the higher-order counterterm $\tilde{c}$ can be detected in data,
or if it modifies cosmological parameter inference in a significant way,
it should be included in the prediction. This is indeed the case
for luminous red galaxies of BOSS and DESI surveys~\cite{Ivanov:2019pdj,Chudaykin:2020hbf,Ivanov:2024xgb,Chudaykin:2025aux}.

\medskip\noindent
\textbf{Summary of free parameters at one loop in redshift space:}
\begin{tcolorbox}[colback=red!5!white, colframe=red!50!black, title=Parameter Count]
  \textbf{Deterministic bias:} $b_1$, $b_2$, $b_{\mathcal{G}_2}$, $b_{\Gamma_3}$ (4 parameters)\\
  \textbf{Counterterms:} $C_0$ ($= b_{\nabla^2\delta}$, from real space), $\tilde{C}_1$, $\tilde{C}_2$ (2 new RSD parameters), and optionally a higher derivative counterterm $\tilde{c}$ multiplying $k^4 \mu^4 P_{11}(k)$ shape \\
  \textbf{Stochastic:} $\tilde\alpha_0$, $\tilde\alpha_1$, $\tilde\alpha_2$ (3 parameters)\\[4pt]
  \textbf{Total:} $ 11$ free parameters for the full one-loop galaxy power spectrum in redshift space.
\end{tcolorbox}

\begin{figure}
    \centering
    \includegraphics[width=0.49\linewidth]{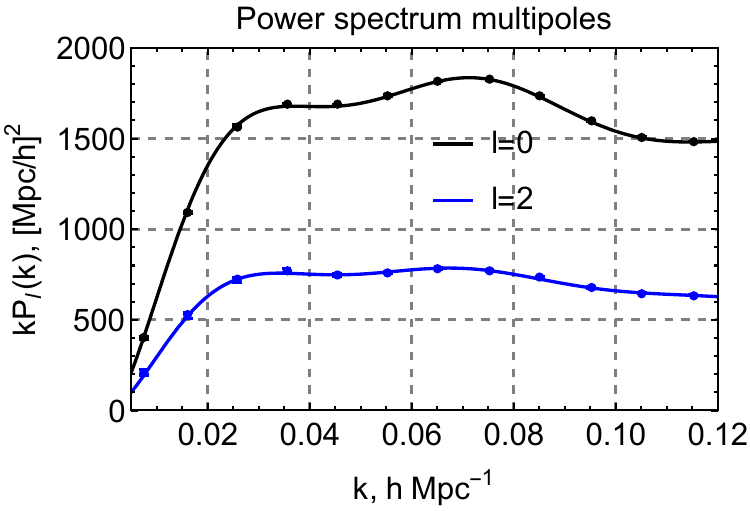}
    \includegraphics[width=0.49\linewidth]{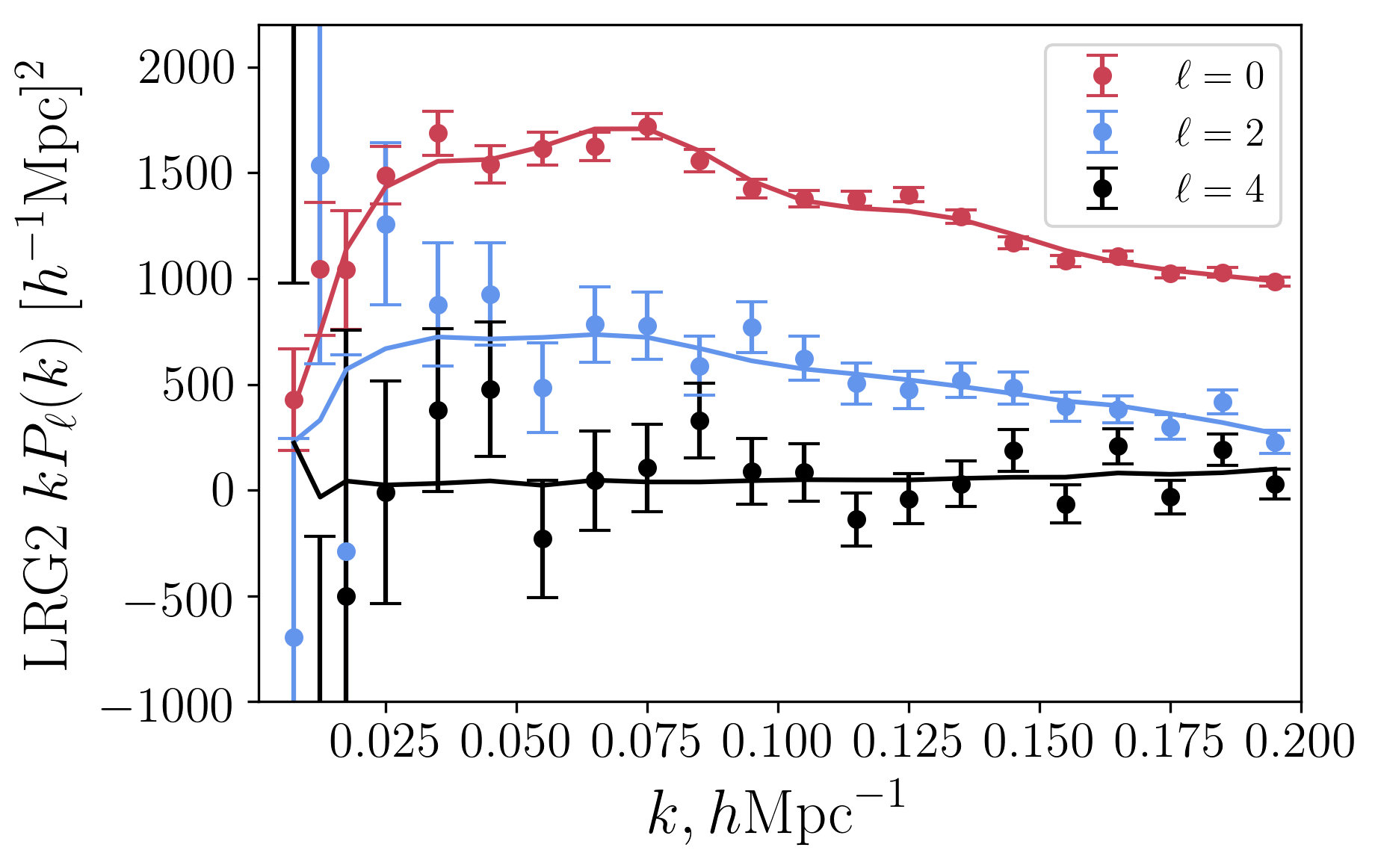}
    \caption{One loop EFT predictions for redshift-space power spectrum multipoles of the luminous red galaxies at $z=0.61$
    of the PT Challenge simulation (\textbf{left panel}) and  
    the DESI DR1 LRG2 data (\textbf{right panel}) at $z=0.71$. These are adapted from~\cite{Nishimichi:2020tvu,Chudaykin:2025vdh}, respectively.
    PT Challenge simulation has a gigantic 
    volume of 566~[$h^{-1}$Gpc]$^3$, which is why the error bars 
    are invisible in the plot. Since the two-loop corrections are 
    greater than the error-bars in this case, the fit is performed at 
    more conservative scale cuts than the actual DESI data.
     }
    \label{fig:prsd_fit}
\end{figure}

\medskip\noindent
\textbf{Comparison to simulations and data.}
The EFT prediction for galaxies in redshift space
matches both simulated and 
observed data well up to $k_{\rm max}\simeq 0.2~\hMpc$, 
see Fig.~\ref{fig:prsd_fit}.
Perfect recovery of cosmological parameters in various
masked data challenges such as~\cite{Nishimichi:2020tvu,Beyond-2pt:2024mqz} demonstrated 
that this computation is not in the over-fitting regime.

An even stronger test of EFT is a comparison to simulations 
at the field level~\cite{Schmittfull:2018yuk,Schmittfull:2020trd,Obuljen:2022cjo,Ivanov:2024hgq,Ivanov:2024xgb,Chen:2025jnr,Ivanov:2024dgv}. 
To  that end one applies the EFT prediction to the known initial
condition of the simulation $\delta_1(\kv)$ and use this prediction
as a \textit{forward model} to be compared to the actual simulated map.
The point it that the power spectrum is a relatively smooth function
of $k$ that can be overfit easily. In contrast, the field level comparison
requires matching not only the amplitudes of the modes, but also their phases,
i.e. we need to fit each pixel of the simulated galaxy map. 
This comparison is nicely illustrated by the following
example. A forward model $\delta^{(s)}_{g~\rm model}=\sqrt{\frac{P_{gg,NL}(k)}{P_{11}(k)}}\delta_1(\k)$ (where $P_{gg,NL}$ is the non-linear galaxy power spectrum
from the simulation) by construction leads to a perfect match 
to the galaxy power spectrum, but it fails completely at the field level because
it misses the phase contributions from higher order operators~\cite{Baldauf:2015tla,Baldauf:2015zga,Schmittfull:2018yuk,Schmittfull:2020trd}.  
Thus, the field-level test allows one to address a key concern of the EFT
skeptics: is the success of EFT entirely due to the free parameters?
The answer is obviously ``no.''
Even if one has infinitely many free parameters, one can't fit every pixel in a
galaxy map without the correct bias model, i.e. operators and the displacement structure predicted by EFT.

The field level comparison allows one
to determine the stochastic power spectrum, see Fig.~\ref{fig:perr_rsd},
which shows its measurement for DESI-like luminous red galaxies (LRG)
in both \texttt{MTNG} hydrodynamical simulation 
and \texttt{Abacus} N-body simulation
from~\cite{Ivanov:2024xgb,Ivanov:2024dgv} (see~\cite{Springel:2017tpz,Maksimova:2021ynf}).
$P_{\rm stoch}\bar n$ is white (orientation and scale-independent)
on large scales. The onset of scale and $\mu$-dependence in redshift 
space is accurately captured by the EFT prediction~\eqref{eq:Pstoch_rsd}.
The product $P_{\rm stoch}\bar n\sim \tilde{\alpha}_0$
is different from one on large scales due to the halo exclusion effect mentioned in the previous chapter.
The correct predictions for the stochastic power spectrum
is another important consistency test of EFT. In addition, let us 
mention that $\tilde{\alpha}_2>0$ for DESI-like LRGs, as clearly 
seen in Fig.~\ref{fig:perr_rsd}. To recover this value from the 
direct power spectrum fits it is crucial to include
the higher-order counterterm $P_{\rm RSD,~\nabla^4_{\hat z}\delta}$,
in the theory prediction.

Note that just like in the case of the matter fields, 
$P_{\rm stoch}$ becomes shallower around $k\sim 0.25~\hMpc$, 
which signals the breakdown of the 
EFT expansion for $P_{\rm stoch}$ around this scale (for this galaxy sample).
In principle, EFT is really useful only for the deterministic 
part of the power spectrum, whilst for the stochastic part it can't offer anything other than the Taylor series in $k^2$. This implies that
once the stochastic power spectrum becomes non-perturbative, 
no higher EFT order corrections can possibly improve the modeling 
of the galaxy power spectrum in redshift space.~\footnote{This is particularly important now as some new ``resummation'' 
techniques resurfaced in the literature. 
In contrast to IR resummation, which resums
large-scale modes that are fully described by the EFT, 
these new ``resummation''  approaches try to resumm the UV contributions, 
paralleling the logic of the unsuccessful program to resum
SPT loop diagrams.
There are two problems with such ``resummations'' in general: 
(a) we don't have a parametric control over the UV contributions, 
i.e. the ``resummed'' terms are of the same order as the unresummed ones;
and (b) we don't know the UV theory. Even if these ``resummation'' 
formulas can formally fit the galaxy power spectrum data at $k\sim 0.3~\hMpc$,
the field level EFT analyses imply that this is achieved only due 
to a massive over-fitting.
No resummation can fix the non-trivial scaling of 
$P_{\rm stoch}$ which all simulations consistently predict 
at $k\sim 0.3~\hMpc$. 
At the same time, simulations suggest that $P_{\rm stoch}$
can be a factor of ten greater than the deterministic part
of $P_{\rm gg}$ on these scale, rendering any UV resummation  
completely useless even if they were correct, see the lower panel of Fig.~\ref{fig:perr_rsd}.}

\begin{figure}%[htb!]
    \centering
    \includegraphics[width=0.99\linewidth]{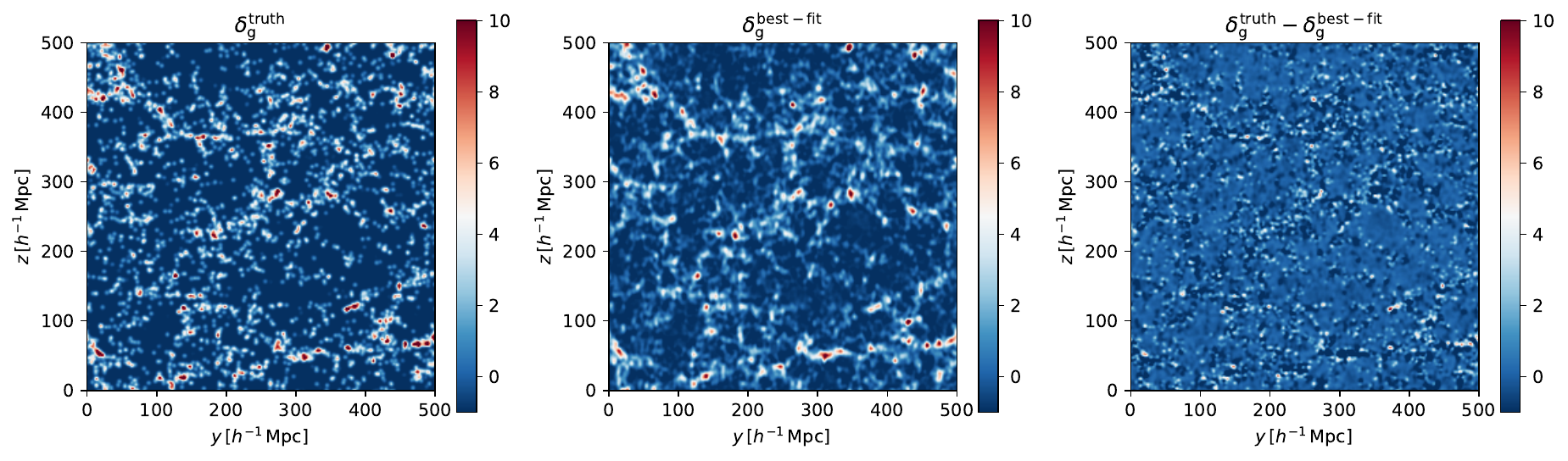}
    \caption{The galaxy density field of DESI-like luminous red galaxies 
    from the \texttt{MTNG} simulation (\textbf{left panel}), the EFT forward model (\textbf{middle panel}), and the residuals between them (\textbf{right panel}). 
    Redshift-space distortions are implemented along the $z$ axis.
     }
    \label{fig:mtng_field_desi}
\end{figure}
\begin{figure}[htb!]
    \centering
    \includegraphics[width=0.49\linewidth]{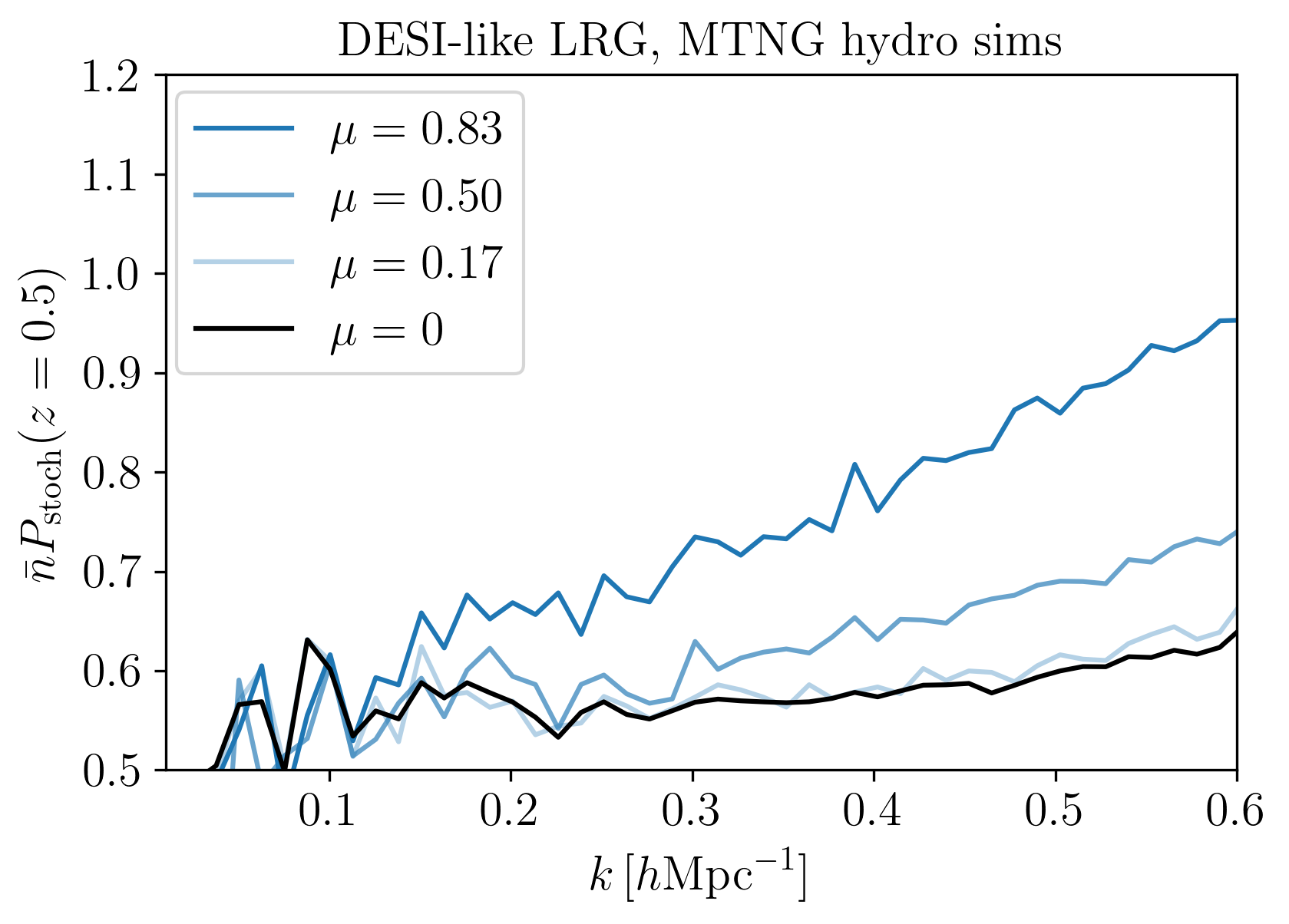}
    \includegraphics[width=0.49\linewidth]{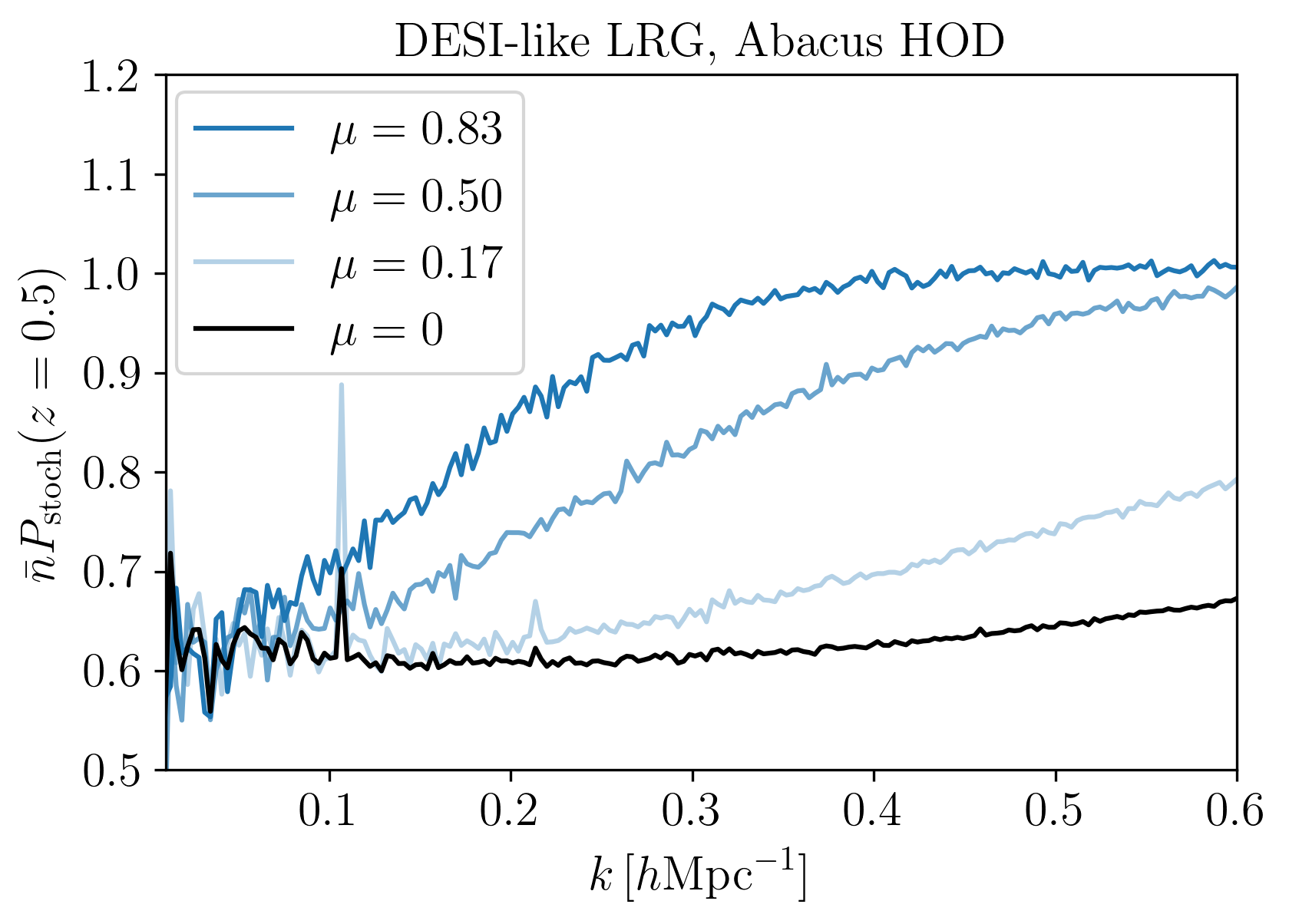}
        \includegraphics[width=0.49\linewidth]{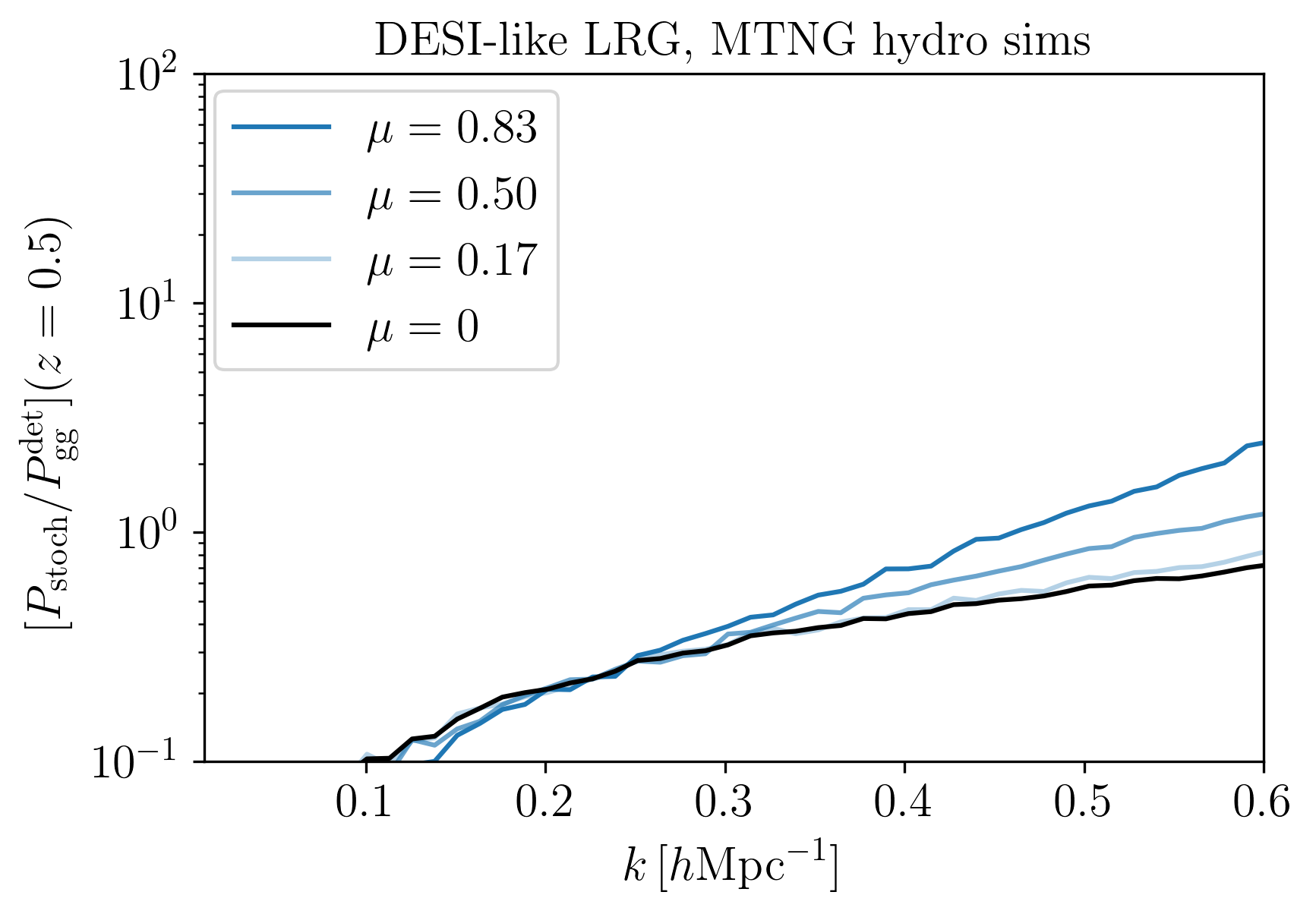}
    \includegraphics[width=0.49\linewidth]{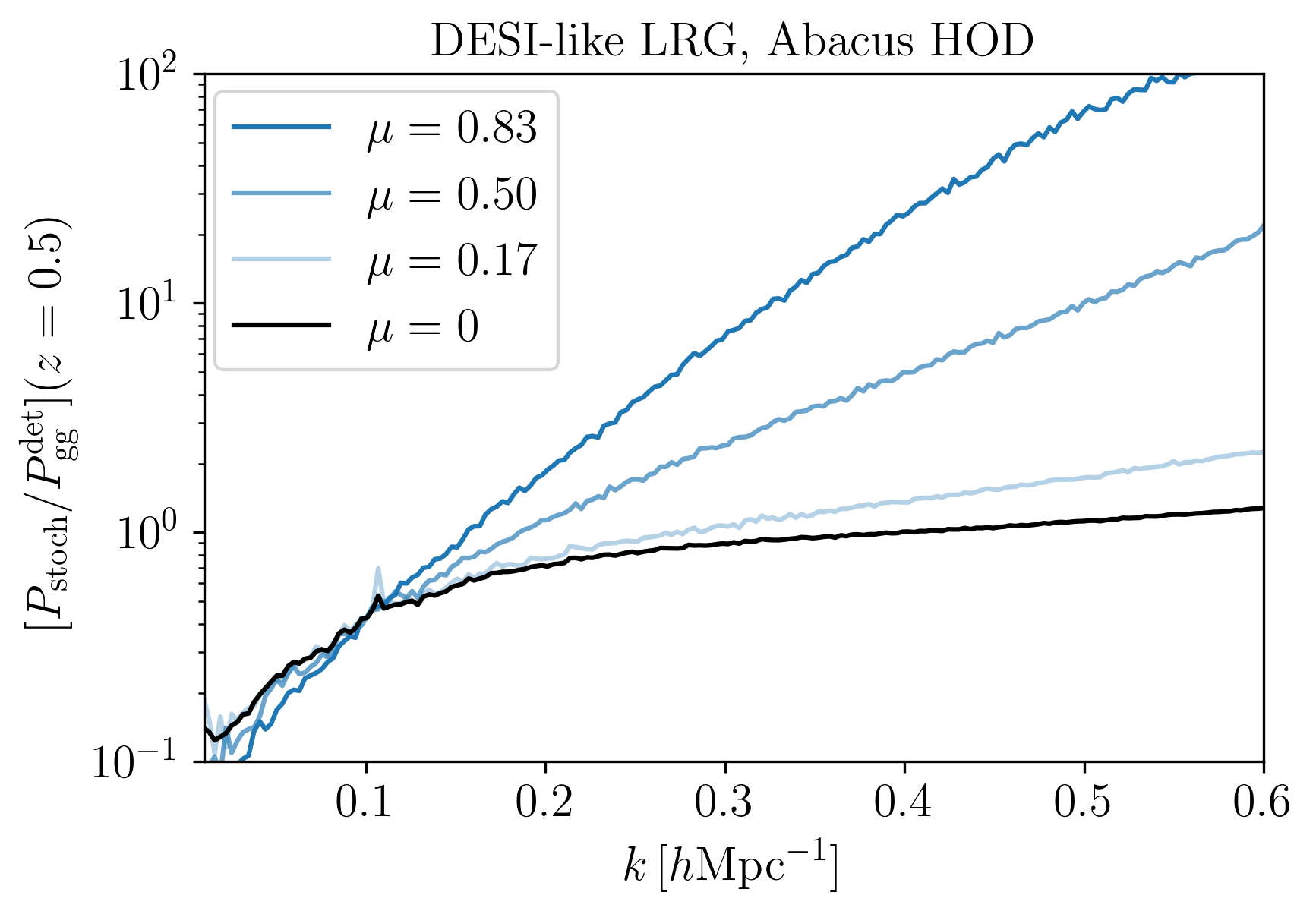}
    \caption{The stochastic power spectra $P_{\rm stoch}$ of DESI-like 
    Luminous Red Galaxies (LRG) from field-level EFT
    fits to \texttt{MTNG} hydrodynamical simulation (\textbf{upper left panel})
    and \texttt{Abacus} halo occupation distribution (HOD) models (\textbf{upper right panel}). $\bar n$ is the number density of the simulated galaxies. $P_{\rm stoch}\bar n$
    is different from unity on large scales due to halo exclusion effects. It is scale and orientation-independent (white) in the $k\to 0$ limit in agreement with EFT. The leading corrections beyond the white noise behavior match the EFT prediction $\alpha_1k^2+\alpha_2 \mu^2 k^2$. $P_{\rm stoch}$ becomes shallower around $k\sim 0.25~\hMpc$, which sets the limit of the EFT reach in redshift space for these galaxy samples. No resummation of deterministic contributions 
    can improve the modeling of the stochastic power spectrum,
    whose amplitude is comparable or exceeding the deterministic one at $k\sim 0.3~\hMpc$, see the~\textbf{lower panels}.
     }
    \label{fig:perr_rsd}
\end{figure}

All in all, while the regime of validity of EFT is somewhat
limited in redshift space, it is nevertheless a powerful tool 
that can be used for precision measurements of cosmological
parameters in the quasi-linear regime. At the same time, simulations suggest
that small galaxies hosted by light halos, which are typically
identified by their strong emission lines, 
have a higher cutoff
than 
more massive galaxies~\cite{Ivanov:2021zmi,Ivanov:2024dgv,Sullivan:2025eei}.
In addition, Fingers-of-God can be suppressed
by a careful data-driven selection~\cite{BaleatoLizancos:2025wdg}.
Therefore,  
the perturbativity of the sample (i.e. the EFT cutoff), 
is a highly sample- and selection-dependent property. 
Finally, simulations suggest
that high-redshift data may be more 
perturbative than the current samples~\cite{Sullivan:2025eei},
suggesting a bright future for the EFT!

%==========================================================================
\section{Chapter 7: Lagrangian Perturbation Theory}
%==========================================================================

In Chapter 3 we saw that the IR-enhanced effects of large-scale
flows on the BAO must be resummed, and we listed Lagrangian
perturbation theory (LPT) as one of the two ways to do it: a
formulation in which the displacement is never Taylor-expanded, so
that the boost symmetry and the associated IR structure are manifest
at every step~\cite{Bernardeau:2001qr,Matsubara:2007wj,Matsubara:2008wx,Matsubara:2015ipa,Sato:2011qr,Porto:2013qua,Vlah:2015sea,Vlah:2018ygt,Chen:2020zjt,Chen:2020fxs}. In this chapter we develop LPT systematically: the
displacement field and the exact expression for the power spectrum,
the Zel'dovich approximation~\cite{Zeldovich:1969sb}, the higher-order Lagrangian kernels
and their matching to SPT, and finally the position-space picture of
why bulk flows are harmless for the broadband power spectrum but
fatal for the perturbative treatment of the BAO peak.

A remark on notation: throughout this chapter $\qv$ denotes the
\emph{Lagrangian coordinate} (the initial position of a fluid
element), not a loop momentum; momenta are denoted $\kv$, $\pv$,
$\vv{l}$, and we abbreviate
$\int_\pv \equiv \int \frac{d^3p}{(2\pi)^3}$, as before.

%------------------------------------------------------------------
\subsection{The Displacement Field}
%------------------------------------------------------------------

In the Lagrangian formulation of hydrodynamics one follows the
trajectories of the particles (fluid elements). A particle that
started at the initial position $\qv$ --- equivalently, $\qv$ is
simply a label --- is found at time $\tau$ at the comoving position
\begin{equation}\label{eq:displ_def}
  \xv(\qv,\tau) = \qv + \Psiv(\qv,\tau)\,,
\end{equation}
which defines the fundamental dynamical variable of LPT, the
\textbf{displacement field} $\Psiv(\qv,\tau)$.

Let us derive its equation of motion directly from Newtonian
cosmology. A particle at physical position $\rv(t)$ obeys Newton's
law in the full gravitational potential,
\begin{equation}\label{eq:newton_traj}
  \frac{d^2\rv}{dt^2} = -\gradr\Phi(\rv)\,.
\end{equation}
We now repeat, for a single trajectory, the steps that led to the
fluid equations of Chapter~3: decouple the Hubble flow by writing
$\rv = a(\tau)\,\xv$, and split the potential into background and
perturbation as in Eq.~\eqref{eq:bg_split},
$\Phi = -\frac{1}{2}\ddot a\,a\,|\xv|^2 + \phi(\xv,\tau)$. With
$dt = a\,d\tau$, the first time derivative of the trajectory is the
familiar Hubble-flow-plus-peculiar-velocity decomposition,
$\dot\rv = \Hubble\,\xv + \xv'$, and one more derivative gives the
left-hand side of Eq.~\eqref{eq:newton_traj},
\begin{equation}
  \frac{d^2\rv}{dt^2}
  = \frac{1}{a}\frac{d}{d\tau}\left(\Hubble\,\xv+\xv'\right)
  = \frac{1}{a}\left(\xv'' + \Hubble\,\xv' + \Hubble'\,\xv\right)\,.
\end{equation}
For the right-hand side, use $\gradr = a^{-1}\gradx$ together with
$\ddot a\,a = \Hubble'$ (which follows from
$\ddot a = \frac{1}{a}(a'/a)'$), so that
\begin{equation}
  -\gradr\Phi
  = \frac{1}{a}\left(\Hubble'\,\xv - \gradx\phi\right)\,.
\end{equation}
The two $\Hubble'\xv$ terms cancel \emph{identically} --- the
background potential is precisely what supports the pure Hubble flow
$\xv=\mathrm{const}$ as a solution:\footnote{Equivalently, on the matter-dominated background
one may substitute the Einstein--de Sitter value of the background
potential, $\Phi = \frac{1}{4}\Hubble^2|\xv|^2+\phi$. The equation
of motion then reads
$\xv''+\Hubble\xv'+\Hubble'\xv = -\frac{1}{2}\Hubble^2\xv
-\gradx\phi$, and the terms without time derivatives cancel by the
Friedmann equation $\Hubble' = -\frac{1}{2}\Hubble^2$ of EdS.}
\begin{equation}\label{eq:comoving_traj}
  \xv'' + \Hubble\,\xv' = -\gradx\phi(\xv)\,.
\end{equation}
Finally, evaluate this along the trajectory
$\xv(\qv,\tau)=\qv+\Psiv(\qv,\tau)$: the label $\qv$ carries no time
dependence, so $\xv'=\Psiv'$ and $\xv''=\Psiv''$, while the force is
evaluated at the displaced position,
\begin{equation}\label{eq:LPT_eom}
  \boxed{\;\Psiv'' + \Hubble\,\Psiv' = -\left.\gradx\phi\right|_{\xv=\qv+\Psiv}\,,}
  \qquad
  \Delta_x\phi = \frac{3}{2}\,\Hubble^2\Omega_m\,\delta\,.
\end{equation}
Note that the force is evaluated at the \emph{displaced} position ---
this is the only place where the non-linearity enters. Equation~\eqref{eq:LPT_eom} is of course nothing
but the characteristic form of the Euler equation of Chapter~3 with
$\tau_{ij}=0$: the peculiar velocity along the trajectory is
$\vvv=\xv'=\Psiv'$, and $\Psiv''$ is its convective derivative
$\left(\partial_\tau+\vvv\cdot\gradx\right)\vvv$, since the partial
derivatives in Eq.~\eqref{eq:LPT_eom} are taken at fixed $\qv$,
i.e.\ following the particle.

The density field is not an independent variable: it is fixed by
mass conservation. There are two ways to see this. First, counting particles:
the number of particles inside an Eulerian volume element centered
at $\xv$ is obtained by summing over all $\qv$ whose trajectories
end up there, so
\begin{equation}\label{eq:LPT_density}
  1+\delta(\xv,\tau) = \int d^3q\;\delta_D^{(3)}\big(\xv-\qv-\Psiv(\qv,\tau)\big)
  = \frac{1}{\det\left(\delta_{ij} + \dfrac{\partial\Psi^i}{\partial q^j}\right)}
  \Bigg|_{\qv:\,\xv(\qv,\tau)=\xv}\,.
\end{equation}
Second, directly from mass conservation: the volume elements get
distorted but carry their mass with them,
$\rhob\,d^3q = \rho(\xv,\tau)\,d^3x$, which gives the same Jacobian
formula. When the map $\qv\to\xv$ ceases to be single-valued
(which happens at shell crossing, $\det\to 0$), the sum over streams in the first
expression must be kept --- this is where the perturbative
description ultimately breaks down.

\medskip\noindent
\textbf{Homework:} show that Eq.~\eqref{eq:LPT_density} solves the
continuity equation.

%------------------------------------------------------------------
\subsection{The Exact Power Spectrum: Cumulant Expansion}
%------------------------------------------------------------------

The great advantage of the Lagrangian formulation is that the
density field can be written in closed form. Fourier transforming
Eq.~\eqref{eq:LPT_density} (from now on we suppress the time
argument and consider $\kv\neq 0$),
\begin{equation}\label{eq:delta_LPT_exact}
  \delta(\kv) = \int d^3q\;e^{-i\kv\cdot\qv}
  \left(e^{-i\kv\cdot\Psiv(\qv)}-1\right)\,:
\end{equation}
the displacement enters only through the phase
$e^{-i\kv\cdot\Psiv}$ --- the same exponential structure that we
encountered in the toy model and in the boost-symmetry argument of
Chapter~3, except that here it is exact and fully non-linear. For
the two-point function, the terms involving a single expectation
value $\avg{e^{-i\kv\cdot\Psiv}}$ are proportional to
$\delta_D^{(3)}(\kv)$ and drop for $\kv\neq 0$; statistical
homogeneity then implies that the remaining average depends only on
the separation $\qv=\qv_1-\qv_2$, and we obtain the \emph{exact}
result
\begin{equation}\label{eq:P_LPT_exact}
  \boxed{\;P(k) = \int d^3q\;e^{-i\kv\cdot\qv}
  \left(\avg{e^{-i\kv\cdot\left[\Psiv(\qv)-\Psiv(0)\right]}}-1\right)\,.\;}
\end{equation}
Note that only the \emph{relative} displacement of the two points
appears: a uniform shift of all particles cancels between the two
exponentials. This is the Lagrangian-space incarnation of the
equivalence-principle protection derived in Chapter~3, and it holds
here non-perturbatively.

To evaluate the average we use the cumulant expansion theorem,
\begin{equation}
  \avg{e^{-iX}} = \exp\left(\sum_{N=1}^{\infty}\frac{(-i)^N}{N!}\,
  \avg{X^N}_c\right)\,,
\end{equation}
with $X = \kv\cdot[\Psiv(\qv)-\Psiv(0)]$. Parity implies
$\avg{\Psiv}=0$, so the expansion starts at $N=2$. Separating the
zero-lag (one-point) cumulants from the $\qv$-dependent ones with
the help of the binomial theorem, one can organize the result as
\begin{equation}\label{eq:P_AB}
  P(k)=\int d^3q\;e^{-i\kv\cdot\qv}
  \left(e^{-2\sum_{n=1}^\infty \frac{k_{i_1}\dots k_{i_{2n}}}{(2n)!}\,
  A^{(2n)}_{i_1\dots i_{2n}}}\;
  e^{\sum_{N=2}^\infty \frac{k_{i_1}\dots k_{i_N}}{N!}\,
  B^{(N)}_{i_1\dots i_N}(\qv)}-1\right)\,,
\end{equation}
with
\begin{equation}\label{eq:AB_def}
\begin{split}
  A^{(2n)}_{i_1\dots i_{2n}} &= (-1)^{n-1}\,
  \avg{\Psi_{i_1}(0)\cdots\Psi_{i_{2n}}(0)}_c\,,\\
  B^{(N)}_{i_1\dots i_N}(\qv) &= i^N \sum_{j=1}^{N-1}\binom{N}{j}\,
  \avg{\Psi_{(i_1}(\qv_1)\cdots\Psi_{i_j}(\qv_1)\,
  \Psi_{i_{j+1}}(\qv_2)\cdots\Psi_{i_N)}(\qv_2)}_c\,,
\end{split}
\end{equation}
where the indices are symmetrized and $\qv = \qv_1 - \qv_2$. The
$A$'s encode the (large) one-point displacement variances, the $B$'s
the correlations between the two points. Equation~\eqref{eq:P_AB} is
still exact --- all of Eulerian perturbation theory is contained in
the low-order cumulants of $\Psiv$.

%------------------------------------------------------------------
\subsection{The Zel'dovich Approximation}
%------------------------------------------------------------------

The Zel'dovich approximation (ZA) is the \emph{linear} solution for
the displacement --- which, via the non-linear
map~\eqref{eq:LPT_density}, still generates a fully non-linear
density field. Linearizing Eq.~\eqref{eq:LPT_eom} (so that
$\gradx\approx\grad_q$) and taking the divergence, the variable
$\psi\equiv\grad_q\cdot\Psiv$ obeys
\begin{equation}
  \psi'' + \Hubble\,\psi' - \frac{3}{2}\,\Hubble^2\Omega_m\,\psi = 0\,,
\end{equation}
i.e.\ exactly the linear growth equation (with $\delta_1=-\psi$, as
follows from expanding the determinant in
Eq.~\eqref{eq:LPT_density}). Selecting the growing mode and fixing
the initial conditions by $\delta_0 = -\grad_q\cdot\Psiv_0$, we find
in Fourier space
\begin{equation}\label{eq:ZA_displacement}
  \Psiv_\mathrm{ZA}(\kv,\tau) = \frac{i\kv}{k^2}\,\delta_1(\kv,\tau)\,,
\end{equation}
which is precisely the linear displacement~\eqref{eq:displacement}
introduced in Chapter~2. Since $\Psiv_\mathrm{ZA}$ is linear in the
Gaussian field $\delta_0$, all its cumulants beyond the second
vanish, and the second one is
\begin{equation}\label{eq:ZA_corr}
  \avg{\Psi^i_\mathrm{ZA}(\qv_1)\,\Psi^j_\mathrm{ZA}(\qv_2)}
  = \int_{\vv{l}}\;\frac{l_i l_j}{l^4}\;e^{i\vv{l}\cdot\qv}\,P_{11}(l)\,,
  \qquad \qv=\qv_1-\qv_2\,.
\end{equation}
The average in Eq.~\eqref{eq:P_LPT_exact} is then a Gaussian
integral,
\begin{equation}
  \avg{e^{-i\kv\cdot[\Psiv(\qv)-\Psiv(0)]}}
  = \exp\left\{-\frac{k_ik_j}{2}
  \avg{\left[\Psi^i(\qv)-\Psi^i(0)\right]\left[\Psi^j(\qv)-\Psi^j(0)\right]}
  \right\}\,,
\end{equation}
and using Eq.~\eqref{eq:ZA_corr} (the term odd in $\vv{l}$
integrates to zero by isotropy) we obtain the
\textbf{Zel'dovich power spectrum}
\begin{equation}\label{eq:P_ZA}
  \boxed{\;P_\mathrm{ZA}(k)=\int d^3q\;e^{-i\kv\cdot\qv}
  \left(\exp\left\{-2\,k_ik_j\int_{\vv{l}}\frac{l_il_j}{l^4}\,P_{11}(l)\,
  \sin^2\!\left(\frac{\vv{l}\cdot\qv}{2}\right)\right\}-1\right)\,.\;}
\end{equation}
Two features are worth noting. First, the combination
$\sin^2(\vv{l}\cdot\qv/2) = \tfrac12[1-\cos(\vv{l}\cdot\qv)]$
vanishes as $l^2q^2$ for $lq\to 0$: very long modes drop out of the
relative displacement --- the same suppression that appeared in the
bracket of the BAO damping factor~\eqref{eq:Sigma2_def}. Second, the
exponent is kept unexpanded: the ZA resums \emph{all} powers of the
displacement, which is exactly what the IR resummation of Chapter~3
requires (indeed, evaluating Eq.~\eqref{eq:P_ZA} for the wiggly part
of the spectrum reproduces the damping $e^{-k^2\Sigma^2}$).

\medskip\noindent
\textbf{Homework:} show from Eq.~\eqref{eq:P_ZA} that
$P_\mathrm{ZA}(k)\to P_{11}(k)$ for $k\to 0$.

%------------------------------------------------------------------
\subsection{Higher Orders: the $L_n$ Kernels}
%------------------------------------------------------------------

To go beyond the ZA systematically, we need the equations of motion
in a form suitable for iteration. For $\kv\neq 0$ the density and
the potential read, from Eq.~\eqref{eq:LPT_density},
\begin{equation}
  \delta(\kv) = \int d^3q\;e^{-i\kv\cdot\xv(\qv)}\,,\qquad
  \phi(\kv) = -\frac{3}{2}\,\frac{\Hubble^2\Omega_m}{k^2}
  \int d^3q\;e^{-i\kv\cdot\xv(\qv)}\,,
\end{equation}
so the equation of motion~\eqref{eq:LPT_eom} becomes
\begin{equation}
  \Psiv''(\qv) + \Hubble\,\Psiv'(\qv)
  = \frac{3}{2}\,\Hubble^2\Omega_m
  \int_\kv \frac{i\kv}{k^2}\int d^3q'\;
  e^{i\kv\cdot\left(\xv(\qv)-\xv(\qv')\right)}\,.
\end{equation}
We now Fourier transform with respect to the \emph{Lagrangian}
coordinate, $\Psiv(\qv)=\int_\pv e^{i\pv\cdot\qv}\,\Psiv_\pv$, and
pass to the time variable $\eta=\ln\Dp$ in the EdS approximation
$\Omega_m=f^2$ (in exact EdS this is the same
as $\eta=\ln a$):
\begin{equation}\label{eq:LPT_master}
  \partial^2_\eta{\Psiv}_\pv + \frac{1}{2}\,\partial_\eta{\Psiv}_\pv
  = \frac{3}{2}\int d^3q\,d^3q'\;e^{-i\pv\cdot\qv}
  \int_\kv \frac{i\kv}{k^2}\;e^{i\kv\cdot(\qv-\qv')}\;
  \exp\left\{i k_j \int_{\pv'} \Psi^j_{\pv'}
  \left(e^{i\pv'\cdot\qv}-e^{i\pv'\cdot\qv'}\right)\right\}\,.
\end{equation}
This single equation contains all of LPT. We solve it with the
ansatz
\begin{equation}\label{eq:Ln_ansatz}
  \Psi^i_\pv = \sum_{n=1}^{\infty}\frac{i\,\Dp^n(\eta)}{n!}
  \int_{\pv_1}\!\!\cdots\!\int_{\pv_n}
  (2\pi)^3\delta_D^{(3)}\Big(\pv-\sum_a\pv_a\Big)\,
  L^i_n(\pv_1,\ldots,\pv_n)\;\delta_0(\pv_1)\cdots\delta_0(\pv_n)\,,
\end{equation}
which defines the \textbf{Lagrangian kernels} $L_n$ --- the analogs
of the $F_n$, $G_n$ of SPT. Expanding the exponential in
Eq.~\eqref{eq:LPT_master} and collecting powers of $\delta_0$, the
time derivatives act on $\Dp^n=e^{n\eta}$ and the \emph{linear}
(longitudinal) part of the right-hand side can be moved to the left,
leaving at each order
\begin{equation}\label{eq:Ln_master}
  \left(n^2+\frac{n}{2}-\frac{3}{2}\right)\frac{L^i_n}{n!}
  = \frac{(n-1)(2n+3)}{2}\,\frac{L^i_n}{n!}
  = \big(\text{source built from } L_{m<n}\big)^i\,,
\end{equation}
with the same characteristic denominators $(n-1)(2n+3)$ as in the
SPT recursion relations below Eq.~\eqref{eq:F2} --- not a
coincidence, since both descend from the same growth operator (cf.\
Appendix~\ref{app:green-eta}).

At $n=1$ the left-hand side vanishes: the linear kernel is a zero
mode of the growth operator (it \emph{is} the growing mode) and is
fixed by the initial conditions~\eqref{eq:ZA_displacement} instead:
\begin{equation}
  L^i_1(\pv) = \frac{p^i}{p^2}\,.
\end{equation}
At $n=2$, expanding the exponential in Eq.~\eqref{eq:LPT_master} to
second order in $\Psiv^{(1)}$, the $\qv,\qv'$ integrals produce
momentum-conserving delta functions and the source evaluates to
$\frac{3}{2}\cdot\frac{1}{2}\left(1-\mu_{12}^2\right)p^i/p^2$, where
$\mu_{12}\equiv\hat{\pv}_1\cdot\hat{\pv}_2$ and $\pv=\pv_1+\pv_2$.
With the ansatz $L^i_2\propto p^i$, Eq.~\eqref{eq:Ln_master} with
$(n-1)(2n+3)/2 = 7/2$ gives
\begin{equation}\label{eq:L2}
  \boxed{\;L^i_2(\pv_1,\pv_2) = \frac{3}{7}\,\frac{p^i}{p^2}
  \left(1-\frac{(\pv_1\cdot\pv_2)^2}{p_1^2\,p_2^2}\right)\,.\;}
\end{equation}
Note that the second-order displacement is controlled by exactly
(minus) the tidal kernel $\mathcal{S}(\pv_1,\pv_2)=\mu_{12}^2-1$ of
the galaxy bias expansion: it vanishes for aligned momenta, i.e.\
for one-dimensional (plane-parallel) collapse the ZA is exact.

\medskip\noindent
\textbf{Homework:} applying the SPT expansion
$\delta = \Dp\,\delta_0 + \Dp^2 F_2\,\delta_0^2+\dots$ to the
left-hand side of Eq.~\eqref{eq:LPT_density} and the
ansatz~\eqref{eq:Ln_ansatz} to the right-hand side, derive the
general relation between the $L_n$ and $F_n$ kernels. (The first two
orders are worked out in the next section.)

%------------------------------------------------------------------
\subsection{Matching to SPT and the Curl Displacement}
%------------------------------------------------------------------

Expanding the exponential in Eq.~\eqref{eq:delta_LPT_exact} in
powers of $\kv\cdot\Psiv$ and inserting the perturbative
displacement connects the two expansions. It is convenient to split
the displacement into its longitudinal and transverse (curl) parts,
\begin{equation}\label{eq:helicity_dec}
  \Psiv(\qv) = \int_\pv e^{i\pv\cdot\qv}
  \left(i\pv\,S^L_\pv + i\pv\times\vv{S}^T_\pv\right)\,,
\end{equation}
with kernels $L_n$ and $\vv{T}_n$ defined as in
Eq.~\eqref{eq:Ln_ansatz}.

\medskip\noindent
\textbf{Zel'dovich kernels.} Keeping the full series in
$\kv\cdot\Psiv$ but only the linear solution
$\Psiv_\mathrm{ZA}$, one finds
\begin{equation}
  \delta^\mathrm{ZA}_\kv = \sum_{n=1}^\infty \Dp^n \int_{\pv_1}\!\!\cdots\!\int_{\pv_n}
  (2\pi)^3\delta_D^{(3)}(\kv-\pv_{1\dots n})\;
  \frac{1}{n!}\,\frac{(\kv\cdot\pv_1)}{p_1^2}\cdots
  \frac{(\kv\cdot\pv_n)}{p_n^2}\;\delta_0(\pv_1)\cdots\delta_0(\pv_n)\,,
\end{equation}
i.e.\ the SPT kernels of the Zel'dovich approximation are simply
\begin{equation}\label{eq:FnZA}
  F_n^\mathrm{ZA}(\pv_1,\ldots,\pv_n)
  = \frac{1}{n!}\,\frac{(\kv\cdot\pv_1)}{p_1^2}\cdots
  \frac{(\kv\cdot\pv_n)}{p_n^2}\,,\qquad \kv = \pv_{1\dots n}\,.
\end{equation}

\medskip\noindent
\textbf{Full kernels.} In general, each way of distributing the $n$
linear fields among the factors of $(-i\kv\cdot\Psiv^{(m)})$
contributes, and one finds order by order (all expressions
symmetrized over the momenta):
\begin{align}
  F_2 &= F_2^\mathrm{ZA} + \frac{1}{2!}\,\kv\cdot\vv{L}_2(\pv_1,\pv_2)\,,
  \label{eq:F2_L2}\\
  F_3 &= F_3^\mathrm{ZA} + \frac{1}{3!}\,\kv\cdot\vv{L}_3
  + \left[\frac{(\kv\cdot\pv_1)}{p_1^2}\,
  \frac{1}{2!}\,\kv\cdot\vv{L}_2(\pv_2,\pv_3)\right]_\mathrm{symm}\,,
  \label{eq:F3_L3}
\end{align}
and at fourth order
\begin{equation}\label{eq:F4_L4}
\begin{split}
  F_4 = F_4^\mathrm{ZA}
  &+ \frac{1}{4!}\,\kv\cdot\vv{L}_4
  + \frac{1}{2}\left[\frac{\kv\cdot\vv{L}_2(\pv_1,\pv_2)}{2!}\,
  \frac{\kv\cdot\vv{L}_2(\pv_3,\pv_4)}{2!}\right]_\mathrm{symm}
  + \left[\frac{(\kv\cdot\pv_1)}{p_1^2}\,
  \frac{\kv\cdot\vv{L}_3(\pv_2,\pv_3,\pv_4)}{3!}\right]_\mathrm{symm}\\
  &+ \left[\frac{(\kv\cdot\pv_1)}{p_1^2}\,
  \frac{\kv\cdot\left(\pv_{234}\times\vv{T}_3(\pv_2,\pv_3,\pv_4)\right)}{3!}
  \right]_\mathrm{symm}
  + \frac{1}{2}\left[\frac{(\kv\cdot\pv_1)}{p_1^2}\,
  \frac{(\kv\cdot\pv_2)}{p_2^2}\,
  \frac{\kv\cdot\vv{L}_2(\pv_3,\pv_4)}{2!}\right]_\mathrm{symm}\,.
\end{split}
\end{equation}
As a consistency check, inserting $L_2$ from Eq.~\eqref{eq:L2} into
Eq.~\eqref{eq:F2_L2} gives
$F_2 = F_2^\mathrm{ZA} + \frac{3}{14}(1-\mu_{12}^2)$, which
reproduces exactly the kernel~\eqref{eq:F2}.

Note the pattern of the transverse contributions: in the term linear
in $\Psiv^{(n)}$, the curl part drops out identically, because there
the Fourier argument of the displacement is the \emph{total}
momentum, $\kv\cdot(\kv\times\vv{T}_n)=0$. The curl displacement
first appears in $F_4$, through the cross term with $\Psiv^{(1)}$
above, where it is evaluated at the partial momentum
$\pv_{234}\neq\kv$.

\medskip\noindent
\textbf{The curl part and the Cauchy invariant.} The transverse
kernels are fixed by a beautiful exact statement. Contract the
equation of motion~\eqref{eq:LPT_eom} with the Jacobian matrix
$\partial x^i/\partial q^j$ and use
$\frac{\partial x^i}{\partial q^j}\,x''^{\,i}
= \frac{d}{d\tau}\!\left(\frac{\partial x^i}{\partial q^j}\,x'^{\,i}\right)
- \frac{\partial x'^{\,i}}{\partial q^j}\,x'^{\,i}$
(primes denote $d/d\tau$ at fixed $\qv$, as everywhere in these
notes) to bring it to the form
\begin{equation}
  \left(\frac{d}{d\tau}+\Hubble\right)
  \frac{\partial x^i}{\partial q^j}\,x'^{\,i}
  = \partial_{q^j}\!\left(\frac{|\xv'|^2}{2}-\phi\right)\,.
\end{equation}
Taking the Lagrangian curl kills the right-hand side:
\begin{equation}
  \left(\frac{d}{d\tau}+\Hubble\right)
  \epsilon_{lmj}\,\frac{\partial x^i}{\partial q^j}\,
  \frac{\partial x'^{\,i}}{\partial q^m} = 0\,,
\end{equation}
so the quantity in brackets (the cosmological version of the
\textbf{Cauchy invariant}) decays, and since it vanishes in the
initial conditions it vanishes at all times:
\begin{equation}\label{eq:cauchy}
  \epsilon_{lmj}\,\frac{\partial x^i}{\partial q^j}\,
  \frac{\partial x'^{\,i}}{\partial q^m} = 0\,.
\end{equation}
This is the Lagrangian-space counterpart of the statement that no
Eulerian vorticity is generated in the single-stream regime ---
note, however, that it does \emph{not} mean that $\Psiv$ is
curl-free: expanding $\xv=\qv+\Psiv$ in Eq.~\eqref{eq:cauchy},
\begin{equation}
  \epsilon_{lmj}\,\partial_{q_m}\Psi'^{\,j}
  = -\,\epsilon_{lmj}\,\partial_{q_j}\Psi^i\,\partial_{q_m}\Psi'^{\,i}\,,
\end{equation}
the curl of $\Psiv$ is sourced at second order by the longitudinal
part. Inserting the EdS expansion
$\Psiv = \sum_n \Psiv_n(\qv)\,\Dp^n$ and symmetrizing the sum
(each term carries exactly one time derivative, so the overall
factor of $\Dp'$ drops out and the resulting relation holds in any
time variable), one finds
\begin{equation}
  \epsilon_{lmj}\,\partial_{q_m}\Psi^{(n)}_j
  = \frac{1}{2n}\sum_{l=1}^{n-1}(n-2l)\,
  \epsilon_{ljm}\,\partial_{q_j}\Psi^{(l)}_i\,\partial_{q_m}\Psi^{(n-l)}_i\,,
\end{equation}
which vanishes for $n=2$ (the two terms cancel): the first
non-trivial curl appears at \emph{third} order. In Fourier space,
using the decomposition~\eqref{eq:helicity_dec},
\begin{equation}
  \vv{T}_3(\pv)\,p^2 = \frac{1}{3}
  \Big[\left(\pv_1\times(\pv_2+\pv_3)\right)
  \left(\pv_1\cdot(\pv_2+\pv_3)\right)
  L_1(\pv_1)\,L_2(\pv_2,\pv_3)\Big]_\mathrm{symm}\,,
\end{equation}
where $L_1$, $L_2$ denote the scalar parts of the kernels above.
This is the object that enters the $F_4$
relation~\eqref{eq:F4_L4} --- i.e.\ the transverse displacement
first affects the density power spectrum at two loops.

\subsection{Displacement Cumulants and Equivalence to SPT}
%------------------------------------------------------------------

To use the master formula~\eqref{eq:P_AB} beyond the ZA we need the
low-order cumulants of the displacement. From the
expansion~\eqref{eq:Ln_ansatz}, the displacement power spectrum,
$\avg{\Psi^i_{\pv}\Psi^j_{\pv'}} = (2\pi)^3\delta_D^{(3)}(\pv+\pv')\,C^{ij}(\pv)$,
receives through one-loop order the contributions
$C^{ij} = C^{ij}_{(11)}+C^{ij}_{(22)}+C^{ij}_{(13)}+C^{ij}_{(31)}$,
with
\begin{equation}\label{eq:Cij_pieces}
\begin{split}
  C^{ij}_{(11)}(\pv) &= \frac{p^ip^j}{p^4}\,P_{11}(p)\,,\qquad
  C^{ij}_{(22)}(\pv) = \frac{1}{2}\int_{\pv'}
  L^i_2(\pv',\pv-\pv')\,L^j_2(\pv',\pv-\pv')\,
  P_{11}(p')\,P_{11}(|\pv-\pv'|)\,,\\
  C^{ij}_{(13)}(\pv) &= C^{ji}_{(31)}(\pv)
  = -\frac{1}{2}\,L^i_1(\pv)\,P_{11}(p)
  \int_{\pv'} L^j_3(-\pv,\pv',-\pv')\,P_{11}(p')\,,
\end{split}
\end{equation}
while the leading (tree-level) third cumulant is
\begin{equation}\label{eq:Cijl_def}
  \avg{\Psi^i_{\pv_1}\Psi^j_{\pv_2}\Psi^l_{\pv_3}}_c
  = i\,(2\pi)^3\delta_D^{(3)}(\pv_{123})\,C^{ijl}(\pv_1,\pv_2,\pv_3)\,,
\end{equation}
\begin{equation}\label{eq:Cijl_expr}
  C^{ijl}(\pv_1,\pv_2,\pv_3)
  = L_1^i(\pv_1)\,L_1^j(\pv_2)\,L_2^l(\pv_1,\pv_2)\,
  P_{11}(p_1)\,P_{11}(p_2) + \text{2 cyclic perms}\,.
\end{equation}
(The overall factors of $i$ come from the $i$'s in the
ansatz~\eqref{eq:Ln_ansatz} together with the parity of the kernels,
$\vv{L}_n(-\pv_1,\ldots,-\pv_n)=-\vv{L}_n(\pv_1,\ldots,\pv_n)$.)

\medskip\noindent
\textbf{Equivalence to SPT.} If we now \emph{expand} the
exponentials in Eq.~\eqref{eq:P_AB} in the cumulants --- undoing the
Lagrangian resummation --- we must recover the Eulerian loop
expansion: LPT and SPT solve the same equations, and differ only in
how the perturbative series is organized. At one loop this can be
verified explicitly; the bookkeeping is carried out in
Appendix~\ref{app:LPT_SPT}, with the result
\begin{equation}
  \boxed{\;P(k) = P_{11}(k) + P_{22}(k) + 2P_{13}(k)\,.\;}
\end{equation}
So order by order in $P_{11}$, LPT and SPT are the \emph{same}
perturbative expansion, as they must be --- they solve the same
equations. The difference is entirely in what is kept resummed, and
here one must be careful: neither the zero-lag cumulants $A^{(2n)}$
nor the correlations $B^{(N)}$ are physical by themselves. Their
large IR parts cancel in the combination in which they enter
Eq.~\eqref{eq:P_AB} --- this is the $1-\cos(\vv{l}\cdot\qv)$
structure of Eq.~\eqref{eq:P_ZA} --- leaving the variance of the
\emph{relative} displacement across the separation $\qv$, as
dictated by the equivalence principle. It is this combination that
LPT keeps in the exponent and that SPT Taylor-expands. For the
smooth part of the spectrum the expansion is harmless; for the BAO
wiggles the surviving exponent is of order unity,
$k^2\Sigma^2 = O(1)$, and the expansion fails --- as we saw in
momentum space in Chapter 3, and as we will shortly see in position space.

%------------------------------------------------------------------
\subsection{EFT Corrections to LPT}
%------------------------------------------------------------------

Since LPT and SPT are the same expansion order by order, LPT
inherits all the UV diseases of Chapter 3: the loop integrals probe
modes beyond $\knl$, where the perturbative displacement is
meaningless. The Lagrangian description also makes the physical
origin of the breakdown transparent: after \emph{shell crossing} the
map $\qv\to\xv$ becomes multi-valued, and the true trajectory of a
mass element bears no resemblance to its perturbative extrapolation.
The remedy is the same as in the Eulerian theory: coarse-grain and
parameterize the effect of the short modes by effective terms in the
equation of motion. This is implemented in Lagrangian EFT~\cite{Porto:2013qua,Vlah:2015sea}. The trajectory equation~\eqref{eq:LPT_eom}
becomes
\begin{equation}\label{eq:LPT_eom_EFT}
  \Psiv'' + \Hubble\,\Psiv'
  = -\left.\gradx\phi\right|_{\xv(\qv)}
  + \vv{F}_\mathrm{eff}\big[\partial_i\partial_j\phi\big](\xv_\mathrm{fl};\tau)
  + \vv{F}_\mathrm{stoch}\,,
\end{equation}
where the effective force $\vv{F}_\mathrm{eff}$ is built, by the
equivalence principle, from second derivatives of the potential
(and their gradients), evaluated along the past trajectory --- it is
non-local in time, exactly as the effective stress tensor of
Chapter~4, and becomes local order by order in perturbation theory
by the same mechanism. At leading order in fields and gradients,
$\vv{F}_\mathrm{eff}\propto \grad\,\Delta\phi \propto \grad\delta$,
and solving with the Green's function of
Appendix~\ref{app:green-eta} produces a \textbf{counterterm
displacement} which is a pure gradient,
\begin{equation}\label{eq:Psi_ctr}
  \Delta\Psiv_\mathrm{ctr}(\kv,\eta) = -\,i\,\gamma(\eta)\,\kv\;\delta_1(\kv,\eta)\,,
\end{equation}
with the very same coefficient $\gamma$ as in Chapter 4: indeed,
$-i\kv\cdot\Delta\Psiv_\mathrm{ctr} = -\gamma k^2\delta_1$
reproduces the Eulerian counterterm~\eqref{eq:ddelta_cs}, as it must
--- the two descriptions are related by a change of variables, so
they share the same Wilson coefficients. Similarly, the stochastic
force generates a \textbf{stochastic displacement}
$\vv{\varepsilon}_\Psi$; momentum conservation (Peebles' argument of
Chapter 4) forces the short-scale dynamics to preserve the center
of mass of each patch, so $\vv{\varepsilon}_\Psi(\kv)\propto\kv$ at
small $k$, its power spectrum starts at $O(k^2)$, and the induced
density noise $-i\kv\cdot\vv{\varepsilon}_\Psi$ again produces
$P_\mathrm{stoch}\propto k^4$~\cite{Baldauf:2015tla}.

One practical remark. In evaluating the master
formula~\eqref{eq:P_AB} one has a choice of what to keep
exponentiated. The standard scheme is: keep the
\emph{linear} (Zel'dovich) displacement --- more precisely, its
long-wavelength part --- resummed in the exponent, and expand
everything else (higher-order cumulants, counterterms, stochastic
terms) to the desired loop order. The result is precisely the
IR-resummed one-loop EFT spectrum of Eq.~\eqref{eq:IR_resum}: the
Lagrangian formulation delivers the IR resummation automatically,
while the EFT corrections restore the correct UV behavior. Expanding
the exponent completely would take us back to SPT; keeping
\emph{all} of it resummed would keep spurious UV pieces of the
displacement in the exponent, which makes the UV convergence
properties even worse.

%------------------------------------------------------------------
\subsection{Lagrangian Bias}
%------------------------------------------------------------------

Galaxy bias also takes a particularly transparent form in Lagrangian
space. Galaxies form out of special initial patches, so it is
natural to weight the fluid elements by a functional of the
\emph{initial} fields and then let gravity transport them:
\begin{equation}\label{eq:Lag_bias_def}
  1+\delta_g(\xv,\tau) = \int d^3q\;F\big[\delta_1(\qv),\ldots\big]\;
  \delta_D^{(3)}\big(\xv-\qv-\Psiv(\qv,\tau)\big)\,,
\end{equation}
i.e.\ the tracers are painted on the initial conditions and then
\emph{advected} by the same displacement field as the matter. By the
equivalence principle the weight $F$ can depend on the initial
density, the initial tidal field, higher derivatives, and a
stochastic component:
\begin{equation}\label{eq:Lag_bias_exp}
  F = 1 + b_1^L\,\delta_1(\qv)
  + \frac{b_2^L}{2}\left(\delta_1^2(\qv)-\avg{\delta_1^2}\right)
  + b_{\mathcal{G}_2}^L \mathcal{G}_2(\qv)
  + b_{\nabla^2}^L\,\Delta\delta_1(\qv) + \varepsilon(\qv) + \cdots,
\end{equation}
where $\mathcal{G}_2(\qv)$
is the Galileon operator built from
the initial tidal field ($\delta_1$ here is the linear field at the
time of observation, so the $b_n^L$ are time-dependent numbers, as
usual). In Fourier space,
\begin{equation}
  \delta_g(\kv) = \int d^3q\;e^{-i\kv\cdot\qv}
  \left(F(\qv)\,e^{-i\kv\cdot\Psiv(\qv)}-1\right)\,,
\end{equation}
and the galaxy spectra are given by the same cumulant machinery as
before, with the weights $F(\qv_1)$, $F(\qv_2)$ inserted under the
average.

Two structural lessons follow immediately. First, expanding to
linear order and using number conservation of the tracers,
$\delta_g = \delta + b_1^L\,\delta_1$, i.e.
\begin{equation}\label{eq:b1_lag}
  \boxed{\;b_1 = 1 + b_1^L\;}
\end{equation}
--- the Eulerian linear bias is the Lagrangian one plus the
contribution of the flow itself. Second, and more importantly, the
Lagrangian bias basis is \emph{non-local in time by construction}: all
the operators are evaluated on the initial slice, and the time
non-locality that plagued the Eulerian discussion of Chapter 5 is
entirely carried by the advection, i.e.\ by $\Psiv$. Expanding
Eq.~\eqref{eq:Lag_bias_def} in perturbations, the advected initial
operators automatically generate the full Eulerian
basis~\eqref{eq:bias_expansion}, so the two expansions
are equivalent at the one-loop order. At the two-loop order,
however the Lagrangian expansion will generate more terms than
the naive Eulerian one due to the curl-displacement contributions~\cite{Ivanov:2026zos}.
However, these can be accounted for in a more systematic version
of the Eulerian bias~\cite{Bakx:2026rmd}.  

%------------------------------------------------------------------
\subsection{RSD in Lagrangian Space}
%------------------------------------------------------------------

Redshift-space distortions are arguably at their simplest in the
Lagrangian language: the mapping of Chapter 6 acts on
\emph{positions}, and in LPT the position and the velocity are both
carried by the displacement, $\vvv = \Psiv'$. The redshift-space
trajectory is therefore
\begin{equation}
  \xv_s(\qv) = \qv + \Psiv(\qv)
  + \frac{\hat{z}\cdot\Psiv'(\qv)}{\Hubble}\,\hat{z}
  \equiv \qv + \Psiv_s(\qv)\,.
\end{equation}
For the EdS-approximated time dependence $\Psiv^{(n)}\propto\Dp^n$
we have $\Psiv'^{(n)} = n f\Hubble\,\Psiv^{(n)}$, so the
redshift-space displacement is obtained order by order by a simple
\emph{linear map},
\begin{equation}\label{eq:RSD_matrix}
  \Psi^{(n)}_{s,i} = R^{(n)}_{ij}\,\Psi^{(n)}_j\,,\qquad
  R^{(n)}_{ij} = \delta_{ij} + n\,f\,\hat{z}_i\hat{z}_j\,.
\end{equation}
All the formulas of this chapter then go through with
$\Psiv\to\Psiv_s$ (the bias weights $F(\qv)$ are untouched --- the
mapping only reshuffles positions):
\begin{equation}\label{eq:Ps_LPT}
  P^{(s)}(k,\mu)=\int d^3q\;e^{-i\kv\cdot\qv}
  \left(\avg{F_1F_2\,e^{-i\kv\cdot[\Psiv_s(\qv)-\Psiv_s(0)]}}-1\right)\,.
\end{equation}
At linear order,
$-i\kv\cdot\Psi^{(1)}_s = -i\,(k_i + f k_z\hat z_i)\Psi^{(1)}_i
= (1+f\mu^2)\,\delta_1$, and together with $b_1^L$ from the weight
we recover the Kaiser formula~\eqref{eq:Kaiser}.

The payoff comes in the infrared. The exponent in
Eq.~\eqref{eq:Ps_LPT} now involves the \emph{redshift-space}
displacement, whose linear variance is anisotropic: contracting
$\kv R^{(1)}=\kv+f k_z\hat z$ with the isotropic
correlator~\eqref{eq:ZA_corr} gives
$|\kv+fk_z\hat{z}|^2 = k^2\left[1+f(f+2)\,\mu^2\right]$ for the
leading (isotropic) part. The BAO damping of Chapter~3 therefore
generalizes to
\begin{equation}\label{eq:Sigma_RSD}
  \boxed{\;k^2\Sigma^2 \;\longrightarrow\;
  k^2\left[1+f(f+2)\,\mu^2\right]\Sigma^2\;}
\end{equation}
(plus subleading anisotropic corrections): the wiggles are damped
more strongly along the line of sight, because the same bulk flows
that displace the pairs also Doppler-shift them. This anisotropic
damping factor is what is used in the IR-resummed redshift-space
power spectrum in practical analyses.

Finally, the EFT corrections. The redshift-space
map~\eqref{eq:RSD_matrix} feeds the \emph{velocity} counterterms
into the displacement: the corrections to $\Psiv'$ (cf.\ the
velocity response $\Delta\Theta^{(c_s^2)} = (m+1)\Delta\delta^{(c_s^2)}$
of Appendix~\ref{app:green-eta}) enter multiplied by
$\hat{z}_i\hat{z}_j$, so the counterterm
displacement~\eqref{eq:Psi_ctr} acquires line-of-sight components
with independent coefficients. Projecting onto the density, this
reproduces exactly the $\mu^2$ and $\mu^4$ counterterm structure of
Eq.~\eqref{eq:RSD_ctr}, truncating at $\mu^4$ for the same
symmetry reasons; and since the short-scale velocity dispersion is
larger than the short-scale density response (Fingers of God, cf.\
Chapter 6), the line-of-sight coefficients are parametrically
enhanced. The Lagrangian and Eulerian EFTs in redshift space are,
once again, the same theory in different variables.

%------------------------------------------------------------------
\subsection{Bulk Flows and the BAO in Position Space}
%------------------------------------------------------------------

We conclude with the position-space picture of the IR story of Chapter~3,
which in the Lagrangian language becomes almost trivial. Consider two
clumps of matter at separation $\xv_1$, with linear correlation function
$\xi(x_1)=\avg{\delta(0)\delta(\xv_1)}$, and plunge them into a
long-wavelength flow. The flow is a linear-theory mode, statistically
independent of the short-wavelength modes that make up the clumps, so in
all correlators below the averages over the flow and over the clumps
factorize.

Suppose first that the flow is \emph{coherent} across the pair: both
clumps are displaced by the same $\Psiv_L$, and their trajectories are
parallel. Then \emph{both} arguments of the correlator are shifted, and
\begin{equation}
\avg{\delta(\Psiv_L)\,\delta(\xv_1+\Psiv_L)}
= \int_{\kv_1}\!\int_{\kv_2} e^{i\kv_2\cdot\xv_1}\,
\avg{e^{i(\kv_1+\kv_2)\cdot\Psiv_L}}\,
\avg{\delta_{\kv_1}\delta_{\kv_2}}
= \int_{\kv} e^{i\kv\cdot\xv_1}\,P(k) = \xi(x_1)\,,
\end{equation}
because $\avg{\delta_{\kv_1}\delta_{\kv_2}}\propto
\delta_D^{(3)}(\kv_1+\kv_2)$ forces $\kv_2=-\kv_1$, upon which the phase
$e^{i(\kv_1+\kv_2)\cdot\Psiv_L}$ equals unity \emph{identically}, before
any averaging. A coherent flow, however large, leaves the correlation
function untouched --- the position-space version of the
equivalence-principle cancellation~\eqref{eq:soft_translation}.

What \emph{is} physical is the \textbf{relative displacement} of the
pair, $\Delta\Psiv_L \equiv \Psiv_L(\xv_1)-\Psiv_L(0)$, generated by the
modes that vary across the separation. Shifting only the separation by
$\Delta\Psiv_L$ and averaging over the (Gaussian, linear-theory) flow,
\begin{equation}\label{eq:xi_smeared}
\avg{\delta(0)\,\delta(\xv_1+\Delta\Psiv_L)}
= \int_{\kv} e^{i\kv\cdot\xv_1}\,
\avg{e^{i\kv\cdot\Delta\Psiv_L}}\,P(k)
= e^{\frac{1}{2}\avg{\Delta\Psi^i_L\,\Delta\Psi^j_L}\,
\partial_i\partial_j}\;\xi(x_1)\,:
\end{equation}
the bulk flows act on the correlation function as a Gaussian
\emph{smearing} operator. Now recall the two components of $\xi$: a
smooth, featureless broadband and the BAO peak,
\begin{equation}
\xi(r) = \xi_s(r) + \xi_w(r)\,,\qquad
\xi_s(r)=\left(\frac{r_0}{r}\right)^{\gamma}\,,\qquad
\xi_w(r) = A_w\,e^{-\frac{(r-\ell_\mathrm{BAO})^2}{2\sigma_w^2}}\,,
\end{equation}
with $\ell_\mathrm{BAO}=1/\kosc\simeq 110\Mpch$
(cf.~Eq.~\eqref{eq:Pw_decomp}) and a peak width $\sigma_w\sim 10\Mpch$
set by Silk damping. Since the displacement correlations decay across
$\ell_\mathrm{BAO}$, the relative displacement is of the order of the
one-point r.m.s.\ displacement estimated in Chapter~2,
$|\Delta\Psiv_L|\sim 10\Mpch$. For the smooth part the smearing is then
negligible --- each derivative in Eq.~\eqref{eq:xi_smeared} costs $1/r$:
\begin{equation}
\avg{\Delta\Psi_L^2}\,\partial^2\xi_s \sim
\frac{\avg{\Delta\Psi_L^2}}{r^2}\,\xi_s \sim 10^{-2}\,\xi_s\,.
\end{equation}
For the peak, however, the derivatives are enhanced by its width, not by
the separation:
\begin{equation}
\avg{\Delta\Psi_L^2}\,\partial^2\xi_w \sim
\frac{\avg{\Delta\Psi_L^2}}{\sigma_w^2}\,\xi_w \sim \xi_w\,.
\end{equation}
The width of the BAO peak is of the same order as the large-scale
relative displacements, so Taylor expanding the exponent in
Eq.~\eqref{eq:xi_smeared} makes no sense. But this Taylor expansion is
exactly what SPT does: the Eulerian expansion is a power series in
$P_{11}$, and the displacement variance is itself
$\avg{\Delta\Psi_L^2}\sim\int_\pv P_{11}(p)/p^2 = O(P_{11})$. This is
the position-space counterpart of the momentum-space breakdown of
Chapter~3 ($k^2\Sigma^2\sim 1$ for the wiggly part), and it makes the
resolution self-evident: keep the displacements in the exponent, as LPT
does automatically.

The quantitative connection to the damping factor of Chapter~3 is
sharp. The Gaussian peak in $\xi_w$ is a \emph{radial} feature at
$|\rv|=\ell_\mathrm{BAO}$, so its smearing --- and hence the damping of
the wiggles in Fourier space --- is controlled by the component of
$\Delta\Psiv_L$ \emph{along} the separation, evaluated at the BAO scale:
\begin{equation}\label{eq:Sigma_long}
\Sigma^2 = \frac{1}{2}\,
\avg{\big(\hat{\qv}\cdot\Delta\Psiv_L(\qv)\big)^2}
\Big|_{q\,=\,\ell_\mathrm{BAO}}\,,
\end{equation}
which reproduces \emph{exactly} Eq.~\eqref{eq:Sigma2_def}, including the
spherical-Bessel bracket (see the Homework below). The IR-resummed
spectrum~\eqref{eq:IR_resum} is thus nothing but the Fourier transform
of the smeared correlation function~\eqref{eq:xi_smeared}.

\medskip\noindent
\textbf{Homework:} using
$\avg{\Delta\Psi_L^i\,\Delta\Psi_L^j}(\qv)
= 2\int_\pv \frac{p^ip^j}{p^4}\,P_{11}(p)
\left[1-\cos(\pv\cdot\qv)\right]$
and the angular integral
$\int\frac{d\Omega_p}{4\pi}\,\hat{p}^i\hat{p}^j\,e^{i\pv\cdot\qv}
= \frac{j_1(pq)}{pq}\,\delta^{ij} - j_2(pq)\,\hat{q}^i\hat{q}^j$,
decompose the relative-displacement tensor into components parallel and
orthogonal to the separation, and show, with the help of the identity
$j_1(x)/x = \left[j_0(x)+j_2(x)\right]/3$, that the longitudinal
dispersion~\eqref{eq:Sigma_long} equals
\begin{equation}
\frac{1}{6\pi^2}\int dp\;P_{11}(p)
\left[1-j_0(p\,\ell_\mathrm{BAO})+2\,j_2(p\,\ell_\mathrm{BAO})\right]\,,
\end{equation}
i.e.\ precisely the damping factor $\Sigma^2$ of
Eq.~\eqref{eq:Sigma2_def} (with the integral restricted to the IR modes
$p\leq\kIR$, as there). Compute also the transverse dispersion and show
that it involves the combination $1-j_0-j_2$.

%==========================================================================
\section{Chapter 8: Outlook}
%==========================================================================

There are many important topics that are not covered in these notes. Let
us mention some of them.

\paragraph{Bispectrum.}
The  bispectrum is the Fourier-space three-point function:
\begin{equation}
  \avg{\delta(\kv_1)\,\delta(\kv_2)\,\delta(\kv_3)} = (2\pi)^3\,\delta_D^{(3)}(\kv_1+\kv_2+\kv_3)\;B(k_1,k_2,k_3)\,.
\end{equation}
It vanishes for Gaussian fields, but 
it is generated by non-linearity. In EFT
at \textit{tree level} (leading order):
\begin{equation}\label{eq:B_tree}
  \boxed{B_{112}(k_1,k_2,k_3) = 2\,F_2(\kv_1,\kv_2)\,P_{11}(k_1)\,P_{11}(k_2) + \text{2 cyclic perms}\,.}
\end{equation}
This uses the same $F_2$ kernel from eq.~\eqref{eq:F2}.
One can generalize it to one-loop order and also include
galaxy bias, redshift space distortions, IR resummation, etc.~\cite{Baldauf:2014qfa,Philcox:2022frc,Chen:2024pyp,Bakx:2025pop}. 
A conceptually interesting 
novelty is the appearance of \textbf{mixed stochastic} contributions
like $\delta_g\supset \varepsilon \delta$ etc., which 
are fully redundant at the power spectrum level.  

\paragraph{Primordial Non-Gaussianity.}
If the initial conditions in our Universe satisfy
\mbox{$\langle \delta_1\delta_1\delta_1\rangle\neq 0$}, many 
interesting modifications to EFT should happen~\cite{Dalal:2007cu,Baumann:2012bc,Assassi:2015fma,Assassi:2015jqa,Welling:2016dng,Schmidt:2012ys,Desjacques:2016bnm,MoradinezhadDizgah:2020whw,Cabass:2022wjy,Cabass:2022epm,Cabass:2024wob,
Green:2023uyz,
Green:2026yev,Sharma:2025xss,Chudaykin:2025vdh}. 
For instance, the $\langle \delta^{(1)} \delta^{(2)}\rangle$
power spectrum corrections are not zero anymore. 
However, most importantly, there will be new terms 
in the galaxy bias expansion. Their origin is similar 
to that of bias itself: the modifications of statistics
of the initial distributions can be effectively captured by 
adding new fields in the deterministic field expansion of $\delta_g$.
In particular, in the 
presence of the \textit{local} primordial non-Gaussianity 
one can add a new operator $b_\phi \phi(\qv)$ ($\phi(\qv)$ is the gravitational Bardeen potential on the Lagrangian time slice), giving rise
to \textit{scale-dependent bias}, whose measurement is 
one of the main
targets of ongoing and upcoming galaxy surveys. Importantly, 
$\phi(\qv)$ should be evaluated on the Lagrangian coordinates
in order to respect the equivalence principle! 

\paragraph{Time-sliced perturbation theory (TSPT).}
This is a formulation of EFT in the language of 
generating functionals
familiar to particle
physicists. The relevant objects in this framework are equal-time 
$n$-point functions,
which are more natural statistical 
observables than the cosmological fields used in other EFT formulations. 
The advantage of this approach is 
that one makes the symmetries related to the equivalence 
principle manifest, which is beneficial for IR resummation.
This allows one to develop IR resummation for an arbitrary
$n$-point functions, which is very difficult in alternative
formulations, such as LPT. 
TSPT also provides a natural framework to implement 
Wilson-Polchinski renormalization group ideas 
at the level of the effective action~\cite{Ivanov:2022mrd}.

\paragraph{Backreaction.}
One of the original motivations to develop EFT for LSS is to show
the absence of backreacion of small scale non-linear 
structure onto large scales.
In the Newtonian settings this is guaranteed by Peebles' argument
we discussed in Chapter 4. EFT for LSS can be formulated
in a fully relativistic setting and used to upgrade 
Peebles' argument to the level of full general relativity~\cite{Baumann:2010tm}. 

\paragraph{Lyman-$\alpha$ forest.}
An interesting example of a tracer with a different symmetry
structure is the Lyman-$\alpha$ forest. The observed 
Lyman-$\alpha$ forest
flux decrement can be modeled 
as a non-linear transformation of a 
conserved redshift-space tracer field,
whose normalization requires 
line-of-sight operators. The corresponding bias expansion
only has an SO(2) symmetry of rotations around the line-of-sight, as opposed
to the full SO(3) rotation symmetry of galaxies.
As a result, at the linear level one has an extra bias parameter
$b_\eta$
in the Kaiser formula $\delta_{F}=(b_1-b_\eta f\mu^2)\delta_1$~\cite{Givans:2020sez,Chen:2021rnb,Ivanov:2023yla,Ivanov:2024jtl,deBelsunce:2024rvv,Chudaykin:2025gsh,deBelsunce:2025bqc,deBelsunce:2025gci,deBelsunce:2026tks}. 

\paragraph{Field-level EFT.} In these notes we focused 
on the galaxy power spectrum calculation, but EFT is a theory for the entire
field. This has several implications. First, one can use EFT as a forward model to produce galaxy maps at the pixel level. 
This is helpful in order to get precision measurements of 
EFT parameters by virtue of cosmic variance cancellation~\cite{Schmittfull:2014tca,Abidi:2018eyd,Lazeyras:2017hxw,Ivanov:2024hgq,Ivanov:2024xgb,Ivanov:2024dgv}. 
Second, one can use EFT to measure cosmological
parameters from the whole density map, 
which can avoid a potential loss of information 
due to compression of the galaxy field into $n$-point functions~\cite{Schmidt:2018bkr,Schmidt:2020viy,Nguyen:2020hxe,Nguyen:2024yth,
cabass_pt_field_level,Akitsu:2025boy}. 
These topics are a subject of intense investigation at the moment. 

\paragraph{EFT for galaxy shapes.} The shapes of galaxies $\gamma_{ab}$,
which is a trace-less spin-2 field, correlates
with the underlying cosmological fields. These correlations,
called \textit{intrinsic alignments} (IA)
play an important role in cosmic shear surveys, where 
they represent a major source of contamination of the weak
lensing signal. One can use EFT arguments, i.e. symmetries
and dimensional analysis, to model IA. This problem is a generalization
of the galaxy bias expansion to a spin-2 field. In real space
at the lowest (linear) order there is only one spin-2 treeless field
which one can write, 
\[
\gamma_{ab}=\mathcal{C}_s \left(\partial_a\partial_b\hat{\Phi}-\frac{\delta_{ab}}{3}\Delta\hat{\Phi}\right)\,.
\]
EFT thinking allows one to systematically go beyond the linear order~\cite{Vlah:2019byq,Bakx:2023mld,Chen:2023yyb}.  

\paragraph{Counts-in-cells statistics.} The path integral 
formulation of EFT can be used to predict the one-point
probability distribution function of the spherically averaged
density field. This is one of the most basic large-scale 
structure observables, which is quite hard to predict
from first principles. The proper
calculation parallels that of tunneling in 
quantum field theory~\cite{Ivanov:2018lcg,Chudaykin:2022sdl}. 
The result is an
EFT for density fluctuations 
in a non-trivial spherical collapse background. 
This formulation is non-perturbative as the spherical
collapse background is a fully non-linear solution.

\paragraph{Acknowledgments.}
We are grateful to the organizers of the ``New Physics from Galaxy Clustering'' institute, and especially Marko Simonovi\'c, 
Emanuele Castorina, and Diego Redigolo 
for the opportunity to deliver these lectures.
An earlier version of these lectures were delivered at the UNAM School 
``Effective Field Theories Across the Universe.''
The notes underling these lectures were written down by MI, 
and partly transcribed by Claude Code for this manuscript. 
Claude Code generated Appendices B and C 
with some straightforward computations that have been independently
checked by MI.

\appendix

%==========================================================================
\section{Gaussian Random Fields and Wick's Theorem}\label{app:GRF}
%==========================================================================

In this appendix we collect the basic formalism of Gaussian random
fields and derive Wick's theorem, which is stated and heavily used
in the main text (Chapters 2 and 3). The same generating-functional
technique is used in the Press--Schechter computation of Chapter 5.

\subsection{Warm-Up: the Multivariate Gaussian}

Consider first a finite collection of random variables
$x=(x_1,\ldots,x_n)$ with the Gaussian probability density
\be
  P(x) = \frac{1}{\mathcal{N}}\;e^{-\frac{1}{2}\,x^T M\, x}\,,\qquad
  \mathcal{N} = \int d^n x\;e^{-\frac{1}{2}\,x^T M\, x}
  = \frac{(2\pi)^{n/2}}{\sqrt{\det M}}\,,
\ee
where $M$ is a symmetric, positive-definite $n\times n$ matrix. All
the moments are encoded in the generating function
\be
  Z(J) \equiv \avg{e^{J^T x}}
  = \frac{1}{\mathcal{N}}\int d^n x\;
  e^{-\frac{1}{2}\,x^T M\, x + J^T x}\,,\qquad
  \avg{x_{i_1}\cdots x_{i_m}} =
  \frac{\partial^m Z(J)}{\partial J_{i_1}\cdots \partial J_{i_m}}
  \bigg|_{J=0}\,.
\ee
The integral is evaluated by completing the square: the shift
$x \to x + M^{-1}J$ removes the source term,
\be
  -\frac{1}{2}\,x^T M\, x + J^T x \;\longrightarrow\;
  -\frac{1}{2}\,x^T M\, x + \frac{1}{2}\,J^T M^{-1} J\,,
\ee
and the remaining integral over the shifted variable cancels against
$\mathcal{N}$, leaving
\be\label{eq:ZJ_finite}
  \boxed{\;Z(J) = \exp\left\{\frac{1}{2}\,J^T M^{-1} J\right\}\,.\;}
\ee
Differentiating twice, $\avg{x_i x_j} = (M^{-1})_{ij} \equiv C_{ij}$:
the covariance matrix is the \emph{inverse} of the matrix in the
exponent. Differentiating four times, each derivative must either
produce a factor $(M^{-1}J)_i$ from the exponent or remove one
already produced, so at $J=0$ the result is a sum over complete
pairings,
\be
  \avg{x_i x_j x_k x_l} = C_{ij}C_{kl} + C_{ik}C_{jl} + C_{il}C_{jk}\,,
\ee
--- Wick's theorem, already at finite $n$.

Nothing in this computation refers to $n$ being finite. To describe
a random \emph{field}, promote the discrete label to a continuous
one, $i \to \xv\in\mathbb{R}^3$, with the dictionary
\be
  x_i \to \phi(\xv)\,,\quad
  J_i \to J(\xv)\,,\quad
  M_{ij} \to M(\xv,\vv{y})\,,\quad
  \sum_i \to \int d^3x\,,\quad
  d^n x \to \mathcal{D}\phi \equiv \lim \prod_{\xv} d\phi(\xv)\,,
\ee
so that $J^T x \to \int d^3x\,J(\xv)\phi(\xv)$, matrix
multiplication becomes integration, and the inverse matrix becomes
the inverse kernel, $\int d^3y\, M(\xv,\vv{y})\,\xi(\vv{y},\vv{z}) =
\delta_D^{(3)}(\xv-\vv{z})$. The only genuinely new feature of the
continuum is that the determinant in $\mathcal{N}$ becomes formally
infinite --- but it cancels in every \emph{normalized} expectation
value, which is why the generating functional below carries the
factor $\mathcal{N}^{-1}$ from the start.

\subsection{Gaussian Random Fields}

With this dictionary at hand, let $\phi(\xv)$ be a classical random field in 3d Euclidean space.
Its statistics are fully specified by the \textbf{generating
functional}
\begin{equation}\label{eq:Zphi}
  Z[J]=\mathcal{N}^{-1}\int \mathcal{D}\phi\;
  e^{-W[\phi]}\,e^{\int d^3x\, J(\xv)\phi(\xv)}\,,\qquad
  \mathcal{N}=\int \mathcal{D}\phi\; e^{-W[\phi]}\,,
\end{equation}
where $W[\phi]$ is the statistical weight (``effective action'').
The field is \textbf{Gaussian} if the weight is quadratic,
\begin{equation}
  W[\phi] = \frac{1}{2}\int d^3x\, d^3y\;
  M(\xv,\vv{y})\,\phi(\xv)\,\phi(\vv{y})\,,
\end{equation}
with some positive kernel $M$. All $n$-point correlation functions
follow from functional derivatives,
\begin{equation}\label{eq:Z_derivs}
  \avg{\phi(\xv_1)\cdots\phi(\xv_n)}
  = \frac{\delta^n Z[J]}{\delta J(\xv_1)\cdots\delta J(\xv_n)}
  \bigg|_{J=0}\,.
\end{equation}
The integral~\eqref{eq:Zphi} is computed by completing the square
exactly as in the warm-up, Eq.~\eqref{eq:ZJ_finite}: shifting
$\phi(\xv)\to\phi(\xv)+\int d^3y\,\xi(\xv,\vv{y})J(\vv{y})$ with
$\xi \equiv M^{-1}$, the source term is eliminated and the
normalization cancels, leaving
\begin{equation}\label{eq:ZJ_result}
  \boxed{\;Z[J]=\exp\left\{\frac{1}{2}\int d^3x_1\,d^3x_2\;
  J(\xv_1)\,\xi(\xv_1,\xv_2)\,J(\xv_2)\right\}\,.\;}
\end{equation}
Everything follows from this expression:
\begin{enumerate}[nosep]
  \item Differentiating twice, $\avg{\phi(\xv_1)\phi(\xv_2)} =
  \xi(\xv_1,\xv_2)\equiv\xi_{12}$: the kernel of the weight is the
  inverse of the two-point function.
  \item $Z[J]$ is even under $J\to-J$, so all correlators with an
  \emph{odd} number of fields vanish.
  \item Differentiating four times, the three ways of pairing up the
  derivatives give
  \begin{equation}\label{eq:wick4}
    \avg{\phi(\xv_1)\phi(\xv_2)\phi(\xv_3)\phi(\xv_4)}
    = \xi_{12}\,\xi_{34}+\xi_{13}\,\xi_{24}+\xi_{14}\,\xi_{23}\,,
  \end{equation}
  which is exactly the four-point contraction identity used for the
  one-loop power spectrum in Chapter 3.
  \item In general, each derivative must either hit the exponent
  (producing a new $\xi J$) or an already-produced $J$; at $J=0$
  only completely paired terms survive, giving
  \textbf{Wick's theorem}:
  \begin{equation}\label{eq:wick_general}
    \avg{\phi(\xv_1)\cdots\phi(\xv_{2N})}
    = \sum_{\text{pairings}}\;\prod_{\text{pairs }(ij)}\xi_{ij}\,,
  \end{equation}
  with $(2N-1)!!$ terms in the sum --- precisely the statement of
  Chapter 2.
\end{enumerate}

Two useful corollaries. First, taking the logarithm of
Eq.~\eqref{eq:ZJ_result}: $\ln Z[J]$ --- the generator of
\emph{connected} correlators (cumulants) --- is exactly quadratic,
so for a Gaussian field all cumulants beyond the second vanish.
Written for a linear combination $X=\int J\phi$, this is the
statement
\begin{equation}
  \avg{e^{X}} = e^{\frac{1}{2}\avg{X^2}_c}\,,
\end{equation}
i.e.\ the cumulant expansion theorem used in Chapter 7 truncates at
$N=2$ --- this is what made the Zel'dovich power
spectrum~\eqref{eq:P_ZA} computable in closed form. Second,
\emph{linear} time evolution preserves Gaussianity: if
$\phi(\kv,\tau) = g(k,\tau)\,\phi_0(\kv)$ is a linear map of
Gaussian initial conditions, all the correlators above, including
unequal-time ones, still obey Wick's theorem. Non-Gaussianity in
the late Universe is generated precisely by the \emph{non-linear}
mode coupling of Chapter 3.

\medskip\noindent
\textbf{Symmetries and Fourier space.} Cosmological ensembles are
statistically homogeneous and isotropic, so
$\xi(\xv_1,\xv_2)=\xi(|\xv_1-\xv_2|)$. In Fourier space (with the
conventions of Eq.~\eqref{eq:deltak}, and the reality condition
$\phi^*(\kv)=\phi(-\kv)$),
\begin{equation}
\begin{split}
  \avg{\phi(\kv)\phi(\kv')} &= \int d^3x_1\,d^3x_2\;
  e^{-i\kv\cdot\xv_1-i\kv'\cdot\xv_2}\;\xi(|\xv_1-\xv_2|)\\
  &= (2\pi)^3\,\delta_D^{(3)}(\kv+\kv')\int d^3z\;
  \xi(z)\,e^{-i\kv\cdot\vv{z}}
  \equiv (2\pi)^3\,\delta_D^{(3)}(\kv+\kv')\,P(k)\,,
\end{split}
\end{equation}
recovering the power spectrum of Eq.~\eqref{eq:pk}: homogeneity
produces the momentum-conserving delta function, isotropy makes
$P$ a function of $k=|\kv|$ only. Wick's theorem in Fourier space
reads
\begin{equation}
  \avg{\phi(\kv_1)\cdots\phi(\kv_{2N+1})} = 0\,,\qquad
  \avg{\phi(\kv_1)\cdots\phi(\kv_{2N})}
  = \sum_{\text{pairings}}\;\prod_{\text{pairs }(ij)}
  \avg{\phi(\kv_i)\,\phi(\kv_j)}\,,
\end{equation}
which is the form used throughout the loop computations of the main
text.

\medskip\noindent
\textbf{Relation to the Press--Schechter computation.} The
generating functional also underlies the halo statistics of
Chapter 5: the one-point PDF of the smoothed density,
Eq.~\eqref{eq:PS_lambda}, is nothing but $Z[J]$ evaluated on the
localized source $J(\xv)=\lambda\,W_R(\xv)$, and the joint
two-point PDF~\eqref{eq:pdf2joint} corresponds to a source with two
components, $J = \lambda_1 W_{R_1} + \lambda_2 W_{R_2}$ shifted by
the separation. Constrained Gaussian statistics of this kind are
the bread and butter of analytic halo models.

%==========================================================================
\section{Green's Functions in $\eta=\ln D_+$}\label{app:green-eta}
%==========================================================================

In this appendix we derive the retarded Green's functions of the
linearized fluid system in the time variable $\eta=\ln\Dp$. They are
used in the main text to compute the response of the density field
to the effective stress tensor (Chapters 3, 4). The punchline is
that in the EdS approximation $\Omega_m=f^2$ the linear system has
\emph{constant} coefficients, so the equations acquire a
time-translation symmetry in $\eta$: the Green's functions depend
only on the difference $s\equiv\eta-\eta'$ and can be obtained in
closed form.

\subsection{The linear system}

It is convenient to collect the fields into a doublet (to avoid confusion, in this appendix we do not use the displacement field which is denoted by $\Psi$ as well),
\begin{equation}
  \Psi_a(\kv,\eta) \equiv
  \begin{pmatrix} \delta_{\kv} \\[2pt] \Theta_{\kv} \end{pmatrix},
  \qquad a = 1,2\,,
\end{equation}
with $\Theta=-\partial_i v^i/(f\Hubble)$ as in the main text; the
normalization is chosen so that on the linear growing mode
$\Psi_1^{(1)}=\Psi_2^{(1)}=e^\eta\,\delta_0(\kv)$. The equations of
motion \eqref{eq:SPT_cont}--\eqref{eq:SPT_euler} then take the form
\begin{equation}\label{eq:doublet_system}
  \partial_\eta\Psi_a(\kv,\eta) + \Omega_{ab}\,\Psi_b(\kv,\eta)
  = \int_{\qv_1}\int_{\qv_2}(2\pi)^3\,\delta_D^{(3)}(\kv-\qv_{12})\;
  \gamma_{abc}(\qv_1,\qv_2)\;\Psi_b(\qv_1,\eta)\,\Psi_c(\qv_2,\eta)
  + J_a(\kv,\eta)\,,
\end{equation}
where $J_a=(0,J_2)^T$ collects possible linear sources --- such as
the counterterm sources of Chapter 4, which enter the Euler equation
only. The symmetrized vertex $\gamma_{abc}(\qv_1,\qv_2)=\gamma_{acb}(\qv_2,\qv_1)$
encodes the mode-coupling kernels $\alpha$ and $\beta$ of
Eq.~\eqref{eq:alpha_beta}; its only non-vanishing components are
\begin{equation}\label{eq:vertex_app}
  \gamma_{121}(\qv_1,\qv_2) = \frac{\alpha(\qv_1,\qv_2)}{2}\,,\qquad
  \gamma_{112}(\qv_1,\qv_2) = \frac{\alpha(\qv_2,\qv_1)}{2}\,,\qquad
  \gamma_{222}(\qv_1,\qv_2) = \beta(\qv_1,\qv_2)\,.
\end{equation}
Indeed, for $a=1$ the two $\alpha$-components add up under the
symmetric integral to reproduce the right-hand side of the
continuity equation,
$\gamma_{121}\Psi_2(\qv_1)\Psi_1(\qv_2)+\gamma_{112}\Psi_1(\qv_1)\Psi_2(\qv_2)
\to \alpha(\qv_1,\qv_2)\,\Theta_{\qv_1}\delta_{\qv_2}$,
while for $a=2$ the single component $\gamma_{222}$ directly
reproduces $\beta(\qv_1,\qv_2)\,\Theta_{\qv_1}\Theta_{\qv_2}$
(no factor of $1/2$ is needed here, since $\beta$ is already
symmetric and carries an explicit $1/2$ in its definition).
The matrix of linear coefficients is
\begin{equation}
  \Omega_{ab} =
  \begin{pmatrix}
    0 & -1 \\[4pt]
    -\dfrac{3\Omega_m}{2f^2} & \dfrac{3\Omega_m}{2f^2}-1
  \end{pmatrix}
  \;\xrightarrow{\;\Omega_m=f^2\;}\;
  \begin{pmatrix}
    0 & -1 \\[4pt]
    -\dfrac{3}{2} & \dfrac{1}{2}
  \end{pmatrix}\,.
\end{equation}
In the EdS approximation the matrix is constant, which is the source
of all the simplifications below.

\subsection{Second-Order Equation, Modes and Wronskian}

Eliminating $\Theta$ from the linearized system (the first equation
gives $\Theta=\partial_\eta\delta$), we obtain a single second-order
equation with constant coefficients,
\begin{equation}\label{eq:growth_eta_app}
  \ddot\delta + \frac{1}{2}\,\dot\delta - \frac{3}{2}\,\delta = j(\eta)\,,
  \qquad \dot{\phantom{x}} \equiv \frac{d}{d\eta}\,,
\end{equation}
where $j$ is a generic linear source. Substituting
$\delta=e^{\lambda\eta}$ into the homogeneous equation gives the
characteristic polynomial
\begin{equation}
  \lambda^2 + \frac{1}{2}\,\lambda - \frac{3}{2} = 0
  \quad\Longleftrightarrow\quad
  (2\lambda+3)(\lambda-1) = 0
  \quad\Longrightarrow\quad
  \lambda_+ = 1\,,\quad \lambda_- = -\frac{3}{2}\,,
\end{equation}
so the two independent solutions are the growing and decaying modes,
\begin{equation}
  D_+(\eta) = e^{\eta}\,,\qquad
  D_-(\eta) = e^{-3\eta/2} = D_+^{-3/2}\,.
\end{equation}
The first is tautological ($\eta=\ln D_+$ by definition), while the
second is the familiar EdS relation $D_-\propto D_+^{-3/2}$, here
valid for general $\Lambda$CDM within the $\Omega_m=f^2$
approximation. Their Wronskian is
\begin{equation}\label{eq:W_eta}
  W(\eta) = D_+\dot D_- - D_-\dot D_+
  = -\frac{3}{2}\,e^{-\eta/2} - e^{-\eta/2}
  = -\frac{5}{2}\,e^{-\eta/2}\,,
\end{equation}
consistent with Abel's identity,
$W\propto\exp[-\int^\eta \tfrac{1}{2}\,d\eta'] = e^{-\eta/2}$.

\subsection{The Scalar Green's Function}

The retarded Green's function of Eq.~\eqref{eq:growth_eta_app}
satisfies
\begin{equation}
  \ddot g_\eta + \frac{1}{2}\,\dot g_\eta - \frac{3}{2}\,g_\eta
  = \delta_D(\eta-\eta')\,,\qquad
  g_\eta = 0 \;\;\text{for}\;\; \eta<\eta'\,,
\end{equation}
and is given by the standard two-solution formula
\begin{equation}
  g_\eta(\eta,\eta') = \Theta_H(\eta-\eta')\,
  \frac{D_+(\eta')\,D_-(\eta) - D_-(\eta')\,D_+(\eta)}{W(\eta')}\,.
\end{equation}
The numerator equals
$e^{-\eta'/2}\big[e^{-3s/2}-e^{s}\big]$ with $s=\eta-\eta'$, and
dividing by $W(\eta')=-\tfrac{5}{2}e^{-\eta'/2}$ the
$\eta'$-dependence cancels completely --- the promised translation
invariance:
\begin{equation}\label{eq:g_eta_app}
  \boxed{\;g_\eta(s) = \frac{2}{5}\,\Theta_H(s)
  \left[e^{s} - e^{-3s/2}\right]\,,\qquad s = \eta - \eta'\;}
\end{equation}
Basic checks: continuity, $g_\eta(0)=0$; unit derivative jump,
$\dot g_\eta(0^+) = \tfrac{2}{5}\big(1+\tfrac{3}{2}\big) = 1$;
positivity for $s>0$ (a positive source enhances the growth of
$\delta$). At late times the growing mode dominates,
$g_\eta(s)\approx \tfrac{2}{5}\,e^s = \tfrac{2}{5}\,\Dp(\eta)/\Dp(\eta')$,
while at early separations $g_\eta(s)\approx s$, the short-``time''
linear-response regime.

\subsection{The Matrix Propagator}

For the first-order system~\eqref{eq:doublet_system} the retarded
matrix propagator satisfies
\begin{equation}
  \partial_\eta\,g_{ab}(\eta,\eta') + \Omega_{ac}\,g_{cb}(\eta,\eta')
  = \delta_{ab}\,\delta_D(\eta-\eta')\,,\qquad
  g_{ab} = 0 \;\;\text{for}\;\;\eta<\eta'\,,
\end{equation}
whose solution is the matrix exponential
$g_{ab}(s) = \Theta_H(s)\,[e^{-{\bm\Omega}s}]_{ab}$. The eigenvalues
of ${\bm\Omega}$ follow from
$\det({\bm\Omega}-\mu\,{\bm 1}) = \mu^2-\tfrac{1}{2}\mu-\tfrac{3}{2}
= (\mu-\tfrac{3}{2})(\mu+1)$, i.e.\ $\mu = 3/2$ and $\mu=-1$, so
$e^{-{\bm\Omega}s}$ contains $e^{-3s/2}$ (decaying) and $e^{s}$
(growing). The corresponding eigenvectors are $(2,-3)^T$ and
$(1,1)^T$, and the associated projectors,
\begin{equation}
  P_+ = \frac{1}{5}\begin{pmatrix} 3 & 2 \\ 3 & 2\end{pmatrix}\,,\qquad
  P_- = \frac{1}{5}\begin{pmatrix} 2 & -2 \\ -3 & 3\end{pmatrix}\,,
\end{equation}
satisfy $P_++P_-={\bm 1}$, $P_\pm^2=P_\pm$, $P_+P_-=0$, and project
any doublet onto its growing and decaying components. The propagator
is therefore
\begin{equation}\label{eq:gab_eta_app}
  \boxed{\;g_{ab}(s) = \Theta_H(s)\left[
  e^{s}\,P_+ + e^{-3s/2}\,P_-\right]_{ab}\,,\qquad s=\eta-\eta'\;}
\end{equation}
or, component by component,
\begin{equation}
  g_{11} = \tfrac{1}{5}\big(3e^s+2e^{-3s/2}\big),\quad
  g_{12} = \tfrac{2}{5}\big(e^s-e^{-3s/2}\big),\quad
  g_{21} = \tfrac{3}{5}\big(e^s-e^{-3s/2}\big),\quad
  g_{22} = \tfrac{1}{5}\big(2e^s+3e^{-3s/2}\big).
\end{equation}
One easily checks that $g_{ab}(0)=\delta_{ab}$ and that each column
solves the homogeneous system for $s>0$. Note that
\begin{equation}
  g_{12}(s) = g_\eta(s)\,:
\end{equation}
a source inserted in the \emph{Euler} equation (the second slot of
the doublet, which is where the stress tensor enters) produces a
$\delta$-response given precisely by the scalar Green's
function~\eqref{eq:g_eta_app}, while its $\Theta$-response is given
by $g_{22}$. The general solution of Eq.~\eqref{eq:doublet_system}
reads
\begin{equation}\label{eq:formal_solution_app}
  \Psi_a(\kv,\eta) = g_{ab}(\eta-\eta_0)\,\Psi_b(\kv,\eta_0)
  + \int_{\eta_0}^{\eta} d\eta'\;g_{ab}(\eta-\eta')
  \left[\big(\text{quadratic}\big)_b + J_b\right](\kv,\eta')\,.
\end{equation}
Discarding the decaying mode, the linear solution is the
growing-mode doublet
$\Psi^{(1)}_a = e^{\eta}\,\delta_0(\kv)\,(1,1)^T$ --- indeed,
$P_+(1,1)^T=(1,1)^T$.

\subsection{Rederivation of $F_2$ and $G_2$}

As a consistency check --- and to expose the origin of the famous
$5/7$-type coefficients --- let us recompute the second-order SPT
kernels from Eq.~\eqref{eq:formal_solution_app}. Evaluating the
mode-coupling term of Eq.~\eqref{eq:doublet_system} on the linear
solution $\Psi^{(1)}_a=e^{\eta'}\delta_0\,(1,1)^T$, i.e.\ contracting
the vertex~\eqref{eq:vertex_app} with $(1,1)^T$ in both slots,
$S_b = \sum_{c,d}\gamma_{bcd}$, the source doublet for the mode
$\kv=\qv_1+\qv_2$ is
\begin{equation}
  S_b(\qv_1,\qv_2;\eta') = e^{2\eta'}
  \begin{pmatrix} \alpha^{(s)}(\qv_1,\qv_2) \\[2pt] \beta(\qv_1,\qv_2)\end{pmatrix}_b
  \delta_0(\qv_1)\,\delta_0(\qv_2)\,,\qquad
  \alpha^{(s)}(\qv_1,\qv_2) \equiv
  \frac{\alpha(\qv_1,\qv_2)+\alpha(\qv_2,\qv_1)}{2}\,,
\end{equation}
where the symmetrized $\alpha^{(s)}$ arises as
$\gamma_{121}+\gamma_{112}$.
Convolving with the propagator and using
$e^{2\eta'}=e^{2\eta}e^{-2s}$, the time integral factorizes into a
pure matrix of numbers:
\begin{equation}
  \int_0^{\infty} ds\;g_{ab}(s)\,e^{-2s}
  = P_+\int_0^\infty ds\,e^{-s} + P_-\int_0^\infty ds\,e^{-7s/2}
  = P_+ + \frac{2}{7}\,P_-
  = \frac{1}{7}\begin{pmatrix} 5 & 2 \\ 3 & 4 \end{pmatrix}\,.
\end{equation}
Reading off the $a=1$ and $a=2$ rows,
\begin{equation}
  F_2 = \frac{5}{7}\,\alpha^{(s)} + \frac{2}{7}\,\beta\,,\qquad
  G_2 = \frac{3}{7}\,\alpha^{(s)} + \frac{4}{7}\,\beta\,,
\end{equation}
which reproduces exactly the kernels of Eq.~\eqref{eq:F2}: the
coefficients $5/7$, $2/7$, $3/7$, $4/7$ are nothing but the entries
of the time-integral matrix above.

\subsection{Response to a Counterterm Source}

Finally, consider the linear counterterm source of Chapter 4,
$J_a = (0,\,J_2)^T$ with
$J_2(\kv,\eta') = -k^2\ell_0^2\,e^{(m+1)\eta'}\delta_0(\kv)$
(cf.\ Eq.~\eqref{eq:cs_ansatz}). Since only the second slot is
sourced, the density and velocity responses are governed by $g_{12}$
and $g_{22}$ respectively:
\begin{align}
  \Delta\delta^{(c_s^2)} &= \int d\eta'\,g_{12}(s)\,J_2(\eta')
  = -\frac{k^2\ell_0^2\,e^{(m+1)\eta}}{m\left(m+\tfrac{5}{2}\right)}\,\delta_0(\kv)\,,\\[4pt]
  \Delta\Theta^{(c_s^2)} &= \int d\eta'\,g_{22}(s)\,J_2(\eta')
  = (m+1)\,\Delta\delta^{(c_s^2)}\,,
\end{align}
reproducing the result used in \S4.1. The constant ratio
$\Delta\Theta^{(c_s^2)}/\Delta\delta^{(c_s^2)} = m+1$ (as opposed to
$\Theta^{(1)}/\delta^{(1)}=1$ for the pure growing mode) is a direct
measure of the decaying-mode admixture generated by the effective
stress tensor.

%==========================================================================
\section{Equivalence of LPT and SPT at One Loop}\label{app:LPT_SPT}
%==========================================================================

In this appendix we start from the exact Lagrangian
formula~\eqref{eq:P_LPT_exact} and show, by direct expansion, that
at one-loop order it reproduces the SPT result,
$P = P_{11} + P_{22} + 2P_{13}$, with $P_{22}$ and $P_{13}$ given by
Eqs.~\eqref{eq:P22}--\eqref{eq:P13}. Besides confirming the
statement made in Chapter 7, the computation is instructive: it
shows explicitly how the IR-sensitive pieces are distributed
differently between the terms of the two expansions.

Throughout the appendix $P\equiv P_{11}$, and it is convenient to
abbreviate the longitudinal kernel contractions
\begin{equation}\label{eq:ab_shorthand}
  a(\pv) \equiv \kv\cdot\vv{L}_1(\pv) = \frac{\kv\cdot\pv}{p^2}\,,\qquad
  b(\pv_1,\pv_2) \equiv \frac{1}{2}\,\kv\cdot\vv{L}_2(\pv_1,\pv_2)\,,
\end{equation}
so that the kernel relation~\eqref{eq:F2_L2} reads
$F_2 = \tfrac{1}{2}a(\pv_1)a(\pv_2) + b(\pv_1,\pv_2)$. We will
repeatedly use three elementary properties, all following from the
parity of the kernels: (i) $a(-\pv)=-a(\pv)$ and
$b(-\pv_1,-\pv_2)=-b(\pv_1,\pv_2)$; (ii) $b(\pv,-\pv)=0$ (the
second-order displacement vanishes for collinear momenta, cf.\ the
tidal structure of Eq.~\eqref{eq:L2}); and (iii) for any fixed
$\kv$,
\begin{equation}\label{eq:b_flip}
  b(-\kv,\pv) = -\,b(\kv,-\pv)\,,
\end{equation}
since flipping both arguments flips $\vv{L}_2$ while the tidal
factor $1-\mu^2$ is even.

\subsection{Expansion of the Master Formula}

Let $X \equiv \kv\cdot\left[\Psiv(\qv_1)-\Psiv(\qv_2)\right]$ and
$\qv=\qv_1-\qv_2$. The cumulant expansion theorem gives
\begin{equation}
  \avg{e^{-iX}} = \exp\left[-\frac{1}{2}\avg{X^2}_c
  + \frac{i}{6}\avg{X^3}_c + \frac{1}{24}\avg{X^4}_c + \cdots\right]\,.
\end{equation}
Power counting in $P_{11}$: $\avg{X^2}_c$ starts at $O(P)$ and
receives one-loop corrections at $O(P^2)$; $\avg{X^3}_c$ starts at
$O(P^2)$ (it needs one second-order displacement); $\avg{X^4}_c$
starts at $O(P^3)$ and can be dropped. Expanding the exponential to
$O(P^2)$,
\begin{equation}\label{eq:exp_expansion}
  \avg{e^{-iX}}-1 = -\frac{1}{2}\avg{X^2}_c + \frac{i}{6}\avg{X^3}_c
  + \frac{1}{8}\left(\avg{X^2}_c\right)^2 + O(P^3)\,.
\end{equation}
Using statistical homogeneity, the second cumulant is
\begin{equation}\label{eq:X2}
  \avg{X^2}_c = 2\,k_ik_j\int_\pv C^{ij}(\pv)
  \left[1-\cos(\pv\cdot\qv)\right]\,,
\end{equation}
with $C^{ij}$ from Eq.~\eqref{eq:Cij_pieces} --- note that at linear
order this is exactly the exponent of the Zel'dovich
formula~\eqref{eq:P_ZA}. For the third cumulant, expanding
$X^3=(\kv\cdot\Psiv_1-\kv\cdot\Psiv_2)^3$, the zero-lag pieces
$\avg{(\kv\cdot\Psiv)^3}$ cancel between the two ends, and using
Eq.~\eqref{eq:Cijl_def},
\begin{equation}\label{eq:X3}
  \avg{X^3}_c = 3i\,k_ik_jk_l\int_{\pv_1}\!\int_{\pv_2}
  C^{ijl}(\pv_1,\pv_2,-\pv_{12})
  \left[e^{i\pv_1\cdot\qv}-e^{i\pv_{12}\cdot\qv}\right]\,,
\end{equation}
where the first (second) phase comes from the configurations with
one (two) displacement(s) at the point $\qv_1$.

\subsection{Fourier Transform}

We now insert Eq.~\eqref{eq:exp_expansion} into
Eq.~\eqref{eq:P_LPT_exact} and integrate over $\qv$. All
$\qv$-independent terms produce $\delta_D^{(3)}(\kv)$ and drop for
$\kv\neq 0$; the oscillating terms lock the internal momenta to
$\kv$. The three terms of Eq.~\eqref{eq:exp_expansion} give:

\medskip\noindent
\emph{(i) Second cumulant.} The surviving piece of
$-\tfrac12\avg{X^2}_c$ is $+k_ik_j\,C^{ij}(\kv)$, evaluated through
one loop. From Eq.~\eqref{eq:Cij_pieces}, using
$k_iL^i_1(\kv)=a(\kv)=1$ and flipping the arguments of $L_3$ by
parity in the $(13)$ piece,
\begin{equation}\label{eq:term_i}
  k_ik_jC^{ij}(\kv)
  = P(k) + 2\int_\pv b^2(\pv,\kv-\pv)\,P(p)\,P(|\kv-\pv|)
  + P(k)\int_\pv \kv\cdot\vv{L}_3(\kv,\pv,-\pv)\,P(p)\,.
\end{equation}

\medskip\noindent
\emph{(ii) Third cumulant.} Using Eq.~\eqref{eq:X3},
$\frac{i}{6}\avg{X^3}_c$ contributes
\begin{equation}\label{eq:term_ii}
  T = \frac{1}{2}\,k_ik_jk_l\left[
  \int_\pv C^{ijl}(\pv,\kv-\pv,-\kv)
  - \int_\pv C^{ijl}(\kv,\pv,-\kv-\pv)\right]\,.
\end{equation}
From Eq.~\eqref{eq:Cijl_expr}, the fully contracted third cumulant
is a sum over the three pairs,
\begin{equation}
  k_ik_jk_l\,C^{ijl}(\pv_1,\pv_2,\pv_3)
  = 2\left[a_1a_2\,b(\pv_1,\pv_2)P_1P_2
  + a_2a_3\,b(\pv_2,\pv_3)P_2P_3
  + a_1a_3\,b(\pv_1,\pv_3)P_1P_3\right]\,,
\end{equation}
with $a_i\equiv a(\pv_i)$, $P_i\equiv P(p_i)$. Substituting the two
momentum configurations of Eq.~\eqref{eq:term_ii} and using the
parity properties (i)--(iii) to align the integration variables
(e.g.\ $\pv\to-\pv$ or $\pv\to\kv-\pv$ where convenient), the pair
that does not contain the external leg assembles into a
$P_{22}$-type structure, while the pairs containing it are
proportional to $P(k)$:
\begin{equation}\label{eq:term_ii_result}
  T = 2\int_\pv a(\pv)\,a(\kv-\pv)\,b(\pv,\kv-\pv)\,P(p)\,P(|\kv-\pv|)
  + P(k)\int_\pv a(\pv)\left[3\,b(\kv,-\pv)-b(\kv,\pv)\right]P(p)\,.
\end{equation}

\medskip\noindent
\emph{(iii) Square of the second cumulant.} At $O(P^2)$ only the
linear pieces $C_{(11)}$ are needed, with
$k_ik_jC_{(11)}^{ij}(\pv)=a^2(\pv)P(p)$. Writing
$1-\cos(\pv\cdot\qv)$ for each factor and collecting the phases that
add up to $\kv$, two structures survive: the cross term between a
zero-lag piece and an oscillating piece, and the product of two
oscillating pieces:
\begin{equation}\label{eq:term_iii}
  \frac{1}{8}\left(\avg{X^2}_c\right)^2 \;\longrightarrow\;
  -\,k^2\sigma_d^2\;P(k)
  \;+\;\frac{1}{2}\int_\pv a^2(\pv)\,a^2(\kv-\pv)\,P(p)\,P(|\kv-\pv|)\,,
\end{equation}
where we used
$k_ik_j\int_\pv C_{(11)}^{ij}(\pv) = k^2\sigma_d^2$ with the
displacement dispersion $\sigma_d^2$ of Chapter 2.

\subsection{Assembly}

\noindent
\textbf{The $P_{22}$ part.} Collecting the terms with two
loop-momentum spectra from
Eqs.~\eqref{eq:term_i},~\eqref{eq:term_ii_result}
and~\eqref{eq:term_iii}:
\begin{equation}
  \int_\pv\left[\frac{1}{2}\,a_1^2a_2^2 + 2\,a_1a_2\,b + 2\,b^2\right]
  P_1P_2
  = 2\int_\pv\left[\frac{a_1a_2}{2}+b\right]^2 P_1P_2
  = 2\int_\pv F_2^2(\pv,\kv-\pv)\,P_1P_2 = P_{22}(k)\,,
\end{equation}
with $(\pv_1,\pv_2)=(\pv,\kv-\pv)$: the three contributions are
precisely the $(F_2^\mathrm{ZA})^2$, cross, and $(\kv\cdot\vv{L}_2)^2$
pieces of the squared kernel relation~\eqref{eq:F2_L2}.

\medskip\noindent
\textbf{The $P_{13}$ part.} The remaining terms are all proportional
to $P(k)$:
\begin{equation}\label{eq:P13_collect}
  P(k)\int_\pv\Big[\kv\cdot\vv{L}_3(\kv,\pv,-\pv)
  + a(\pv)\big(3\,b(\kv,-\pv)-b(\kv,\pv)\big)\Big]P(p)
  \;-\;k^2\sigma_d^2\,P(k)\,.
\end{equation}
Two massages are needed. First, split
$3b(\kv,-\pv)-b(\kv,\pv) =
2\left[b(\kv,-\pv)-b(\kv,\pv)\right] +
\left[b(\kv,-\pv)+b(\kv,\pv)\right]$
and note that $a(\pv)\left[b(\kv,-\pv)+b(\kv,\pv)\right]$ is
\emph{odd} under $\pv\to-\pv$, so it integrates to zero. Second,
recognize the zero-lag term as the Zel'dovich part of the $13$
configuration:
$k^2\sigma_d^2 P(k) = P(k)\int_\pv a^2(\pv)P(p) =
-6P(k)\int_\pv F_3^\mathrm{ZA}(\kv,\pv,-\pv)P(p)$,
since $F_3^\mathrm{ZA}(\kv,\pv,-\pv) =
\tfrac{1}{3!}a(\kv)a(\pv)a(-\pv) = -\tfrac{1}{6}a^2(\pv)$. On the
other hand, the symmetrized kernel relation~\eqref{eq:F3_L3},
evaluated in the $(\kv,\pv,-\pv)$ configuration and using
$b(\pv,-\pv)=0$, reads
\begin{equation}
  F_3(\kv,\pv,-\pv) = -\frac{1}{6}\,a^2(\pv)
  + \frac{1}{3}\,a(\pv)\left[b(\kv,-\pv)-b(\kv,\pv)\right]
  + \frac{1}{3!}\,\kv\cdot\vv{L}_3(\kv,\pv,-\pv)\,.
\end{equation}
Comparing term by term, the expression~\eqref{eq:P13_collect} is
exactly
\begin{equation}
  6\,P(k)\int_\pv F_3(\kv,\pv,-\pv)\,P(p) = 2P_{13}(k)\,.
\end{equation}

\medskip\noindent
Adding the tree-level piece from Eq.~\eqref{eq:term_i}, we obtain
\begin{equation}
  \boxed{\;P(k) = P_{11}(k) + P_{22}(k) + 2P_{13}(k)\,,\;}
\end{equation}
completing the proof of the one-loop equivalence.

\subsection{Remarks}

First, note where the IR-sensitive pieces reside. In the Lagrangian
organization, \emph{both} the $+k^2\sigma^2_{d,\mathrm{IR}}$ piece
of $P_{22}$ (the soft limit of the
$\tfrac12 a^2a^2$ term) and the $-k^2\sigma^2_{d,\mathrm{IR}}$ piece
of $2P_{13}$ (the zero-lag cross term) originate from one and the
same object --- the square of the second
cumulant~\eqref{eq:term_iii}, whose $[1-\cos][1-\cos]$ structure
guarantees their cancellation before any expansion is performed.
The equal-time IR cancellation of Chapter 3, which in SPT requires a
conspiracy between different diagrams, is manifest term by term in
LPT: the exponent only ever contains the relative displacement.

Second, the derivation used nothing but the parity of the kernels
and the kernel relations~\eqref{eq:F2_L2}--\eqref{eq:F3_L3}, which
are themselves consequences of expanding the same exponential at the
field level. The pattern therefore persists at all orders: expanding
the cumulant representation of Eq.~\eqref{eq:P_LPT_exact} reproduces
the Eulerian loop expansion order by order, with the partitions of
the displacement cumulants reassembling into products of SPT
kernels.

\bibliographystyle{JHEP}
\bibliography{short}

\end{document}